# Tables of practical invariants for distinguishing multiplicity-free fusion categories up to rank 7

G. Vercleyen[*]

July 22, 2025

**Abstract**

This paper discusses to what extent the census of multiplicity-free fusion categories up to rank 7, proposed by the software package Anyonica and the anyonwiki website, can be regarded as a proper classification. The questions of correctness and completeness are briefly discussed, and the question of inequivalence of the provided categories is resolved under the assumption that all proposed data is correct. This is done by providing tables of small sets of invariants for these categories, for which (a) their invariance under automorphisms can be checked manually, and (b) the (in)equivalence of two skeletal fusion categories can be checked manually. These invariants can also be used to identify the category on the anyonwiki that corresponds to one for which the skeletal data is known.

## Contents

[*]Purdue University, Mathematical Sciences Bldg, 150 N University St, West Lafayette, IN 47907, USA.

# 1 Introduction

Fusion categories play a central role in bridging the gap between various research domains in mathematics and theoretical physics [Eti+15] such as topological quantum computation [Fre+02; Kit06; Row16; RW18], knot theory [Kas95; RT90; RT91; Tur10], representation theory of weak Hopf algebras and quantum groups [Kas95; BK01], low-dimensional topology and topological field theory [Tur92; CKY97; BK01; CW17; TV17], subfactor theory and planar algebras [JMS13; Gro+18; CMS11; EG12], vertex operator algebras, and conformal field theory [MS89; Cre19; EG12]. Sadly, due to their wide range of applications, it is challenging to obtain an overview of which fusion categories are known and what their properties are. Many examples are scattered throughout the literature, and written down from a vast number of different perspectives. This is an obvious consequence of the wide applicability of fusion categories but it provides frustration to researchers who need explicit data on fusion categories. In order to tackle this problem, the author, under the supervision of J. Slingerland, constructed the Anyonica[Ver24a] software package and a website called the anyonwiki[VS]. The goal of the anyonwiki is to collect data on known fusion categories and make it available to the public. The goal of Anyonica is to provide a computational platform to work with fusion rings and categories. It provides tools for solving large sparse systems of polynomial equations with certain symmetries. It is particularly optimized to solve the pentagon, hexagon, and pivotal equations that arise when categorifying fusion rings. Besides providing tools to find fusion categories, it also provides tools to work with fusion categories and a dataset of what the author proposed as a census of skeletal data of all multiplicity-free pivotal braided fusion categories (MFPBFCs) and all multiplicity-free pivotal non-braided fusion categories (MFPNBFCs) up to rank 7 [Ver24b].

Although the code of Anyonica is open-source, its evaluation relies on Mathematica, which is proprietary software. This implies that it is impossible to verify the correctness of the package formally, and one may, and should, wonder to what extent the presented census is could be regarded as an actual classification. For this to be true, three statements need to be verified

1. Each provided set of skeletal data does correspond to a pivotal and possibly braided multiplicity-free fusion category. I.e., each set of skeletal data solves the pentagon equations, the pivotal equations, and, if applicable, the hexagon equations.

2. The census contains the skeletal data of all MFPBCFs and all MFPNBFCs up to rank 7.

3. Each one of the MFP(N)BCFs, up to equivalence, is listed only once in the census.

In this paper, we will briefly discuss to what extent these statements can be regarded as true. Rather than providing fully rigorous proofs, we will provide techniques to obtain evidence of their correctness. The reason these techniques provide evidence rather than rigorous proof is because they rely on the use of computer algebra systems. Therefore, as long as no rigorous mathematical proof is given, the correctness of the proposed classification becomes a matter of faith. Some of the statements require more faith than others. We will see that it is possible to provide convincing and ever-increasing evidence for the correctness of the first statement. If the first statement holds, the last statement follows rigorously from the results of this paper. In particular, those who trust Mathematica's symbolic algorithms may regard the provided evidence of the first and last statements in this paper as a proof. We will see that, from a computational viewpoint, evidence for the second statement is much harder to obtain.

The paper is organized as follows. In Section 2, we provide a minimal introduction to multiplicity-free fusion categories and their related structures. In Section 3, we briefly discuss the three statements required for the proposed census to be a classification. We provide a method, based on '*gauge-split bases*' to construct small sets of invariants of multiplicity-free fusion categories. In Section 4 we provide tables of small sets of invariants that distinguish between all MFPBFCs and MFPNBFCs up to rank 7.

## 2 Fusion categories and fusion systems

### 2.1 Multiplicity-free fusion systems

We start this section by presenting the definition of a multiplicity-free fusion system. A fusion system is a term coined in [DHW13] to refer to the data that determines a skeletal fusion category. Then, we will add extra structures to arrive at the notions of pivotal, spherical, braided, ribbon, and modular fusion systems. The notion of unitarity of fusion systems and the link with unitarity of its data is discussed as well. While fusion categories can be defined over any field, we will only consider categories defined over the complex numbers. Most of the material in this section comes straight from [DHW13] and its presentation comes from [Ver24b].

**Definitions 2.1**. • A **multiplicity-free fusion system** is a collection of data $(\mathbf{L}, *, \mathbf{N}, \mathbf{F})$ where

1. $\mathbf{L} = \{1, \ldots, r\}, r \in \mathbb{N}$.
2. $* : \mathbf{L} \to \mathbf{L}$ is a map for which $1^* = 1$ and $(a^*)^* = a$.
3. $\mathbf{N} = \left\{ N_{a,b}^c \,\middle|\, a, b, c = 1, \ldots, r \right\}$ is a finite set of natural numbers that satisfy

$$N_{a,b}^c \in \{0, 1\}, \tag{1}$$

$$N_{a,1}^b = N_{1,a}^b = \delta_b^a, \tag{2}$$

$$N_{a,b}^1 = \delta_{b^*}^a, \tag{3}$$

$$N_{a,b,c}^d := \sum_e N_{a,b}^e N_{e,c}^d = \sum_f N_{a,f}^d N_{b,c}^f. \tag{4}$$

4. $\mathbf{F} = \left\{ [F_d^{abc}] \in \mathrm{Mat}_{N_{a,b,c}^d \times N_{a,b,c}^d}(\mathbb{C}) \,\middle|\, a, b, c, d = 1, \ldots, r \right\}$ is a finite set of finite-dimensional invertible matrices, with inverses $\left\{ \left[ \tilde{F}_d^{abc} \right] \right\}$, that satisfy

$$[F_a^{aa^*a}]_1^1 \neq 0, \tag{5}$$

$$[F_d^{1bc}]_f^e = 1, \tag{6}$$

$$[F_d^{a1c}]_f^e = 1, \tag{7}$$

$$[F_d^{ab1}]_f^e = 1, \tag{8}$$

$$[F_e^{fcd}]_l^g [F_e^{abl}]_k^f = \sum_h [F_g^{abc}]_h^f [F_e^{ahd}]_k^g [F_k^{bcd}]_l^h, \tag{9}$$

where whenever a zero dimensional matrix is encountered in a term, it is automatically 0.

- Equations (7) are called the **triangle equations** while equations (9) are called the **pentagon equations**.

- The matrices $[F_d^{abc}]$ are called $F$-matrices, their elements are called $F$-symbols and if $a = 1$ or $b = 1$ or $c = 1$ then we call such the $F$-matrix resp. $F$-symbol a vacuum $F$-matrix resp. vacuum $F$-symbol.

- **N** provides the structure constants for a fusion ring. Since, by assumption, this ring is part of a fusion system, it is also called the **Gröthendieck ring** of the fusion system.

*Remark* 2.2. We did not demand that $N_{a,b}^c = N_{b,a}^c$, i.e., that fusion is commutative. Neither did we demand that the $F$-matrices are unitary, as is often assumed in physics literature.

Fusion systems are closely related to fusion categories. In [DHW13], it is shown that each fusion system gives rise to a unique fusion category. On the other hand, each fusion category gives rise to an infinite number of fusion systems. The reason is two-fold.

1. First, a choice of bases, also called a choice of the gauge, is required to determine the $F$-matrices. From a categorical perspective, the $F$-matrices correspond to the pull-back of the associator $\alpha : a \otimes (b \otimes c) \to (a \otimes b) \otimes c$ as a map on objects to a map on the (finite-dimensional) morphism spaces. In particular $[F_d^{abc}] : \bigoplus_e V_e^{a,b} \otimes V_d^{e,c} \to V_d^{a,f} \otimes V_f^{b,c}$ where $V_c^{a,b} = \hom(a \otimes b, c)$ and $\dim(V_c^{a,b}) = N_{a,b}^c \in \{0,1\}$. Let the variables $\{g_c^{ab} | a, b, c \in \mathbf{L} \text{ such that } N_{a,b}^c \neq 0\}$, represent basis transformations (also called gauge transformations) of the various one-dimensional $V_c^{a,b}$. The induced transformations on the $F$-symbols are of the form

$$[F_d^{abc}]_f^e \mapsto \frac{g_e^{ab} g_d^{ec}}{g_d^{af} g_f^{bc}} [F_d^{abc}]_f^e. \tag{10}$$

   For any solution $\left\{ [F_d^{abc}]_f^e \right\}$ to the pentagon equations,

$$\left\{ \frac{g_e^{ab} g_d^{ec}}{g_d^{af} g_f^{bc}} [F_d^{abc}]_f^e \right\} \tag{11}$$

   is also a solution to the pentagon equations. So as long as the gauge transforms are chosen such that (5) (6), (7) and (8) are satisfied for the new set of $F$-symbols (which is always possible), we obtain a fusion system that corresponds to the same fusion category.

2. Second, any permutation $\sigma$ of the elements of $\mathbf{L}$ in such a way that $N_{\sigma(a),\sigma(b)}^{\sigma(c)} = N_{a,b}^c, \forall a, b, c \in \mathbf{L}$ leaves the pentagon equations invariant and thus for any solution $\left\{ [F_d^{abc}]_f^e \right\}$ to the pentagon equations,

$$\left\{ [F_{\sigma(d)}^{\sigma(a)\sigma(b)\sigma(c)}]_{\sigma(f)}^{\sigma(e)} \right\} \tag{12}$$

   is also a solution to the pentagon equations. In this paper, we will not refer to such permutations as gauge transforms.

*Remark* 2.3. In contrast with [DHW13], definition 2.1 demands that any vacuum $F$-symbol equals 1. Out of the three demands (6), (7), and (8) only the triangle equations, i.e. (7) are necessary. It is, however, without loss of generality, always possible to choose a gauge in which all vacuum $F$-symbols are 1 (see, e.g., [Ver24b] for a non-categorical proof). The main difference between the triangle equations and demands (6) and (8) is that the former is a necessary consistency condition while the latter two are unnecessary and only there to simplify calculations. The latter two also restrict the choice of gauge, while the first does not.

The terms fusion system and fusion category will often be used interchangeably in what follows. The latter is the more common term, while the former is, as far as we are aware, only introduced in [DHW13] to point out the relationship between the abstract structure and the more down-to-earth, numerical description of a fusion category.

Fusion systems can have additional structure. The most common structures are a *pivotal structure*, a *braiding*, and a *Hermitian structure*. We start by discussing the *pivotal structure*.

## 2.2 Multiplicity-free pivotal and spherical fusion systems

**Definition 2.4.**

- A **multiplicity-free pivotal fusion system** $(\mathbf{L}, *, \mathbf{N}, \mathbf{F}, \mathbf{P})$ is a multiplicity-free fusion system together with a list of phases $\mathbf{P} = \{p_a \in \mathrm{U}(1) \,|\, a \in \mathbf{L}\}$, called **pivotal coefficitents**, for which

$$p_1 = 1, \tag{13}$$

$$p_a = p_{a^*}^{-1}, \tag{14}$$

$$\frac{p_a p_b}{p_c} = [F_1^{abc^*}]_{a^*}^c [F_1^{bc^*a}]_{b^*}^{a^*} [F_1^{c^*ba}]_c^{b^*}. \tag{15}$$

- The **left quantum dimensions** $\{d_a^L \in \mathbb{C} \,|\, a \in L\}$ of a multiplicity-free pivotal fusion system are defined as

$$d_a^L := \frac{p_a}{[F_{a^*}^{a^*aa^*}]_1^1}. \tag{16}$$

- The pivotal structure **P** is called **spherical** if $d_a^R := d_{a^*}^L = d_a^L$ for all $a \in \mathbf{L}$. In this case, the multiplicity-free pivotal fusion system is called a **multiplicity-free spherical fusion system** and the left quantum dimensions are called the quantum dimensions and denoted by $d_a$.

*Remarks* 2.5.   • While the pivotal coefficients are gauge-dependent, the left quantum dimensions are not. Therefore, in Section 3.3, we will express the invariants for a category using the left quantum dimensions rather than the pivotal coefficients.

- A pivotal structure is often not regarded as an extra structure in the physics literature. This is because physicists most often work with a unitary fusion system. We will expand more unitary fusion systems in Section 2.4.

## 2.3 Multiplicity-free braided, ribbon, and modular fusion systems

We can also add a *braided structure* to a multiplicity-free fusion system. This structure is independent of a pivotal structure but when combined they can lead to interesting invariants.

**Definition 2.6.**

- A **multiplicity-free braided fusion system** $(\mathbf{L}, *, \mathbf{N}, \mathbf{F}, \mathbf{R})$ is a fusion system for which together with a finite list $\mathbf{R}$ of non-zero complex numbers $\left\{ R_c^{ab} \in \mathbb{C} \setminus \{0\} \,\middle|\, a, b, c \in \mathbf{L} \text{ such that } N_{a,b}^c \neq 0 \right\}$,

    1. $N_{a,b}^c = N_{b,a}^c$ for all $a, b, c \in \mathbf{L}$, and

2. the symbols $R_c^{ab}$ satisfy

$$R_e^{ca}[F_d^{acb}]_g^e R_g^{cb} = \sum_f [F_d^{cab}]_f^e R_d^{cf} [F_d^{abc}]_g^f, \tag{17}$$

$$(R_e^{ac})^{-1} [F_d^{acb}]_g^e \left(R_g^{bc}\right)^{-1} = \sum_f [F_d^{cab}]_f^e (R_d^{fc})^{-1} [F_d^{abc}]_g^f, \tag{18}$$

- The variables $R_c^{ab}$ are called *R-symbols* and equations (17) and (18) are called the **hexagon equations**.

- A **multiplicity-free ribbon fusion system** is a multiplicity-free spherical braided fusion system.

- A **multiplicity-free modular fusion system** (**L**, ∗, **N**, **F**, **P**, **R**) is a multipicity-free ribbon fusion system for which the matrix $\hat{S} \in \text{Mat}_{r\times r}(\mathbb{C})$, whose entries are given by

$$[\hat{S}]_b^a = \sum_{c=1}^r [\tilde{F}_a^{ab^*b}]_c^1 R_c^{b^*a} R_c^{ab^*} [F_a^{ab^*b}]_1^c. \tag{19}$$

is invertible.

In [DHW13], it is shown that each of the systems above defines a fusion category, with the corresponding adjectives, that is unique up to equivalence.

Just like the *F*-symbols and pivotal coefficients, the *R*-symbols are gauge-dependent. A gauge-transform, with variables $\{g_c^{ab} \mid a, b, c \in \mathbf{L} \text{ such that } N_{a,b}^c \neq 0\}$, has the following effect

$$R_c^{ab} \mapsto \frac{g_c^{ab}}{g_c^{ba}} R_c^{ab}. \tag{20}$$

## 2.4 Multiplicity-free unitary, unitary ribbon, and unitary modular fusion systems

Categorically, a unitary fusion category is a fusion category that comes with a *Hermitian structure*, †, that satisfies several properties (see, e.g., [Gal14; HP16] for the full definition). The Hermitian structure defines the notion of unitary maps and can, in particular, be used to define evaluation and coevaluation morphisms that can be combined to calculate the left quantum dimensions $d_a^L$ [JP17; Yam02; Yam04]. The left quantum dimensions, arising this way, satisfy $d_a^L = \text{FPDim}(a)$ where $\text{FPDim}(a)$ is the largest real eigenvalue of the matrix $[N_a]$ with entries $[N_a]_b^c = N_{a,b}^c$. Due to the Frobenius-Perron theorem, this value is unique, real, and $\text{FPDim}(a) \geq 1$ for all $a \in \mathbf{L}$. The Hermitian structure also fixes a spherical pivotal structure **P**, which, for the corresponding fusion system, is given by $\left\{ p_a = d_a^L [F_{a^*}^{a^*aa^*}]_1^1 \,\middle|\, a \in \mathbf{L} \right\}$. This pivotal structure is also called the canonical spherical structure of the unitary fusion category.

It was recently shown in [Reu23] that any unitarizable fusion category admits a unique unitary structure (up to unitary monoidal equivalence). Therefore, the canonical spherical structure can be implicitly included in the definition of a unitary fusion category.

In [Yam02] (part 4) it is shown that if a fusion system has a gauge for which its *F*-matrices are unitary, then its category can always be made into a unitary category via the choice of an appropriate Hermitian structure. Therefore, we define a unitary fusion system as follows

**Definition 2.7**. A **multiplicity-free unitary fusion system** is a multiplicity-free spherical fusion system (**L**, ∗, **N**, **F**, **P**) for which the *F*-matrices are unitary and the spherical structure is such that the quantum dimensions satisfy $d_a = \text{FPDim}(a)$.

For a unitary category, there always exists a gauge choice for which the *F*-symbols of the corresponding spherical fusion system are unitary. So, the notion of a unitary fusion system is equivalent to that of a unitary fusion category.

*Remark* 2.8. It is important to note that if the *F*-matrices of a fusion system are unitary, one could still choose a (spherical) pivotal structure that is not canonical. For example, any fusion system with a fusion ring corresponding to the group ring $\mathbb{Z}_3$ has unitary *F*-symbols, but there are multiple non-spherical pivotal structures for which the left quantum dimensions are not positive. We will not refer to these fusion systems as unitary.

The definition of unitarity of a fusion category can be extended to braided, ribbon, and modular fusion categories by demanding that the braided and ribbon structures are *compatible with the unitary structure*. In the paper [Gal14], these demands are written out, and the following statements are proven:

- If a unitary fusion category admits a braiding, it is automatically a unitary braided fusion category.

- Every unitary braided fusion category admits a unique unitary ribbon structure. This ribbon structure is the canonical ribbon structure derived from from the canonical spherical structure combined with the braiding.

In particular, a unitary braided fusion category is immediately a ribbon fusion category. We will therefore use the following definitions

**Definitions 2.9**.

- A **multiplicity-free unitary braided fusion system**, or equivalently a **multiplicity-free unitary ribbon fusion system**, is a multiplicity-free ribbon fusion system (**L**, ∗, **N**, **F**, **P**, **R**) for which (**L**, ∗, **N**, **F**, **P**) form a multiplicity-free unitary fusion system and the *R*-matrices are unitary matrices.

- A **multiplicity-free unitary modular fusion system** is a multiplicity-free unitary ribbon fusion system for which the matrix $\hat{S}$ is invertible.

*Remark* 2.10. The demand that the *F*- and *R*-matrices of a unitary fusion system are unitary is made for convenience rather than necessity. Since such a basis always exists and it is not difficult to find (see [Ver24b], for an algorithm) this should cause no issues in practice.

## 3 The census of multiplicity-free fusion categories up to rank 7

In [Ver24b] a census of all multiplicity-free fusion categories up to rank 7 and all their braided, and pivotal structures is made. This result was obtained using computational algebra. The package used to obtain these results is quite extensive, so it is possible that the results may contain errors due to bugs or mistakes. As a matter of fact, while writing this paper, the author discovered that 4 categories with Rep($D_7$) fusion rules were actually equivalent to other ones. This was caused by a combination of a software bug at that time and the author forgetting to redo the calculations for that specific ring. It is, therefore, important that further evidence is provided to support the claim that the presented data is indeed a proper classification. In particular, there needs to be evidence for each of the following claims

1. **Correctness.** Each fusion system corresponds to a category. I.e., each set of skeletal data solves the pentagon, hexagon, and pivotal equations.

2. **Completeness.** The census contains all categories. I.e., the skeletal data of all multiplicity-free categories up to rank 7 and all compatible braided and pivotal structures have been found.

3. **Uniqueness.** None of the MFP(N)BFCs are equivalent to each other.

In this section, we will briefly discuss to what extent these statements can be regarded as true. We will mainly focus on the last statement since, under the assumption of the first, this statement can actually be proven formally.

## 3.1 Correctness of the skeletal data

One can verify the first statement by checking that skeletal data of the provided categories solves the pentagon, hexagon, and pivotal equations. Since some of the systems of equations contain more than a hundred thousand equations, a brute force proof by hand is not feasible. However, the skeletal data provided by, and also stored as part of, Anyonica is exact. This means that the first statement can be tested computationally by substituting the solutions into the pentagon, hexagon, and pivotal equations and verifying that these are satisfied. By using different methods from different software packages to verify the data's correctness, the chance of obtaining false positives is reduced. While such computations don't provide a rigorous proof of exactness, they can provide strong evidence that the data should be taken seriously.

The correctness of the data has been tested using exact methods provided by Mathematica. In particular, the correctness of the data has been verified by using the functions

- `RootReduce`, which expresses algebraic combinations of roots of polynomials as a root of a single polynomial, and
- `N[#,{Infinity,1000}]&`, which calculates the exact numerical value of its argument with 1000 correct digits and uses an arbitrarily big extra internal precision to guarantee correctness of those digits.

The data has also been independently verified by the TensorCategories.jl package[MT] with symbolic algebra provided by the Oscar package[25]. This package is developed independently of Anyonica and uses different methods to verify the correctness of the data. Therefore, we think it is safe to say that the data provided by the census is correct.

## 3.2 Completeness of the skeletal data

Due to the size of Anyonica and the complexity of its methods, it is harder to argue that the obtained sets of solutions to the pentagon-, hexagon- and pivotal equations are complete. In contrast to verifying the correctness of the data (which is relatively easy to test with other programming languages and/or packages), verifying completeness suffers from the fact that one cannot easily export a whole software package to another programming language. Therefore, the completeness of the data relies on the assumption that there are no silent critical bugs in Anyonica and/or Mathematica and that no human error was made while processing the data.

To reduce the risk of critical bugs and enhance the transparency of calculations, Anyonica provided log files with calculations for all the categories found. The most time-intensive steps, consisting of matrix decompositions and substitutions of well-chosen variables, and their results have been logged and can therefore be checked using other languages as well. Therefore, the log files provide a way to check the completeness of the set of fusion categories belonging to a single Gröthendieck ring. However, there are too many log files to check if one wants evidence for the correctness of all data.

Currently, the author has begun exporting some of Anyonica's functionality to Julia, but there is no clear timeline for completion.

## 3.3 Inequivalence of the skeletal data

In this section, we will examine a method of creating such a complete invariant of an MFP(N)BFC. We will also show how one can construct a weaker invariant based on the complete one, which is more practical to work with.

### 3.3.1 Invariants of MFPBFCs and MFPNBFCs

Two MFPBFCs or MFPNBFCs, $\mathscr{C}_1$ and $\mathscr{C}_2$ are equivalent to each other if and only if the skeletal data of $\mathscr{C}_1$ equals the skeletal data of $\mathscr{C}_2$ after applying a combination of

1. a, possibly trivial, gauge transform with variables $\{g_c^{ab} \,|\, a, b, c \in \mathbf{L} \text{ such that } N_{a,b}^c \neq 0\}$ that transforms the skeletal data as follows

$$[F_d^{abc}]_f^e \mapsto \frac{g_e^{ab} g_d^{ec}}{g_d^{af} g_f^{bc}} [F_d^{abc}]_f^e, \tag{21}$$

$$R_c^{ab} \mapsto \frac{g_c^{ab}}{g_c^{ba}} R_c^{ab}, \tag{22}$$

$$p_a \mapsto \frac{g_1^{aa^*}}{g_1^{a^*a}} p_a, \tag{23}$$

and

2. a, possibly trivial, permutation $\sigma$ of the elements of $\mathbf{L}$ in such a way that $N_{\sigma(a),\sigma(b)}^{\sigma(c)} = N_{a,b}^c$, $\forall a, b, c \in \mathbf{L}$ that transforms the skeletal data as follows

$$[F_d^{abc}]_f^e \mapsto [F_{\sigma(d)}^{\sigma(a)\sigma(b)\sigma(c)}]_{\sigma(f)}^{\sigma(e)}, \tag{24}$$

$$R_c^{ab} \mapsto R_{\sigma(c)}^{\sigma(a)\sigma(b)}, \tag{25}$$

$$p_a \mapsto p_{\sigma(a)}. \tag{26}$$

Consider a pivotal (and possibly braided) fusion category $\mathscr{C}$ with Gröthendieck ring $\mathscr{R}$. Let $S = \mathrm{Aut}(\mathscr{R})$ be the group of fusion-ring automorphisms of $\mathscr{R}$. An example of a complete invariant of an MFP(N)BFC is the orbit under $S$ of the set of gauge-invariants from a 'gauge-split basis' of a fusion system $\Phi$ corresponding to $\mathscr{C}$.

To define a gauge-split basis the following definitions are useful

**Definitions 3.1.** Let $\Phi = (\mathbf{L}, *, \mathbf{N}, \mathbf{F}, \mathbf{P}, \mathbf{R})$ be a multiplicity-free pivotal braided fusion system. Let $\mathscr{L}_\Phi$ be the ring of rational functions that is generated by formal $F$-symbols, formal $R$-symbols, and formal left quantum dimensions of $\Phi$. The formal $F$-symbols, $R$-symbols, and left quantum dimensions are regarded as functions from respectively $\mathbf{L}^6 \to \mathbb{C}$, $\mathbf{L}^3 \to \mathbb{C}$, and $\mathbf{L} \to \mathbb{C}$ without assigned values or constraints. The multiplication and addition in $\mathscr{L}_\Phi$ are the usual multiplication and addition of functions.

- An element $\rho \in \mathscr{L}_\Phi$ is called **de jure gauge-invariant** if it is invariant under any gauge transform $\varphi$, i.e. $\varphi(\rho) = \rho$ for all $\varphi$.

- Let $\rho \in \mathscr{L}_\Phi$ and let $[\rho]_\Phi$ be $\rho$ where its $F$-symbols, $R$-symbols, and left quantum dimensions have been replaced by their values from $\Phi$. $\rho$ is called **de facto gauge-invariant** if $[\rho]_\Phi$ is well defined and $[\varphi(\rho)]_\Phi = [\rho]_\Phi$ for all gauge-transforms $\varphi$.

*Example* 3.2. For the multiplicity-free category $[\mathrm{Adj}(SO(16))]_{1,1,1}$ the $F$-symbol $[F_5^{555}]_6^6$ is not de jure gauge-invariant since $\varphi([F_5^{555}]_6^6) = \frac{g_5^{65}}{g_5^{56}}[F_5^{555}]_6^6$ but it is de facto gauge-invariant since its value is 0 and therefore unaffected by any gauge transform.

*Notes* 3.3. 
- We are restricting ourselves to fusion systems that are braided at the moment. This is because we will use the same gauge-split bases for braided fusion systems in the non-braided case as well.

- It is not necessary to know the values of the $F$-symbols, $R$-symbols, or left quantum dimensions to determine whether an element $\rho$ of $\mathscr{L}_\Phi$ is de jure gauge-invariant. To determine whether $\rho$ is de facto gauge-invariant, however, certain extra information about the values of the $F$-symbols and $R$-symbols could be required. Which information depends on the form of $\rho$. If, for example, $\rho$ is a rational monomial in formal $F$- and $R$-symbols, we only need to know which $F$-symbols are 0.

- Since the values of the left quantum dimensions are gauge-invariant, any left quantum dimension is always both de jure and de facto gauge-invariant.

A gauge-split basis of an MFPBFC can then be defined as follows.

**Definition 3.4.** Let $\Phi = (\mathbf{L}, *, \mathbf{N}, \mathbf{F}, \mathbf{P}, \mathbf{R})$ be a multiplicity-free pivotal braided fusion system and $\mathscr{M}_\Phi \subset \mathscr{L}_\Phi$ be the subset of rational monomials (RMs) in formal $F$-symbols, formal $R$-symbols, and formal left quantum dimensions. A **gauge split basis** of $\Phi$ is a tuple $(I, D)$ where $I = (\iota_1, \ldots, \iota_m) \in \mathscr{M}_\Phi^m$, $D = (\delta_1, \ldots, \delta_n) \in \mathscr{M}_\Phi^n$ and for which the following properties hold.

1. For each RM $\rho$ in $I$ and $D$, $[\rho]_\Phi$ is well-defined.

2. Each element of $I$ is de facto gauge-invariant.

3. Each element of $D$ is gauge-dependent and moreover the values of each of the $\delta_i$ can be independently set to any non-zero value by performing a suitable gauge transform.

4. For any RM $\rho \in \mathscr{M}_\Phi$ there exist $m + n$ unique integers $a_1, \ldots, a_m, b_1, \ldots, b_n$ such that

$$\rho = \prod_{i=1}^m \iota_i^{a_i} \prod_{i=1}^n \delta_i^{b_i} \tag{27}$$

*Remark* 3.5. A similar definition can be given for a gauge-split basis of a multiplicity free pivotal fusion system, a multiplicity-free braided fusion system, and a fusion system by leaving out the formal symbols that are not available.

In [Ver24b], it is shown that any fusion system (pivotal or non-pivotal, braided or non-braided) has an infinite number of gauge-split bases[1], and there exists an efficient method for constructing such bases. The technique used in [Ver24b] for removing duplicate solutions comes down to constructing a complete invariant for a fusion system $\mathscr{C}$ as follows. Let $(I, D)$ be a gauge-split basis of a fusion system $\Phi$ belonging to $\mathscr{C}$ and $S(I)$ the orbit of the tuple of gauge invariants $I$ under the automorphisms of $\mathscr{R}$. Let $[S(I)]_\Phi$ be $S(I)$ where every formal $F$-symbol, quantum dimension, and, if applicable, $R$-symbol has been replaced by its value from $\Phi$. By construction the set $[S(I)]_\Phi$ is invariant under gauge-transforms and fusion ring automorphisms. From properties 4 in definition 3.4, it also follows that the original set of skeletal data can be recovered from the tuple $(S(I), [S(I)]_\Phi)$. So $(S(I), [S(I)]_\Phi)$ is a complete invariant of $\mathscr{C}$.

*Example* 3.6. The category $[\text{Rep}(D_3)]_{1,1,1}$ has Gröthendieck ring $\mathscr{R} = \text{Rep}(D_3)$ with $S = \{()\}$ the trivial group and $[F_3^{333}]_3^3$ as its only $F$-symbol that equals 0. The following provides an invariant that uniquely identifies this category.

$$\left( S(I), [S(I)]_{[\text{Rep}(D_3)]_{1,1,1}} \right) \tag{28}$$

---

[1] Called gauge-split sets in that paper, but since all its components are tuples and because of property 4 in the definition, we will call these gauge-split bases

where

$$
\begin{aligned}
S(I) = \Bigg\{\Bigg( & [F_3^{333}]_3^3, [F_1^{111}]_1^1, [F_2^{121}]_2^2, [F_3^{131}]_3^3, [F_3^{132}]_3^3, [F_3^{133}]_3^3, [F_3^{231}]_3^3, [F_3^{232}]_3^3, [F_3^{233}]_3^3, [F_3^{331}]_3^3, [F_3^{332}]_3^3, [F_1^{333}]_3^3, \\
& [F_2^{333}]_3^3, R_1^{11}, R_1^{22}, R_1^{33}, R_2^{33}, R_3^{33}, [F_2^{112}]_2^1[F_1^{122}]_1^2, [F_3^{113}]_3^1[F_1^{133}]_1^3, [F_2^{211}]_1^2[F_1^{221}]_1^1, [F_3^{311}]_1^3[F_1^{331}]_1^3, \\
& [F_3^{333}]_2^3 R_3^{32}, R_3^{23} R_3^{32}, \frac{[F_2^{112}]_2^1[F_3^{123}]_3^2}{[F_3^{113}]_3^1}, \frac{[F_3^{113}]_3^1[F_2^{133}]_2^3}{[F_2^{112}]_2^1}, \frac{[F_1^{212}]_1^2}{[F_2^{112}]_2^1[F_2^{211}]_1^2}, \frac{[F_3^{213}]_3^2}{[F_3^{113}]_3^1[F_2^{211}]_1^2}, \\
& [F_2^{112}]_2^1[F_2^{211}]_1^2[F_2^{222}]_1^1, \frac{[F_3^{312}]_3^2}{[F_2^{112}]_2^1[F_3^{311}]_1^3}, \frac{[F_1^{313}]_3^3}{[F_3^{113}]_3^1[F_3^{311}]_1^3}, \frac{[F_3^{313}]_3^3}{[F_3^{113}]_3^1[F_3^{311}]_1^3}, \frac{[F_3^{313}]_3^3}{[F_3^{113}]_3^1[F_3^{311}]_1^3}, \\
& \frac{[F_2^{211}]_1^2[F_3^{321}]_2^3}{[F_3^{311}]_1^3}, \frac{[F_3^{311}]_1^3[F_2^{331}]_3^2}{[F_2^{211}]_1^2}, [F_3^{113}]_3^1[F_3^{311}]_1^3[F_3^{333}]_1^1, [F_2^{112}]_2^1[F_2^{211}]_1^2 R_2^{12}, [F_3^{113}]_3^1[F_3^{311}]_1^3 R_3^{13}, \\
& [F_1^{233}]_2^3[F_1^{332}]_3^2 R_3^{32}, \frac{[F_3^{223}]_3^1[F_1^{233}]_2^3[F_3^{333}]_3^2}{[F_3^{333}]_3^1}, \frac{R_2^{21}}{[F_2^{112}]_2^1[F_2^{211}]_1^2}, \frac{R_3^{31}}{[F_3^{113}]_3^1[F_3^{311}]_1^3}, \\
& [F_3^{113}]_3^1[F_3^{311}]_1^3[F_3^{333}]_3^1[F_3^{333}]_1^3, \frac{[F_3^{333}]_2^1 R_3^{32}}{[F_3^{223}]_3^1[F_1^{233}]_2^3}, [F_3^{113}]_3^1[F_2^{211}]_1^2[F_3^{223}]_3^1[F_1^{233}]_2^3 \\
& [F_2^{233}]_1^3, [F_3^{113}]_3^1[F_2^{223}]_3^1[F_1^{233}]_2^3[F_3^{311}]_1^3[F_3^{333}]_1^2, \frac{[F_3^{333}]_3^1[F_3^{333}]_2^3 R_3^{32}}{[F_3^{223}]_3^1[F_1^{233}]_2^3}, \frac{[F_1^{323}]_3^3 R_3^{32}}{[F_3^{113}]_3^1[F_3^{223}]_3^1[F_3^{311}]_1^3[F_3^{322}]_1^3}, \\
& \frac{[F_3^{323}]_3^3 R_3^{32}}{[F_3^{113}]_3^1[F_3^{223}]_3^1[F_3^{311}]_1^3[F_3^{322}]_1^3}, \frac{[F_2^{112}]_2^1[F_2^{332}]_3^1 R_3^{32}}{[F_3^{113}]_3^1[F_3^{223}]_3^1[F_1^{233}]_2^3}, \frac{[F_2^{323}]_3^3 R_3^{32}}{[F_3^{113}]_3^1[F_3^{223}]_3^1[F_3^{311}]_1^3[F_3^{322}]_1^3}, \\
& \frac{\left(R_3^{32}\right)^2}{[F_3^{113}]_3^1[F_3^{223}]_3^1[F_3^{311}]_1^3[F_3^{322}]_1^3}, d_1^L, d_2^L, d_3^L \Bigg)\Bigg\}.
\end{aligned}
\tag{29}
$$

and

$$
[S(I)]_{[\text{Rep}(D_3)]_{1,1}^1} = \Bigg\{\Bigg(0, 1, 1, 1, 1, 1, 1, 1, -1, 1, -1, 1, -1, 1, 1, e^{\frac{2i\pi}{3}}, -e^{\frac{2i\pi}{3}}, e^{-\frac{2i\pi}{3}}, 1, 1, 1, 1, -\frac{1}{2}, 1, 1, 1,
$$
$$
1, 1, 1, 1, 1, 1, 1, 1, \frac{1}{2}, 1, 1, 1, 1, -1, -1, \frac{1}{2}, -\frac{1}{2}, 1, \frac{1}{2}, \frac{1}{2}, -1, -1, 1, -1, 1, 1, 2\Bigg)\Bigg\}
$$

A few remarks are in order

*Remarks* 3.7. 
- It is possible to reconstruct the fusion ring from the invariant $(S(I), [S(I)]_\Phi)$ by looking at all the vacuum $F$-symbols. Indeed, an $F$-symbol of the form $[F_d^{1bc}]_d^b$ exists if and only if $N_{b,c}^d \neq 0$ so the $F$-symbols if this form give direct access to the structure constants $N_{b,c}^d$.

- The invariant in 3.6 contains vacuum $F$-symbols, vacuum $R$-symbols and $d_1^L$. For our definition of a fusion system, those will always assumed to be 1, and in what follows, we will only be concerned with gauge transforms that leave the vacuum $F$- and $R$-symbols invariant.

### 3.3.2 Reduction of invariants

While the invariant $(S(I), [S(I)]_\Phi)$ is complete, it is an inpractical invariant for distinguishing categories by hand. If we restrict the use case of the invariants to distinguishing fusion categories, we can create invariants that are much smaller. As can be seen from Section 4.5.5, the values of $[F_1^{333}]_3^3$ and $R_1^{33}$ suffice to distinguish between all categories with $\text{Rep}(D_3)$ fusion rules. As a matter of fact, most multiplicity-free categories with rank up to 7 can be identified using only a handful of well-chosen $F$-symbols, $R$-symbols, and left quantum dimensions.

In example 3.6, the construction of $S(I)$ was only based on

1. the fusion ring,

2. the assumption that $\Phi$ is braided, and

3. the knowledge of which $F$-symbols are 0.

It turns out that, for MFP(N)BFCs up to rank 7, every Gröthendieck ring $\mathcal{R}$ provides a list of tuples of formal symbols $S(I)$ that (1) uses no $F$-symbols that evaluate to 0, (2) works for braided and non-braided fusion systems, and (3) distinguishes between all fusion systems with Gröthendieck ring $\mathcal{R}$.

The fact that no $F$-symbol that evaluates to 0 is necessecary is not trivial but it just turns out to be the case. Whether this is still the case for higher ranks is unknown to the author.

That it is possible to build a single invariant that works for both braided and non-braided categories is rather easy to see. The set $S(I)$ can be constructed the same way for a braided as for an unbraided fusion system $\Phi$ since it only consists of formal symbols. We can then define $[S(I)]_\Phi$ for a non-braided fusion system as $S(I)$ where any RM that contains a formal $R$ symbol has been removed, and any other RM has its formal $F$-symbols and left quantum dimensions replaced by their values from $\Phi$. When using the same $S(I)$ for all fusion systems with the same Gröthendieck ring, it is clear that many elements of the tuples in $S(I)$ evaluate to the same value for all fusion systems. These can be removed without losing any power in distinguishing fusion systems.

*Example* 3.8. For the Rep($D_3$) fusion ring we can use $S(I)$ as in equation (29) for all fusion systems. When computing $[S(I)]_\Phi$ for each MFP(N)BFC with Rep($D_3$) fusion rules, it turns out that many elements of the tuple in $[S(I)]_\Phi$ don't provide any power in distinguishing the different categories. For example, all vacuum $F$- and $R$-symbols are always 1, $[F_3^{333}]_3^3$ is always 0, and various other combinations of $F$- and $R$-symbols have equal values as well. After removing the elements of each tuple in $S(I)$ whose values are equal for all categories, we obtain a smaller invariant. The values for this invariant for each category are given in Table 1.

From Example 3.8, it is clear that there could still be some redundant elements of the tuples in $S(I)$ left. For example, from Table 1 it is clear that some gauge invariants, such as $R_1^{33}$ and $R_3^{33}$, have the same power in distinguishing categories.

In order to simplify the invariant furter we will need to choose which gauge-invariants are prefered. The following are the, sometimes incompatible, criteria that were considered when deciding which invariants to use.

- Whenever possible, de gauge-invariants were set up that allow one to distinguish between inequivalent sets of
  - $F$-symbols by only using $F$-symbols,
  - $R$-symbols by only using $R$-symbols, and
  - pivotal structures by only using left quantum dimensions.

  This allows researchers with only partial information, e.g. only the $F$-symbols of a pivotal possibly braided fusion category, to identify their structure with the ones provided by Anyonica or the anyonwiki. In some cases, e.g. for Fibonacci fusion categories that are completely distinguished by the values of a single $R$-symbol, this means that provided invariant is bigger than strictly necessary.

- Gauge-invariants containing $F$-symbols that evaluate to 0 are left out whenever possible. This is because these might be de facto gauge-invariant for some fusion systems, but not for others.

- In general, invariants containing a small number of symbols were prioritized. E.g. for Rep($D_3$) categories all invariants that contain only $F$-symbols have the same power and therefore we chose to use $[F_1^{333}]_3^3$ in favour of some of the bigger ones.

- Ideally, we try to keep the tuples of gauge-invariants as small as possible. To do so we sometimes have to choose power over simplicity and a bigger gauge-invariant might appear as part of a tuple, rather than multiple smaller ones. This criterion sometimes clashes with the previous one, in which case the choice of invariant was rather add-hoc.

| Elements of the tuple in $S(I)$ | $[\text{Rep}(D_3)]_{1,1,1}$ | $[\text{Rep}(D_3)]_{1,2,1}$ | $[\text{Rep}(D_3)]_{1,3,1}$ | $[\text{Rep}(D_3)]_{2,0,1}$ | $[\text{Rep}(D_3)]_{3,0,1}$ |
|---|---|---|---|---|---|
| $[F_1^{333}]_3^3$ | 1 | 1 | 1 | $e^{\frac{2i\pi}{3}}$ | $e^{-\frac{2i\pi}{3}}$ |
| $[F_2^{333}]_3^3$ | $-1$ | $-1$ | $-1$ | $e^{-\frac{i\pi}{3}}$ | $e^{\frac{i\pi}{3}}$ |
| $[F_3^{333}]_3^1[F_3^{333}]_1^3$ | $\frac{1}{2}$ | $\frac{1}{2}$ | $\frac{1}{2}$ | $\frac{1}{2}e^{-\frac{2i\pi}{3}}$ | $\frac{1}{2}e^{\frac{2i\pi}{3}}$ |
| $\frac{[F_1^{323}]_3^3[F_3^{333}]_3^1[F_3^{333}]_2^3}{[F_3^{223}]_3^1[F_1^{233}]_2^3}$ | $-\frac{1}{2}$ | $-\frac{1}{2}$ | $-\frac{1}{2}$ | $-\frac{1}{2}e^{-\frac{2i\pi}{3}}$ | $-\frac{1}{2}e^{\frac{2i\pi}{3}}$ |
| $\frac{[F_2^{323}]_3^3[F_3^{333}]_3^1[F_3^{333}]_2^3}{[F_3^{223}]_3^1[F_1^{233}]_2^3}$ | $-\frac{1}{2}$ | $-\frac{1}{2}$ | $-\frac{1}{2}$ | $-\frac{1}{2}e^{-\frac{2i\pi}{3}}$ | $-\frac{1}{2}e^{\frac{2i\pi}{3}}$ |
| $\frac{[F_3^{323}]_3^3[F_3^{333}]_3^1[F_3^{333}]_2^3}{[F_3^{223}]_3^1[F_1^{233}]_2^3}$ | $\frac{1}{2}$ | $\frac{1}{2}$ | $\frac{1}{2}$ | $\frac{1}{2}e^{-\frac{2i\pi}{3}}$ | $\frac{1}{2}e^{\frac{2i\pi}{3}}$ |
| $[F_3^{322}]_1^3[F_1^{332}]_3^2[F_3^{333}]_3^1[F_3^{333}]_2^3$ | $-\frac{1}{2}$ | $-\frac{1}{2}$ | $-\frac{1}{2}$ | $-\frac{1}{2}e^{-\frac{2i\pi}{3}}$ | $-\frac{1}{2}e^{\frac{2i\pi}{3}}$ |
| $\frac{[F_3^{322}]_1^3[F_3^{333}]_3^1[F_3^{333}]_2^2[F_3^{333}]_2^3}{[F_1^{233}]_2^3}$ | $-\frac{1}{4}$ | $-\frac{1}{4}$ | $-\frac{1}{4}$ | $-\frac{1}{4}e^{-\frac{2i\pi}{3}}$ | $-\frac{1}{4}e^{\frac{2i\pi}{3}}$ |
| $\frac{[F_3^{322}]_1^3[F_2^{332}]_3^1[F_3^{333}]_3^1[F_3^{333}]_2^3}{[F_3^{223}]_3^1\left([F_1^{233}]_2^3\right)^2}$ | $-\frac{1}{2}$ | $-\frac{1}{2}$ | $-\frac{1}{2}$ | $-\frac{1}{2}e^{-\frac{2i\pi}{3}}$ | $-\frac{1}{2}e^{\frac{2i\pi}{3}}$ |
| $\frac{[F_3^{322}]_1^3[F_3^{333}]_3^1[F_3^{333}]_3^1[F_3^{333}]_2^3}{[F_3^{223}]_3^1\left([F_1^{233}]_2^3\right)^2}$ | $-\frac{1}{4}$ | $-\frac{1}{4}$ | $-\frac{1}{4}$ | $-\frac{1}{4}e^{-\frac{2i\pi}{3}}$ | $-\frac{1}{4}e^{\frac{2i\pi}{3}}$ |
| $\frac{[F_3^{322}]_1^3\left([F_3^{333}]_3^1[F_3^{333}]_2^3\right)^2}{[F_3^{223}]_3^1\left([F_1^{233}]_2^3\right)^2}$ | $\frac{1}{4}$ | $\frac{1}{4}$ | $\frac{1}{4}$ | $\frac{1}{4}e^{\frac{2i\pi}{3}}$ | $\frac{1}{4}e^{-\frac{2i\pi}{3}}$ |
| $R_1^{22}$ | 1 | 1 | 1 | - | - |
| $R_1^{33}$ | 1 | $e^{-\frac{2i\pi}{3}}$ | $e^{\frac{2i\pi}{3}}$ | - | - |
| $R_2^{33}$ | $-1$ | $-e^{-\frac{2i\pi}{3}}$ | $-e^{\frac{2i\pi}{3}}$ | - | - |
| $R_3^{33}$ | 1 | $e^{\frac{2i\pi}{3}}$ | $e^{-\frac{2i\pi}{3}}$ | - | - |
| $[F_3^{333}]_2^2 R_3^{32}$ | $-\frac{1}{2}$ | $-\frac{1}{2}$ | $-\frac{1}{2}$ | - | - |
| $R_3^{23} R_3^{32}$ | 1 | 1 | 1 | - | - |
| $[F_1^{233}]_2^3[F_1^{332}]_3^2 R_3^{32}$ | $-1$ | $-1$ | $-1$ | - | - |
| $\frac{[F_3^{333}]_2^1 R_3^{32}}{[F_3^{223}]_3^1[F_1^{233}]_2^3}$ | $-\frac{1}{2}$ | $-\frac{1}{2}$ | $-\frac{1}{2}$ | - | - |
| $\frac{[F_3^{333}]_3^1[F_3^{333}]_2^3 R_3^{32}}{[F_3^{223}]_3^1[F_1^{233}]_2^3}$ | $\frac{1}{2}$ | $\frac{1}{2}$ | $\frac{1}{2}$ | - | - |
| $\frac{[F_1^{323}]_3^3 R_3^{32}}{[F_3^{223}]_3^1[F_3^{322}]_1^3}$ | $-1$ | $-1$ | $-1$ | - | - |
| $\frac{[F_2^{323}]_3^3 R_3^{32}}{[F_3^{223}]_3^1[F_3^{322}]_1^3}$ | $-1$ | $-1$ | $-1$ | - | - |
| $\frac{[F_3^{323}]_3^3 R_3^{32}}{[F_3^{223}]_3^1[F_3^{322}]_1^3}$ | 1 | 1 | 1 | - | - |
| $\frac{[F_2^{332}]_3^1 R_3^{32}}{[F_3^{223}]_3^1[F_1^{233}]_2^3}$ | $-1$ | $-1$ | $-1$ | - | - |
| $\frac{\left(R_3^{32}\right)^2}{[F_3^{223}]_3^1[F_3^{322}]_1^3}$ | 1 | 1 | 1 | - | - |

Table 1: Values of elements of the reduced $S(I)$ for the Rep($D_3$) pivotal (braided) fusion categories. The symbol '-' means that the category is not braided

- Lastly, between several gauge-invariants with equal power and simplicity, we opted to choose the ones with the most elegant values. E.g. the SU(2)$_k$ and PSU(2)$_k$ categories often contain an object $a$ for which $[F_a^{aaa}]_a^a$ distinguishes between all sets of inequivalent $F$-symbols and has a nicer value than other $F$-symbols with the same properties.

In case of $\text{Rep}(D_3)$, a reduced invariant could look like $(S(I))_{\text{reduced}} = \left\{([F_1^{333}]_3^3, R_1^{33})\right\}$. Since $S$ is trivial, there is no need to look at its orbit and we obtain the couple $c = \left([F_1^{333}]_3^3, R_1^{33}\right)$ as a distinguishing invariant. Note that each elemement of $c$ is independent and has its own role in distinguishing between the various categories. $[F_1^{333}]_3^3$ distinguishes between all inequivalent sets of $F$-symbols and $R_1^{33}$ distinguishes between different sets of $R$-symbols. The use of an ordered couple is thus not necessary in this case and we might as well use $\left\{[F_1^{333}]_3^3, R_1^{33}\right\}$ as a distinguishing invariant.

One of the reasons the final invariant for fusion systems with $\mathcal{R} = \text{Rep}(D_3)$ is so nice, is that $S(R)$ is trivial. If $S = \text{Aut}(\mathcal{R})$ is not trivial, two scenarios can occur.

1. It is possible to write the invariant as a set of orbits of gauge-invariants under $S$. This occurs, for example, for $\mathcal{R} = \mathbb{Z}_3$.

2. The invariant can only be written as an orbit of ordered tuples of gauge-invariants. This occurs, for example for $\mathcal{R} = \mathbb{Z}_2 \times \mathbb{Z}_2$. In this case the fusion ring has so much symmetry, and the invariants attain so few different values, that it is impossible to simplify the invariant $(S(I), [S(I)]_\Phi)$ in a way that doesn't take account of the order of the gauge invariants.

In Section 4 we will see how invariants for each of these scenarios are represented.

# 4 Tables of invariants

In this section, we provide a table of invariants for each multiplicity-free Gröthendieck ring, $\mathcal{R}$, up to rank 7 that distinguishes between all MFPBFCs and MFPNBFCs with Gröthendieck ring $\mathcal{R}$. Each subsection that covers a Gröthendieck ring also provides some extra practical information. In particular, each such subsection contains

- A multiplication table of the fusion ring whose categories are distinguished. **Be careful:** The order of the elements of the fusion rings in this paper may differ from certain conventions. The elements are sorted, first by Perron-Frobenius dimension and second by whether they are self-dual or not. For example, the order of the elements of $\mathbb{Z}_4$ in this paper is different from that of the common representation as the abelian group of integers under addition modulo 4.

- The automorphism group $S$ of the fusion ring as a set of permutations in cycle notation.

- A table that lists the values of the invariant for all MFPRFCs and MFPNRFCs, as provided by Anyonica, with fusion ring $\mathcal{R}$.

Before presenting the tables, it is important to know how they are constructed. In Section 4.1 the layout of the tables is discussed, while Section 4.2 discusses the content of the tables. Sections 4.3 and 4.4 respectively present some tips for effectively using the tables and some important remarks before interpreting the data. Section 4.5 provides the tables of invariants for all MFPBFCs and MFPNBFCs up to rank 7.

## 4.1 Layout of the tables

The following conventions and visual guides are used in the tables.

- The naming scheme of the fusion categories is a slight adaptation from [Ver24b] and has the general form $[\text{RingName}]_{n_F, n_R, n_P}$. Here, RingName is the name of the fusion ring for which we use the formal naming scheme set up in [VS23], while $n_F, n_R, n_P$ are indices indicating which categories have equivalent $F$-symbols, $R$-symbols, and pivotal structures. Categories with the same Gröthendieck ring are

- equivalent as fusion categories (without extra structure) if and only if they have the same value of $n_F$,
- equivalent as braided fusion categories (without pivotal structure) if and only if they have the same value for both $n_F$ and $n_R$,
- equivalent as pivotal fusion categories (without braided structure) if and only if they have the same value for both $n_F$ and $n_p$,
- equivalent as pivotal braided fusion categories if and only if they have the same value for $n_F$, $n_R$, and $n_p$.

If $n_R = 0$, it means the fusion category cannot be braided.

- For non-braided categories, the values of the invariants containing $R$-symbols will be denoted by -.

- The following are visual guidelines to distinguish between the various categories.
  - A horizontal dashed line separates the categories with different values for $n_F$.
  - Categories with the same value for $n_R$ are grouped together in colored bands. The colors themselves have no meaning besides separating consequtive groups of categories that are braided equivalent.
  - Each category is followed by a list of symbols , , , ,  that respectively denotes that the category is braided, unitary, spherical, ribbon, modular.

- Some numbers appear so often that it is useful to define the following shorthand notation

$$\zeta_n := \exp\left(\frac{2\pi i}{n}\right), \quad \phi := \frac{1+\sqrt{5}}{2}, \quad \overline{\phi} := \frac{1-\sqrt{5}}{2}.$$

- For each ring, a table with symbolic invariants and one with numeric invariants is given. We made an exception in the case where the symbolic invariants contain only rational complex numbers. The numerical invariants are printed by default with 3 decimal digits but trailing 0's have been removed for rational numbers.

- Sets will sometimes list their elements vertically if the content of the table would otherwise not fit on the page.

## 4.2 Content of the tables

A table that lists the values of the invariant for all categories with fusion ring $R$ can have three different formats.

### 4.2.1 Case 1: The invariant is a set of gauge-invariants

If the invariant can be written as a set of gauge-invariants then all such gauge invariants are listed as column headers and their values for each category make up the rows.

*Example* 4.1. For, e.g. $\mathscr{R} = \mathbb{Z}_2$ this looks like

Table 2: Symbolic invariants for categories with $\mathscr{R} = \mathbb{Z}_2$

| Name | Properties | $[F_2^{222}]_1^1$ | $R_1^{22}$ | $d_2^L$ |
|---|---|---|---|---|
| $[FR_1^{2,1,0}]_{1,1,1}$ | ✗ † 🌐 🎗 ⚭ | $-1$ | $i$ | $1$ |
| $[FR_1^{2,1,0}]_{1,1,2}$ | ✗ 🌐 🎗 ⚭ | $-1$ | $i$ | $-1$ |
| $[FR_1^{2,1,0}]_{1,2,1}$ | ✗ † 🌐 🎗 ⚭ | $-1$ | $-i$ | $1$ |
| $[FR_1^{2,1,0}]_{1,2,2}$ | ✗ 🌐 🎗 ⚭ | $-1$ | $-i$ | $-1$ |
| $[FR_1^{2,1,0}]_{2,1,1}$ | ✗ † 🌐 🎗 | $1$ | $1$ | $1$ |
| $[FR_1^{2,1,0}]_{2,1,2}$ | ✗ 🌐 🎗 | $1$ | $1$ | $-1$ |
| $[FR_1^{2,1,0}]_{2,2,1}$ | ✗ † 🌐 🎗 | $1$ | $-1$ | $1$ |
| $[FR_1^{2,1,0}]_{2,2,2}$ | ✗ 🌐 🎗 | $1$ | $-1$ | $-1$ |

From this table it is clear that the value of $[F_2^{222}]_1^1$ distinguishes the first 4 categories from the last 4. Each colored band contains the consecutive categories with equal values of $R_1^{22}$ and the categories within the bands are distinguished by their value of $d_2^L$. Since $[F_2^{222}]_1^1$, $R_1^{22}$, and $d_2^L$ are gauge-invariant and $\text{Aut}(\mathbb{Z}_2)$ is trivial, it is clear that all the proposed braided pivotal categories are indeed inequivalent.

### 4.2.2 Case 2: The invariant is a set of orbits of gauge-invariants

In this case the header of the table contains a combination of orbits of formal gauge-invariants under $S$ and possibly formal gauge-invariants that are invariant under $S$. Orbits of gauge-invariants under $S$ are unsorted sets and can, after substituting the $F$-symbols, $R$-symbols, and left quantum dimensions, differ in size between categories.

*Example* 4.2. For $\mathscr{R} = \mathbb{Z}_3$, $S = \{(), (2\ 3)\}$. Let

$$X_1 = S\left(R_3^{22}\right), X_2 = S\left(d_2^L\right).$$

The following table lists a small set of invariants whose values completely distinguish between all MFPBFCs and MFPNBFCs with the given fusion rules.

Table 3: Symbolic invariants for categories with $\mathscr{R} = \mathbb{Z}_3$

| Name | Properties | $[F_1^{222}]_3^3[F_1^{333}]_2^2$ | $X_1$ | $X_2$ |
|---|---|---|---|---|
| $[FR_1^{3,1,2}]_{1,1,1}$ | ✗ † 🌐 🎗 ⚭ | $1$ | $\{\zeta_3\}$ | $\{1\}$ |
| $[FR_1^{3,1,2}]_{1,1,2}$ | ✗ | $1$ | $\{\zeta_3\}$ | $\{\zeta_3^2, \zeta_3\}$ |
| $[FR_1^{3,1,2}]_{1,2,1}$ | ✗ † 🌐 🎗 ⚭ | $1$ | $\{\zeta_3^2\}$ | $\{1\}$ |
| $[FR_1^{3,1,2}]_{1,2,2}$ | ✗ | $1$ | $\{\zeta_3^2\}$ | $\{\zeta_3^2, \zeta_3\}$ |
| $[FR_1^{3,1,2}]_{1,3,1}$ | ✗ † 🌐 🎗 | $1$ | $\{1\}$ | $\{1\}$ |
| $[FR_1^{3,1,2}]_{1,3,2}$ | ✗ | $1$ | $\{1\}$ | $\{\zeta_3^2, \zeta_3\}$ |
| $[FR_1^{3,1,2}]_{2,1,1}$ | † 🌐 | $\zeta_3$ | $-$ | $\{1\}$ |
| $[FR_1^{3,1,2}]_{2,1,2}$ |  | $\zeta_3$ | $-$ | $\{\zeta_3^2, \zeta_3\}$ |

Table 3: Symbolic invariants for categories with $\mathscr{R} = \mathbb{Z}_3$ (Continued)

| Name | Properties | $[F_1^{222}]_3^3[F_1^{333}]_2^2$ | $X_1$ | $X_2$ |
|---|---|---|---|---|
| $[\text{FR}_1^{3,1,2}]_{3,1,1}$ |  | $\zeta_3^2$ | - | $\{1\}$ |
| $[\text{FR}_1^{3,1,2}]_{3,1,2}$ | | $\zeta_3^2$ | - | $\{\zeta_3^2, \zeta_3\}$ |

Since the gauge-invariant $[F_1^{222}]_3^3[F_1^{333}]_2^2$ is invariant under $S$, there is no need to look at its orbit. While this is not the case for $R_3^{22}$ and $d_2^L$, it is clear that both for $R_3^{22}$ and $d_2^L$ one only needs to compute one value in their orbits. Indeed, it turns out that $R_3^{22} = R_2^{33}$ for the given categories and $\{d_2^L, d_3^L\}$, after substitution of the left quantum dimensions, either contains only the element 1 or no elements equal to 1. So, even though $S$ is non-trivial, only one value of each gauge-invariant is required to distinguish all categories.

#### 4.2.3 Case 3: The invariant is an orbit of couples

In case the invariant is an orbit of couples, the table contains only one invariant.

*Example* 4.3. For $\mathscr{R} = \mathbb{Z}_5$, $S = \{(), (2\ 4\ 3\ 5), (2\ 5\ 3\ 4), (2\ 3)(4\ 5)\}$ and the invariant

$$X_1 = S\left(\left([F_4^{324}]_3^1[F_3^{332}]_1^4[F_3^{444}]_2^2, R_4^{33}, d_3^L\right)\right) \tag{30}$$

distinguishes between all categories.

Table 4: Symbolic invariants for categories with $\mathscr{R} = \mathbb{Z}_5$

| Name | Properties | $X_1$ |
|---|---|---|
| $[\text{FR}_1^{5,1,4}]_{1,1,1}$ |  | $\{(1, \zeta_5^3, 1), (1, \zeta_5^2, 1)\}$ |
| $[\text{FR}_1^{5,1,4}]_{1,1,2}$ |  | $\{(1, \zeta_5^3, \zeta_5^4), (1, \zeta_5^3, \zeta_5), (1, \zeta_5^2, \zeta_5^3), (1, \zeta_5^2, \zeta_5^2)\}$ |
| $[\text{FR}_1^{5,1,4}]_{1,1,3}$ |  | $\{(1, \zeta_5^3, \zeta_5^3), (1, \zeta_5^3, \zeta_5^2), (1, \zeta_5^2, \zeta_5^4), (1, \zeta_5^2, \zeta_5)\}$ |
| $[\text{FR}_1^{5,1,4}]_{1,2,1}$ |  | $\{(1, \zeta_5^4, 1), (1, \zeta_5, 1)\}$ |
| $[\text{FR}_1^{5,1,4}]_{1,2,2}$ |  | $\{(1, \zeta_5^4, \zeta_5^4), (1, \zeta_5^4, \zeta_5), (1, \zeta_5, \zeta_5^3), (1, \zeta_5, \zeta_5^2)\}$ |
| $[\text{FR}_1^{5,1,4}]_{1,2,3}$ |  | $\{(1, \zeta_5^4, \zeta_5^3), (1, \zeta_5^4, \zeta_5^2), (1, \zeta_5, \zeta_5^4), (1, \zeta_5, \zeta_5)\}$ |
| $[\text{FR}_1^{5,1,4}]_{1,3,1}$ |  | $\{(1, 1, 1)\}$ |
| $[\text{FR}_1^{5,1,4}]_{1,3,2}$ |  | $\{(1, 1, \zeta_5^3), (1, 1, \zeta_5^2), (1, 1, \zeta_5^4), (1, 1, \zeta_5)\}$ |
| $[\text{FR}_1^{5,1,4}]_{2,1,1}$ |  | $\{(\zeta_5^3, -, 1), (\zeta_5^2, -, 1)\}$ |
| $[\text{FR}_1^{5,1,4}]_{2,1,2}$ | | $\{(\zeta_5^3, -, \zeta_5^3), (\zeta_5^3, -, \zeta_5^2), (\zeta_5^2, -, \zeta_5), (\zeta_5^2, -, \zeta_5^4)\}$ |
| $[\text{FR}_1^{5,1,4}]_{2,1,3}$ | | $\{(\zeta_5^3, -, \zeta_5^4), (\zeta_5^3, -, \zeta_5), (\zeta_5^2, -, \zeta_5^3), (\zeta_5^2, -, \zeta_5^2)\}$ |
| $[\text{FR}_1^{5,1,4}]_{3,1,1}$ |  | $\{(\zeta_5^4, -, 1), (\zeta_5, -, 1)\}$ |
| $[\text{FR}_1^{5,1,4}]_{3,1,2}$ | | $\{(\zeta_5^4, -, \zeta_5^3), (\zeta_5^4, -, \zeta_5^2), (\zeta_5, -, \zeta_5), (\zeta_5, -, \zeta_5^4)\}$ |
| $[\text{FR}_1^{5,1,4}]_{3,1,3}$ | | $\{(\zeta_5^4, -, \zeta_5^4), (\zeta_5^4, -, \zeta_5), (\zeta_5, -, \zeta_5^3), (\zeta_5, -, \zeta_5^2)\}$ |

To identify a category with $\mathbb{Z}_5$ fusion rules it is possible to apply some clever tricks to avoid computations. If the value of $[F_4^{324}]_3^1[F_3^{332}]_1^4[F_3^{444}]_2^2$ equals 1 it is clear that the category is of the form $[\text{FR}_1^{5,1,4}]_{1,n_R,n_p}$, if it equals $\zeta_5^3$ or $\zeta_5^2$ the category is of the form $[\text{FR}_1^{5,1,4}]_{2,n_R,n_p}$ and otherwise it is of the form $[\text{FR}_1^{5,1,4}]_{3,n_R,n_p}$. There is, therefore, no need to calculate all values of the symbols in the orbit $S([F_4^{324}]_3^1[F_3^{332}]_1^4[F_3^{444}]_2^2)$, only one suffices. The same trick applies to identifying $n_R$. The fact that an ordered tuple is required to set up the invariant is because otherwise the pivotal structure can not be identified. If additional information about the pivotal structure is known, e.g. if the category one wants to identify is known to be spherical or unitary, then one can often identify the pivotal structure by examining the list of properties for each category in the table.

## 4.3 Using the tables effectively

Here are some tips for using the tables effectively

- Often, all categories belonging to some ring are distinguished by the values of a given $R$-symbol. If this is the case and all skeletal data of an unidentified category is known, such an $R$-symbol provides an immediate lookup table.

- Often, the values of all algebraic expressions in formal symbols in an orbit have the same value. In that case, only one of the values of such an algebraic expression needs to be calculated.

## 4.4 Important remarks

Several important remarks should be noted before reviewing the tables. Some of these remarks were already made in the previous part of the paper but will still be revised here.

- **The order of the basis elements of the fusion rings in this paper may differ from certain conventions.** When trying to identify a fusion category, the first step should always be to make sure the fusion ring of the category-to-be-identified has the same multiplication table as presented in this paper.

- **The presented invariants are not invariant under the full group of automorphisms of the category.** They are only invariant under the subgroup of automorphisms that leave the values of the vacuum $F$-symbols fixed. We also assume that all vacuum $F$-symbols and $R$-symbols equal 1. To identify a fusion system by using the tables, one should always check that the vacuum $F$- and $R$-symbols of that category equal 1.

- **The tables assume that the pivotal and (if it exists) braided structure is part of the category.** For example, two fusion systems, say $\Phi_1$ and $\Phi_2$, with equivalent $F$- and $R$-symbols will be regarded as inequivalent if their combination of $F$-symbols, $R$-symbols and pivotal structure are not equivalent, i.e. if there exists no combination of a fusion ring automorphism and gauge-transform that maps the $F$-symbols, $R$-symbols, and pivotal structure of $\Phi_1$ to those of $\Phi_2$. Likewise, two fusion categories with equivalent $F$-symbols will not be regarded as equivalent if either their $R$-symbols or their pivotal structures are not equivalent.

- Each category is followed by a list of symbols 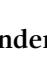, 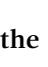, 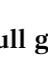, 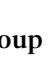, 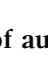 that respectively denote that the category is braided, unitary, spherical, ribbon, modular. Some authors identify unitary categories with categories whose $F$-matrices are unitary in an appropriate gauge. In this paper, as in [Ver24b], **a category is called unitary and denoted by 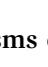 if it has the canonical spherical structure for which all left quantum dimensions are positive and there exists a gauge for which the $F$-matrices are unitary.** If one prefers the definition purely based on the unitarity of the $F$-matrices then any fusion category $[\mathscr{R}]_{n_F,n_R,n_p}$ for which there is another unitary category $[\mathscr{R}]_{n_F,m_R,m_p}$ can be regarded as unitary.

## 4.5 The tables of invariants

### 4.5.1 $FR_1^{1,1,0}$: Trivial

For the fusion ring, the following multiplication table is used.

$$\boxed{\begin{array}{c} 1 \end{array}}$$

Only the trivial permutation leaves the fusion rules invariant.

The following table lists a small set of invariants whose values completely distinguish between all MFPBFCs and MFPNBFCs with the given fusion rules.

Table 5: Symbolic invariants.

| Name | Properties | $[F_1^{111}]_1^1$ |
|---|---|---|
| $[FR_1^{1,1,0}]_{1,1,1}$ | 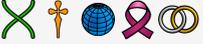 | 1 |

### 4.5.2 $FR_1^{2,1,0}$: $\mathbb{Z}_2$

For the fusion ring, the following multiplication table is used.

$$\boxed{\begin{array}{cc} 1 & 2 \\ 2 & 1 \end{array}}$$

Only the trivial permutation leaves the fusion rules invariant.

The following table lists a small set of invariants whose values completely distinguish between all MFPBFCs and MFPNBFCs with the given fusion rules.

Table 6: Symbolic invariants

| Name | Properties | $[F_2^{222}]_1^1$ | $R_1^{22}$ | $d_2^L$ |
|---|---|---|---|---|
| $[FR_1^{2,1,0}]_{1,1,1}$ | 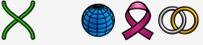 | $-1$ | $i$ | 1 |
| $[FR_1^{2,1,0}]_{1,1,2}$ | 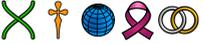 | $-1$ | $i$ | $-1$ |
| $[FR_1^{2,1,0}]_{1,2,1}$ | 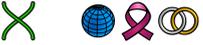 | $-1$ | $-i$ | 1 |
| $[FR_1^{2,1,0}]_{1,2,2}$ | 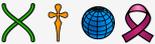 | $-1$ | $-i$ | $-1$ |
| $[FR_1^{2,1,0}]_{2,1,1}$ | 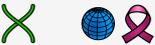 | 1 | 1 | 1 |
| $[FR_1^{2,1,0}]_{2,1,2}$ | 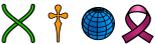 | 1 | 1 | $-1$ |
| $[FR_1^{2,1,0}]_{2,2,1}$ | 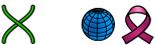 | 1 | $-1$ | 1 |
| $[FR_1^{2,1,0}]_{2,2,2}$ |  | 1 | $-1$ | $-1$ |

### 4.5.3 $FR_2^{2,1,0}$: Fib

For the fusion ring, the following multiplication table is used.

$$\boxed{\begin{array}{cc} 1 & 2 \\ 2 & 1+2 \end{array}}$$

Only the trivial permutation leaves the fusion rules invariant.

The following table lists a small set of invariants whose values completely distinguish between all MFPBFCs and MFPNBFCs with the given fusion rules.

Table 7: Symbolic invariants

| Name | Properties | $[F_2^{222}]_2^2$ | $R_1^{22}$ |
|---|---|---|---|
| $[FR_2^{2,1,0}]_{1,1,1}$ | ✕ † 🌐 🎗 ⚭ | $\overline{\phi}$ | $\zeta_5^3$ |
| $[FR_2^{2,1,0}]_{1,2,1}$ | ✕ † 🌐 🎗 ⚭ | $\overline{\phi}$ | $\zeta_5^2$ |
| $[FR_2^{2,1,0}]_{2,1,1}$ | ✕ 🌐 🎗 ⚭ | $\phi$ | $\zeta_5$ |
| $[FR_2^{2,1,0}]_{2,2,1}$ | ✕ 🌐 🎗 ⚭ | $\phi$ | $\zeta_5^4$ |

Table 8: Numeric invariants

| Name | Properties | $[F_2^{222}]_2^2$ | $R_1^{22}$ |
|---|---|---|---|
| $[FR_2^{2,1,0}]_{1,1,1}$ | ✕ † 🌐 🎗 ⚭ | $-0.618$ | $-0.809 - 0.588i$ |
| $[FR_2^{2,1,0}]_{1,2,1}$ | ✕ † 🌐 🎗 ⚭ | $-0.618$ | $-0.809 + 0.588i$ |
| $[FR_2^{2,1,0}]_{2,1,1}$ | ✕ 🌐 🎗 ⚭ | $1.618$ | $0.309 + 0.951i$ |
| $[FR_2^{2,1,0}]_{2,2,1}$ | ✕ 🌐 🎗 ⚭ | $1.618$ | $0.309 - 0.951i$ |

### 4.5.4 $FR_1^{3,1,0}$: Ising

For the fusion ring, the following multiplication table is used.

| **1** | **2** | **3** |
|---|---|---|
| **2** | **1** | **3** |
| **3** | **3** | **1 + 2** |

Only the trivial permutation leaves the fusion rules invariant.

The following table lists a small set of invariants whose values completely distinguish between all MFPBFCs and MFPNBFCs with the given fusion rules.

Table 9: Symbolic invariants

| Name | Properties | $[F_1^{323}]_3^3 [F_3^{333}]_2^2$ | $R_1^{33}$ | $d_3^L$ |
|---|---|---|---|---|
| $[FR_1^{3,1,0}]_{1,1,1}$ | ✕ † 🌐 🎗 ⚭ | $-\frac{1}{\sqrt{2}}$ | $\zeta_{16}^{15}$ | $\sqrt{2}$ |
| $[FR_1^{3,1,0}]_{1,1,2}$ | ✕ 🌐 🎗 ⚭ | $-\frac{1}{\sqrt{2}}$ | $\zeta_{16}^{15}$ | $-\sqrt{2}$ |
| $[FR_1^{3,1,0}]_{1,2,1}$ | ✕ † 🌐 🎗 ⚭ | $-\frac{1}{\sqrt{2}}$ | $\zeta_{16}^{7}$ | $\sqrt{2}$ |
| $[FR_1^{3,1,0}]_{1,2,2}$ | ✕ 🌐 🎗 ⚭ | $-\frac{1}{\sqrt{2}}$ | $\zeta_{16}^{7}$ | $-\sqrt{2}$ |
| $[FR_1^{3,1,0}]_{1,3,1}$ | ✕ † 🌐 🎗 ⚭ | $-\frac{1}{\sqrt{2}}$ | $\zeta_{16}^{9}$ | $\sqrt{2}$ |
| $[FR_1^{3,1,0}]_{1,3,2}$ | ✕ 🌐 🎗 ⚭ | $-\frac{1}{\sqrt{2}}$ | $\zeta_{16}^{9}$ | $-\sqrt{2}$ |
| $[FR_1^{3,1,0}]_{1,4,1}$ | ✕ † 🌐 🎗 ⚭ | $-\frac{1}{\sqrt{2}}$ | $\zeta_{16}$ | $\sqrt{2}$ |
| $[FR_1^{3,1,0}]_{1,4,2}$ | ✕ 🌐 🎗 ⚭ | $-\frac{1}{\sqrt{2}}$ | $\zeta_{16}$ | $-\sqrt{2}$ |
| $[FR_1^{3,1,0}]_{2,1,1}$ | ✕ † 🌐 🎗 ⚭ | $\frac{1}{\sqrt{2}}$ | $\zeta_{16}^{3}$ | $\sqrt{2}$ |

Table 9: Symbolic invariants (Continued)

| Name | Properties | $[F_1^{323}]_3^3[F_3^{333}]_2^2$ | $R_1^{33}$ | $d_3^L$ |
|---|---|---|---|---|
| $[FR_1^{3,1,0}]_{2,1,2}$ | | $\frac{1}{\sqrt{2}}$ | $\zeta_{16}^{3}$ | $-\sqrt{2}$ |
| $[FR_1^{3,1,0}]_{2,2,1}$ | | $\frac{1}{\sqrt{2}}$ | $\zeta_{16}^{11}$ | $\sqrt{2}$ |
| $[FR_1^{3,1,0}]_{2,2,2}$ | | $\frac{1}{\sqrt{2}}$ | $\zeta_{16}^{11}$ | $-\sqrt{2}$ |
| $[FR_1^{3,1,0}]_{2,3,1}$ | | $\frac{1}{\sqrt{2}}$ | $\zeta_{16}^{5}$ | $\sqrt{2}$ |
| $[FR_1^{3,1,0}]_{2,3,2}$ | | $\frac{1}{\sqrt{2}}$ | $\zeta_{16}^{5}$ | $-\sqrt{2}$ |
| $[FR_1^{3,1,0}]_{2,4,1}$ | | $\frac{1}{\sqrt{2}}$ | $\zeta_{16}^{13}$ | $\sqrt{2}$ |
| $[FR_1^{3,1,0}]_{2,4,2}$ | | $\frac{1}{\sqrt{2}}$ | $\zeta_{16}^{13}$ | $-\sqrt{2}$ |

Table 10: Numeric invariants

| Name | Properties | $[F_1^{323}]_3^3[F_3^{333}]_2^2$ | $R_1^{33}$ | $d_3^L$ |
|---|---|---|---|---|
| $[FR_1^{3,1,0}]_{1,1,1}$ | | $-0.707$ | $0.924 - 0.383i$ | $1.414$ |
| $[FR_1^{3,1,0}]_{1,1,2}$ | | $-0.707$ | $0.924 - 0.383i$ | $-1.414$ |
| $[FR_1^{3,1,0}]_{1,2,1}$ | | $-0.707$ | $-0.924 + 0.383i$ | $1.414$ |
| $[FR_1^{3,1,0}]_{1,2,2}$ | | $-0.707$ | $-0.924 + 0.383i$ | $-1.414$ |
| $[FR_1^{3,1,0}]_{1,3,1}$ | | $-0.707$ | $-0.924 - 0.383i$ | $1.414$ |
| $[FR_1^{3,1,0}]_{1,3,2}$ | | $-0.707$ | $-0.924 - 0.383i$ | $-1.414$ |
| $[FR_1^{3,1,0}]_{1,4,1}$ | | $-0.707$ | $0.924 + 0.383i$ | $1.414$ |
| $[FR_1^{3,1,0}]_{1,4,2}$ | | $-0.707$ | $0.924 + 0.383i$ | $-1.414$ |
| $[FR_1^{3,1,0}]_{2,1,1}$ | | $0.707$ | $0.383 + 0.924i$ | $1.414$ |
| $[FR_1^{3,1,0}]_{2,1,2}$ | | $0.707$ | $0.383 + 0.924i$ | $-1.414$ |
| $[FR_1^{3,1,0}]_{2,2,1}$ | | $0.707$ | $-0.383 - 0.924i$ | $1.414$ |
| $[FR_1^{3,1,0}]_{2,2,2}$ | | $0.707$ | $-0.383 - 0.924i$ | $-1.414$ |
| $[FR_1^{3,1,0}]_{2,3,1}$ | | $0.707$ | $-0.383 + 0.924i$ | $1.414$ |
| $[FR_1^{3,1,0}]_{2,3,2}$ | | $0.707$ | $-0.383 + 0.924i$ | $-1.414$ |
| $[FR_1^{3,1,0}]_{2,4,1}$ | | $0.707$ | $0.383 - 0.924i$ | $1.414$ |
| $[FR_1^{3,1,0}]_{2,4,2}$ | | $0.707$ | $0.383 - 0.924i$ | $-1.414$ |

### 4.5.5 $FR_2^{3,1,0}$: $\mathbf{Rep}(D_3)$

For the fusion ring, the following multiplication table is used.

| **1** | **2** | **3** |
|---|---|---|
| **2** | **1** | **3** |
| **3** | **3** | $1+2+3$ |

Only the trivial permutation leaves the fusion rules invariant.

The following table lists a small set of invariants whose values completely distinguish between all MFPBFCs and MFPNBFCs with the given fusion rules.

Table 11: Symbolic invariants

| Name | Properties | $[F_1^{333}]_3^3$ | $R_1^{33}$ |
|---|---|---|---|
| $[FR_2^{3,1,0}]_{1,1,1}$ | ✂ † 🌐 🎗 | 1 | 1 |
| $[FR_2^{3,1,0}]_{1,2,1}$ | ✂ † 🌐 🎗 | 1 | $\zeta_3^2$ |
| $[FR_2^{3,1,0}]_{1,3,1}$ | ✂ † 🌐 🎗 | 1 | $\zeta_3$ |
| $[FR_2^{3,1,0}]_{2,0,1}$ | † 🌐 | $\zeta_3$ | - |
| $[FR_2^{3,1,0}]_{3,0,1}$ | † 🌐 | $\zeta_3^2$ | - |

Table 12: Numeric invariants

| Name | Properties | $[F_1^{333}]_3^3$ | $R_1^{33}$ |
|---|---|---|---|
| $[FR_2^{3,1,0}]_{1,1,1}$ | ✂ † 🌐 🎗 | 1 | 1 |
| $[FR_2^{3,1,0}]_{1,2,1}$ | ✂ † 🌐 🎗 | 1 | $-0.5 - 0.866i$ |
| $[FR_2^{3,1,0}]_{1,3,1}$ | ✂ † 🌐 🎗 | 1 | $-0.5 + 0.866i$ |
| $[FR_2^{3,1,0}]_{2,0,1}$ | † 🌐 | $-0.5 + 0.866i$ | - |
| $[FR_2^{3,1,0}]_{3,0,1}$ | † 🌐 | $-0.5 - 0.866i$ | - |

### 4.5.6  $FR_3^{3,1,0}$: $PSU(2)_5$

For the fusion ring, the following multiplication table is used.

| | 1 | 2 | 3 |
|---|---|---|---|
| 1 | | | |
| 2 | | $1+3$ | $2+3$ |
| 3 | | $2+3$ | $1+2+3$ |

Only the trivial permutation leaves the fusion rules invariant.

The following table lists a small set of invariants whose values completely distinguish between all MFPBFCs and MFPNBFCs with the given fusion rules.

Table 13: Symbolic invariants

| Name | Properties | $[F_3^{333}]_3^3$ | $R_1^{33}$ |
|---|---|---|---|
| $[FR_3^{3,1,0}]_{1,1,1}$ | ✂ † 🌐 🎗 ⭕ | $\zeta_7^4 + \zeta_7^3 + 2$ | $\zeta_7^2$ |
| $[FR_3^{3,1,0}]_{1,2,1}$ | ✂ † 🌐 🎗 ⭕ | $\zeta_7^4 + \zeta_7^3 + 2$ | $\zeta_7^5$ |
| $[FR_3^{3,1,0}]_{2,1,1}$ | ✂ 🌐 🎗 ⭕ | $\zeta_7^5 + \zeta_7^2 + 2$ | $\zeta_7^6$ |
| $[FR_3^{3,1,0}]_{2,2,1}$ | ✂ 🌐 🎗 ⭕ | $\zeta_7^5 + \zeta_7^2 + 2$ | $\zeta_7$ |
| $[FR_3^{3,1,0}]_{3,1,1}$ | ✂ 🌐 🎗 ⭕ | $\zeta_7^6 + \zeta_7 + 2$ | $\zeta_7^3$ |

Table 13: Symbolic invariants (Continued)

| Name | Properties | $[F_3^{333}]_3^3$ | $R_1^{33}$ |
|---|---|---|---|
| $[FR_3^{3,1,0}]_{3,2,1}$ | 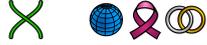 | $\zeta_7^6 + \zeta_7 + 2$ | $\zeta_7^4$ |

Table 14: Numeric invariants

| Name | Properties | $[F_3^{333}]_3^3$ | $R_1^{33}$ |
|---|---|---|---|
| $[FR_3^{3,1,0}]_{1,1,1}$ | | 0.198 | $-0.223 + 0.975i$ |
| $[FR_3^{3,1,0}]_{1,2,1}$ | | 0.198 | $-0.223 - 0.975i$ |
| $[FR_3^{3,1,0}]_{2,1,1}$ | | 1.555 | $0.623 - 0.782i$ |
| $[FR_3^{3,1,0}]_{2,2,1}$ | | 1.555 | $0.623 + 0.782i$ |
| $[FR_3^{3,1,0}]_{3,1,1}$ | | 3.247 | $-0.901 + 0.434i$ |
| $[FR_3^{3,1,0}]_{3,2,1}$ | | 3.247 | $-0.901 - 0.434i$ |

### 4.5.7 $FR_1^{3,1,2}$: $\mathbb{Z}_3$

For the fusion ring, the following multiplication table is used.

| 1 | 2 | 3 |
|---|---|---|
| 2 | 3 | 1 |
| 3 | 1 | 2 |

The following is the group of all non-trivial permutations that leave the fusion rules invariant:

$$S = \{(), (2\ 3)\}.$$

Let

$$X_1 = S\left(R_3^{22}\right), \tag{31}$$
$$X_2 = S\left(d_2^L\right). \tag{32}$$

The following table lists a small set of invariants whose values completely distinguish between all MFPBFCs and MFPNBFCs with the given fusion rules.

Table 15: Symbolic invariants

| Name | Properties | $[F_1^{222}]_3^3 [F_1^{333}]_2^2$ | $X_1$ | $X_2$ |
|---|---|---|---|---|
| $[FR_1^{3,1,2}]_{1,1,1}$ | | 1 | $\{\zeta_3\}$ | $\{1\}$ |
| $[FR_1^{3,1,2}]_{1,1,2}$ | | 1 | $\{\zeta_3\}$ | $\{\zeta_3^2, \zeta_3\}$ |
| $[FR_1^{3,1,2}]_{1,2,1}$ | | 1 | $\{\zeta_3^2\}$ | $\{1\}$ |
| $[FR_1^{3,1,2}]_{1,2,2}$ | | 1 | $\{\zeta_3^2\}$ | $\{\zeta_3^2, \zeta_3\}$ |

Table 15: Symbolic invariants (Continued)

| Name | Properties | $[F_1^{222}]_3^3[F_1^{333}]_2^2$ | $X_1$ | $X_2$ |
|---|---|---|---|---|
| $[FR_1^{3,1,2}]_{1,3,1}$ | ✗ † ● ♀ | 1 | {1} | {1} |
| $[FR_1^{3,1,2}]_{1,3,2}$ | ✗ | 1 | {1} | $\{\zeta_3^2, \zeta_3\}$ |
| $[FR_1^{3,1,2}]_{2,0,1}$ | † ● | $\zeta_3^2$ | - | {1} |
| $[FR_1^{3,1,2}]_{2,0,2}$ |  | $\zeta_3^2$ | - | $\{\zeta_3^2, \zeta_3\}$ |
| $[FR_1^{3,1,2}]_{3,0,1}$ | † ● | $\zeta_3$ | - | {1} |
| $[FR_1^{3,1,2}]_{3,0,2}$ |  | $\zeta_3$ | - | $\{\zeta_3^2, \zeta_3\}$ |

Table 16: Numeric invariants

| Name | Properties | $[F_1^{222}]_3^3[F_1^{333}]_2^2$ | $X_1$ | $X_2$ |
|---|---|---|---|---|
| $[FR_1^{3,1,2}]_{1,1,1}$ | ✗ † ● ♀ ⊙ | 1 | $\{-0.5 + 0.866i\}$ | {1} |
| $[FR_1^{3,1,2}]_{1,1,2}$ | ✗ | 1 | $\{-0.5 + 0.866i\}$ | $\left\{\begin{array}{l}-0.5 - 0.866i, \\ -0.5 + 0.866i\end{array}\right\}$ |
| $[FR_1^{3,1,2}]_{1,2,1}$ | ✗ † ● ♀ ⊙ | 1 | $\{-0.5 - 0.866i\}$ | {1} |
| $[FR_1^{3,1,2}]_{1,2,2}$ | ✗ | 1 | $\{-0.5 - 0.866i\}$ | $\left\{\begin{array}{l}-0.5 - 0.866i, \\ -0.5 + 0.866i\end{array}\right\}$ |
| $[FR_1^{3,1,2}]_{1,3,1}$ | ✗ † ● ♀ | 1 | {1} | {1} |
| $[FR_1^{3,1,2}]_{1,3,2}$ | ✗ | 1 | {1} | $\left\{\begin{array}{l}-0.5 - 0.866i, \\ -0.5 + 0.866i\end{array}\right\}$ |
| $[FR_1^{3,1,2}]_{2,0,1}$ | † ● | $-0.5 - 0.866i$ | - | {1, 1} |
| $[FR_1^{3,1,2}]_{2,0,2}$ |  | $-0.5 - 0.866i$ | - | $\left\{\begin{array}{l}-0.5 - 0.866i, \\ -0.5 + 0.866i\end{array}\right\}$ |
| $[FR_1^{3,1,2}]_{3,0,1}$ | † ● | $-0.5 + 0.866i$ | - | {1, 1} |
| $[FR_1^{3,1,2}]_{3,0,2}$ |  | $-0.5 + 0.866i$ | - | $\left\{\begin{array}{l}-0.5 - 0.866i, \\ -0.5 + 0.866i\end{array}\right\}$ |

### 4.5.8  $FR_1^{4,1,0}$: $\mathbb{Z}_2 \otimes \mathbb{Z}_2$

For the fusion ring, the following multiplication table is used.

| 1 | 2 | 3 | 4 |
|---|---|---|---|
| 2 | 1 | 4 | 3 |
| 3 | 4 | 1 | 2 |
| 4 | 3 | 2 | 1 |

The following is the group of all non-trivial permutations that leave the fusion rules invariant:

$$S = \{(), (2\ 3), (2\ 4), (3\ 4), (2\ 3\ 4), (2\ 4\ 3)\}.$$

Let
$$X_1 = S\left(\left([F_3^{232}]_4^4 [F_4^{242}]_3^3, [F_2^{323}]_4^4 [F_4^{343}]_2^2, R_1^{22}, R_1^{33}, d_2^L\right)\right). \tag{33}$$

The following table lists a small set of invariants whose values completely distinguish between all MFPBFCs and MFPNBFCs with the given fusion rules.

Table 17: Symbolic invariants

| Name | Properties | $X_1$ |
|---|---|---|
| $[\mathrm{FR}_1^{4,1,0}]_{1,1,1}$ | ✕ ✝ 🌐 🎗 ⚭ | $\{(-1,-1,-i,i,1),(-1,-1,i,-i,1),(-1,1,-i,1,1),(-1,1,i,1,1),(1,-1,1,-i,1),(1,-1,1,i,1)\}$ |
| $[\mathrm{FR}_1^{4,1,0}]_{1,1,2}$ | ✕ 🌐 🎗 ⚭ | $\{(-1,-1,-i,i,1),(-1,-1,i,-i,-1),(-1,1,-i,1,1),(-1,1,i,1,-1),(1,-1,1,-i,-1),(1,-1,1,i,-1)\}$ |
| $[\mathrm{FR}_1^{4,1,0}]_{1,1,3}$ | ✕ 🌐 🎗 ⚭ | $\{(-1,-1,-i,i,-1),(-1,-1,i,-i,-1),(-1,1,-i,1,-1),(-1,1,i,1,-1),(1,-1,1,-i,1),(1,-1,1,i,1)\}$ |
| $[\mathrm{FR}_1^{4,1,0}]_{1,2,1}$ | ✕ 🌐 🎗 ⚭ | $\{(-1,-1,-i,i,-1),(-1,-1,i,-i,1),(-1,1,-i,1,-1),(-1,1,i,1,1),(1,-1,1,-i,-1),(1,-1,1,i,-1)\}$ |
| $[\mathrm{FR}_1^{4,1,0}]_{1,2,2}$ | ✕ ✝ 🌐 🎗 ⚭ | $\{(-1,-1,-i,-i,1),(-1,1,-i,-1,1),(1,-1,-1,-i,1)\}$ |
| $[\mathrm{FR}_1^{4,1,0}]_{1,3,1}$ | ✕ 🌐 🎗 ⚭ | $\{(-1,-1,-i,-i,-1),(-1,-1,-i,-i,1),(-1,1,-i,-1,-1),(-1,1,-i,-1,1),(1,-1,-1,-i,-1)\}$ |
| $[\mathrm{FR}_1^{4,1,0}]_{1,3,2}$ | ✕ 🌐 🎗 ⚭ | $\{(-1,-1,-i,-i,-1),(-1,1,-i,-1,-1),(1,-1,-1,-i,1)\}$ |
| $[\mathrm{FR}_1^{4,1,0}]_{1,4,1}$ | ✕ ✝ 🌐 🎗 ⚭ | $\{(-1,-1,i,i,1),(-1,1,i,-1,1),(1,-1,-1,i,1)\}$ |
| $[\mathrm{FR}_1^{4,1,0}]_{1,4,2}$ | ✕ 🌐 🎗 ⚭ | $\{(-1,-1,i,i,-1),(-1,-1,i,i,1),(-1,1,i,-1,-1),(-1,1,i,-1,1),(1,-1,-1,i,-1)\}$ |
| $[\mathrm{FR}_1^{4,1,0}]_{1,4,3}$ | ✕ 🌐 🎗 ⚭ | $\{(-1,-1,i,i,-1),(-1,1,i,-1,-1),(1,-1,-1,i,1)\}$ |
| $[\mathrm{FR}_1^{4,1,0}]_{2,1,1}$ | ✕ ✝ 🌐 🎗 | $\{(-1,-1,-i,i,1),(-1,-1,i,-i,1),(-1,1,-i,-1,1),(-1,1,i,-1,1),(1,-1,-1,-i,1),(1,-1,-1,i,1)\}$ |
| $[\mathrm{FR}_1^{4,1,0}]_{2,1,2}$ | ✕ 🌐 🎗 | $\{(-1,-1,-i,i,-1),(-1,-1,i,-i,1),(-1,1,-i,-1,-1),(-1,1,i,-1,1),(1,-1,-1,-i,-1),(1,-1,-1,i,-1)\}$ |
| $[\mathrm{FR}_1^{4,1,0}]_{2,1,3}$ | ✕ 🌐 🎗 | $\{(-1,-1,-i,i,-1),(-1,-1,i,-i,-1),(-1,1,-i,-1,-1),(-1,1,i,-1,-1),(1,-1,-1,-i,1),(1,-1,-1,i,1)\}$ |
| $[\mathrm{FR}_1^{4,1,0}]_{2,2,1}$ | ✕ 🌐 🎗 | $\{(-1,-1,-i,i,1),(-1,-1,i,-i,-1),(-1,1,-i,-1,1),(-1,1,i,-1,-1),(1,-1,-1,-i,-1),(1,-1,-1,i,-1)\}$ |
| $[\mathrm{FR}_1^{4,1,0}]_{2,2,2}$ | ✕ ✝ 🌐 🎗 | $\{(-1,-1,-i,-i,1),(-1,1,-i,1,1),(1,-1,1,-i,1)\}$ |
| $[\mathrm{FR}_1^{4,1,0}]_{2,2,3}$ | ✕ 🌐 🎗 | $\{(-1,-1,-i,-i,-1),(-1,-1,-i,-i,1),(-1,1,-i,1,-1),(-1,1,-i,1,1),(1,-1,1,-i,-1)\}$ |
| $[\mathrm{FR}_1^{4,1,0}]_{2,2,4}$ | ✕ 🌐 🎗 | $\{(-1,-1,-i,-i,-1),(-1,1,-i,1,-1),(1,-1,1,-i,1)\}$ |
| $[\mathrm{FR}_1^{4,1,0}]_{2,3,1}$ | ✕ ✝ 🌐 🎗 | $\{(-1,-1,i,i,1),(-1,1,i,1,1),(1,-1,1,i,1)\}$ |

| Name | Properties | $X_1$ |
|---|---|---|
| $[FR_1^{4,1,0}]_{2,3,2}$ | ✕ 🌐 🎗 | $\{(-1,-1,i,i,-1),(-1,-1,i,i,1),(-1,1,i,1,-1),$ $(-1,1,i,1,1),(1,-1,1,i,-1)\}$ |
| $[FR_1^{4,1,0}]_{2,3,3}$ | ✕ 🌐 🎗 | $\{(-1,-1,i,i,-1),(-1,1,i,1,-1),(1,-1,1,i,1)\}$ |
| $[FR_1^{4,1,0}]_{2,4,1}$ | ✕ † 🌐 🎗 ⭕ | $\{(1,1,-1,1,1),(1,1,1,-1,1),(1,1,1,1,1)\}$ |
| $[FR_1^{4,1,0}]_{2,4,2}$ | ✕ 🌐 🎗 ⭕ | $\{(1,1,-1,1,1),(1,1,1,-1,-1),(1,1,1,1,-1)\}$ |
| $[FR_1^{4,1,0}]_{2,4,3}$ | ✕ 🌐 🎗 ⭕ | $\{(1,1,-1,1,-1),(1,1,1,-1,-1),(1,1,1,-1,1),$ $(1,1,1,1,-1),(1,1,1,1,1)\}$ |
| $[FR_1^{4,1,0}]_{2,5,1}$ | ✕ † 🌐 🎗 ⭕ | $\{(1,1,-1,-1,1)\}$ |
| $[FR_1^{4,1,0}]_{2,5,2}$ | ✕ 🌐 🎗 ⭕ | $\{(1,1,-1,-1,-1),(1,1,-1,-1,1)\}$ |
| $[FR_1^{4,1,0}]_{2,5,3}$ | ✕ † 🌐 🎗 | $\{(1,1,1,1,1)\}$ |
| $[FR_1^{4,1,0}]_{2,5,4}$ | ✕ 🌐 🎗 | $\{(1,1,1,1,-1),(1,1,1,1,1)\}$ |
| $[FR_1^{4,1,0}]_{2,6,1}$ | ✕ † 🌐 🎗 | $\{(1,1,-1,-1,1),(1,1,-1,1,1),(1,1,1,-1,1)\}$ |
| $[FR_1^{4,1,0}]_{2,6,2}$ | ✕ 🌐 🎗 | $\{(1,1,-1,-1,-1),(1,1,-1,1,-1),(1,1,1,-1,1)\}$ |
| $[FR_1^{4,1,0}]_{2,6,3}$ | ✕ 🌐 🎗 | $\{(1,1,-1,-1,-1),(1,1,-1,-1,1),(1,1,-1,1,-1),$ $(1,1,-1,1,1),(1,1,1,-1,-1)\}$ |
| $[FR_1^{4,1,0}]_{3,0,1}$ | † 🌐 | $\{(-1,1,-,-,1),(1,-1,-,-,1),(1,1,-,-,1)\}$ |
| $[FR_1^{4,1,0}]_{3,0,2}$ | 🌐 | $\{(-1,1,-,-,1),(1,-1,-,-,-1),(1,1,-,-,-1)\}$ |
| $[FR_1^{4,1,0}]_{3,0,3}$ | 🌐 | $\{(-1,1,-,-,-1),(1,-1,-,-,-1),(1,-1,-,-,1),$ $(1,1,-,-,-1),(1,1,-,-,1)\}$ |
| $[FR_1^{4,1,0}]_{4,0,1}$ | † 🌐 | $\{(-1,-1,-,-,1)\}$ |
| $[FR_1^{4,1,0}]_{4,0,2}$ | 🌐 | $\{(-1,-1,-,-,1),(-1,-1,-,-,-1),(-1,-1,-,-,-1)\}$ |

### 4.5.9 $FR_2^{4,1,0}$: $SU(2)_3$

For the fusion ring, the following multiplication table is used.

| **1** | **2** | **3** | **4** |
|---|---|---|---|
| **2** | 1 | 4 | 3 |
| **3** | 4 | 1+4 | 2+3 |
| **4** | 3 | 2+3 | 1+4 |

Only the trivial permutation leaves the fusion rules invariant.

The following table lists a small set of invariants whose values completely distinguish between all MFPBFCs and MFPNBFCs with the given fusion rules.

Table 18: Symbolic invariants

| Name | Properties | $[F_2^{323}]_4^4[F_4^{343}]_2^2$ | $R_1^{33}$ | $d_3^L$ |
|---|---|---|---|---|
| $[\text{FR}_2^{4,1,0}]_{1,1,1}$ | | | $\overline{\phi}$ | $\zeta_{20}^7$ | $\phi$ |
| $[\text{FR}_2^{4,1,0}]_{1,1,2}$ | | | $\overline{\phi}$ | $\zeta_{20}^7$ | $-\phi$ |
| $[\text{FR}_2^{4,1,0}]_{1,2,1}$ | | | $\overline{\phi}$ | $\zeta_{20}^3$ | $\phi$ |
| $[\text{FR}_2^{4,1,0}]_{1,2,2}$ | | | $\overline{\phi}$ | $\zeta_{20}^3$ | $-\phi$ |
| $[\text{FR}_2^{4,1,0}]_{1,3,1}$ | | | $\overline{\phi}$ | $\zeta_{20}^{17}$ | $\phi$ |
| $[\text{FR}_2^{4,1,0}]_{1,3,2}$ | | | $\overline{\phi}$ | $\zeta_{20}^{17}$ | $-\phi$ |
| $[\text{FR}_2^{4,1,0}]_{1,4,1}$ | | | $\overline{\phi}$ | $\zeta_{20}^{13}$ | $\phi$ |
| $[\text{FR}_2^{4,1,0}]_{1,4,2}$ | | | $\overline{\phi}$ | $\zeta_{20}^{13}$ | $-\phi$ |
| $[\text{FR}_2^{4,1,0}]_{2,1,1}$ | | | $-\overline{\phi}$ | $\zeta_5^3$ | $\phi$ |
| $[\text{FR}_2^{4,1,0}]_{2,1,2}$ | | | $-\overline{\phi}$ | $\zeta_5^3$ | $-\phi$ |
| $[\text{FR}_2^{4,1,0}]_{2,2,1}$ | | | $-\overline{\phi}$ | $\zeta_5^2$ | $\phi$ |
| $[\text{FR}_2^{4,1,0}]_{2,2,2}$ | | | $-\overline{\phi}$ | $\zeta_5^2$ | $-\phi$ |
| $[\text{FR}_2^{4,1,0}]_{2,3,1}$ | | | $-\overline{\phi}$ | $\zeta_{10}$ | $\phi$ |
| $[\text{FR}_2^{4,1,0}]_{2,3,2}$ | | | $-\overline{\phi}$ | $\zeta_{10}$ | $-\phi$ |
| $[\text{FR}_2^{4,1,0}]_{2,4,1}$ | | | $-\overline{\phi}$ | $\zeta_{10}^9$ | $\phi$ |
| $[\text{FR}_2^{4,1,0}]_{2,4,2}$ | | | $-\overline{\phi}$ | $\zeta_{10}^9$ | $-\phi$ |
| $[\text{FR}_2^{4,1,0}]_{3,1,1}$ | | | $\phi$ | $\zeta_{20}$ | $\overline{\phi}$ |
| $[\text{FR}_2^{4,1,0}]_{3,1,2}$ | | | $\phi$ | $\zeta_{20}$ | $-\overline{\phi}$ |
| $[\text{FR}_2^{4,1,0}]_{3,2,1}$ | | | $\phi$ | $\zeta_{20}^9$ | $\overline{\phi}$ |
| $[\text{FR}_2^{4,1,0}]_{3,2,2}$ | | | $\phi$ | $\zeta_{20}^9$ | $-\overline{\phi}$ |
| $[\text{FR}_2^{4,1,0}]_{3,3,1}$ | | | $\phi$ | $\zeta_{20}^{11}$ | $\overline{\phi}$ |
| $[\text{FR}_2^{4,1,0}]_{3,3,2}$ | | | $\phi$ | $\zeta_{20}^{11}$ | $-\overline{\phi}$ |
| $[\text{FR}_2^{4,1,0}]_{3,4,1}$ | | | $\phi$ | $\zeta_{20}^{19}$ | $\overline{\phi}$ |
| $[\text{FR}_2^{4,1,0}]_{3,4,2}$ | | | $\phi$ | $\zeta_{20}^{19}$ | $-\overline{\phi}$ |
| $[\text{FR}_2^{4,1,0}]_{4,1,1}$ | | | $-\phi$ | $\zeta_5^4$ | $\overline{\phi}$ |
| $[\text{FR}_2^{4,1,0}]_{4,1,2}$ | | | $-\phi$ | $\zeta_5^4$ | $-\overline{\phi}$ |
| $[\text{FR}_2^{4,1,0}]_{4,2,1}$ | | | $-\phi$ | $\zeta_5$ | $\overline{\phi}$ |
| $[\text{FR}_2^{4,1,0}]_{4,2,2}$ | | | $-\phi$ | $\zeta_5$ | $-\overline{\phi}$ |
| $[\text{FR}_2^{4,1,0}]_{4,3,1}$ | | | $-\phi$ | $\zeta_{10}^3$ | $\overline{\phi}$ |
| $[\text{FR}_2^{4,1,0}]_{4,3,2}$ | | | $-\phi$ | $\zeta_{10}^3$ | $-\overline{\phi}$ |
| $[\text{FR}_2^{4,1,0}]_{4,4,1}$ | | | $-\phi$ | $\zeta_{10}^7$ | $\overline{\phi}$ |
| $[\text{FR}_2^{4,1,0}]_{4,4,2}$ | | | $-\phi$ | $\zeta_{10}^7$ | $-\overline{\phi}$ |

Table 19: Numeric invariants

| Name | Properties | $[F_2^{323}]_4^4[F_4^{343}]_2^2$ | $R_1^{33}$ | $d_3^L$ |
|---|---|---|---|---|
| $[FR_2^{4,1,0}]_{1,1,1}$ | ✗ † 🌐 🎗 ⊚ | −0.618 | −0.588 + 0.809$i$ | 1.618 |
| $[FR_2^{4,1,0}]_{1,1,2}$ | ✗ 🌐 🎗 ⊚ | −0.618 | −0.588 + 0.809$i$ | −1.618 |
| $[FR_2^{4,1,0}]_{1,2,1}$ | ✗ † 🌐 🎗 ⊚ | −0.618 | 0.588 + 0.809$i$ | 1.618 |
| $[FR_2^{4,1,0}]_{1,2,2}$ | ✗ 🌐 🎗 ⊚ | −0.618 | 0.588 + 0.809$i$ | −1.618 |
| $[FR_2^{4,1,0}]_{1,3,1}$ | ✗ † 🌐 🎗 ⊚ | −0.618 | 0.588 − 0.809$i$ | 1.618 |
| $[FR_2^{4,1,0}]_{1,3,2}$ | ✗ 🌐 🎗 ⊚ | −0.618 | 0.588 − 0.809$i$ | −1.618 |
| $[FR_2^{4,1,0}]_{1,4,1}$ | ✗ † 🌐 🎗 ⊚ | −0.618 | −0.588 − 0.809$i$ | 1.618 |
| $[FR_2^{4,1,0}]_{1,4,2}$ | ✗ 🌐 🎗 ⊚ | −0.618 | −0.588 − 0.809$i$ | −1.618 |
| $[FR_2^{4,1,0}]_{2,1,1}$ | ✗ † 🌐 🎗 | 0.618 | −0.809 − 0.588$i$ | 1.618 |
| $[FR_2^{4,1,0}]_{2,1,2}$ | ✗ 🌐 🎗 | 0.618 | −0.809 − 0.588$i$ | −1.618 |
| $[FR_2^{4,1,0}]_{2,2,1}$ | ✗ † 🌐 🎗 | 0.618 | −0.809 + 0.588$i$ | 1.618 |
| $[FR_2^{4,1,0}]_{2,2,2}$ | ✗ 🌐 🎗 | 0.618 | −0.809 + 0.588$i$ | −1.618 |
| $[FR_2^{4,1,0}]_{2,3,1}$ | ✗ † 🌐 🎗 | 0.618 | 0.809 + 0.588$i$ | 1.618 |
| $[FR_2^{4,1,0}]_{2,3,2}$ | ✗ 🌐 🎗 | 0.618 | 0.809 + 0.588$i$ | −1.618 |
| $[FR_2^{4,1,0}]_{2,4,1}$ | ✗ † 🌐 🎗 | 0.618 | 0.809 − 0.588$i$ | 1.618 |
| $[FR_2^{4,1,0}]_{2,4,2}$ | ✗ 🌐 🎗 | 0.618 | 0.809 − 0.588$i$ | −1.618 |
| $[FR_2^{4,1,0}]_{3,1,1}$ | ✗ 🌐 🎗 ⊚ | 1.618 | 0.951 + 0.309$i$ | −0.618 |
| $[FR_2^{4,1,0}]_{3,1,2}$ | ✗ 🌐 🎗 ⊚ | 1.618 | 0.951 + 0.309$i$ | 0.618 |
| $[FR_2^{4,1,0}]_{3,2,1}$ | ✗ 🌐 🎗 ⊚ | 1.618 | −0.951 + 0.309$i$ | −0.618 |
| $[FR_2^{4,1,0}]_{3,2,2}$ | ✗ 🌐 🎗 ⊚ | 1.618 | −0.951 + 0.309$i$ | 0.618 |
| $[FR_2^{4,1,0}]_{3,3,1}$ | ✗ 🌐 🎗 ⊚ | 1.618 | −0.951 − 0.309$i$ | −0.618 |
| $[FR_2^{4,1,0}]_{3,3,2}$ | ✗ 🌐 🎗 ⊚ | 1.618 | −0.951 − 0.309$i$ | 0.618 |
| $[FR_2^{4,1,0}]_{3,4,1}$ | ✗ 🌐 🎗 ⊚ | 1.618 | 0.951 − 0.309$i$ | −0.618 |
| $[FR_2^{4,1,0}]_{3,4,2}$ | ✗ 🌐 🎗 ⊚ | 1.618 | 0.951 − 0.309$i$ | 0.618 |
| $[FR_2^{4,1,0}]_{4,1,1}$ | ✗ 🌐 🎗 | −1.618 | 0.309 − 0.951$i$ | −0.618 |
| $[FR_2^{4,1,0}]_{4,1,2}$ | ✗ 🌐 🎗 | −1.618 | 0.309 − 0.951$i$ | 0.618 |
| $[FR_2^{4,1,0}]_{4,2,1}$ | ✗ 🌐 🎗 | −1.618 | 0.309 + 0.951$i$ | −0.618 |
| $[FR_2^{4,1,0}]_{4,2,2}$ | ✗ 🌐 🎗 | −1.618 | 0.309 + 0.951$i$ | 0.618 |
| $[FR_2^{4,1,0}]_{4,3,1}$ | ✗ 🌐 🎗 | −1.618 | −0.309 + 0.951$i$ | −0.618 |
| $[FR_2^{4,1,0}]_{4,3,2}$ | ✗ 🌐 🎗 | −1.618 | −0.309 + 0.951$i$ | 0.618 |
| $[FR_2^{4,1,0}]_{4,4,1}$ | ✗ 🌐 🎗 | −1.618 | −0.309 − 0.951$i$ | −0.618 |
| $[FR_2^{4,1,0}]_{4,4,2}$ | ✗ 🌐 🎗 | −1.618 | −0.309 − 0.951$i$ | 0.618 |

### 4.5.10 $FR_3^{4,1,0}$: $\mathbf{Rep}(D_5)$

For the fusion ring, the following multiplication table is used.

| 1 | 2 | 3         | 4         |
|---|---|-----------|-----------|
| 2 | 1 | 3         | 4         |
| 3 | 3 | 1 + 2 + 4 | 3 + 4     |
| 4 | 4 | 3 + 4     | 1 + 2 + 3 |

The following is the group of all non-trivial permutations that leave the fusion rules invariant:

$$S = \{(), (3\ 4)\}.$$

Let

$$X_1 = S\left(\frac{[F_1^{434}]_4^4}{[F_4^{333}]_4^4}\right), \tag{34}$$

$$X_2 = S\left(R_1^{33}\right). \tag{35}$$

The following table lists a small set of invariants whose values completely distinguish between all MFPBFCs and MFPNBFCs with the given fusion rules.

Table 20: Symbolic invariants

| Name | Properties | $X_1$ | $X_2$ |
|---|---|---|---|
| $[FR_3^{4,1,0}]_{1,1,1}$ | 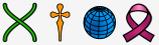 | {1} | $\{\zeta_5^4, \zeta_5\}$ |
| $[FR_3^{4,1,0}]_{1,2,1}$ | 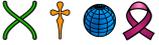 | {1} | $\{\zeta_5^3, \zeta_5^2\}$ |
| $[FR_3^{4,1,0}]_{1,3,1}$ | 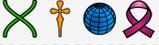 | {1} | {1} |
| $[FR_3^{4,1,0}]_{2,0,1}$ | 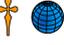 | $\{\zeta_5^4, \zeta_5\}$ | - |
| $[FR_3^{4,1,0}]_{3,0,1}$ | 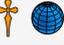 | $\{\zeta_5^3, \zeta_5^2\}$ | - |

Table 21: Numeric invariants

| Name | Properties | $X_1$ | $X_2$ |
|---|---|---|---|
| $[FR_3^{4,1,0}]_{1,1,1}$ | 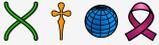 | {1} | $\{0.309 - 0.951i,\ 0.309 + 0.951i\}$ |
| $[FR_3^{4,1,0}]_{1,2,1}$ | 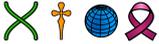 | {1} | $\{-0.809 - 0.588i,\ -0.809 + 0.588i\}$ |
| $[FR_3^{4,1,0}]_{1,3,1}$ | 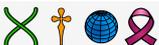 | {1} | {1} |
| $[FR_3^{4,1,0}]_{2,0,1}$ | 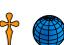 | $\{0.309 - 0.951i,\ 0.309 + 0.951i\}$ | - |
| $[FR_3^{4,1,0}]_{3,0,1}$ | 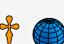 | $\{-0.809 - 0.588i,\ -0.809 + 0.588i\}$ | - |

### 4.5.11 $FR_4^{4,1,0}$: $PSU(2)_6$

For the fusion ring, the following multiplication table is used.

| 1 | 2 | 3 | 4 |
|---|---|---|---|
| 2 | 1 | 4 | 3 |
| 3 | 4 | 1+3+4 | 2+3+4 |
| 4 | 3 | 2+3+4 | 1+3+4 |

The following is the group of all non-trivial permutations that leave the fusion rules invariant:

$$S = \{(), (3\ 4)\}.$$

Let

$$X_1 = S\left([F_3^{333}]_3^3\right). \tag{36}$$

The following table lists a small set of invariants whose values completely distinguish between all MFPBFCs and MFPNBFCs with the given fusion rules.

Table 22: Symbolic invariants

| Name | Properties | $X_1$ |
|---|---|---|
| $[FR_4^{4,1,0}]_{1,1,1}$ | 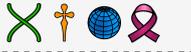 | $\left\{\frac{1}{2}\left(2-\sqrt{2}\right)\right\}$ |
| $[FR_4^{4,1,0}]_{2,1,1}$ | 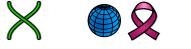 | $\left\{\frac{1}{2}\left(2+\sqrt{2}\right)\right\}$ |

Table 23: Numeric invariants

| Name | Properties | $X_1$ |
|---|---|---|
| $[FR_4^{4,1,0}]_{1,1,1}$ | 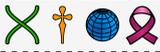 | $\{0.293\}$ |
| $[FR_4^{4,1,0}]_{2,1,1}$ | 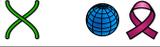 | $\{1.707\}$ |

### 4.5.12 $FR_5^{4,1,0}$: Fib ⊗ Fib

For the fusion ring, the following multiplication table is used.

| 1 | 2 | 3 | 4 |
|---|---|---|---|
| 2 | 1+2 | 4 | 3+4 |
| 3 | 4 | 1+3 | 2+4 |
| 4 | 3+4 | 2+4 | 1+2+3+4 |

The following is the group of all non-trivial permutations that leave the fusion rules invariant:

$$S = \{(), (2\ 3)\}.$$

Let

$$X_1 = S\left([F_2^{222}]_2^2\right). \tag{37}$$

The following table lists a small set of invariants whose values completely distinguish between all MFPBFCs and MFPNBFCs with the given fusion rules.

Table 24: Symbolic invariants

| Name | Properties | $X_1$ | $R_1^{44}$ |
|---|---|---|---|
| $[\text{FR}_5^{4,1,0}]_{1,1,1}$ | ✕ † 🌐 🎗 ⭕ | $\{\overline{\phi}\}$ | $\zeta_5$ |
| $[\text{FR}_5^{4,1,0}]_{1,2,1}$ | ✕ † 🌐 🎗 ⭕ | $\{\overline{\phi}\}$ | $1$ |
| $[\text{FR}_5^{4,1,0}]_{1,3,1}$ | ✕ † 🌐 🎗 ⭕ | $\{\overline{\phi}\}$ | $\zeta_5^4$ |
| $[\text{FR}_5^{4,1,0}]_{2,1,1}$ | ✕ 🌐 🎗 ⭕ | $\{\overline{\phi}, \phi\}$ | $\zeta_5^2$ |
| $[\text{FR}_5^{4,1,0}]_{2,2,1}$ | ✕ 🌐 🎗 ⭕ | $\{\overline{\phi}, \phi\}$ | $\zeta_5^4$ |
| $[\text{FR}_5^{4,1,0}]_{2,3,1}$ | ✕ 🌐 🎗 ⭕ | $\{\overline{\phi}, \phi\}$ | $\zeta_5$ |
| $[\text{FR}_5^{4,1,0}]_{2,4,1}$ | ✕ 🌐 🎗 ⭕ | $\{\overline{\phi}, \phi\}$ | $\zeta_5^3$ |
| $[\text{FR}_5^{4,1,0}]_{3,1,1}$ | ✕ 🌐 🎗 ⭕ | $\{\phi\}$ | $\zeta_5^3$ |
| $[\text{FR}_5^{4,1,0}]_{3,2,1}$ | ✕ 🌐 🎗 ⭕ | $\{\phi\}$ | $1$ |
| $[\text{FR}_5^{4,1,0}]_{3,3,1}$ | ✕ 🌐 🎗 ⭕ | $\{\phi\}$ | $\zeta_5^2$ |

Table 25: Numeric invariants

| Name | Properties | $X_1$ | $R_1^{44}$ |
|---|---|---|---|
| $[\text{FR}_5^{4,1,0}]_{1,1,1}$ | ✕ † 🌐 🎗 ⭕ | $\{-0.618\}$ | $0.309 + 0.951i$ |
| $[\text{FR}_5^{4,1,0}]_{1,2,1}$ | ✕ † 🌐 🎗 ⭕ | $\{-0.618\}$ | $1$ |
| $[\text{FR}_5^{4,1,0}]_{1,3,1}$ | ✕ † 🌐 🎗 ⭕ | $\{-0.618\}$ | $0.309 - 0.951i$ |
| $[\text{FR}_5^{4,1,0}]_{2,1,1}$ | ✕ 🌐 🎗 ⭕ | $\{-0.618, 1.618\}$ | $-0.809 + 0.588i$ |
| $[\text{FR}_5^{4,1,0}]_{2,2,1}$ | ✕ 🌐 🎗 ⭕ | $\{-0.618, 1.618\}$ | $0.309 - 0.951i$ |
| $[\text{FR}_5^{4,1,0}]_{2,3,1}$ | ✕ 🌐 🎗 ⭕ | $\{-0.618, 1.618\}$ | $0.309 + 0.951i$ |
| $[\text{FR}_5^{4,1,0}]_{2,4,1}$ | ✕ 🌐 🎗 ⭕ | $\{-0.618, 1.618\}$ | $-0.809 - 0.588i$ |
| $[\text{FR}_5^{4,1,0}]_{3,1,1}$ | ✕ 🌐 🎗 ⭕ | $\{1.618\}$ | $-0.809 - 0.588i$ |
| $[\text{FR}_5^{4,1,0}]_{3,2,1}$ | ✕ 🌐 🎗 ⭕ | $\{1.618\}$ | $1$ |
| $[\text{FR}_5^{4,1,0}]_{3,3,1}$ | ✕ 🌐 🎗 ⭕ | $\{1.618\}$ | $-0.809 + 0.588i$ |

### 4.5.13 $\text{FR}_6^{4,1,0}$: $\text{PSU}(2)_7$

For the fusion ring, the following multiplication table is used.

| | 1 | 2 | 3 | 4 |
|---|---|---|---|---|
| 2 | | $1+4$ | $3+4$ | $2+3$ |
| 3 | | $3+4$ | $1+2+3+4$ | $2+3+4$ |
| 4 | | $2+3$ | $2+3+4$ | $1+3+4$ |

Only the trivial permutation leaves the fusion rules invariant.

The following table lists a small set of invariants whose values completely distinguish between all MFPBFCs and MFPNBFCs with the given fusion rules.

Table 26: Symbolic invariants

| Name | Properties | $[F_3^{333}]_3^3$ | $R_1^{33}$ |
|---|---|---|---|
| $[\text{FR}_6^{4,1,0}]_{1,1,1}$ | ✕ † 🌐 🎗 ⦾ | $-\zeta_9^4 + \zeta_9^2 - \zeta_9$ | $\zeta_9^2$ |
| $[\text{FR}_6^{4,1,0}]_{1,2,1}$ | ✕ † 🌐 🎗 ⦾ | $-\zeta_9^4 + \zeta_9^2 - \zeta_9$ | $\zeta_9^7$ |
| $[\text{FR}_6^{4,1,0}]_{2,1,1}$ | ✕ 🌐 🎗 ⦾ | $-\zeta_9^5 - \zeta_9^2 + \zeta_9$ | $\zeta_9$ |
| $[\text{FR}_6^{4,1,0}]_{2,2,1}$ | ✕ 🌐 🎗 ⦾ | $-\zeta_9^5 - \zeta_9^2 + \zeta_9$ | $\zeta_9^8$ |
| $[\text{FR}_6^{4,1,0}]_{3,1,1}$ | ✕ 🌐 🎗 ⦾ | $\zeta_9^5 + \zeta_9^4$ | $\zeta_9^4$ |
| $[\text{FR}_6^{4,1,0}]_{3,2,1}$ | ✕ 🌐 🎗 ⦾ | $\zeta_9^5 + \zeta_9^4$ | $\zeta_9^5$ |

Table 27: Numeric invariants

| Name | Properties | $[F_3^{333}]_3^3$ | $R_1^{33}$ |
|---|---|---|---|
| $[\text{FR}_6^{4,1,0}]_{1,1,1}$ | ✕ † 🌐 🎗 ⦾ | 0.347 | $0.174 + 0.985i$ |
| $[\text{FR}_6^{4,1,0}]_{1,2,1}$ | ✕ † 🌐 🎗 ⦾ | 0.347 | $0.174 - 0.985i$ |
| $[\text{FR}_6^{4,1,0}]_{2,1,1}$ | ✕ 🌐 🎗 ⦾ | 1.532 | $0.766 + 0.643i$ |
| $[\text{FR}_6^{4,1,0}]_{2,2,1}$ | ✕ 🌐 🎗 ⦾ | 1.532 | $0.766 - 0.643i$ |
| $[\text{FR}_6^{4,1,0}]_{3,1,1}$ | ✕ 🌐 🎗 ⦾ | $-1.879$ | $-0.940 + 0.342i$ |
| $[\text{FR}_6^{4,1,0}]_{3,2,1}$ | ✕ 🌐 🎗 ⦾ | $-1.879$ | $-0.940 - 0.342i$ |

### 4.5.14 $\text{FR}_1^{4,1,2}$: $\mathbb{Z}_4$

For the fusion ring, the following multiplication table is used.

| 1 | 2 | 3 | 4 |
|---|---|---|---|
| 2 | 1 | 4 | 3 |
| 3 | 4 | 2 | 1 |
| 4 | 3 | 1 | 2 |

The following is the group of all non-trivial permutations that leave the fusion rules invariant:

$$S = \{(), (3\ 4)\}.$$

Let

$$X_1 = S\left(\frac{[F_3^{232}]_4^4 [F_1^{323}]_4^4 [F_2^{342}]_3^1}{[F_3^{223}]_4^1 [F_1^{233}]_2^4}\right), \tag{38}$$

$$X_2 = S\left(R_2^{33}\right), \tag{39}$$

$$X_3 = \left(d_3^L\right). \tag{40}$$

The following table lists a small set of invariants whose values completely distinguish between all MFPBFCs and MFPNBFCs with the given fusion rules.

Table 28: Symbolic invariants

| Name | Properties | $X_1$ | $X_2$ | $X_3$ |
|---|---|---|---|---|
| $[FR_1^{4,1,2}]_{1,1,1}$ | ✂ † 🌐 🎗 ⚭ | $\{-1\}$ | $\{\zeta_8^3\}$ | $\{1\}$ |
| $[FR_1^{4,1,2}]_{1,1,2}$ | ✂ 🌐 🎗 ⚭ | $\{-1\}$ | $\{\zeta_8^3\}$ | $\{-1\}$ |
| $[FR_1^{4,1,2}]_{1,1,3}$ | ✂ | $\{-1\}$ | $\{\zeta_8^3\}$ | $\{-i, i\}$ |
| $[FR_1^{4,1,2}]_{1,2,1}$ | ✂ † 🌐 🎗 ⚭ | $\{-1\}$ | $\{\zeta_8^5\}$ | $\{1\}$ |
| $[FR_1^{4,1,2}]_{1,2,2}$ | ✂ 🌐 🎗 ⚭ | $\{-1\}$ | $\{\zeta_8^5\}$ | $\{-1\}$ |
| $[FR_1^{4,1,2}]_{1,2,3}$ | ✂ | $\{-1\}$ | $\{\zeta_8^5\}$ | $\{-i, i\}$ |
| $[FR_1^{4,1,2}]_{1,3,1}$ | ✂ † 🌐 🎗 ⚭ | $\{-1\}$ | $\{\zeta_8\}$ | $\{1\}$ |
| $[FR_1^{4,1,2}]_{1,3,2}$ | ✂ 🌐 🎗 ⚭ | $\{-1\}$ | $\{\zeta_8\}$ | $\{-1\}$ |
| $[FR_1^{4,1,2}]_{1,3,3}$ | ✂ | $\{-1\}$ | $\{\zeta_8\}$ | $\{-i, i\}$ |
| $[FR_1^{4,1,2}]_{1,4,1}$ | ✂ † 🌐 🎗 ⚭ | $\{-1\}$ | $\{\zeta_8^7\}$ | $\{1\}$ |
| $[FR_1^{4,1,2}]_{1,4,2}$ | ✂ 🌐 🎗 ⚭ | $\{-1\}$ | $\{\zeta_8^7\}$ | $\{-1\}$ |
| $[FR_1^{4,1,2}]_{1,4,3}$ | ✂ | $\{-1\}$ | $\{\zeta_8^7\}$ | $\{-i, i\}$ |
| $[FR_1^{4,1,2}]_{2,1,1}$ | ✂ † 🌐 🎗 | $\{1\}$ | $\{1\}$ | $\{1\}$ |
| $[FR_1^{4,1,2}]_{2,1,2}$ | ✂ 🌐 🎗 | $\{1\}$ | $\{1\}$ | $\{-1\}$ |
| $[FR_1^{4,1,2}]_{2,1,3}$ | ✂ | $\{1\}$ | $\{1\}$ | $\{-i, i\}$ |
| $[FR_1^{4,1,2}]_{2,2,1}$ | ✂ † 🌐 🎗 | $\{1\}$ | $\{-1\}$ | $\{1\}$ |
| $[FR_1^{4,1,2}]_{2,2,2}$ | ✂ 🌐 🎗 | $\{1\}$ | $\{-1\}$ | $\{-1\}$ |
| $[FR_1^{4,1,2}]_{2,2,3}$ | ✂ | $\{1\}$ | $\{-1\}$ | $\{-i, i\}$ |
| $[FR_1^{4,1,2}]_{2,3,1}$ | ✂ † 🌐 🎗 | $\{1\}$ | $\{i\}$ | $\{1\}$ |
| $[FR_1^{4,1,2}]_{2,3,2}$ | ✂ 🌐 🎗 | $\{1\}$ | $\{i\}$ | $\{-1\}$ |
| $[FR_1^{4,1,2}]_{2,3,3}$ | ✂ | $\{1\}$ | $\{i\}$ | $\{-i, i\}$ |
| $[FR_1^{4,1,2}]_{2,4,1}$ | ✂ † 🌐 🎗 | $\{1\}$ | $\{-i\}$ | $\{1\}$ |
| $[FR_1^{4,1,2}]_{2,4,2}$ | ✂ 🌐 🎗 | $\{1\}$ | $\{-i\}$ | $\{-1\}$ |
| $[FR_1^{4,1,2}]_{2,4,3}$ | ✂ | $\{1\}$ | $\{-i\}$ | $\{-i, i\}$ |
| $[FR_1^{4,1,2}]_{3,0,1}$ | † 🌐 | $\{-i\}$ | - | $\{1\}$ |
| $[FR_1^{4,1,2}]_{3,0,2}$ | 🌐 | $\{-i\}$ | - | $\{-1\}$ |
| $[FR_1^{4,1,2}]_{3,0,3}$ |  | $\{-i\}$ | - | $\{-i, i\}$ |
| $[FR_1^{4,1,2}]_{4,0,1}$ | † 🌐 | $\{i\}$ | - | $\{1\}$ |
| $[FR_1^{4,1,2}]_{4,0,2}$ | 🌐 | $\{i\}$ | - | $\{-1\}$ |
| $[FR_1^{4,1,2}]_{4,0,3}$ |  | $\{i\}$ | - | $\{-i, i\}$ |

Table 29: Numeric invariants

| Name | Properties | $X_1$ | $X_2$ | $X_3$ |
|---|---|---|---|---|
| $[FR_1^{4,1,2}]_{1,1,1}$ | ✕ † 🌐 🎗 ⚭ | $\{-1\}$ | $\{-0.707 + 0.707i\}$ | $\{1\}$ |
| $[FR_1^{4,1,2}]_{1,1,2}$ | ✕ 🌐 🎗 ⚭ | $\{-1\}$ | $\{-0.707 + 0.707i\}$ | $\{-1\}$ |
| $[FR_1^{4,1,2}]_{1,1,3}$ | ✕ | $\{-1\}$ | $\{-0.707 + 0.707i\}$ | $\{-i, i\}$ |
| $[FR_1^{4,1,2}]_{1,2,1}$ | ✕ † 🌐 🎗 ⚭ | $\{-1\}$ | $\{-0.707 - 0.707i\}$ | $\{1\}$ |
| $[FR_1^{4,1,2}]_{1,2,2}$ | ✕ 🌐 🎗 ⚭ | $\{-1\}$ | $\{-0.707 - 0.707i\}$ | $\{-1\}$ |
| $[FR_1^{4,1,2}]_{1,2,3}$ | ✕ | $\{-1\}$ | $\{-0.707 - 0.707i\}$ | $\{-i, i\}$ |
| $[FR_1^{4,1,2}]_{1,3,1}$ | ✕ † 🌐 🎗 ⚭ | $\{-1\}$ | $\{0.707 + 0.707i\}$ | $\{1\}$ |
| $[FR_1^{4,1,2}]_{1,3,2}$ | ✕ 🌐 🎗 ⚭ | $\{-1\}$ | $\{0.707 + 0.707i\}$ | $\{-1\}$ |
| $[FR_1^{4,1,2}]_{1,3,3}$ | ✕ | $\{-1\}$ | $\{0.707 + 0.707i\}$ | $\{-i, i\}$ |
| $[FR_1^{4,1,2}]_{1,4,1}$ | ✕ † 🌐 🎗 ⚭ | $\{-1\}$ | $\{0.707 - 0.707i\}$ | $\{1\}$ |
| $[FR_1^{4,1,2}]_{1,4,2}$ | ✕ 🌐 🎗 ⚭ | $\{-1\}$ | $\{0.707 - 0.707i\}$ | $\{-1\}$ |
| $[FR_1^{4,1,2}]_{1,4,3}$ | ✕ | $\{-1\}$ | $\{0.707 - 0.707i\}$ | $\{-i, i\}$ |
| $[FR_1^{4,1,2}]_{2,1,1}$ | ✕ † 🌐 🎗 | $\{1\}$ | $\{1\}$ | $\{1\}$ |
| $[FR_1^{4,1,2}]_{2,1,2}$ | ✕ 🌐 🎗 | $\{1\}$ | $\{1\}$ | $\{-1\}$ |
| $[FR_1^{4,1,2}]_{2,1,3}$ | ✕ | $\{1\}$ | $\{1\}$ | $\{-i, i\}$ |
| $[FR_1^{4,1,2}]_{2,2,1}$ | ✕ † 🌐 🎗 | $\{1\}$ | $\{-1\}$ | $\{1\}$ |
| $[FR_1^{4,1,2}]_{2,2,2}$ | ✕ 🌐 🎗 | $\{1\}$ | $\{-1\}$ | $\{-1\}$ |
| $[FR_1^{4,1,2}]_{2,2,3}$ | ✕ | $\{1\}$ | $\{-1\}$ | $\{-i, i\}$ |
| $[FR_1^{4,1,2}]_{2,3,1}$ | ✕ † 🌐 🎗 | $\{1\}$ | $\{i\}$ | $\{1\}$ |
| $[FR_1^{4,1,2}]_{2,3,2}$ | ✕ 🌐 🎗 | $\{1\}$ | $\{i\}$ | $\{-1\}$ |
| $[FR_1^{4,1,2}]_{2,3,3}$ | ✕ | $\{1\}$ | $\{i\}$ | $\{-i, i\}$ |
| $[FR_1^{4,1,2}]_{2,4,1}$ | ✕ † 🌐 🎗 | $\{1\}$ | $\{-i\}$ | $\{1\}$ |
| $[FR_1^{4,1,2}]_{2,4,2}$ | ✕ 🌐 🎗 | $\{1\}$ | $\{-i\}$ | $\{-1\}$ |
| $[FR_1^{4,1,2}]_{2,4,3}$ | ✕ | $\{1\}$ | $\{-i\}$ | $\{-i, i\}$ |
| $[FR_1^{4,1,2}]_{3,0,1}$ | † 🌐 | $\{-i\}$ | - | $\{1\}$ |
| $[FR_1^{4,1,2}]_{3,0,2}$ | 🌐 | $\{-i\}$ | - | $\{-1\}$ |
| $[FR_1^{4,1,2}]_{3,0,3}$ |  | $\{-i\}$ | - | $\{-i, i\}$ |
| $[FR_1^{4,1,2}]_{4,0,1}$ | † 🌐 | $\{i\}$ | - | $\{1\}$ |
| $[FR_1^{4,1,2}]_{4,0,2}$ | 🌐 | $\{i\}$ | - | $\{-1\}$ |
| $[FR_1^{4,1,2}]_{4,0,3}$ |  | $\{i\}$ | - | $\{-i, i\}$ |

## 4.5.15 $FR_2^{4,1,2}$: $TY(\mathbb{Z}_3)$

| 1 | 2 | 3 | 4 |
| --- | --- | --- | --- |
| 2 | 3 | 1 | 4 |
| 3 | 1 | 2 | 4 |
| 4 | 4 | 4 | 1+2+3 |

The following is the group of all non-trivial permutations that leave the fusion rules invariant:

$$S = \{(), (2\ 3)\}.$$

Let

$$X_1 = S\left([F_4^{242}]_4^4\right), \tag{41}$$

$$X_2 = S\left([F_1^{424}]_4^4 [F_4^{444}]_2^2\right). \tag{42}$$

The following table lists a small set of invariants whose values completely distinguish between all MFPBFCs and MFPNBFCs with the given fusion rules.

Table 30: Symbolic invariants

| Name | Properties | $X_1$ | $X_2$ | $d_4^L$ |
| --- | --- | --- | --- | --- |
| $[FR_2^{4,1,2}]_{1,0,1}$ |   | $\{\zeta_3^2\}$ | $\{\frac{\zeta_3}{\sqrt{3}}\}$ | $\sqrt{3}$ |
| $[FR_2^{4,1,2}]_{1,0,2}$ |  | $\{\zeta_3^2\}$ | $\{\frac{\zeta_3}{\sqrt{3}}\}$ | $-\sqrt{3}$ |
| $[FR_2^{4,1,2}]_{2,0,1}$ |   | $\{\zeta_3^2\}$ | $\{\frac{\zeta_6^5}{\sqrt{3}}\}$ | $\sqrt{3}$ |
| $[FR_2^{4,1,2}]_{2,0,2}$ |  | $\{\zeta_3^2\}$ | $\{\frac{\zeta_6^5}{\sqrt{3}}\}$ | $-\sqrt{3}$ |
| $[FR_2^{4,1,2}]_{3,0,1}$ |   | $\{\zeta_3\}$ | $\{\frac{\zeta_6}{\sqrt{3}}\}$ | $\sqrt{3}$ |
| $[FR_2^{4,1,2}]_{3,0,2}$ |  | $\{\zeta_3\}$ | $\{\frac{\zeta_6}{\sqrt{3}}\}$ | $-\sqrt{3}$ |
| $[FR_2^{4,1,2}]_{4,0,1}$ |   | $\{\zeta_3\}$ | $\{\frac{\zeta_3^2}{\sqrt{3}}\}$ | $\sqrt{3}$ |
| $[FR_2^{4,1,2}]_{4,0,2}$ |  | $\{\zeta_3\}$ | $\{\frac{\zeta_3^2}{\sqrt{3}}\}$ | $-\sqrt{3}$ |

Table 31: Numeric invariants

| Name | Properties | $X_1$ | $X_2$ | $d_4^L$ |
| --- | --- | --- | --- | --- |
| $[FR_2^{4,1,2}]_{1,0,1}$ |   | $\{-0.5 - 0.866i\}$ | $\{-0.289 + 0.5i\}$ | $1.732$ |
| $[FR_2^{4,1,2}]_{1,0,2}$ |  | $\{-0.5 - 0.866i\}$ | $\{-0.289 + 0.5i\}$ | $-1.732$ |
| $[FR_2^{4,1,2}]_{2,0,1}$ |   | $\{-0.5 - 0.866i\}$ | $\{0.289 - 0.5i\}$ | $1.732$ |
| $[FR_2^{4,1,2}]_{2,0,2}$ |  | $\{-0.5 - 0.866i\}$ | $\{0.289 - 0.5i\}$ | $-1.732$ |
| $[FR_2^{4,1,2}]_{3,0,1}$ |   | $\{-0.5 + 0.866i\}$ | $\{0.289 + 0.5i\}$ | $1.732$ |

Table 31: Numeric invariants (Continued)

| Name | Properties | $X_1$ | $X_2$ | $d_4^L$ |
|---|---|---|---|---|
| $[FR_2^{4,1,2}]_{3,0,2}$ |  | $\{-0.5 + 0.866i\}$ | $\{0.289 + 0.5i\}$ | $-1.732$ |
| $[FR_2^{4,1,2}]_{4,0,1}$ |  | $\{-0.5 + 0.866i\}$ | $\{-0.289 - 0.5i\}$ | $1.732$ |
| $[FR_2^{4,1,2}]_{4,0,2}$ |  | $\{-0.5 + 0.866i\}$ | $\{-0.289 - 0.5i\}$ | $-1.732$ |

### 4.5.16 $FR_4^{4,1,2}$: Pseudo $PSU(2)_6$

For the fusion ring, the following multiplication table is used.

| | | | |
|---|---|---|---|
| **1** | **2** | **3** | **4** |
| **2** | **1** | **4** | **3** |
| **3** | **4** | $2+3+4$ | $1+3+4$ |
| **4** | **3** | $1+3+4$ | $2+3+4$ |

The following is the group of all non-trivial permutations that leave the fusion rules invariant:

$$S = \{(), (3\ 4)\}.$$

Let

$$X_1 = S\left([F_2^{333}]_3^3\right). \tag{43}$$

The following table lists a small set of invariants whose values completely distinguish between all MFPBFCs and MFPNBFCs with the given fusion rules.

Table 32: Symbolic invariants

| Name | Properties | $X_1$ |
|---|---|---|
| $[FR_4^{4,1,2}]_{1,0,1}$ |  | $\{\zeta_8^3\}$ |
| $[FR_4^{4,1,2}]_{2,0,1}$ |  | $\{\zeta_8^5\}$ |
| $[FR_4^{4,1,2}]_{3,0,1}$ |  | $\{\zeta_8\}$ |
| $[FR_4^{4,1,2}]_{4,0,1}$ |  | $\{\zeta_8^7\}$ |

Table 33: Numeric invariants

| Name | Properties | $X_1$ |
|---|---|---|
| $[FR_4^{4,1,2}]_{1,0,1}$ |  | $\{-0.707 - 0.707i\}$ |
| $[FR_4^{4,1,2}]_{2,0,1}$ |  | $\{-0.707 + 0.707i\}$ |
| $[FR_4^{4,1,2}]_{3,0,1}$ |  | $\{0.707 - 0.707i\}$ |
| $[FR_4^{4,1,2}]_{4,0,1}$ |  | $\{0.707 + 0.707i\}$ |

### 4.5.17 $FR_1^{5,1,0}$: **Rep**($D_4$)

For the fusion ring, the following multiplication table is used.

| 1 | 2 | 3 | 4 | 5 |
|---|---|---|---|---|
| 2 | 1 | 4 | 3 | 5 |
| 3 | 4 | 1 | 2 | 5 |
| 4 | 3 | 2 | 1 | 5 |
| 5 | 5 | 5 | 5 | $1+2+3+4$ |

The following is the group of all non-trivial permutations that leave the fusion rules invariant:

$$S = \{(), (2\ 3), (2\ 4), (3\ 4), (2\ 3\ 4), (2\ 4\ 3)\}.$$

Let

$$X_1 = S\left([F_5^{252}]_5^5\right), \tag{44}$$

$$X_2 = S\left([F_1^{535}]_5^5 [F_5^{555}]_3^3\right), \tag{45}$$

$$X_3 = S\left(R_1^{55}\right), \tag{46}$$

$$X_4 = S\left(d_5^L\right) \tag{47}$$

Table 34: Distinguishing invariants for categories with $FR_1^{5,1,0}$ fusion rules.

| Name | Properties | $X_1$ | $X_2$ | $R_1^{55}$ | $d_5^L$ |
|---|---|---|---|---|---|
| $[FR_1^{5,1,0}]_{1,1,1}$ | ✕ † ● ✄ | $\{-1, 1\}$ | $\left\{-\frac{1}{2}, \frac{1}{2}\right\}$ | $\zeta_8^7$ | 2 |
| $[FR_1^{5,1,0}]_{1,1,2}$ | ✕ ● ✄ | $\{-1, 1\}$ | $\left\{-\frac{1}{2}, \frac{1}{2}\right\}$ | $\zeta_8^7$ | $-2$ |
| $[FR_1^{5,1,0}]_{1,2,1}$ | ✕ † ● ✄ | $\{-1, 1\}$ | $\left\{-\frac{1}{2}, \frac{1}{2}\right\}$ | $\zeta_8^3$ | 2 |
| $[FR_1^{5,1,0}]_{1,2,2}$ | ✕ ● ✄ | $\{-1, 1\}$ | $\left\{-\frac{1}{2}, \frac{1}{2}\right\}$ | $\zeta_8^3$ | $-2$ |
| $[FR_1^{5,1,0}]_{1,3,1}$ | ✕ † ● ✄ | $\{-1, 1\}$ | $\left\{-\frac{1}{2}, \frac{1}{2}\right\}$ | 1 | 2 |
| $[FR_1^{5,1,0}]_{1,3,2}$ | ✕ ● ✄ | $\{-1, 1\}$ | $\left\{-\frac{1}{2}, \frac{1}{2}\right\}$ | 1 | $-2$ |
| $[FR_1^{5,1,0}]_{1,4,1}$ | ✕ † ● ✄ | $\{-1, 1\}$ | $\left\{-\frac{1}{2}, \frac{1}{2}\right\}$ | $-1$ | 2 |
| $[FR_1^{5,1,0}]_{1,4,2}$ | ✕ ● ✄ | $\{-1, 1\}$ | $\left\{-\frac{1}{2}, \frac{1}{2}\right\}$ | $-1$ | $-2$ |
| $[FR_1^{5,1,0}]_{2,1,1}$ | ✕ † ● ✄ | $\{-1, 1\}$ | $\left\{-\frac{1}{2}, \frac{1}{2}\right\}$ | $\zeta_8^5$ | 2 |
| $[FR_1^{5,1,0}]_{2,1,2}$ | ✕ ● ✄ | $\{-1, 1\}$ | $\left\{-\frac{1}{2}, \frac{1}{2}\right\}$ | $\zeta_8^5$ | $-2$ |
| $[FR_1^{5,1,0}]_{2,2,1}$ | ✕ † ● ✄ | $\{-1, 1\}$ | $\left\{-\frac{1}{2}, \frac{1}{2}\right\}$ | $\zeta_8$ | 2 |
| $[FR_1^{5,1,0}]_{2,2,2}$ | ✕ ● ✄ | $\{-1, 1\}$ | $\left\{-\frac{1}{2}, \frac{1}{2}\right\}$ | $\zeta_8$ | $-2$ |
| $[FR_1^{5,1,0}]_{2,3,1}$ | ✕ † ● ✄ | $\{1\}$ | $\left\{\frac{1}{2}\right\}$ | $-1$ | 2 |
| $[FR_1^{5,1,0}]_{2,3,2}$ | ✕ ● ✄ | $\{1\}$ | $\left\{\frac{1}{2}\right\}$ | $-1$ | $-2$ |
| $[FR_1^{5,1,0}]_{2,4,1}$ | ✕ † ● ✄ | $\{1\}$ | $\left\{\frac{1}{2}\right\}$ | 1 | 2 |

Table 34: Distinguishing invariants for categories with $FR_1^{5,1,0}$ fusion rules. (Continued)

| Name | Properties | $X_1$ | $X_2$ | $R_1^{55}$ | $d_5^L$ |
|---|---|---|---|---|---|
| $[FR_1^{5,1,0}]_{2,4,2}$ | ✗ 🌐 🎗 | $\{1\}$ | $\left\{\frac{1}{2}\right\}$ | $1$ | $-2$ |
| $[FR_1^{5,1,0}]_{3,1,1}$ | ✗ † 🌐 🎗 | $\{1\}$ | $\left\{\frac{1}{2}\right\}$ | $-i$ | $2$ |
| $[FR_1^{5,1,0}]_{3,1,2}$ | ✗ 🌐 🎗 | $\{1\}$ | $\left\{\frac{1}{2}\right\}$ | $-i$ | $-2$ |
| $[FR_1^{5,1,0}]_{3,2,1}$ | ✗ † 🌐 🎗 | $\{1\}$ | $\left\{\frac{1}{2}\right\}$ | $i$ | $2$ |
| $[FR_1^{5,1,0}]_{3,2,2}$ | ✗ 🌐 🎗 | $\{1\}$ | $\left\{\frac{1}{2}\right\}$ | $i$ | $-2$ |
| $[FR_1^{5,1,0}]_{3,3,1}$ | ✗ † 🌐 🎗 | $\{-1,1\}$ | $\left\{-\frac{1}{2},\frac{1}{2}\right\}$ | $\zeta_8$ | $2$ |
| $[FR_1^{5,1,0}]_{3,3,2}$ | ✗ 🌐 🎗 | $\{-1,1\}$ | $\left\{-\frac{1}{2},\frac{1}{2}\right\}$ | $\zeta_8$ | $-2$ |
| $[FR_1^{5,1,0}]_{3,4,1}$ | ✗ † 🌐 🎗 | $\{-1,1\}$ | $\left\{-\frac{1}{2},\frac{1}{2}\right\}$ | $\zeta_8^5$ | $2$ |
| $[FR_1^{5,1,0}]_{3,4,2}$ | ✗ 🌐 🎗 | $\{-1,1\}$ | $\left\{-\frac{1}{2},\frac{1}{2}\right\}$ | $\zeta_8^5$ | $-2$ |
| $[FR_1^{5,1,0}]_{3,5,1}$ | ✗ † 🌐 🎗 | $\{-1,1\}$ | $\left\{-\frac{1}{2},\frac{1}{2}\right\}$ | $-i$ | $2$ |
| $[FR_1^{5,1,0}]_{3,5,2}$ | ✗ 🌐 🎗 | $\{-1,1\}$ | $\left\{-\frac{1}{2},\frac{1}{2}\right\}$ | $-i$ | $-2$ |
| $[FR_1^{5,1,0}]_{3,6,1}$ | ✗ † 🌐 🎗 | $\{-1,1\}$ | $\left\{-\frac{1}{2},\frac{1}{2}\right\}$ | $i$ | $2$ |
| $[FR_1^{5,1,0}]_{3,6,2}$ | ✗ 🌐 🎗 | $\{-1,1\}$ | $\left\{-\frac{1}{2},\frac{1}{2}\right\}$ | $i$ | $-2$ |
| $[FR_1^{5,1,0}]_{4,1,1}$ | ✗ † 🌐 🎗 | $\{-1,1\}$ | $\left\{-\frac{1}{2},\frac{1}{2}\right\}$ | $\zeta_8^3$ | $2$ |
| $[FR_1^{5,1,0}]_{4,1,2}$ | ✗ 🌐 🎗 | $\{-1,1\}$ | $\left\{-\frac{1}{2},\frac{1}{2}\right\}$ | $\zeta_8^3$ | $-2$ |
| $[FR_1^{5,1,0}]_{4,2,1}$ | ✗ † 🌐 🎗 | $\{-1,1\}$ | $\left\{-\frac{1}{2},\frac{1}{2}\right\}$ | $\zeta_8^7$ | $2$ |
| $[FR_1^{5,1,0}]_{4,2,2}$ | ✗ 🌐 🎗 | $\{-1,1\}$ | $\left\{-\frac{1}{2},\frac{1}{2}\right\}$ | $\zeta_8^7$ | $-2$ |
| $[FR_1^{5,1,0}]_{4,3,1}$ | ✗ † 🌐 🎗 | $\{1\}$ | $\left\{-\frac{1}{2}\right\}$ | $-i$ | $2$ |
| $[FR_1^{5,1,0}]_{4,3,2}$ | ✗ 🌐 🎗 | $\{1\}$ | $\left\{-\frac{1}{2}\right\}$ | $-i$ | $-2$ |
| $[FR_1^{5,1,0}]_{4,4,1}$ | ✗ † 🌐 🎗 | $\{1\}$ | $\left\{-\frac{1}{2}\right\}$ | $i$ | $2$ |
| $[FR_1^{5,1,0}]_{4,4,2}$ | ✗ 🌐 🎗 | $\{1\}$ | $\left\{-\frac{1}{2}\right\}$ | $i$ | $-2$ |
| $[FR_1^{5,1,0}]_{4,5,1}$ | ✗ † 🌐 🎗 | $\{1\}$ | $\left\{-\frac{1}{2}\right\}$ | $-1$ | $2$ |
| $[FR_1^{5,1,0}]_{4,5,2}$ | ✗ 🌐 🎗 | $\{1\}$ | $\left\{-\frac{1}{2}\right\}$ | $-1$ | $-2$ |
| $[FR_1^{5,1,0}]_{4,6,1}$ | ✗ † 🌐 🎗 | $\{1\}$ | $\left\{-\frac{1}{2}\right\}$ | $1$ | $2$ |
| $[FR_1^{5,1,0}]_{4,6,2}$ | ✗ 🌐 🎗 | $\{1\}$ | $\left\{-\frac{1}{2}\right\}$ | $1$ | $-2$ |

Table 35: Numeric invariants

| Name | Properties | $X_1$ | $X_2$ | $R_1^{55}$ | $d_5^L$ |
|---|---|---|---|---|---|
| $[FR_1^{5,1,0}]_{1,1,1}$ | ✗ † 🌐 🎗 | $\{-1,1\}$ | $\{-0.5, 0.5\}$ | $0.707 - 0.707i$ | $2$ |
| $[FR_1^{5,1,0}]_{1,1,2}$ | ✗ 🌐 🎗 | $\{-1,1\}$ | $\{-0.5, 0.5\}$ | $0.707 - 0.707i$ | $-2$ |

Table 35: Numeric invariants (Continued)

| Name | Properties | $X_1$ | $X_2$ | $R_1^{55}$ | $d_5^L$ |
|---|---|---|---|---|---|
| $[FR_1^{5,1,0}]_{1,2,1}$ | ✖ ✝ 🌐 🎗 | $\{-1, 1\}$ | $\{-0.5, 0.5\}$ | $-0.707 + 0.707i$ | 2 |
| $[FR_1^{5,1,0}]_{1,2,2}$ | ✖ 🌐 🎗 | $\{-1, 1\}$ | $\{-0.5, 0.5\}$ | $-0.707 + 0.707i$ | $-2$ |
| $[FR_1^{5,1,0}]_{1,3,1}$ | ✖ ✝ 🌐 🎗 | $\{-1, 1\}$ | $\{-0.5, 0.5\}$ | $1$ | $2$ |
| $[FR_1^{5,1,0}]_{1,3,2}$ | ✖ 🌐 🎗 | $\{-1, 1\}$ | $\{-0.5, 0.5\}$ | $1$ | $-2$ |
| $[FR_1^{5,1,0}]_{1,4,1}$ | ✖ ✝ 🌐 🎗 | $\{-1, 1\}$ | $\{-0.5, 0.5\}$ | $-1$ | $2$ |
| $[FR_1^{5,1,0}]_{1,4,2}$ | ✖ 🌐 🎗 | $\{-1, 1\}$ | $\{-0.5, 0.5\}$ | $-1$ | $-2$ |
| $[FR_1^{5,1,0}]_{2,1,1}$ | ✖ ✝ 🌐 🎗 | $\{-1, 1\}$ | $\{-0.5, 0.5\}$ | $-0.707 - 0.707i$ | $2$ |
| $[FR_1^{5,1,0}]_{2,1,2}$ | ✖ 🌐 🎗 | $\{-1, 1\}$ | $\{-0.5, 0.5\}$ | $-0.707 - 0.707i$ | $-2$ |
| $[FR_1^{5,1,0}]_{2,2,1}$ | ✖ ✝ 🌐 🎗 | $\{-1, 1\}$ | $\{-0.5, 0.5\}$ | $0.707 + 0.707i$ | $2$ |
| $[FR_1^{5,1,0}]_{2,2,2}$ | ✖ 🌐 🎗 | $\{-1, 1\}$ | $\{-0.5, 0.5\}$ | $0.707 + 0.707i$ | $-2$ |
| $[FR_1^{5,1,0}]_{2,3,1}$ | ✖ ✝ 🌐 🎗 | $\{1\}$ | $\{0.5\}$ | $-1$ | $2$ |
| $[FR_1^{5,1,0}]_{2,3,2}$ | ✖ 🌐 🎗 | $\{1\}$ | $\{0.5\}$ | $-1$ | $-2$ |
| $[FR_1^{5,1,0}]_{2,4,1}$ | ✖ ✝ 🌐 🎗 | $\{1\}$ | $\{0.5\}$ | $1$ | $2$ |
| $[FR_1^{5,1,0}]_{2,4,2}$ | ✖ 🌐 🎗 | $\{1\}$ | $\{0.5\}$ | $1$ | $-2$ |
| $[FR_1^{5,1,0}]_{3,1,1}$ | ✖ ✝ 🌐 🎗 | $\{1\}$ | $\{0.5\}$ | $-i$ | $2$ |
| $[FR_1^{5,1,0}]_{3,1,2}$ | ✖ 🌐 🎗 | $\{1\}$ | $\{0.5\}$ | $-i$ | $-2$ |
| $[FR_1^{5,1,0}]_{3,2,1}$ | ✖ ✝ 🌐 🎗 | $\{1\}$ | $\{0.5\}$ | $i$ | $2$ |
| $[FR_1^{5,1,0}]_{3,2,2}$ | ✖ 🌐 🎗 | $\{1\}$ | $\{0.5\}$ | $i$ | $-2$ |
| $[FR_1^{5,1,0}]_{3,3,1}$ | ✖ ✝ 🌐 🎗 | $\{-1, 1\}$ | $\{-0.5, 0.5\}$ | $0.707 + 0.707i$ | $2$ |
| $[FR_1^{5,1,0}]_{3,3,2}$ | ✖ 🌐 🎗 | $\{-1, 1\}$ | $\{-0.5, 0.5\}$ | $0.707 + 0.707i$ | $-2$ |
| $[FR_1^{5,1,0}]_{3,4,1}$ | ✖ ✝ 🌐 🎗 | $\{-1, 1\}$ | $\{-0.5, 0.5\}$ | $-0.707 - 0.707i$ | $2$ |
| $[FR_1^{5,1,0}]_{3,4,2}$ | ✖ 🌐 🎗 | $\{-1, 1\}$ | $\{-0.5, 0.5\}$ | $-0.707 - 0.707i$ | $-2$ |
| $[FR_1^{5,1,0}]_{3,5,1}$ | ✖ ✝ 🌐 🎗 | $\{-1, 1\}$ | $\{-0.5, 0.5\}$ | $-i$ | $2$ |
| $[FR_1^{5,1,0}]_{3,5,2}$ | ✖ 🌐 🎗 | $\{-1, 1\}$ | $\{-0.5, 0.5\}$ | $-i$ | $-2$ |
| $[FR_1^{5,1,0}]_{3,6,1}$ | ✖ ✝ 🌐 🎗 | $\{-1, 1\}$ | $\{-0.5, 0.5\}$ | $i$ | $2$ |
| $[FR_1^{5,1,0}]_{3,6,2}$ | ✖ 🌐 🎗 | $\{-1, 1\}$ | $\{-0.5, 0.5\}$ | $i$ | $-2$ |
| $[FR_1^{5,1,0}]_{4,1,1}$ | ✖ ✝ 🌐 🎗 | $\{-1, 1\}$ | $\{-0.5, 0.5\}$ | $-0.707 + 0.707i$ | $2$ |
| $[FR_1^{5,1,0}]_{4,1,2}$ | ✖ 🌐 🎗 | $\{-1, 1\}$ | $\{-0.5, 0.5\}$ | $-0.707 + 0.707i$ | $-2$ |
| $[FR_1^{5,1,0}]_{4,2,1}$ | ✖ ✝ 🌐 🎗 | $\{-1, 1\}$ | $\{-0.5, 0.5\}$ | $0.707 - 0.707i$ | $2$ |
| $[FR_1^{5,1,0}]_{4,2,2}$ | ✖ 🌐 🎗 | $\{-1, 1\}$ | $\{-0.5, 0.5\}$ | $0.707 - 0.707i$ | $-2$ |
| $[FR_1^{5,1,0}]_{4,3,1}$ | ✖ ✝ 🌐 🎗 | $\{1\}$ | $\{-0.5\}$ | $-i$ | $2$ |
| $[FR_1^{5,1,0}]_{4,3,2}$ | ✖ 🌐 🎗 | $\{1\}$ | $\{-0.5\}$ | $-i$ | $-2$ |
| $[FR_1^{5,1,0}]_{4,4,1}$ | ✖ ✝ 🌐 🎗 | $\{1\}$ | $\{-0.5\}$ | $i$ | $2$ |

Table 35: Numeric invariants (Continued)

| Name | Properties | $X_1$ | $X_2$ | $R_1^{55}$ | $d_5^L$ |
|---|---|---|---|---|---|
| $[FR_1^{5,1,0}]_{4,4,2}$ | ✕ 🌐🎗 | $\{1\}$ | $\{-0.5\}$ | $i$ | $-2$ |
| $[FR_1^{5,1,0}]_{4,5,1}$ | ✕†🌐🎗 | $\{1\}$ | $\{-0.5\}$ | $-1$ | $2$ |
| $[FR_1^{5,1,0}]_{4,5,2}$ | ✕ 🌐🎗 | $\{1\}$ | $\{-0.5\}$ | $-1$ | $-2$ |
| $[FR_1^{5,1,0}]_{4,6,1}$ | ✕†🌐🎗 | $\{1\}$ | $\{-0.5\}$ | $1$ | $2$ |
| $[FR_1^{5,1,0}]_{4,6,2}$ | ✕ 🌐🎗 | $\{1\}$ | $\{-0.5\}$ | $1$ | $-2$ |

### 4.5.18 $FR_3^{5,1,0}$: $SU(2)_4$

For the fusion ring, the following multiplication table is used.

| 1 | 2 | 3 | 4 | 5 |
|---|---|---|---|---|
| 2 | 1 | 4 | 3 | 5 |
| 3 | 4 | 1+5 | 2+5 | 3+4 |
| 4 | 3 | 2+5 | 1+5 | 3+4 |
| 5 | 5 | 3+4 | 3+4 | 1+2+5 |

The following is the group of all non-trivial permutations that leave the fusion rules invariant:

$$S = \{(), (3\ 4)\}.$$

Let

$$X_1 = S\left([F_2^{323}]_4^4 [F_4^{343}]_2^2\right), \tag{48}$$

$$X_2 = S\left(d_3^L\right). \tag{49}$$

The following table lists a small set of invariants whose values completely distinguish between all MFPBFCs and MFPNBFCs with the given fusion rules.

Table 36: Symbolic invariants

| Name | Properties | $X_1$ | $R_1^{55}$ | $X_2$ |
|---|---|---|---|---|
| $[FR_3^{5,1,0}]_{1,1,1}$ | ✕†🌐🎗⊚ | $\{\frac{1}{\sqrt{3}}\}$ | $\zeta_3$ | $\{\sqrt{3}\}$ |
| $[FR_3^{5,1,0}]_{1,1,2}$ | ✕ 🌐🎗⊚ | $\{\frac{1}{\sqrt{3}}\}$ | $\zeta_3$ | $\{-\sqrt{3}\}$ |
| $[FR_3^{5,1,0}]_{1,2,1}$ | ✕†🌐🎗⊚ | $\{\frac{1}{\sqrt{3}}\}$ | $\zeta_3^2$ | $\{\sqrt{3}\}$ |
| $[FR_3^{5,1,0}]_{1,2,2}$ | ✕ 🌐🎗⊚ | $\{\frac{1}{\sqrt{3}}\}$ | $\zeta_3^2$ | $\{-\sqrt{3}\}$ |
| $[FR_3^{5,1,0}]_{2,1,1}$ | ✕†🌐🎗⊚ | $\{-\frac{1}{\sqrt{3}}\}$ | $\zeta_3$ | $\{\sqrt{3}\}$ |
| $[FR_3^{5,1,0}]_{2,1,2}$ | ✕ 🌐🎗⊚ | $\{-\frac{1}{\sqrt{3}}\}$ | $\zeta_3$ | $\{-\sqrt{3}\}$ |
| $[FR_3^{5,1,0}]_{2,2,1}$ | ✕†🌐🎗⊚ | $\{-\frac{1}{\sqrt{3}}\}$ | $\zeta_3^2$ | $\{\sqrt{3}\}$ |
| $[FR_3^{5,1,0}]_{2,2,2}$ | ✕ 🌐🎗⊚ | $\{-\frac{1}{\sqrt{3}}\}$ | $\zeta_3^2$ | $\{-\sqrt{3}\}$ |

Table 37: Numeric invariants

| Name | Properties | $X_1$ | $R_1^{55}$ | $X_2$ |
|---|---|---|---|---|
| $[FR_3^{5,1,0}]_{1,1,1}$ | 🧬 🗡 🌐 🎗 ⭕ | {0.577} | $-0.5 + 0.866i$ | {1.732} |
| $[FR_3^{5,1,0}]_{1,1,2}$ | 🧬 🌐 🎗 ⭕ | {0.577} | $-0.5 + 0.866i$ | {−1.732} |
| $[FR_3^{5,1,0}]_{1,2,1}$ | 🧬 🗡 🌐 🎗 ⭕ | {0.577} | $-0.5 - 0.866i$ | {1.732} |
| $[FR_3^{5,1,0}]_{1,2,2}$ | 🧬 🌐 🎗 ⭕ | {0.577} | $-0.5 - 0.866i$ | {−1.732} |
| $[FR_3^{5,1,0}]_{2,1,1}$ | 🧬 🗡 🌐 🎗 ⭕ | {−0.577} | $-0.5 + 0.866i$ | {1.732} |
| $[FR_3^{5,1,0}]_{2,1,2}$ | 🧬 🌐 🎗 ⭕ | {−0.577} | $-0.5 + 0.866i$ | {−1.732} |
| $[FR_3^{5,1,0}]_{2,2,1}$ | 🧬 🗡 🌐 🎗 ⭕ | {−0.577} | $-0.5 - 0.866i$ | {1.732} |
| $[FR_3^{5,1,0}]_{2,2,2}$ | 🧬 🌐 🎗 ⭕ | {−0.577} | $-0.5 - 0.866i$ | {−1.732} |

### 4.5.19  $FR_4^{5,1,0}$: $\mathbf{Rep}(D_7)$

For the fusion ring, the following multiplication table is used.

| | 1 | 2 | 3 | 4 | 5 |
|---|---|---|---|---|---|
| 1 | 2 | 3 | | 4 | 5 |
| 2 | 1 | 3 | | 4 | 5 |
| 3 | 3 | $1 + 2 + 5$ | $4 + 5$ | $3 + 4$ |
| 4 | 4 | $4 + 5$ | $1 + 2 + 3$ | $3 + 5$ |
| 5 | 5 | $3 + 4$ | $3 + 5$ | $1 + 2 + 4$ |

The following is the group of all non-trivial permutations that leave the fusion rules invariant:

$$S = \{(), (3\ 4\ 5), (3\ 5\ 4)\}.$$

Let

$$X_1 = S\left(\frac{[F_5^{334}]_4^5 [F_3^{345}]_5^4 [F_3^{454}]_5^5}{[F_3^{334}]_5^5}\right), \tag{50}$$

$$X_2 = S\left(R_1^{33}\right). \tag{51}$$

The following table lists a small set of invariants whose values completely distinguish between all MFPBFCs and MFPNBFCs with the given fusion rules.

Table 38: Symbolic invariants

| Name | Properties | $X_1$ | $X_2$ |
|---|---|---|---|
| $[FR_4^{5,1,0}]_{1,1,1}$ | 🧬 🗡 🌐 🎗 | {1} | {1} |
| $[FR_4^{5,1,0}]_{1,2,1}$ | 🧬 🗡 🌐 🎗 | {1} | $\{\zeta_7^3, \zeta_7^5, \zeta_7^6\}$ |
| $[FR_4^{5,1,0}]_{1,3,1}$ | 🧬 🗡 🌐 🎗 | {1} | $\{\zeta_7, \zeta_7^2, \zeta_7^4\}$ |
| $[FR_4^{5,1,0}]_{2,0,1}$ | 🗡 🌐 | $\{\zeta_7, \zeta_7^2, \zeta_7^4\}$ | - |
| $[FR_4^{5,1,0}]_{3,0,1}$ | 🗡 🌐 | $\{\zeta_7^3, \zeta_7^5, \zeta_7^6\}$ | - |

Table 39: Numeric invariants

| Name | Properties | $X_1$ | $X_2$ |
|---|---|---|---|
| $[FR_4^{5,1,0}]_{1,1,1}$ | ✕ ♱ 🌐 🎗 | $\{1.\}$ | $\{1\}$ |
| $[FR_4^{5,1,0}]_{1,2,1}$ | ✕ ♱ 🌐 🎗 | $\{1.\}$ | $\begin{Bmatrix} -0.901 + 0.434i, \\ -0.223 - 0.975i, \\ 0.623 - 0.782i \end{Bmatrix}$ |
| $[FR_4^{5,1,0}]_{1,3,1}$ | ✕ ♱ 🌐 🎗 | $\{1.\}$ | $\begin{Bmatrix} 0.623 + 0.782i, \\ -0.223 + 0.975i, \\ -0.901 - 0.434i \end{Bmatrix}$ |
| $[FR_4^{5,1,0}]_{2,0,1}$ | ♱ 🌐 | $\begin{Bmatrix} 0.623 + 0.782i, \\ -0.223 + 0.975i, \\ -0.901 - 0.434i \end{Bmatrix}$ | - |
| $[FR_4^{5,1,0}]_{3,0,1}$ | ♱ 🌐 | $\begin{Bmatrix} -0.901 + 0.434i, \\ -0.223 - 0.975i, \\ 0.623 - 0.782i \end{Bmatrix}$ | - |

### 4.5.20  $FR_6^{5,1,0}$: $\mathbf{Rep}(S_4)$

For the fusion ring, the following multiplication table is used.

| 1 | 2 | 3 | 4 | 5 |
|---|---|---|---|---|
| 2 | 1 | 3 | 5 | 4 |
| 3 | 3 | 1+2+3 | 4+5 | 4+5 |
| 4 | 5 | 4+5 | 1+3+4+5 | 2+3+4+5 |
| 5 | 4 | 4+5 | 2+3+4+5 | 1+3+4+5 |

The following is the group of all non-trivial permutations that leave the fusion rules invariant:

$$S = \{(), (4\ 5)\}.$$

Let

$$X_1 = S\left([F_4^{444}]_4^4\right). \tag{52}$$

The following table lists a small set of invariants whose values completely distinguish between all MFPBFCs and MFPNBFCs with the given fusion rules.

Table 40: Symbolic invariants

| Name | Properties | $X_1$ |
|---|---|---|
| $[FR_6^{5,1,0}]_{1,1,1}$ | ✕ ♱ 🌐 🎗 | $\{-\frac{1}{2}\}$ |
| $[FR_6^{5,1,0}]_{2,1,1}$ | ✕ ♱ 🌐 🎗 | $\{\frac{1}{2}\}$ |

### 4.5.21  $FR_7^{5,1,0}$: $\mathbf{PSU}(2)_8$

For the fusion ring, the following multiplication table is used.

|   |   |           |           |               |
|---|---|-----------|-----------|---------------|
| 1 | 2 | 3         | 4         | 5             |
| 2 | 1 | 4         | 3         | 5             |
| 3 | 4 | 1 + 4 + 5 | 2 + 3 + 5 | 3 + 4 + 5     |
| 4 | 3 | 2 + 3 + 5 | 1 + 4 + 5 | 3 + 4 + 5     |
| 5 | 5 | 3 + 4 + 5 | 3 + 4 + 5 | 1 + 2 + 3 + 4 + 5 |

Only the trivial permutation leaves the fusion rules invariant.

The following table lists a small set of invariants whose values completely distinguish between all MFPBFCs and MFPNBFCs with the given fusion rules.

Table 41: Symbolic invariants

| Name | Properties | $[F_5^{555}]_5^5$ | $R_1^{55}$ |
|---|---|---|---|
| $[FR_7^{5,1,0}]_{1,1,1}$ | 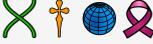 | $\overline{\phi}$ | $\zeta_5^2$ |
| $[FR_7^{5,1,0}]_{1,2,1}$ | 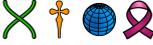 | $\overline{\phi}$ | $\zeta_5^3$ |
| $[FR_7^{5,1,0}]_{2,1,1}$ | 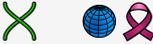 | $\phi$ | $\zeta_5^4$ |
| $[FR_7^{5,1,0}]_{2,2,1}$ | 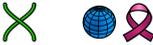 | $\phi$ | $\zeta_5$ |

Table 42: Numeric invariants

| Name | Properties | $[F_5^{555}]_5^5$ | $R_1^{55}$ |
|---|---|---|---|
| $[FR_7^{5,1,0}]_{1,1,1}$ | 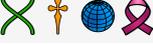 | $-0.618$ | $-0.809 + 0.588i$ |
| $[FR_7^{5,1,0}]_{1,2,1}$ | 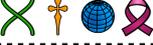 | $-0.618$ | $-0.809 - 0.588i$ |
| $[FR_7^{5,1,0}]_{2,1,1}$ | 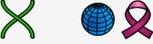 | $1.618$ | $0.309 - 0.951i$ |
| $[FR_7^{5,1,0}]_{2,2,1}$ | 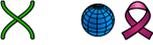 | $1.618$ | $0.309 + 0.951i$ |

### 4.5.22 $FR_{10}^{5,1,0}$: $PSU(2)_9$

For the fusion ring, the following multiplication table is used.

|   |   |   |   |   |
|---|---|---|---|---|
| 1 | 2 | 3 | 4 | 5 |
| 2 | 1 + 3 | 2 + 4 | 3 + 5 | 4 + 5 |
| 3 | 2 + 4 | 1 + 3 + 5 | 2 + 4 + 5 | 3 + 4 + 5 |
| 4 | 3 + 5 | 2 + 4 + 5 | 1 + 3 + 4 + 5 | 2 + 3 + 4 + 5 |
| 5 | 4 + 5 | 3 + 4 + 5 | 2 + 3 + 4 + 5 | 1 + 2 + 3 + 4 + 5 |

Only the trivial permutation leaves the fusion rules invariant.

The following table lists a small set of invariants whose values completely distinguish between all MFPBFCs and MFPNBFCs with the given fusion rules.

Table 43: Symbolic invariants

| Name | Properties | $[F_3^{333}]_3^3$ | $R_1^{33}$ |
|---|---|---|---|
| $[\mathrm{FR}_{10}^{5,1,0}]_{1,1,1}$ | ✗ ✝ 🌐 🎗 ⭕ | $\zeta_{11}^8 + \zeta_{11}^7 + \zeta_{11}^4 + \zeta_{11}^3 + 2$ | $\zeta_{11}^2$ |
| $[\mathrm{FR}_{10}^{5,1,0}]_{1,2,1}$ | ✗ ✝ 🌐 🎗 ⭕ | $\zeta_{11}^8 + \zeta_{11}^7 + \zeta_{11}^4 + \zeta_{11}^3 + 2$ | $\zeta_{11}^9$ |
| $[\mathrm{FR}_{10}^{5,1,0}]_{2,1,1}$ | ✗ 🌐 🎗 ⭕ | $\zeta_{11}^8 + \zeta_{11}^6 + \zeta_{11}^5 + \zeta_{11}^3 + 2$ | $\zeta_{11}^3$ |
| $[\mathrm{FR}_{10}^{5,1,0}]_{2,2,1}$ | ✗ 🌐 🎗 ⭕ | $\zeta_{11}^8 + \zeta_{11}^6 + \zeta_{11}^5 + \zeta_{11}^3 + 2$ | $\zeta_{11}^7$ |
| $[\mathrm{FR}_{10}^{5,1,0}]_{3,1,1}$ | ✗ 🌐 🎗 ⭕ | $-\zeta_{11}^9 - \zeta_{11}^8 - \zeta_{11}^7 - \zeta_{11}^4 - \zeta_{11}^3 - \zeta_{11}^2 + 1$ | $\zeta_{11}^8$ |
| $[\mathrm{FR}_{10}^{5,1,0}]_{3,2,1}$ | ✗ 🌐 🎗 ⭕ | $-\zeta_{11}^9 - \zeta_{11}^8 - \zeta_{11}^7 - \zeta_{11}^4 - \zeta_{11}^3 - \zeta_{11}^2 + 1$ | $\zeta_{11}^3$ |
| $[\mathrm{FR}_{10}^{5,1,0}]_{4,1,1}$ | ✗ 🌐 🎗 ⭕ | $-\zeta_{11}^8 - \zeta_{11}^7 - \zeta_{11}^6 - \zeta_{11}^5 - \zeta_{11}^4 - \zeta_{11}^3 + 1$ | $\zeta_{11}^5$ |
| $[\mathrm{FR}_{10}^{5,1,0}]_{4,2,1}$ | ✗ 🌐 🎗 ⭕ | $-\zeta_{11}^8 - \zeta_{11}^7 - \zeta_{11}^6 - \zeta_{11}^5 - \zeta_{11}^4 - \zeta_{11}^3 + 1$ | $\zeta_{11}^6$ |
| $[\mathrm{FR}_{10}^{5,1,0}]_{5,1,1}$ | ✗ 🌐 🎗 ⭕ | $\zeta_{11}^9 + \zeta_{11}^7 + \zeta_{11}^4 + \zeta_{11}^2 + 2$ | $\zeta_{11}^{10}$ |
| $[\mathrm{FR}_{10}^{5,1,0}]_{5,2,1}$ | ✗ 🌐 🎗 ⭕ | $\zeta_{11}^9 + \zeta_{11}^7 + \zeta_{11}^4 + \zeta_{11}^2 + 2$ | $\zeta_{11}$ |

Table 44: Numeric invariants

| Name | Properties | $[F_3^{333}]_3^3$ | $R_1^{33}$ |
|---|---|---|---|
| $[\mathrm{FR}_{10}^{5,1,0}]_{1,1,1}$ | ✗ ✝ 🌐 🎗 ⭕ | 0.406 | $0.415 + 0.910i$ |
| $[\mathrm{FR}_{10}^{5,1,0}]_{1,2,1}$ | ✗ ✝ 🌐 🎗 ⭕ | 0.406 | $0.415 - 0.910i$ |
| $[\mathrm{FR}_{10}^{5,1,0}]_{2,1,1}$ | ✗ 🌐 🎗 ⭕ | $-0.204$ | $-0.655 + 0.756i$ |
| $[\mathrm{FR}_{10}^{5,1,0}]_{2,2,1}$ | ✗ 🌐 🎗 ⭕ | $-0.204$ | $-0.655 - 0.756i$ |
| $[\mathrm{FR}_{10}^{5,1,0}]_{3,1,1}$ | ✗ 🌐 🎗 ⭕ | 1.764 | $-0.142 - 0.990i$ |
| $[\mathrm{FR}_{10}^{5,1,0}]_{3,2,1}$ | ✗ 🌐 🎗 ⭕ | 1.764 | $-0.142 + 0.990i$ |
| $[\mathrm{FR}_{10}^{5,1,0}]_{4,1,1}$ | ✗ 🌐 🎗 ⭕ | 4.513 | $-0.959 + 0.282i$ |
| $[\mathrm{FR}_{10}^{5,1,0}]_{4,2,1}$ | ✗ 🌐 🎗 ⭕ | 4.513 | $-0.959 - 0.282i$ |
| $[\mathrm{FR}_{10}^{5,1,0}]_{5,1,1}$ | ✗ 🌐 🎗 ⭕ | 1.521 | $0.841 - 0.541i$ |
| $[\mathrm{FR}_{10}^{5,1,0}]_{5,2,1}$ | ✗ 🌐 🎗 ⭕ | 1.521 | $0.841 + 0.541i$ |

### 4.5.23 $\mathrm{FR}_1^{5,1,2}$: $\mathbf{TY}(\mathbb{Z}_4)$

For the fusion ring, the following multiplication table is used.

| 1 | 2 | 3 | 4 | 5 |
|---|---|---|---|---|
| 2 | 1 | 4 | 3 | 5 |
| 3 | 4 | 2 | 1 | 5 |
| 4 | 3 | 1 | 2 | 5 |
| 5 | 5 | 5 | 5 | $1 + 2 + 3 + 4$ |

The following is the group of all non-trivial permutations that leave the fusion rules invariant:

$$S = \{(), (3\ 4)\}.$$

Let

$$X_1 = S\left([F_5^{353}]_5^5\right), \quad (53)$$

$$X_2 = S\left([F_4^{535}]_5^5[F_5^{555}]_3^3\right). \quad (54)$$

The following table lists a small set of invariants whose values completely distinguish between all MFPBFCs and MFPNBFCs with the given fusion rules.

Table 45: Symbolic invariants

| Name | Properties | $X_1$ | $X_2$ | $d_5^L$ |
|---|---|---|---|---|
| $[FR_1^{5,1,2}]_{1,0,1}$ |  | $\{-i\}$ | $\{-\frac{1}{2}\}$ | 2 |
| $[FR_1^{5,1,2}]_{1,0,2}$ |  | $\{-i\}$ | $\{-\frac{1}{2}\}$ | $-2$ |
| $[FR_1^{5,1,2}]_{2,0,1}$ |  | $\{i\}$ | $\{\frac{1}{2}\}$ | 2 |
| $[FR_1^{5,1,2}]_{2,0,2}$ |  | $\{i\}$ | $\{\frac{1}{2}\}$ | $-2$ |
| $[FR_1^{5,1,2}]_{3,0,1}$ |  | $\{i\}$ | $\{-\frac{1}{2}\}$ | 2 |
| $[FR_1^{5,1,2}]_{3,0,2}$ |  | $\{i\}$ | $\{-\frac{1}{2}\}$ | $-2$ |
| $[FR_1^{5,1,2}]_{4,0,1}$ |  | $\{-i\}$ | $\{\frac{1}{2}\}$ | 2 |
| $[FR_1^{5,1,2}]_{4,0,2}$ |  | $\{-i\}$ | $\{\frac{1}{2}\}$ | $-2$ |

### 4.5.24  $FR_3^{5,1,2}$: Pseudo $SU(2)_4$

For the fusion ring, the following multiplication table is used.

| 1 | 2 | 3 | 4 | 5 |
|---|---|---|---|---|
| 2 | 1 | 4 | 3 | 5 |
| 3 | 4 | 2+5 | 1+5 | 3+4 |
| 4 | 3 | 1+5 | 2+5 | 3+4 |
| 5 | 5 | 3+4 | 3+4 | 1+2+5 |

The following is the group of all non-trivial permutations that leave the fusion rules invariant:

$$S = \{(), (3\ 4)\}.$$

Let

$$X_1 = S\left([F_3^{333}]_5^5[F_2^{353}]_3^3\right), \quad (55)$$

$$X_2 = S\left(d_3^L\right). \quad (56)$$

The following table lists a small set of invariants whose values completely distinguish between all MFPBFCs and MFPNBFCs with the given fusion rules.

Table 46: Symbolic invariants

| Name | Properties | $X_1$ | $X_2$ |
|---|---|---|---|
| $[FR_3^{5,1,2}]_{1,0,1}$ | 🗡️🌐 | $\{-i\}$ | $\{\sqrt{3}\}$ |
| $[FR_3^{5,1,2}]_{1,0,2}$ | 🌐 | $\{-i\}$ | $\{-\sqrt{3}\}$ |
| $[FR_3^{5,1,2}]_{2,0,1}$ | 🗡️🌐 | $\{i\}$ | $\{\sqrt{3}\}$ |
| $[FR_3^{5,1,2}]_{2,0,2}$ | 🌐 | $\{i\}$ | $\{-\sqrt{3}\}$ |

Table 47: Numeric invariants

| Name | Properties | $X_1$ | $X_2$ |
|---|---|---|---|
| $[FR_3^{5,1,2}]_{1,0,1}$ | 🗡️🌐 | $\{-i\}$ | $\{1.732\}$ |
| $[FR_3^{5,1,2}]_{1,0,2}$ | 🌐 | $\{-i\}$ | $\{-1.732\}$ |
| $[FR_3^{5,1,2}]_{2,0,1}$ | 🗡️🌐 | $\{i\}$ | $\{1.732\}$ |
| $[FR_3^{5,1,2}]_{2,0,2}$ | 🌐 | $\{i\}$ | $\{-1.732\}$ |

### 4.5.25 $FR_4^{5,1,2}$: Pseudo Rep$(S_4)$

For the fusion ring, the following multiplication table is used.

| 1 | 2 | 3 | 4 | 5 |
|---|---|---|---|---|
| 2 | 1 | 3 | 5 | 4 |
| 3 | 3 | $1+2+3$ | $4+5$ | $4+5$ |
| 4 | 5 | $4+5$ | $2+3+4+5$ | $1+3+4+5$ |
| 5 | 4 | $4+5$ | $1+3+4+5$ | $2+3+4+5$ |

The following is the group of all non-trivial permutations that leave the fusion rules invariant:

$$S = \{(), (4\ 5)\}.$$

Let

$$X_1 = S\left([F_4^{444}]_4^4\right).$$

The following table lists a small set of invariants whose values completely distinguish between all MFPBFCs and MFPNBFCs with the given fusion rules.

Table 48: Symbolic invariants

| Name | Properties | $X_1$ |
|---|---|---|
| $[FR_4^{5,1,2}]_{1,0,1}$ | 🗡️🌐 | $\{\frac{i}{2}\}$ |
| $[FR_4^{5,1,2}]_{2,0,1}$ | 🗡️🌐 | $\{-\frac{i}{2}\}$ |

### 4.5.26 $FR_1^{5,1,4}$: $\mathbb{Z}_5$

For the fusion ring, the following multiplication table is used.

|   |   |   |   |   |
|---|---|---|---|---|
| 1 | 2 | 3 | 4 | 5 |
| 2 | 5 | 1 | 3 | 4 |
| 3 | 1 | 4 | 5 | 2 |
| 4 | 3 | 5 | 2 | 1 |
| 5 | 4 | 2 | 1 | 3 |

The following is the group of all non-trivial permutations that leave the fusion rules invariant:

$$S = \{(), (2\ 4\ 3\ 5), (2\ 5\ 3\ 4), (2\ 3)(4\ 5)\}.$$

Let

$$X_1 = S\left(\left([F_4^{324}]_3^1 [F_3^{332}]_1^4 [F_3^{444}]_2^2, R_4^{33}, d_3^L\right)\right). \tag{57}$$

The following table lists a small set of invariants whose values completely distinguish between all MFPBFCs and MFPNBFCs with the given fusion rules.

Table 49: Symbolic invariants

| Name | Properties | $X_1$ |
|---|---|---|
| $[FR_1^{5,1,4}]_{1,1,1}$ | ✕ † 🌐 🎗 ⦿ | $\{(1, \zeta_5^3, 1), (1, \zeta_5^2, 1)\}$ |
| $[FR_1^{5,1,4}]_{1,1,2}$ | ✕ | $\{(1, \zeta_5^3, \zeta_5^3), (1, \zeta_5^3, \zeta_5^2), (1, \zeta_5^2, \zeta_5^4), (1, \zeta_5^2, \zeta_5)\}$ |
| $[FR_1^{5,1,4}]_{1,1,3}$ | ✕ | $\{(1, \zeta_5^3, \zeta_5^4), (1, \zeta_5^3, \zeta_5), (1, \zeta_5^2, \zeta_5^3), (1, \zeta_5^2, \zeta_5^2)\}$ |
| $[FR_1^{5,1,4}]_{1,2,1}$ | ✕ † 🌐 🎗 ⦿ | $\{(1, \zeta_5^4, 1), (1, \zeta_5, 1)\}$ |
| $[FR_1^{5,1,4}]_{1,2,2}$ | ✕ | $\{(1, \zeta_5^4, \zeta_5^3), (1, \zeta_5^4, \zeta_5^2), (1, \zeta_5, \zeta_5^4), (1, \zeta_5, \zeta_5)\}$ |
| $[FR_1^{5,1,4}]_{1,2,3}$ | ✕ | $\{(1, \zeta_5^4, \zeta_5^4), (1, \zeta_5^4, \zeta_5), (1, \zeta_5, \zeta_5^3), (1, \zeta_5, \zeta_5^2)\}$ |
| $[FR_1^{5,1,4}]_{1,3,1}$ | ✕ † 🌐 🎗 | $\{(1, 1, 1)\}$ |
| $[FR_1^{5,1,4}]_{1,3,2}$ | ✕ | $\{(1, 1, \zeta_5^3), (1, 1, \zeta_5^2), (1, 1, \zeta_5^4), (1, 1, \zeta_5)\}$ |
| $[FR_1^{5,1,4}]_{2,0,1}$ | † 🌐 | $\{(\zeta_5^3, -, 1), (\zeta_5^2, -, 1)\}$ |
| $[FR_1^{5,1,4}]_{2,0,2}$ | | $\{(\zeta_5^3, -, \zeta_5^4), (\zeta_5^3, -, \zeta_5), (\zeta_5^2, -, \zeta_5^3), (\zeta_5^2, -, \zeta_5^2)\}$ |
| $[FR_1^{5,1,4}]_{2,0,3}$ | | $\{(\zeta_5^3, -, \zeta_5^3), (\zeta_5^3, -, \zeta_5^2), (\zeta_5^2, -, \zeta_5), (\zeta_5^2, -, \zeta_5^4)\}$ |
| $[FR_1^{5,1,4}]_{3,0,1}$ | † 🌐 | $\{(\zeta_5^4, -, 1), (\zeta_5, -, 1)\}$ |
| $[FR_1^{5,1,4}]_{3,0,2}$ | | $\{(\zeta_5^4, -, \zeta_5^4), (\zeta_5^4, -, \zeta_5), (\zeta_5, -, \zeta_5^3), (\zeta_5, -, \zeta_5^2)\}$ |
| $[FR_1^{5,1,4}]_{3,0,3}$ | | $\{(\zeta_5^4, -, \zeta_5^3), (\zeta_5^4, -, \zeta_5^2), (\zeta_5, -, \zeta_5), (\zeta_5, -, \zeta_5^4)\}$ |

Table 50: Numeric invariants

| Name | Properties | $X_1$ |
|---|---|---|
| $[FR_1^{5,1,4}]_{1,1,1}$ | ✕ † 🌐 🎗 ⦿ | $\left\{\begin{array}{l}(1, -0.809 - 0.588i, 1),\\(1, -0.809 + 0.588i, 1)\end{array}\right\}$ |

Table 50: Numeric invariants (Continued)

| Name | Properties | $X_1$ |
|---|---|---|
| $[FR_1^{5,1,4}]_{1,1,2}$ | ✗ | $\left\{\begin{array}{l}(1,\ -0.809-0.588i,\ -0.809-0.588i),\\(1,\ -0.809-0.588i,\ -0.809+0.588i),\\(1,\ -0.809+0.588i,\ 0.309-0.951i),\\(1,\ -0.809+0.588i,\ 0.309+0.951i)\end{array}\right\}$ |
| $[FR_1^{5,1,4}]_{1,1,3}$ | ✗ | $\left\{\begin{array}{l}(1,\ -0.809-0.588i,\ 0.309-0.951i),\\(1,\ -0.809-0.588i,\ 0.309+0.951i),\\(1,\ -0.809+0.588i,\ -0.809-0.588i),\\(1,\ -0.809+0.588i,\ -0.809+0.588i)\end{array}\right\}$ |
| $[FR_1^{5,1,4}]_{1,2,1}$ | ✗ † 🌐 🎗 ⚭ | $\left\{\begin{array}{l}(1,\ 0.309-0.951i,\ 1),\\(1,\ 0.309+0.951i,\ 1)\end{array}\right\}$ |
| $[FR_1^{5,1,4}]_{1,2,2}$ | ✗ | $\left\{\begin{array}{l}(1,\ 0.309-0.951i,\ -0.809-0.588i),\\(1,\ 0.309-0.951i,\ -0.809+0.588i),\\(1,\ 0.309+0.951i,\ 0.309-0.951i),\\(1,\ 0.309+0.951i,\ 0.309+0.951i)\end{array}\right\}$ |
| $[FR_1^{5,1,4}]_{1,2,3}$ | ✗ | $\left\{\begin{array}{l}(1,\ 0.309-0.951i,\ 0.309-0.951i),\\(1,\ 0.309-0.951i,\ 0.309+0.951i),\\(1,\ 0.309+0.951i,\ -0.809-0.588i),\\(1,\ 0.309+0.951i,\ -0.809+0.588i)\end{array}\right\}$ |
| $[FR_1^{5,1,4}]_{1,3,1}$ | ✗ † 🌐 🎗 | $\left\{(1,\ 1,\ 1)\right\}$ |
| $[FR_1^{5,1,4}]_{1,3,2}$ | ✗ | $\left\{\begin{array}{l}(1,\ 1,\ -0.809-0.588i),\\(1,\ 1,\ -0.809+0.588i),\\(1,\ 1,\ 0.309-0.951i),\\(1,\ 1,\ 0.309+0.951i)\end{array}\right\}$ |
| $[FR_1^{5,1,4}]_{2,0,1}$ | † 🌐 | $\left\{\begin{array}{l}(-0.809-0.588i,\ \text{-},\ 1),\\(-0.809+0.588i,\ \text{-},\ 1)\end{array}\right\}$ |
| $[FR_1^{5,1,4}]_{2,0,2}$ | | $\left\{\begin{array}{l}(-0.809-0.588i,\ \text{-},\ 0.309-0.951i),\\(-0.809-0.588i,\ \text{-},\ 0.309+0.951i),\\(-0.809+0.588i,\ \text{-},\ -0.809-0.588i),\\(-0.809+0.588i,\ \text{-},\ -0.809+0.588i)\end{array}\right\}$ |
| $[FR_1^{5,1,4}]_{2,0,3}$ | | $\left\{\begin{array}{l}(-0.809-0.588i,\ \text{-},\ -0.809-0.588i),\\(-0.809-0.588i,\ \text{-},\ -0.809+0.588i),\\(-0.809+0.588i,\ \text{-},\ 0.309+0.951i),\\(-0.809+0.588i,\ \text{-},\ 0.309-0.951i)\end{array}\right\}$ |
| $[FR_1^{5,1,4}]_{3,0,1}$ | † 🌐 | $\left\{\begin{array}{l}(0.309-0.951i,\ \text{-},\ 1),\\(0.309+0.951i,\ \text{-},\ 1)\end{array}\right\}$ |
| $[FR_1^{5,1,4}]_{3,0,2}$ | | $\left\{\begin{array}{l}(0.309-0.951i,\ \text{-},\ 0.309-0.951i),\\(0.309-0.951i,\ \text{-},\ 0.309+0.951i),\\(0.309+0.951i,\ \text{-},\ -0.809-0.588i),\\(0.309+0.951i,\ \text{-},\ -0.809+0.588i)\end{array}\right\}$ |

Table 50: Numeric invariants (Continued)

| Name | Properties | $X_1$ |
|---|---|---|
| $[FR_1^{5,1,4}]_{3,0,3}$ | | $\begin{Bmatrix} (0.309 - 0.951i, & -, & -0.809 - 0.588i), \\ (0.309 - 0.951i, & -, & -0.809 + 0.588i), \\ (0.309 + 0.951i, & -, & 0.309 + 0.951i), \\ (0.309 + 0.951i, & -, & 0.309 - 0.951i) \end{Bmatrix}$ |

### 4.5.27 $FR_1^{6,1,0}$: $\mathbb{Z}_2 \otimes$ Ising

For the fusion ring, the following multiplication table is used.

| 1 | 2 | 3 | 4 | 5 | 6 |
|---|---|---|---|---|---|
| 2 | 1 | 4 | 3 | 6 | 5 |
| 3 | 4 | 1 | 2 | 6 | 5 |
| 4 | 3 | 2 | 1 | 5 | 6 |
| 5 | 6 | 6 | 5 | 1+4 | 2+3 |
| 6 | 5 | 5 | 6 | 2+3 | 1+4 |

The following is the group of all non-trivial permutations that leave the fusion rules invariant:

$$S = \{(), (2\ 3), (5\ 6), (2\ 3)(5\ 6)\}.$$

Let

$$X_1 = S\left(\left([F_3^{232}]_4^4 [F_4^{242}]_3^3, [F_2^{525}]_6^6 [F_6^{565}]_2^2, R_1^{55}, d_2^L, d_5^L\right)\right). \tag{58}$$

The following table lists a small set of invariants whose values completely distinguish between all MFPBFCs and MFPNBFCs with the given fusion rules.

Table 51: Symbolic invariants

| Name | Properties | $X_1$ |
|---|---|---|
| $[FR_1^{6,1,0}]_{1,1,1}$ | ✕ † 🌐 🎗 ⊚ | $\left\{\left(-1, -\frac{1}{\sqrt{2}}, \zeta_{16}^5, 1, \sqrt{2}\right), \left(-1, \frac{1}{\sqrt{2}}, \zeta_{16}^9, 1, \sqrt{2}\right)\right\}$ |
| $[FR_1^{6,1,0}]_{1,1,2}$ | ✕ 🌐 🎗 ⊚ | $\left\{\left(-1, -\frac{1}{\sqrt{2}}, \zeta_{16}^5, 1, -\sqrt{2}\right), \left(-1, \frac{1}{\sqrt{2}}, \zeta_{16}^9, 1, -\sqrt{2}\right)\right\}$ |
| $[FR_1^{6,1,0}]_{1,1,3}$ | ✕ 🌐 🎗 ⊚ | $\left\{\left(-1, -\frac{1}{\sqrt{2}}, \zeta_{16}^5, -1, -\sqrt{2}\right), \left(-1, \frac{1}{\sqrt{2}}, \zeta_{16}^9, -1, \sqrt{2}\right)\right\}$ |
| $[FR_1^{6,1,0}]_{1,1,4}$ | ✕ 🌐 🎗 ⊚ | $\left\{\left(-1, -\frac{1}{\sqrt{2}}, \zeta_{16}^5, -1, \sqrt{2}\right), \left(-1, \frac{1}{\sqrt{2}}, \zeta_{16}^9, -1, -\sqrt{2}\right)\right\}$ |
| $[FR_1^{6,1,0}]_{1,2,1}$ | ✕ † 🌐 🎗 ⊚ | $\left\{\left(-1, -\frac{1}{\sqrt{2}}, \zeta_{16}^5, 1, \sqrt{2}\right), \left(-1, \frac{1}{\sqrt{2}}, \zeta_{16}, 1, \sqrt{2}\right)\right\}$ |
| $[FR_1^{6,1,0}]_{1,2,2}$ | ✕ 🌐 🎗 ⊚ | $\left\{\left(-1, -\frac{1}{\sqrt{2}}, \zeta_{16}^5, 1, -\sqrt{2}\right), \left(-1, \frac{1}{\sqrt{2}}, \zeta_{16}, 1, -\sqrt{2}\right)\right\}$ |
| $[FR_1^{6,1,0}]_{1,2,3}$ | ✕ 🌐 🎗 ⊚ | $\left\{\left(-1, -\frac{1}{\sqrt{2}}, \zeta_{16}^5, -1, -\sqrt{2}\right), \left(-1, \frac{1}{\sqrt{2}}, \zeta_{16}, -1, \sqrt{2}\right)\right\}$ |
| $[FR_1^{6,1,0}]_{1,2,4}$ | ✕ 🌐 🎗 ⊚ | $\left\{\left(-1, -\frac{1}{\sqrt{2}}, \zeta_{16}^5, -1, \sqrt{2}\right), \left(-1, \frac{1}{\sqrt{2}}, \zeta_{16}, -1, -\sqrt{2}\right)\right\}$ |
| $[FR_1^{6,1,0}]_{1,3,1}$ | ✕ † 🌐 🎗 ⊚ | $\left\{\left(-1, -\frac{1}{\sqrt{2}}, \zeta_{16}^{13}, 1, \sqrt{2}\right), \left(-1, \frac{1}{\sqrt{2}}, \zeta_{16}, 1, \sqrt{2}\right)\right\}$ |
| $[FR_1^{6,1,0}]_{1,3,2}$ | ✕ 🌐 🎗 ⊚ | $\left\{\left(-1, -\frac{1}{\sqrt{2}}, \zeta_{16}^{13}, 1, -\sqrt{2}\right), \left(-1, \frac{1}{\sqrt{2}}, \zeta_{16}, 1, -\sqrt{2}\right)\right\}$ |

Table 51: Symbolic invariants (Continued)

| Name | Properties | $X_1$ |
|---|---|---|
| $[FR_1^{6,1,0}]_{1,3,3}$ | ✗ 🌐 🎗 ⚭ | $\left\{\left(-1, -\frac{1}{\sqrt{2}}, \zeta_{16}^{13}, -1, -\sqrt{2}\right), \left(-1, \frac{1}{\sqrt{2}}, \zeta_{16}, -1, \sqrt{2}\right)\right\}$ |
| $[FR_1^{6,1,0}]_{1,3,4}$ | ✗ 🌐 🎗 ⚭ | $\left\{\left(-1, -\frac{1}{\sqrt{2}}, \zeta_{16}^{13}, -1, \sqrt{2}\right), \left(-1, \frac{1}{\sqrt{2}}, \zeta_{16}, -1, -\sqrt{2}\right)\right\}$ |
| $[FR_1^{6,1,0}]_{1,4,1}$ | ✗ † 🌐 🎗 ⚭ | $\left\{\left(-1, -\frac{1}{\sqrt{2}}, \zeta_{16}^{13}, 1, \sqrt{2}\right), \left(-1, \frac{1}{\sqrt{2}}, \zeta_{16}^{9}, 1, \sqrt{2}\right)\right\}$ |
| $[FR_1^{6,1,0}]_{1,4,2}$ | ✗ 🌐 🎗 | $\left\{\left(-1, -\frac{1}{\sqrt{2}}, \zeta_{16}^{13}, 1, -\sqrt{2}\right), \left(-1, \frac{1}{\sqrt{2}}, \zeta_{16}^{9}, 1, -\sqrt{2}\right)\right\}$ |
| $[FR_1^{6,1,0}]_{1,4,3}$ | ✗ 🌐 🎗 ⚭ | $\left\{\left(-1, -\frac{1}{\sqrt{2}}, \zeta_{16}^{13}, -1, -\sqrt{2}\right), \left(-1, \frac{1}{\sqrt{2}}, \zeta_{16}^{9}, -1, \sqrt{2}\right)\right\}$ |
| $[FR_1^{6,1,0}]_{1,4,4}$ | ✗ 🌐 🎗 ⚭ | $\left\{\left(-1, -\frac{1}{\sqrt{2}}, \zeta_{16}^{13}, -1, \sqrt{2}\right), \left(-1, \frac{1}{\sqrt{2}}, \zeta_{16}^{9}, -1, -\sqrt{2}\right)\right\}$ |
| $[FR_1^{6,1,0}]_{1,5,1}$ | ✗ † 🌐 🎗 ⚭ | $\left\{\left(-1, -\frac{1}{\sqrt{2}}, \zeta_{16}^{3}, 1, \sqrt{2}\right), \left(-1, \frac{1}{\sqrt{2}}, \zeta_{16}^{7}, 1, \sqrt{2}\right)\right\}$ |
| $[FR_1^{6,1,0}]_{1,5,2}$ | ✗ 🌐 🎗 ⚭ | $\left\{\left(-1, -\frac{1}{\sqrt{2}}, \zeta_{16}^{3}, 1, -\sqrt{2}\right), \left(-1, \frac{1}{\sqrt{2}}, \zeta_{16}^{7}, 1, -\sqrt{2}\right)\right\}$ |
| $[FR_1^{6,1,0}]_{1,5,3}$ | ✗ 🌐 🎗 ⚭ | $\left\{\left(-1, -\frac{1}{\sqrt{2}}, \zeta_{16}^{3}, -1, -\sqrt{2}\right), \left(-1, \frac{1}{\sqrt{2}}, \zeta_{16}^{7}, -1, \sqrt{2}\right)\right\}$ |
| $[FR_1^{6,1,0}]_{1,5,4}$ | ✗ 🌐 🎗 ⚭ | $\left\{\left(-1, -\frac{1}{\sqrt{2}}, \zeta_{16}^{3}, -1, \sqrt{2}\right), \left(-1, \frac{1}{\sqrt{2}}, \zeta_{16}^{7}, -1, -\sqrt{2}\right)\right\}$ |
| $[FR_1^{6,1,0}]_{1,6,1}$ | ✗ † 🌐 🎗 ⚭ | $\left\{\left(-1, -\frac{1}{\sqrt{2}}, \zeta_{16}^{3}, 1, \sqrt{2}\right), \left(-1, \frac{1}{\sqrt{2}}, \zeta_{16}^{15}, 1, \sqrt{2}\right)\right\}$ |
| $[FR_1^{6,1,0}]_{1,6,2}$ | ✗ 🌐 🎗 | $\left\{\left(-1, -\frac{1}{\sqrt{2}}, \zeta_{16}^{3}, 1, -\sqrt{2}\right), \left(-1, \frac{1}{\sqrt{2}}, \zeta_{16}^{15}, 1, -\sqrt{2}\right)\right\}$ |
| $[FR_1^{6,1,0}]_{1,6,3}$ | ✗ 🌐 🎗 | $\left\{\left(-1, -\frac{1}{\sqrt{2}}, \zeta_{16}^{3}, -1, -\sqrt{2}\right), \left(-1, \frac{1}{\sqrt{2}}, \zeta_{16}^{15}, -1, \sqrt{2}\right)\right\}$ |
| $[FR_1^{6,1,0}]_{1,6,4}$ | ✗ 🌐 🎗 | $\left\{\left(-1, -\frac{1}{\sqrt{2}}, \zeta_{16}^{3}, -1, \sqrt{2}\right), \left(-1, \frac{1}{\sqrt{2}}, \zeta_{16}^{15}, -1, -\sqrt{2}\right)\right\}$ |
| $[FR_1^{6,1,0}]_{1,7,1}$ | ✗ † 🌐 🎗 ⚭ | $\left\{\left(-1, -\frac{1}{\sqrt{2}}, \zeta_{16}^{11}, 1, \sqrt{2}\right), \left(-1, \frac{1}{\sqrt{2}}, \zeta_{16}^{15}, 1, \sqrt{2}\right)\right\}$ |
| $[FR_1^{6,1,0}]_{1,7,2}$ | ✗ 🌐 🎗 ⚭ | $\left\{\left(-1, -\frac{1}{\sqrt{2}}, \zeta_{16}^{11}, 1, -\sqrt{2}\right), \left(-1, \frac{1}{\sqrt{2}}, \zeta_{16}^{15}, 1, -\sqrt{2}\right)\right\}$ |
| $[FR_1^{6,1,0}]_{1,7,3}$ | ✗ 🌐 🎗 ⚭ | $\left\{\left(-1, -\frac{1}{\sqrt{2}}, \zeta_{16}^{11}, -1, -\sqrt{2}\right), \left(-1, \frac{1}{\sqrt{2}}, \zeta_{16}^{15}, -1, -\sqrt{2}\right)\right\}$ |
| $[FR_1^{6,1,0}]_{1,7,4}$ | ✗ 🌐 🎗 ⚭ | $\left\{\left(-1, -\frac{1}{\sqrt{2}}, \zeta_{16}^{11}, -1, \sqrt{2}\right), \left(-1, \frac{1}{\sqrt{2}}, \zeta_{16}^{15}, -1, -\sqrt{2}\right)\right\}$ |
| $[FR_1^{6,1,0}]_{1,8,1}$ | ✗ † 🌐 🎗 ⚭ | $\left\{\left(-1, -\frac{1}{\sqrt{2}}, \zeta_{16}^{11}, 1, \sqrt{2}\right), \left(-1, \frac{1}{\sqrt{2}}, \zeta_{16}^{7}, 1, \sqrt{2}\right)\right\}$ |
| $[FR_1^{6,1,0}]_{1,8,2}$ | ✗ 🌐 🎗 ⚭ | $\left\{\left(-1, -\frac{1}{\sqrt{2}}, \zeta_{16}^{11}, 1, -\sqrt{2}\right), \left(-1, \frac{1}{\sqrt{2}}, \zeta_{16}^{7}, 1, -\sqrt{2}\right)\right\}$ |
| $[FR_1^{6,1,0}]_{1,8,3}$ | ✗ 🌐 🎗 ⚭ | $\left\{\left(-1, -\frac{1}{\sqrt{2}}, \zeta_{16}^{11}, -1, -\sqrt{2}\right), \left(-1, \frac{1}{\sqrt{2}}, \zeta_{16}^{7}, -1, \sqrt{2}\right)\right\}$ |
| $[FR_1^{6,1,0}]_{1,8,4}$ | ✗ 🌐 🎗 ⚭ | $\left\{\left(-1, -\frac{1}{\sqrt{2}}, \zeta_{16}^{11}, -1, \sqrt{2}\right), \left(-1, \frac{1}{\sqrt{2}}, \zeta_{16}^{7}, -1, -\sqrt{2}\right)\right\}$ |
| $[FR_1^{6,1,0}]_{2,1,1}$ | ✗ † 🌐 🎗 | $\left\{\left(1, \frac{1}{\sqrt{2}}, \zeta_{16}^{15}, 1, \sqrt{2}\right)\right\}$ |
| $[FR_1^{6,1,0}]_{2,1,2}$ | ✗ 🌐 🎗 | $\left\{\left(1, \frac{1}{\sqrt{2}}, \zeta_{16}^{15}, 1, -\sqrt{2}\right)\right\}$ |
| $[FR_1^{6,1,0}]_{2,1,3}$ | ✗ 🌐 🎗 | $\left\{\left(1, \frac{1}{\sqrt{2}}, \zeta_{16}^{15}, -1, -\sqrt{2}\right), \left(1, \frac{1}{\sqrt{2}}, \zeta_{16}^{15}, -1, \sqrt{2}\right)\right\}$ |
| $[FR_1^{6,1,0}]_{2,1,4}$ | ✗ † 🌐 🎗 | $\left\{\left(1, \frac{1}{\sqrt{2}}, \zeta_{16}^{7}, 1, \sqrt{2}\right), \left(1, \frac{1}{\sqrt{2}}, \zeta_{16}^{15}, 1, \sqrt{2}\right)\right\}$ |
| $[FR_1^{6,1,0}]_{2,2,1}$ | ✗ 🌐 🎗 | $\left\{\left(1, \frac{1}{\sqrt{2}}, \zeta_{16}^{7}, 1, -\sqrt{2}\right), \left(1, \frac{1}{\sqrt{2}}, \zeta_{16}^{15}, 1, -\sqrt{2}\right)\right\}$ |
| $[FR_1^{6,1,0}]_{2,2,2}$ | ✗ 🌐 🎗 | $\left\{\left(1, \frac{1}{\sqrt{2}}, \zeta_{16}^{7}, -1, -\sqrt{2}\right), \left(1, \frac{1}{\sqrt{2}}, \zeta_{16}^{15}, -1, \sqrt{2}\right)\right\}$ |
| $[FR_1^{6,1,0}]_{2,2,3}$ | ✗ 🌐 🎗 | $\left\{\left(1, \frac{1}{\sqrt{2}}, \zeta_{16}^{7}, -1, \sqrt{2}\right), \left(1, \frac{1}{\sqrt{2}}, \zeta_{16}^{15}, -1, -\sqrt{2}\right)\right\}$ |
| $[FR_1^{6,1,0}]_{2,2,4}$ | ✗ † 🌐 🎗 | $\left\{\left(1, \frac{1}{\sqrt{2}}, \zeta_{16}^{7}, 1, \sqrt{2}\right)\right\}$ |
| $[FR_1^{6,1,0}]_{2,3,1}$ | ✗ 🌐 🎗 | $\left\{\left(1, \frac{1}{\sqrt{2}}, \zeta_{16}^{7}, 1, -\sqrt{2}\right)\right\}$ |

Table 51: Symbolic invariants (Continued)

| Name | Properties | $X_1$ |
|---|---|---|
| $[\mathrm{FR}_1^{6,1,0}]_{2,3,2}$ | ✗ 🌐 🎗 | $\{(1,\frac{1}{\sqrt{2}},\zeta_{16}^7,-1,-\sqrt{2}),(1,\frac{1}{\sqrt{2}},\zeta_{16}^7,-1,\sqrt{2})\}$ |
| $[\mathrm{FR}_1^{6,1,0}]_{2,3,3}$ | ✗ † 🌐 🎗 | $\{(1,\frac{1}{\sqrt{2}},\zeta_{16},1,\sqrt{2})\}$ |
| $[\mathrm{FR}_1^{6,1,0}]_{2,4,1}$ | ✗ 🌐 🎗 | $\{(1,\frac{1}{\sqrt{2}},\zeta_{16},1,-\sqrt{2})\}$ |
| $[\mathrm{FR}_1^{6,1,0}]_{2,4,2}$ | ✗ 🌐 🎗 | $\{(1,\frac{1}{\sqrt{2}},\zeta_{16},-1,-\sqrt{2}),(1,\frac{1}{\sqrt{2}},\zeta_{16},-1,\sqrt{2})\}$ |
| $[\mathrm{FR}_1^{6,1,0}]_{2,4,3}$ | ✗ † 🌐 🎗 | $\{(1,\frac{1}{\sqrt{2}},\zeta_{16}^9,1,\sqrt{2}),(1,\frac{1}{\sqrt{2}},\zeta_{16},1,\sqrt{2})\}$ |
| $[\mathrm{FR}_1^{6,1,0}]_{2,5,1}$ | ✗ 🌐 🎗 | $\{(1,\frac{1}{\sqrt{2}},\zeta_{16}^9,1,-\sqrt{2}),(1,\frac{1}{\sqrt{2}},\zeta_{16},1,-\sqrt{2})\}$ |
| $[\mathrm{FR}_1^{6,1,0}]_{2,5,2}$ | ✗ 🌐 🎗 | $\{(1,\frac{1}{\sqrt{2}},\zeta_{16}^9,-1,-\sqrt{2}),(1,\frac{1}{\sqrt{2}},\zeta_{16},-1,\sqrt{2})\}$ |
| $[\mathrm{FR}_1^{6,1,0}]_{2,5,3}$ | ✗ 🌐 🎗 | $\{(1,\frac{1}{\sqrt{2}},\zeta_{16}^9,-1,\sqrt{2}),(1,\frac{1}{\sqrt{2}},\zeta_{16},-1,-\sqrt{2})\}$ |
| $[\mathrm{FR}_1^{6,1,0}]_{2,6,1}$ | ✗ † 🌐 🎗 | $\{(1,\frac{1}{\sqrt{2}},\zeta_{16}^9,1,\sqrt{2})\}$ |
| $[\mathrm{FR}_1^{6,1,0}]_{2,6,2}$ | ✗ 🌐 🎗 | $\{(1,\frac{1}{\sqrt{2}},\zeta_{16}^9,1,-\sqrt{2})\}$ |
| $[\mathrm{FR}_1^{6,1,0}]_{2,6,3}$ | ✗ 🌐 🎗 | $\{(1,\frac{1}{\sqrt{2}},\zeta_{16}^9,-1,-\sqrt{2}),(1,\frac{1}{\sqrt{2}},\zeta_{16}^9,-1,\sqrt{2})\}$ |
| $[\mathrm{FR}_1^{6,1,0}]_{3,1,1}$ | ✗ † 🌐 🎗 | $\{(1,-\frac{1}{\sqrt{2}},\zeta_{16}^{11},1,\sqrt{2}),(1,-\frac{1}{\sqrt{2}},\zeta_{16}^3,1,\sqrt{2})\}$ |
| $[\mathrm{FR}_1^{6,1,0}]_{3,1,2}$ | ✗ 🌐 🎗 | $\{(1,-\frac{1}{\sqrt{2}},\zeta_{16}^{11},1,-\sqrt{2}),(1,-\frac{1}{\sqrt{2}},\zeta_{16}^3,1,-\sqrt{2})\}$ |
| $[\mathrm{FR}_1^{6,1,0}]_{3,1,3}$ | ✗ 🌐 🎗 | $\{(1,-\frac{1}{\sqrt{2}},\zeta_{16}^{11},-1,\sqrt{2}),(1,-\frac{1}{\sqrt{2}},\zeta_{16}^3,-1,-\sqrt{2})\}$ |
| $[\mathrm{FR}_1^{6,1,0}]_{3,1,4}$ | ✗ 🌐 🎗 | $\{(1,-\frac{1}{\sqrt{2}},\zeta_{16}^{11},-1,-\sqrt{2}),(1,-\frac{1}{\sqrt{2}},\zeta_{16}^3,-1,\sqrt{2})\}$ |
| $[\mathrm{FR}_1^{6,1,0}]_{3,2,1}$ | ✗ † 🌐 🎗 | $\{(1,-\frac{1}{\sqrt{2}},\zeta_{16}^3,1,\sqrt{2})\}$ |
| $[\mathrm{FR}_1^{6,1,0}]_{3,2,2}$ | ✗ 🌐 🎗 | $\{(1,-\frac{1}{\sqrt{2}},\zeta_{16}^3,1,-\sqrt{2})\}$ |
| $[\mathrm{FR}_1^{6,1,0}]_{3,2,3}$ | ✗ 🌐 🎗 | $\{(1,-\frac{1}{\sqrt{2}},\zeta_{16}^3,-1,-\sqrt{2}),(1,-\frac{1}{\sqrt{2}},\zeta_{16}^3,-1,\sqrt{2})\}$ |
| $[\mathrm{FR}_1^{6,1,0}]_{3,2,4}$ | ✗ † 🌐 🎗 | $\{(1,-\frac{1}{\sqrt{2}},\zeta_{16}^{11},1,\sqrt{2})\}$ |
| $[\mathrm{FR}_1^{6,1,0}]_{3,3,1}$ | ✗ 🌐 🎗 | $\{(1,-\frac{1}{\sqrt{2}},\zeta_{16}^{11},1,-\sqrt{2})\}$ |
| $[\mathrm{FR}_1^{6,1,0}]_{3,3,2}$ | ✗ 🌐 🎗 | $\{(1,-\frac{1}{\sqrt{2}},\zeta_{16}^{11},-1,-\sqrt{2}),(1,-\frac{1}{\sqrt{2}},\zeta_{16}^{11},-1,\sqrt{2})\}$ |
| $[\mathrm{FR}_1^{6,1,0}]_{3,3,3}$ | ✗ † 🌐 🎗 | $\{(1,-\frac{1}{\sqrt{2}},\zeta_{16}^5,1,\sqrt{2}),(1,-\frac{1}{\sqrt{2}},\zeta_{16}^{13},1,\sqrt{2})\}$ |
| $[\mathrm{FR}_1^{6,1,0}]_{3,4,1}$ | ✗ 🌐 🎗 | $\{(1,-\frac{1}{\sqrt{2}},\zeta_{16}^5,1,-\sqrt{2}),(1,-\frac{1}{\sqrt{2}},\zeta_{16}^{13},1,-\sqrt{2})\}$ |
| $[\mathrm{FR}_1^{6,1,0}]_{3,4,2}$ | ✗ 🌐 🎗 | $\{(1,-\frac{1}{\sqrt{2}},\zeta_{16}^5,-1,\sqrt{2}),(1,-\frac{1}{\sqrt{2}},\zeta_{16}^{13},-1,-\sqrt{2})\}$ |
| $[\mathrm{FR}_1^{6,1,0}]_{3,4,3}$ | ✗ 🌐 🎗 | $\{(1,-\frac{1}{\sqrt{2}},\zeta_{16}^5,-1,-\sqrt{2}),(1,-\frac{1}{\sqrt{2}},\zeta_{16}^{13},-1,\sqrt{2})\}$ |
| $[\mathrm{FR}_1^{6,1,0}]_{3,5,1}$ | ✗ † 🌐 🎗 | $\{(1,-\frac{1}{\sqrt{2}},\zeta_{16}^{13},1,\sqrt{2})\}$ |
| $[\mathrm{FR}_1^{6,1,0}]_{3,5,2}$ | ✗ 🌐 🎗 | $\{(1,-\frac{1}{\sqrt{2}},\zeta_{16}^{13},1,-\sqrt{2})\}$ |
| $[\mathrm{FR}_1^{6,1,0}]_{3,5,3}$ | ✗ 🌐 🎗 | $\{(1,-\frac{1}{\sqrt{2}},\zeta_{16}^{13},-1,-\sqrt{2}),(1,-\frac{1}{\sqrt{2}},\zeta_{16}^{13},-1,\sqrt{2})\}$ |
| $[\mathrm{FR}_1^{6,1,0}]_{3,6,1}$ | ✗ † 🌐 🎗 | $\{(1,-\frac{1}{\sqrt{2}},\zeta_{16}^5,1,\sqrt{2})\}$ |
| $[\mathrm{FR}_1^{6,1,0}]_{3,6,2}$ | ✗ 🌐 🎗 | $\{(1,-\frac{1}{\sqrt{2}},\zeta_{16}^5,1,-\sqrt{2})\}$ |
| $[\mathrm{FR}_1^{6,1,0}]_{3,6,3}$ | ✗ 🌐 🎗 | $\{(1,-\frac{1}{\sqrt{2}},\zeta_{16}^5,-1,-\sqrt{2}),(1,-\frac{1}{\sqrt{2}},\zeta_{16}^5,-1,\sqrt{2})\}$ |

Table 51: Symbolic invariants (Continued)

| Name | Properties | $X_1$ |
|---|---|---|
| $[FR_1^{6,1,0}]_{4,0,1}$ | † 🌐 | $\{(1, -\frac{1}{\sqrt{2}}, -, 1, \sqrt{2}), (1, \frac{1}{\sqrt{2}}, -, 1, \sqrt{2})\}$ |
| $[FR_1^{6,1,0}]_{4,0,2}$ | 🌐 | $\{(1, -\frac{1}{\sqrt{2}}, -, 1, -\sqrt{2}), (1, \frac{1}{\sqrt{2}}, -, 1, -\sqrt{2})\}$ |
| $[FR_1^{6,1,0}]_{4,0,3}$ | 🌐 | $\{(1, -\frac{1}{\sqrt{2}}, -, -1, \sqrt{2}), (1, \frac{1}{\sqrt{2}}, -, -1, -\sqrt{2})\}$ |
| $[FR_1^{6,1,0}]_{4,0,4}$ | 🌐 | $\{(1, -\frac{1}{\sqrt{2}}, -, -1, -\sqrt{2}), (1, \frac{1}{\sqrt{2}}, -, -1, \sqrt{2})\}$ |
| $[FR_1^{6,1,0}]_{5,0,1}$ | † 🌐 | $\{(-1, \frac{1}{\sqrt{2}}, -, 1, \sqrt{2})\}$ |
| $[FR_1^{6,1,0}]_{5,0,2}$ | 🌐 | $\{(-1, \frac{1}{\sqrt{2}}, -, 1, -\sqrt{2})\}$ |
| $[FR_1^{6,1,0}]_{5,0,3}$ | 🌐 | $\{(-1, \frac{1}{\sqrt{2}}, -, -1, \sqrt{2})\}$ |
| $[FR_1^{6,1,0}]_{6,0,1}$ | † 🌐 | $\{(-1, -\frac{1}{\sqrt{2}}, -, 1, \sqrt{2})\}$ |
| $[FR_1^{6,1,0}]_{6,0,2}$ | 🌐 | $\{(-1, -\frac{1}{\sqrt{2}}, -, 1, -\sqrt{2})\}$ |
| $[FR_1^{6,1,0}]_{6,0,3}$ | 🌐 | $\{(-1, -\frac{1}{\sqrt{2}}, -, -1, -\sqrt{2})\}$ |

Table 52: Numeric invariants

| Name | Properties | $X_1$ |
|---|---|---|
| $[FR_1^{6,1,0}]_{1,1,1}$ | ✕ † 🌐 🎗 ⭕ | $\{(-1, -0.707, -0.383 + 0.924i, 1, 1.414), (-1, 0.707, -0.924 - 0.383i, 1, 1.414)\}$ |
| $[FR_1^{6,1,0}]_{1,1,2}$ | ✕ 🌐 🎗 ⭕ | $\{(-1, -0.707, -0.383 + 0.924i, 1, -1.414), (-1, 0.707, -0.924 - 0.383i, 1, -1.414)\}$ |
| $[FR_1^{6,1,0}]_{1,1,3}$ | ✕ 🌐 🎗 ⭕ | $\{(-1, -0.707, -0.383 + 0.924i, -1, -1.414), (-1, 0.707, -0.924 - 0.383i, -1, 1.414)\}$ |
| $[FR_1^{6,1,0}]_{1,1,4}$ | ✕ 🌐 🎗 ⭕ | $\{(-1, -0.707, -0.383 + 0.924i, -1, 1.414), (-1, 0.707, -0.924 - 0.383i, -1, -1.414)\}$ |
| $[FR_1^{6,1,0}]_{1,2,1}$ | ✕ † 🌐 🎗 ⭕ | $\{(-1, -0.707, -0.383 + 0.924i, 1, 1.414), (-1, 0.707, 0.924 + 0.383i, 1, 1.414)\}$ |
| $[FR_1^{6,1,0}]_{1,2,2}$ | ✕ 🌐 🎗 ⭕ | $\{(-1, -0.707, -0.383 + 0.924i, 1, -1.414), (-1, 0.707, 0.924 + 0.383i, 1, -1.414)\}$ |
| $[FR_1^{6,1,0}]_{1,2,3}$ | ✕ 🌐 🎗 ⭕ | $\{(-1, -0.707, -0.383 + 0.924i, -1, -1.414), (-1, 0.707, 0.924 + 0.383i, -1, 1.414)\}$ |
| $[FR_1^{6,1,0}]_{1,2,4}$ | ✕ 🌐 🎗 ⭕ | $\{(-1, -0.707, -0.383 + 0.924i, -1, 1.414), (-1, 0.707, 0.924 + 0.383i, -1, -1.414)\}$ |
| $[FR_1^{6,1,0}]_{1,3,1}$ | ✕ † 🌐 🎗 ⭕ | $\{(-1, -0.707, 0.383 - 0.924i, 1, 1.414), (-1, 0.707, 0.924 + 0.383i, 1, 1.414)\}$ |
| $[FR_1^{6,1,0}]_{1,3,2}$ | ✕ 🌐 🎗 ⭕ | $\{(-1, -0.707, 0.383 - 0.924i, 1, -1.414), (-1, 0.707, 0.924 + 0.383i, 1, -1.414)\}$ |

| Name | Properties | | $X_1$ |
|---|---|---|---|
| $[FR_1^{6,1,0}]_{1,3,3}$ | ✗ | 🌐🎗️⚭ | $\{(-1, -0.707, 0.383 - 0.924i, -1, -1.414),$ $(-1, 0.707, 0.924 + 0.383i, -1, 1.414)\}$ |
| $[FR_1^{6,1,0}]_{1,3,4}$ | ✗ | 🌐🎗️⚭ | $\{(-1, -0.707, 0.383 - 0.924i, -1, 1.414),$ $(-1, 0.707, 0.924 + 0.383i, -1, -1.414)\}$ |
| $[FR_1^{6,1,0}]_{1,4,1}$ | ✗† | 🌐🎗️⚭ | $\{(-1, -0.707, 0.383 - 0.924i, 1, 1.414),$ $(-1, 0.707, -0.924 - 0.383i, 1, 1.414)\}$ |
| $[FR_1^{6,1,0}]_{1,4,2}$ | ✗ | 🌐🎗️⚭ | $\{(-1, -0.707, 0.383 - 0.924i, 1, -1.414),$ $(-1, 0.707, -0.924 - 0.383i, 1, -1.414)\}$ |
| $[FR_1^{6,1,0}]_{1,4,3}$ | ✗ | 🌐🎗️⚭ | $\{(-1, -0.707, 0.383 - 0.924i, -1, -1.414),$ $(-1, 0.707, -0.924 - 0.383i, -1, 1.414)\}$ |
| $[FR_1^{6,1,0}]_{1,4,4}$ | ✗ | 🌐🎗️⚭ | $\{(-1, -0.707, 0.383 - 0.924i, -1, 1.414),$ $(-1, 0.707, -0.924 - 0.383i, -1, -1.414)\}$ |
| $[FR_1^{6,1,0}]_{1,5,1}$ | ✗† | 🌐🎗️⚭ | $\{(-1, -0.707, 0.383 + 0.924i, 1, 1.414),$ $(-1, 0.707, -0.924 + 0.383i, 1, 1.414)\}$ |
| $[FR_1^{6,1,0}]_{1,5,2}$ | ✗ | 🌐🎗️⚭ | $\{(-1, -0.707, 0.383 + 0.924i, 1, -1.414),$ $(-1, 0.707, -0.924 + 0.383i, 1, -1.414)\}$ |
| $[FR_1^{6,1,0}]_{1,5,3}$ | ✗ | 🌐🎗️⚭ | $\{(-1, -0.707, 0.383 + 0.924i, -1, -1.414),$ $(-1, 0.707, -0.924 + 0.383i, -1, 1.414)\}$ |
| $[FR_1^{6,1,0}]_{1,5,4}$ | ✗ | 🌐🎗️⚭ | $\{(-1, -0.707, 0.383 + 0.924i, -1, 1.414),$ $(-1, 0.707, -0.924 + 0.383i, -1, -1.414)\}$ |
| $[FR_1^{6,1,0}]_{1,6,1}$ | ✗† | 🌐🎗️⚭ | $\{(-1, -0.707, 0.383 + 0.924i, 1, 1.414),$ $(-1, 0.707, 0.924 - 0.383i, 1, 1.414)\}$ |
| $[FR_1^{6,1,0}]_{1,6,2}$ | ✗ | 🌐🎗️⚭ | $\{(-1, -0.707, 0.383 + 0.924i, 1, -1.414),$ $(-1, 0.707, 0.924 - 0.383i, 1, -1.414)\}$ |
| $[FR_1^{6,1,0}]_{1,6,3}$ | ✗ | 🌐🎗️⚭ | $\{(-1, -0.707, 0.383 + 0.924i, -1, -1.414),$ $(-1, 0.707, 0.924 - 0.383i, -1, 1.414)\}$ |
| $[FR_1^{6,1,0}]_{1,6,4}$ | ✗ | 🌐🎗️⚭ | $\{(-1, -0.707, 0.383 + 0.924i, -1, 1.414),$ $(-1, 0.707, 0.924 - 0.383i, -1, -1.414)\}$ |
| $[FR_1^{6,1,0}]_{1,7,1}$ | ✗† | 🌐🎗️⚭ | $\{(-1, -0.707, -0.383 - 0.924i, 1, 1.414),$ $(-1, 0.707, 0.924 - 0.383i, 1, 1.414)\}$ |
| $[FR_1^{6,1,0}]_{1,7,2}$ | ✗ | 🌐🎗️⚭ | $\{(-1, -0.707, -0.383 - 0.924i, 1, -1.414),$ $(-1, 0.707, 0.924 - 0.383i, 1, -1.414)\}$ |
| $[FR_1^{6,1,0}]_{1,7,3}$ | ✗ | 🌐🎗️⚭ | $\{(-1, -0.707, -0.383 - 0.924i, -1, -1.414),$ $(-1, 0.707, 0.924 - 0.383i, -1, 1.414)\}$ |
| $[FR_1^{6,1,0}]_{1,7,4}$ | ✗ | 🌐🎗️⚭ | $\{(-1, -0.707, -0.383 - 0.924i, -1, 1.414),$ $(-1, 0.707, 0.924 - 0.383i, -1, -1.414)\}$ |
| $[FR_1^{6,1,0}]_{1,8,1}$ | ✗† | 🌐🎗️⚭ | $\{(-1, -0.707, -0.383 - 0.924i, 1, 1.414),$ $(-1, 0.707, -0.924 + 0.383i, 1, 1.414)\}$ |

Table 52: Numeric invariants (Continued)

| Name | Properties | | | | $X_1$ |
|---|---|---|---|---|---|
| $[FR_1^{6,1,0}]_{1,8,2}$ | ✗ | 🌐🎗️⚭ | | | $\left\{\begin{array}{l}(-1,\ -0.707,\ -0.383-0.924i,\ 1,\ -1.414),\\(-1,\ \phantom{-}0.707,\ -0.924+0.383i,\ 1,\ -1.414)\end{array}\right\}$ |
| $[FR_1^{6,1,0}]_{1,8,3}$ | ✗ | 🌐🎗️⚭ | | | $\left\{\begin{array}{l}(-1,\ -0.707,\ -0.383-0.924i,\ -1,\ -1.414),\\(-1,\ \phantom{-}0.707,\ -0.924+0.383i,\ -1,\ \phantom{-}1.414)\end{array}\right\}$ |
| $[FR_1^{6,1,0}]_{1,8,4}$ | ✗ | 🌐🎗️⚭ | | | $\left\{\begin{array}{l}(-1,\ -0.707,\ -0.383-0.924i,\ -1,\ \phantom{-}1.414),\\(-1,\ \phantom{-}0.707,\ -0.924+0.383i,\ -1,\ -1.414)\end{array}\right\}$ |
| $[FR_1^{6,1,0}]_{2,1,1}$ | ✗† | 🌐🎗️ | | | $\left\{\ (1,\ 0.707,\ \phantom{-}0.924-0.383i,\ 1,\ \phantom{-}1.414)\ \right\}$ |
| $[FR_1^{6,1,0}]_{2,1,2}$ | ✗ | 🌐🎗️ | | | $\left\{\ (1,\ 0.707,\ \phantom{-}0.924-0.383i,\ 1,\ -1.414)\ \right\}$ |
| $[FR_1^{6,1,0}]_{2,1,3}$ | ✗ | 🌐🎗️ | | | $\left\{\begin{array}{l}(1,\ 0.707,\ \phantom{-}0.924-0.383i,\ -1,\ -1.414),\\(1,\ 0.707,\ \phantom{-}0.924-0.383i,\ -1,\ \phantom{-}1.414)\end{array}\right\}$ |
| $[FR_1^{6,1,0}]_{2,1,4}$ | ✗† | 🌐🎗️ | | | $\left\{\begin{array}{l}(1,\ 0.707,\ -0.924+0.383i,\ 1,\ \phantom{-}1.414),\\(1,\ 0.707,\ \phantom{-}0.924-0.383i,\ 1,\ \phantom{-}1.414)\end{array}\right\}$ |
| $[FR_1^{6,1,0}]_{2,2,1}$ | ✗ | 🌐🎗️ | | | $\left\{\begin{array}{l}(1,\ 0.707,\ -0.924+0.383i,\ 1,\ -1.414),\\(1,\ 0.707,\ \phantom{-}0.924-0.383i,\ 1,\ -1.414)\end{array}\right\}$ |
| $[FR_1^{6,1,0}]_{2,2,2}$ | ✗ | 🌐🎗️ | | | $\left\{\begin{array}{l}(1,\ 0.707,\ -0.924+0.383i,\ -1,\ -1.414),\\(1,\ 0.707,\ \phantom{-}0.924-0.383i,\ -1,\ \phantom{-}1.414)\end{array}\right\}$ |
| $[FR_1^{6,1,0}]_{2,2,3}$ | ✗ | 🌐🎗️ | | | $\left\{\begin{array}{l}(1,\ 0.707,\ -0.924+0.383i,\ -1,\ \phantom{-}1.414),\\(1,\ 0.707,\ \phantom{-}0.924-0.383i,\ -1,\ -1.414)\end{array}\right\}$ |
| $[FR_1^{6,1,0}]_{2,2,4}$ | ✗† | 🌐🎗️ | | | $\left\{\ (1,\ 0.707,\ -0.924+0.383i,\ 1,\ 1.414)\ \right\}$ |
| $[FR_1^{6,1,0}]_{2,3,1}$ | ✗ | 🌐🎗️ | | | $\left\{\ (1,\ 0.707,\ -0.924+0.383i,\ 1,\ -1.414)\ \right\}$ |
| $[FR_1^{6,1,0}]_{2,3,2}$ | ✗ | 🌐🎗️ | | | $\left\{\begin{array}{l}(1,\ 0.707,\ -0.924+0.383i,\ -1,\ -1.414),\\(1,\ 0.707,\ -0.924+0.383i,\ -1,\ \phantom{-}1.414)\end{array}\right\}$ |
| $[FR_1^{6,1,0}]_{2,3,3}$ | ✗† | 🌐🎗️ | | | $\left\{\ (1,\ 0.707,\ 0.924+0.383i,\ 1,\ 1.414)\ \right\}$ |
| $[FR_1^{6,1,0}]_{2,4,1}$ | ✗ | 🌐🎗️ | | | $\left\{\ (1,\ 0.707,\ 0.924+0.383i,\ 1,\ -1.414)\ \right\}$ |
| $[FR_1^{6,1,0}]_{2,4,2}$ | ✗ | 🌐🎗️ | | | $\left\{\begin{array}{l}(1,\ 0.707,\ 0.924+0.383i,\ -1,\ -1.414),\\(1,\ 0.707,\ 0.924+0.383i,\ -1,\ \phantom{-}1.414)\end{array}\right\}$ |
| $[FR_1^{6,1,0}]_{2,4,3}$ | ✗† | 🌐🎗️ | | | $\left\{\begin{array}{l}(1,\ 0.707,\ -0.924-0.383i,\ 1,\ 1.414),\\(1,\ 0.707,\ \phantom{-}0.924+0.383i,\ 1,\ 1.414)\end{array}\right\}$ |
| $[FR_1^{6,1,0}]_{2,5,1}$ | ✗ | 🌐🎗️ | | | $\left\{\begin{array}{l}(1,\ 0.707,\ -0.924-0.383i,\ 1,\ -1.414),\\(1,\ 0.707,\ \phantom{-}0.924+0.383i,\ 1,\ -1.414)\end{array}\right\}$ |
| $[FR_1^{6,1,0}]_{2,5,2}$ | ✗ | 🌐🎗️ | | | $\left\{\begin{array}{l}(1,\ 0.707,\ -0.924-0.383i,\ -1,\ -1.414),\\(1,\ 0.707,\ \phantom{-}0.924+0.383i,\ -1,\ \phantom{-}1.414)\end{array}\right\}$ |
| $[FR_1^{6,1,0}]_{2,5,3}$ | ✗ | 🌐🎗️ | | | $\left\{\begin{array}{l}(1,\ 0.707,\ -0.924-0.383i,\ -1,\ \phantom{-}1.414),\\(1,\ 0.707,\ \phantom{-}0.924+0.383i,\ -1,\ -1.414)\end{array}\right\}$ |
| $[FR_1^{6,1,0}]_{2,6,1}$ | ✗† | 🌐🎗️ | | | $\left\{\ (1,\ 0.707,\ -0.924-0.383i,\ 1,\ 1.414)\ \right\}$ |
| $[FR_1^{6,1,0}]_{2,6,2}$ | ✗ | 🌐🎗️ | | | $\left\{\ (1,\ 0.707,\ -0.924-0.383i,\ 1,\ -1.414)\ \right\}$ |

Table 52: Numeric invariants (Continued)

| Name | Properties | $X_1$ |
|---|---|---|
| $[FR_1^{6,1,0}]_{2,6,3}$ | ✗ 🌐 🎀 | $\{(1, 0.707, -0.924 - 0.383i, -1, -1.414),$ $(1, 0.707, -0.924 - 0.383i, -1, 1.414)\}$ |
| $[FR_1^{6,1,0}]_{3,1,1}$ | ✗ † 🌐 🎀 | $\{(1, -0.707, -0.383 - 0.924i, 1, 1.414),$ $(1, -0.707, 0.383 + 0.924i, 1, 1.414)\}$ |
| $[FR_1^{6,1,0}]_{3,1,2}$ | ✗ 🌐 🎀 | $\{(1, -0.707, -0.383 - 0.924i, 1, -1.414),$ $(1, -0.707, 0.383 + 0.924i, 1, -1.414)\}$ |
| $[FR_1^{6,1,0}]_{3,1,3}$ | ✗ 🌐 🎀 | $\{(1, -0.707, -0.383 - 0.924i, -1, 1.414),$ $(1, -0.707, 0.383 + 0.924i, -1, -1.414)\}$ |
| $[FR_1^{6,1,0}]_{3,1,4}$ | ✗ 🌐 🎀 | $\{(1, -0.707, -0.383 - 0.924i, -1, -1.414),$ $(1, -0.707, 0.383 + 0.924i, -1, 1.414)\}$ |
| $[FR_1^{6,1,0}]_{3,2,1}$ | ✗ † 🌐 🎀 | $\{(1, -0.707, 0.383 + 0.924i, 1, 1.414)\}$ |
| $[FR_1^{6,1,0}]_{3,2,2}$ | ✗ 🌐 🎀 | $\{(1, -0.707, 0.383 + 0.924i, 1, -1.414)\}$ |
| $[FR_1^{6,1,0}]_{3,2,3}$ | ✗ 🌐 🎀 | $\{(1, -0.707, 0.383 + 0.924i, -1, -1.414),$ $(1, -0.707, 0.383 + 0.924i, -1, 1.414)\}$ |
| $[FR_1^{6,1,0}]_{3,2,4}$ | ✗ † 🌐 🎀 | $\{(1, -0.707, -0.383 - 0.924i, 1, 1.414)\}$ |
| $[FR_1^{6,1,0}]_{3,3,1}$ | ✗ 🌐 🎀 | $\{(1, -0.707, -0.383 - 0.924i, 1, -1.414)\}$ |
| $[FR_1^{6,1,0}]_{3,3,2}$ | ✗ 🌐 🎀 | $\{(1, -0.707, -0.383 - 0.924i, -1, -1.414),$ $(1, -0.707, -0.383 - 0.924i, -1, 1.414)\}$ |
| $[FR_1^{6,1,0}]_{3,3,3}$ | ✗ † 🌐 🎀 | $\{(1, -0.707, -0.383 + 0.924i, 1, 1.414),$ $(1, -0.707, 0.383 - 0.924i, 1, 1.414)\}$ |
| $[FR_1^{6,1,0}]_{3,4,1}$ | ✗ 🌐 🎀 | $\{(1, -0.707, -0.383 + 0.924i, 1, -1.414),$ $(1, -0.707, 0.383 - 0.924i, 1, -1.414)\}$ |
| $[FR_1^{6,1,0}]_{3,4,2}$ | ✗ 🌐 🎀 | $\{(1, -0.707, -0.383 + 0.924i, -1, 1.414),$ $(1, -0.707, 0.383 - 0.924i, -1, -1.414)\}$ |
| $[FR_1^{6,1,0}]_{3,4,3}$ | ✗ 🌐 🎀 | $\{(1, -0.707, -0.383 + 0.924i, -1, -1.414),$ $(1, -0.707, 0.383 - 0.924i, -1, 1.414)\}$ |
| $[FR_1^{6,1,0}]_{3,5,1}$ | ✗ † 🌐 🎀 | $\{1, -0.707, 0.383 - 0.924i, 1, 1.414)\}$ |
| $[FR_1^{6,1,0}]_{3,5,2}$ | ✗ 🌐 🎀 | $\{(1, -0.707, 0.383 - 0.924i, 1, -1.414)\}$ |
| $[FR_1^{6,1,0}]_{3,5,3}$ | ✗ 🌐 🎀 | $\{(1, -0.707, 0.383 - 0.924i, -1, -1.414),$ $(1, -0.707, 0.383 - 0.924i, -1, 1.414)\}$ |
| $[FR_1^{6,1,0}]_{3,6,1}$ | ✗ † 🌐 🎀 | $\{(1, -0.707, -0.383 + 0.924i, 1, 1.414)\}$ |
| $[FR_1^{6,1,0}]_{3,6,2}$ | ✗ 🌐 🎀 | $\{(1, -0.707, -0.383 + 0.924i, 1, -1.414)\}$ |
| $[FR_1^{6,1,0}]_{3,6,3}$ | ✗ 🌐 🎀 | $\{(1, -0.707, -0.383 + 0.924i, -1, -1.414),$ $(1, -0.707, -0.383 + 0.924i, -1, 1.414)\}$ |
| $[FR_1^{6,1,0}]_{4,0,1}$ | † 🌐 | $\{(1, -0.707, -, 1, 1.414),$ $(1, 0.707, -, 1, 1.414)\}$ |

Table 52: Numeric invariants (Continued)

| Name | Properties | $X_1$ |
|---|---|---|
| $[FR_1^{6,1,0}]_{4,0,2}$ |  | $\{(1, -0.707, -, 1, -1.414), (1, 0.707, -, 1, -1.414)\}$ |
| $[FR_1^{6,1,0}]_{4,0,3}$ |  | $\{(1, -0.707, -, -1, 1.414), (1, 0.707, -, -1, -1.414)\}$ |
| $[FR_1^{6,1,0}]_{4,0,4}$ |  | $\{(1, -0.707, -, -1, -1.414), (1, 0.707, -, -1, 1.414)\}$ |
| $[FR_1^{6,1,0}]_{5,0,1}$ |  | $\{(-1, 0.707, -, 1, 1.414)\}$ |
| $[FR_1^{6,1,0}]_{5,0,2}$ |  | $\{(-1, 0.707, -, 1, -1.414)\}$ |
| $[FR_1^{6,1,0}]_{5,0,3}$ |  | $\{(-1, 0.707, -, -1, 1.414)\}$ |
| $[FR_1^{6,1,0}]_{6,0,1}$ |  | $\{(-1, -0.707, -, 1, 1.414)\}$ |
| $[FR_1^{6,1,0}]_{6,0,2}$ |  | $\{(-1, -0.707, -, 1, -1.414)\}$ |
| $[FR_1^{6,1,0}]_{6,0,3}$ |  | $\{(-1, -0.707, -, -1, -1.414)\}$ |

### 4.5.28  $FR_2^{6,1,0}$: $\mathbb{Z}_2 \otimes \textbf{Rep}(D_3)$

For the fusion ring, the following multiplication table is used.

| 1 | 2 | 3 | 4 | 5 | 6 |
|---|---|---|---|---|---|
| 2 | 1 | 4 | 3 | 6 | 5 |
| 3 | 4 | 1 | 2 | 6 | 5 |
| 4 | 3 | 2 | 1 | 5 | 6 |
| 5 | 6 | 6 | 5 | $1+4+6$ | $2+3+5$ |
| 6 | 5 | 5 | 6 | $2+3+5$ | $1+4+6$ |

The following is the group of all non-trivial permutations that leave the fusion rules invariant:

$$S = \{(), (2\ 3)\}.$$

Let

$$X_1 = S\left([F_3^{232}]_4^4 [F_4^{242}]_3^3\right), \tag{59}$$
$$X_2 = S\left(d_2^L\right). \tag{60}$$

The following table lists a small set of invariants whose values completely distinguish between all MFPBFCs and MFPNBFCs with the given fusion rules.

Table 53: Symbolic invariants

| Name | Properties | $[F_1^{666}]_6^6$ | $X_1$ | $R_1^{55}$ | $X_2$ |
|---|---|---|---|---|---|
| $[FR_2^{6,1,0}]_{1,1,1}$ |  | 1 | $\{1\}$ | $\zeta_3^2$ | $\{1\}$ |
| $[FR_2^{6,1,0}]_{1,1,2}$ |  | 1 | $\{1\}$ | $\zeta_3^2$ | $\{-1\}$ |

Table 53: Symbolic invariants (Continued)

| Name | Properties | $[F_1^{666}]_6^6$ | $X_1$ | $R_1^{55}$ | $X_2$ |
|---|---|---|---|---|---|
| $[FR_2^{6,1,0}]_{1,2,1}$ | ✕ ✝ 🌐 🎗 | 1 | {1} | $\zeta_3$ | {1} |
| $[FR_2^{6,1,0}]_{1,2,2}$ | ✕ 🌐 🎗 | 1 | {1} | $\zeta_3$ | {−1} |
| $[FR_2^{6,1,0}]_{1,3,1}$ | ✕ ✝ 🌐 🎗 | 1 | {1} | 1 | {1} |
| $[FR_2^{6,1,0}]_{1,3,2}$ | ✕ 🌐 🎗 | 1 | {1} | 1 | {−1} |
| $[FR_2^{6,1,0}]_{1,4,1}$ | ✕ ✝ 🌐 🎗 | 1 | {1} | $\zeta_6$ | {1} |
| $[FR_2^{6,1,0}]_{1,4,2}$ | ✕ 🌐 🎗 | 1 | {1} | $\zeta_6$ | {−1} |
| $[FR_2^{6,1,0}]_{1,5,1}$ | ✕ ✝ 🌐 🎗 | 1 | {1} | $\zeta_6^5$ | {1} |
| $[FR_2^{6,1,0}]_{1,5,2}$ | ✕ 🌐 🎗 | 1 | {1} | $\zeta_6^5$ | {−1} |
| $[FR_2^{6,1,0}]_{1,6,1}$ | ✕ ✝ 🌐 🎗 | 1 | {1} | −1 | {1} |
| $[FR_2^{6,1,0}]_{1,6,2}$ | ✕ 🌐 🎗 | 1 | {1} | −1 | {−1} |
| $[FR_2^{6,1,0}]_{2,1,1}$ | ✕ ✝ 🌐 🎗 | 1 | {−1} | $\zeta_{12}^5$ | {1} |
| $[FR_2^{6,1,0}]_{2,1,2}$ | ✕ 🌐 🎗 | 1 | {−1} | $\zeta_{12}^5$ | {−1} |
| $[FR_2^{6,1,0}]_{2,2,1}$ | ✕ ✝ 🌐 🎗 | 1 | {−1} | $-i$ | {1} |
| $[FR_2^{6,1,0}]_{2,2,2}$ | ✕ 🌐 🎗 | 1 | {−1} | $-i$ | {−1} |
| $[FR_2^{6,1,0}]_{2,3,1}$ | ✕ ✝ 🌐 🎗 | 1 | {−1} | $\zeta_{12}$ | {1} |
| $[FR_2^{6,1,0}]_{2,3,2}$ | ✕ 🌐 🎗 | 1 | {−1} | $\zeta_{12}$ | {−1} |
| $[FR_2^{6,1,0}]_{2,4,1}$ | ✕ ✝ 🌐 🎗 | 1 | {−1} | $\zeta_{12}^{11}$ | {1} |
| $[FR_2^{6,1,0}]_{2,4,2}$ | ✕ 🌐 🎗 | 1 | {−1} | $\zeta_{12}^{11}$ | {−1} |
| $[FR_2^{6,1,0}]_{2,5,1}$ | ✕ ✝ 🌐 🎗 | 1 | {−1} | $i$ | {1} |
| $[FR_2^{6,1,0}]_{2,5,2}$ | ✕ 🌐 🎗 | 1 | {−1} | $i$ | {−1} |
| $[FR_2^{6,1,0}]_{2,6,1}$ | ✕ ✝ 🌐 🎗 | 1 | {−1} | $\zeta_{12}^7$ | {1} |
| $[FR_2^{6,1,0}]_{2,6,2}$ | ✕ 🌐 🎗 | 1 | {−1} | $\zeta_{12}^7$ | {−1} |
| $[FR_2^{6,1,0}]_{3,0,1}$ | ✝ 🌐 | $\zeta_3^2$ | {1} | - | {1} |
| $[FR_2^{6,1,0}]_{3,0,2}$ | 🌐 | $\zeta_3^2$ | {1} | - | {−1} |
| $[FR_2^{6,1,0}]_{4,0,1}$ | ✝ 🌐 | $\zeta_3$ | {1} | - | {1} |
| $[FR_2^{6,1,0}]_{4,0,2}$ | 🌐 | $\zeta_3$ | {1} | - | {−1} |
| $[FR_2^{6,1,0}]_{5,0,1}$ | ✝ 🌐 | $\zeta_3^2$ | {−1} | - | {1} |
| $[FR_2^{6,1,0}]_{5,0,2}$ | 🌐 | $\zeta_3^2$ | {−1} | - | {−1} |
| $[FR_2^{6,1,0}]_{6,0,1}$ | ✝ 🌐 | $\zeta_3$ | {−1} | - | {1} |
| $[FR_2^{6,1,0}]_{6,0,2}$ | 🌐 | $\zeta_3$ | {−1} | - | {−1} |

Table 54: Numeric invariants

| Name | Properties | $[F_1^{666}]_6^6$ | $X_1$ | $R_1^{55}$ | $X_2$ |
|---|---|---|---|---|---|
| $[FR_2^{6,1,0}]_{1,1,1}$ | ✗ † 🌐 🎗 | 1 | {1} | $-0.5 - 0.866i$ | {1} |
| $[FR_2^{6,1,0}]_{1,1,2}$ | ✗ 🌐 🎗 | 1 | {1} | $-0.5 - 0.866i$ | {−1} |
| $[FR_2^{6,1,0}]_{1,2,1}$ | ✗ † 🌐 🎗 | 1 | {1} | $-0.5 + 0.866i$ | {1} |
| $[FR_2^{6,1,0}]_{1,2,2}$ | ✗ 🌐 🎗 | 1 | {1} | $-0.5 + 0.866i$ | {−1} |
| $[FR_2^{6,1,0}]_{1,3,1}$ | ✗ † 🌐 🎗 | 1 | {1} | 1 | {1} |
| $[FR_2^{6,1,0}]_{1,3,2}$ | ✗ 🌐 🎗 | 1 | {1} | 1 | {−1} |
| $[FR_2^{6,1,0}]_{1,4,1}$ | ✗ † 🌐 🎗 | 1 | {1} | $0.5 + 0.866i$ | {1} |
| $[FR_2^{6,1,0}]_{1,4,2}$ | ✗ 🌐 🎗 | 1 | {1} | $0.5 + 0.866i$ | {−1} |
| $[FR_2^{6,1,0}]_{1,5,1}$ | ✗ † 🌐 🎗 | 1 | {1} | $0.5 - 0.866i$ | {1} |
| $[FR_2^{6,1,0}]_{1,5,2}$ | ✗ 🌐 🎗 | 1 | {1} | $0.5 - 0.866i$ | {−1} |
| $[FR_2^{6,1,0}]_{1,6,1}$ | ✗ † 🌐 🎗 | 1 | {1} | $-1$ | {1} |
| $[FR_2^{6,1,0}]_{1,6,2}$ | ✗ 🌐 🎗 | 1 | {1} | $-1$ | {−1} |
| $[FR_2^{6,1,0}]_{2,1,1}$ | ✗ † 🌐 🎗 | 1 | {−1} | $-0.866 + 0.5i$ | {1} |
| $[FR_2^{6,1,0}]_{2,1,2}$ | ✗ 🌐 🎗 | 1 | {−1} | $-0.866 + 0.5i$ | {−1} |
| $[FR_2^{6,1,0}]_{2,2,1}$ | ✗ † 🌐 🎗 | 1 | {−1} | $-i$ | {1} |
| $[FR_2^{6,1,0}]_{2,2,2}$ | ✗ 🌐 🎗 | 1 | {−1} | $-i$ | {−1} |
| $[FR_2^{6,1,0}]_{2,3,1}$ | ✗ † 🌐 🎗 | 1 | {−1} | $0.866 + 0.5i$ | {1} |
| $[FR_2^{6,1,0}]_{2,3,2}$ | ✗ 🌐 🎗 | 1 | {−1} | $0.866 + 0.5i$ | {−1} |
| $[FR_2^{6,1,0}]_{2,4,1}$ | ✗ † 🌐 🎗 | 1 | {−1} | $0.866 - 0.5i$ | {1} |
| $[FR_2^{6,1,0}]_{2,4,2}$ | ✗ 🌐 🎗 | 1 | {−1} | $0.866 - 0.5i$ | {−1} |
| $[FR_2^{6,1,0}]_{2,5,1}$ | ✗ † 🌐 🎗 | 1 | {−1} | $i$ | {1} |
| $[FR_2^{6,1,0}]_{2,5,2}$ | ✗ 🌐 🎗 | 1 | {−1} | $i$ | {−1} |
| $[FR_2^{6,1,0}]_{2,6,1}$ | ✗ † 🌐 🎗 | 1 | {−1} | $-0.866 - 0.5i$ | {1} |
| $[FR_2^{6,1,0}]_{2,6,2}$ | ✗ 🌐 🎗 | 1 | {−1} | $-0.866 - 0.5i$ | {−1} |
| $[FR_2^{6,1,0}]_{3,0,1}$ | † 🌐 | $-0.5 - 0.866i$ | {1} | - | {1} |
| $[FR_2^{6,1,0}]_{3,0,2}$ | 🌐 | $-0.5 - 0.866i$ | {1} | - | {−1} |
| $[FR_2^{6,1,0}]_{4,0,1}$ | † 🌐 | $-0.5 + 0.866i$ | {1} | - | {1} |
| $[FR_2^{6,1,0}]_{4,0,2}$ | 🌐 | $-0.5 + 0.866i$ | {1} | - | {−1} |
| $[FR_2^{6,1,0}]_{5,0,1}$ | † 🌐 | $-0.5 - 0.866i$ | {−1} | - | {1} |
| $[FR_2^{6,1,0}]_{5,0,2}$ | 🌐 | $-0.5 - 0.866i$ | {−1} | - | {−1} |
| $[FR_2^{6,1,0}]_{6,0,1}$ | † 🌐 | $-0.5 + 0.866i$ | {−1} | - | {1} |
| $[FR_2^{6,1,0}]_{6,0,2}$ | 🌐 | $-0.5 + 0.866i$ | {−1} | - | {−1} |

## 4.5.29 $FR_4^{6,1,0}$: Fib ⊗ Ising

For the fusion ring, the following multiplication table is used.

| 1 | 2 | 3   | 4     | 5     | 6         |
|---|---|-----|-------|-------|-----------|
| 2 | 1 | 3   | 5     | 4     | 6         |
| 3 | 3 | 1+2 | 6     | 6     | 4+5       |
| 4 | 5 | 6   | 1+5   | 2+4   | 3+6       |
| 5 | 4 | 6   | 2+4   | 1+5   | 3+6       |
| 6 | 6 | 4+5 | 3+6   | 3+6   | 1+2+4+5   |

Only the trivial permutation leaves the fusion rules invariant.

The following table lists a small set of invariants whose values completely distinguish between all MFPBFCs and MFPNBFCs with the given fusion rules.

Table 55: Symbolic invariants

| Name | Properties | $[F_1^{656}]_6^6[F_6^{666}]_5^5$ | $R_1^{66}$ | $d_6^L$ |
|---|---|---|---|---|
| $[FR_4^{6,1,0}]_{1,1,1}$ | ✗ ✝ 🌐 🎗 ⭕ | $-\bar\phi^2$ | $\zeta_{80}^{3}$ | $\sqrt{3+\sqrt{5}}$ |
| $[FR_4^{6,1,0}]_{1,1,2}$ | ✗ 🌐 🎗 ⭕ | $-\bar\phi^2$ | $\zeta_{80}^{3}$ | $-\sqrt{3+\sqrt{5}}$ |
| $[FR_4^{6,1,0}]_{1,2,1}$ | ✗ ✝ 🌐 🎗 ⭕ | $-\bar\phi^2$ | $\zeta_{80}^{67}$ | $\sqrt{3+\sqrt{5}}$ |
| $[FR_4^{6,1,0}]_{1,2,2}$ | ✗ 🌐 🎗 ⭕ | $-\bar\phi^2$ | $\zeta_{80}^{67}$ | $-\sqrt{3+\sqrt{5}}$ |
| $[FR_4^{6,1,0}]_{1,3,1}$ | ✗ ✝ 🌐 🎗 ⭕ | $-\bar\phi^2$ | $\zeta_{80}^{53}$ | $\sqrt{3+\sqrt{5}}$ |
| $[FR_4^{6,1,0}]_{1,3,2}$ | ✗ 🌐 🎗 ⭕ | $-\bar\phi^2$ | $\zeta_{80}^{53}$ | $-\sqrt{3+\sqrt{5}}$ |
| $[FR_4^{6,1,0}]_{1,4,1}$ | ✗ ✝ 🌐 🎗 ⭕ | $-\bar\phi^2$ | $\zeta_{80}^{37}$ | $\sqrt{3+\sqrt{5}}$ |
| $[FR_4^{6,1,0}]_{1,4,2}$ | ✗ 🌐 🎗 ⭕ | $-\bar\phi^2$ | $\zeta_{80}^{37}$ | $-\sqrt{3+\sqrt{5}}$ |
| $[FR_4^{6,1,0}]_{1,5,1}$ | ✗ ✝ 🌐 🎗 ⭕ | $-\bar\phi^2$ | $\zeta_{80}^{13}$ | $\sqrt{3+\sqrt{5}}$ |
| $[FR_4^{6,1,0}]_{1,5,2}$ | ✗ 🌐 🎗 ⭕ | $-\bar\phi^2$ | $\zeta_{80}^{13}$ | $-\sqrt{3+\sqrt{5}}$ |
| $[FR_4^{6,1,0}]_{1,6,1}$ | ✗ ✝ 🌐 🎗 ⭕ | $-\bar\phi^2$ | $\zeta_{80}^{77}$ | $\sqrt{3+\sqrt{5}}$ |
| $[FR_4^{6,1,0}]_{1,6,2}$ | ✗ 🌐 🎗 ⭕ | $-\bar\phi^2$ | $\zeta_{80}^{77}$ | $-\sqrt{3+\sqrt{5}}$ |
| $[FR_4^{6,1,0}]_{1,7,1}$ | ✗ ✝ 🌐 🎗 ⭕ | $-\bar\phi^2$ | $\zeta_{80}^{43}$ | $\sqrt{3+\sqrt{5}}$ |
| $[FR_4^{6,1,0}]_{1,7,2}$ | ✗ 🌐 🎗 ⭕ | $-\bar\phi^2$ | $\zeta_{80}^{43}$ | $-\sqrt{3+\sqrt{5}}$ |
| $[FR_4^{6,1,0}]_{1,8,1}$ | ✗ ✝ 🌐 🎗 ⭕ | $-\bar\phi^2$ | $\zeta_{80}^{27}$ | $\sqrt{3+\sqrt{5}}$ |
| $[FR_4^{6,1,0}]_{1,8,2}$ | ✗ 🌐 🎗 ⭕ | $-\bar\phi^2$ | $\zeta_{80}^{27}$ | $-\sqrt{3+\sqrt{5}}$ |
| $[FR_4^{6,1,0}]_{2,1,1}$ | ✗ ✝ 🌐 🎗 ⭕ | $\bar\phi^2$ | $\zeta_{80}^{73}$ | $\sqrt{3+\sqrt{5}}$ |
| $[FR_4^{6,1,0}]_{2,1,2}$ | ✗ 🌐 🎗 ⭕ | $\bar\phi^2$ | $\zeta_{80}^{73}$ | $-\sqrt{3+\sqrt{5}}$ |
| $[FR_4^{6,1,0}]_{2,2,1}$ | ✗ ✝ 🌐 🎗 ⭕ | $\bar\phi^2$ | $\zeta_{80}^{57}$ | $\sqrt{3+\sqrt{5}}$ |
| $[FR_4^{6,1,0}]_{2,2,2}$ | ✗ 🌐 🎗 ⭕ | $\bar\phi^2$ | $\zeta_{80}^{57}$ | $-\sqrt{3+\sqrt{5}}$ |
| $[FR_4^{6,1,0}]_{2,3,1}$ | ✗ ✝ 🌐 🎗 ⭕ | $\bar\phi^2$ | $\zeta_{80}^{63}$ | $\sqrt{3+\sqrt{5}}$ |

Table 55: Symbolic invariants (Continued)

| Name | Properties | $[F_1^{656}]_6^6[F_6^{666}]_5^5$ | $R_1^{66}$ | $d_6^L$ |
|---|---|---|---|---|
| $[FR_4^{6,1,0}]_{2,3,2}$ | ✕ 🌐 🎗 ⭕ | $\overline{\phi}^2$ | $\zeta_{80}^{63}$ | $-\sqrt{3+\sqrt{5}}$ |
| $[FR_4^{6,1,0}]_{2,4,1}$ | ✕ † 🌐 🎗 ⭕ | $\overline{\phi}^2$ | $\zeta_{80}^{47}$ | $\sqrt{3+\sqrt{5}}$ |
| $[FR_4^{6,1,0}]_{2,4,2}$ | ✕ 🌐 🎗 ⭕ | $\overline{\phi}^2$ | $\zeta_{80}^{47}$ | $-\sqrt{3+\sqrt{5}}$ |
| $[FR_4^{6,1,0}]_{2,5,1}$ | ✕ † 🌐 🎗 ⭕ | $\overline{\phi}^2$ | $\zeta_{80}^{23}$ | $\sqrt{3+\sqrt{5}}$ |
| $[FR_4^{6,1,0}]_{2,5,2}$ | ✕ 🌐 🎗 ⭕ | $\overline{\phi}^2$ | $\zeta_{80}^{23}$ | $-\sqrt{3+\sqrt{5}}$ |
| $[FR_4^{6,1,0}]_{2,6,1}$ | ✕ † 🌐 🎗 ⭕ | $\overline{\phi}^2$ | $\zeta_{80}^{7}$ | $\sqrt{3+\sqrt{5}}$ |
| $[FR_4^{6,1,0}]_{2,6,2}$ | ✕ 🌐 🎗 ⭕ | $\overline{\phi}^2$ | $\zeta_{80}^{7}$ | $-\sqrt{3+\sqrt{5}}$ |
| $[FR_4^{6,1,0}]_{2,7,1}$ | ✕ † 🌐 🎗 ⭕ | $\overline{\phi}^2$ | $\zeta_{80}^{33}$ | $\sqrt{3+\sqrt{5}}$ |
| $[FR_4^{6,1,0}]_{2,7,2}$ | ✕ 🌐 🎗 ⭕ | $\overline{\phi}^2$ | $\zeta_{80}^{33}$ | $-\sqrt{3+\sqrt{5}}$ |
| $[FR_4^{6,1,0}]_{2,8,1}$ | ✕ † 🌐 🎗 ⭕ | $\overline{\phi}^2$ | $\zeta_{80}^{17}$ | $\sqrt{3+\sqrt{5}}$ |
| $[FR_4^{6,1,0}]_{2,8,2}$ | ✕ 🌐 🎗 ⭕ | $\overline{\phi}^2$ | $\zeta_{80}^{17}$ | $-\sqrt{3+\sqrt{5}}$ |
| $[FR_4^{6,1,0}]_{3,1,1}$ | ✕ 🌐 🎗 ⭕ | $-\phi^2$ | $\zeta_{80}^{49}$ | $-\sqrt{3-\sqrt{5}}$ |
| $[FR_4^{6,1,0}]_{3,1,2}$ | ✕ 🌐 🎗 ⭕ | $-\phi^2$ | $\zeta_{80}^{49}$ | $\sqrt{3-\sqrt{5}}$ |
| $[FR_4^{6,1,0}]_{3,2,1}$ | ✕ 🌐 🎗 ⭕ | $-\phi^2$ | $\zeta_{80}$ | $-\sqrt{3-\sqrt{5}}$ |
| $[FR_4^{6,1,0}]_{3,2,2}$ | ✕ 🌐 🎗 ⭕ | $-\phi^2$ | $\zeta_{80}$ | $\sqrt{3-\sqrt{5}}$ |
| $[FR_4^{6,1,0}]_{3,3,1}$ | ✕ 🌐 🎗 ⭕ | $-\phi^2$ | $\zeta_{80}^{39}$ | $-\sqrt{3-\sqrt{5}}$ |
| $[FR_4^{6,1,0}]_{3,3,2}$ | ✕ 🌐 🎗 ⭕ | $-\phi^2$ | $\zeta_{80}^{39}$ | $\sqrt{3-\sqrt{5}}$ |
| $[FR_4^{6,1,0}]_{3,4,1}$ | ✕ 🌐 🎗 ⭕ | $-\phi^2$ | $\zeta_{80}^{71}$ | $-\sqrt{3-\sqrt{5}}$ |
| $[FR_4^{6,1,0}]_{3,4,2}$ | ✕ 🌐 🎗 ⭕ | $-\phi^2$ | $\zeta_{80}^{71}$ | $\sqrt{3-\sqrt{5}}$ |
| $[FR_4^{6,1,0}]_{3,5,1}$ | ✕ 🌐 🎗 ⭕ | $-\phi^2$ | $\zeta_{80}^{79}$ | $-\sqrt{3-\sqrt{5}}$ |
| $[FR_4^{6,1,0}]_{3,5,2}$ | ✕ 🌐 🎗 ⭕ | $-\phi^2$ | $\zeta_{80}^{79}$ | $\sqrt{3-\sqrt{5}}$ |
| $[FR_4^{6,1,0}]_{3,6,1}$ | ✕ 🌐 🎗 ⭕ | $-\phi^2$ | $\zeta_{80}^{31}$ | $-\sqrt{3-\sqrt{5}}$ |
| $[FR_4^{6,1,0}]_{3,6,2}$ | ✕ 🌐 🎗 ⭕ | $-\phi^2$ | $\zeta_{80}^{31}$ | $\sqrt{3-\sqrt{5}}$ |
| $[FR_4^{6,1,0}]_{3,7,1}$ | ✕ 🌐 🎗 ⭕ | $-\phi^2$ | $\zeta_{80}^{9}$ | $-\sqrt{3-\sqrt{5}}$ |
| $[FR_4^{6,1,0}]_{3,7,2}$ | ✕ 🌐 🎗 ⭕ | $-\phi^2$ | $\zeta_{80}^{9}$ | $\sqrt{3-\sqrt{5}}$ |
| $[FR_4^{6,1,0}]_{3,8,1}$ | ✕ 🌐 🎗 ⭕ | $-\phi^2$ | $\zeta_{80}^{41}$ | $-\sqrt{3-\sqrt{5}}$ |
| $[FR_4^{6,1,0}]_{3,8,2}$ | ✕ 🌐 🎗 ⭕ | $-\phi^2$ | $\zeta_{80}^{41}$ | $\sqrt{3-\sqrt{5}}$ |
| $[FR_4^{6,1,0}]_{4,1,1}$ | ✕ 🌐 🎗 ⭕ | $\phi^2$ | $\zeta_{80}^{59}$ | $-\sqrt{3-\sqrt{5}}$ |
| $[FR_4^{6,1,0}]_{4,1,2}$ | ✕ 🌐 🎗 ⭕ | $\phi^2$ | $\zeta_{80}^{59}$ | $\sqrt{3-\sqrt{5}}$ |
| $[FR_4^{6,1,0}]_{4,2,1}$ | ✕ 🌐 🎗 ⭕ | $\phi^2$ | $\zeta_{80}^{11}$ | $-\sqrt{3-\sqrt{5}}$ |
| $[FR_4^{6,1,0}]_{4,2,2}$ | ✕ 🌐 🎗 ⭕ | $\phi^2$ | $\zeta_{80}^{11}$ | $\sqrt{3-\sqrt{5}}$ |

Table 55: Symbolic invariants (Continued)

| Name | Properties | $[F_1^{656}]_6^6[F_6^{666}]_5^5$ | $R_1^{66}$ | $d_6^L$ |
|---|---|---|---|---|
| $[\text{FR}_4^{6,1,0}]_{4,3,1}$ | | $\phi^2$ | $\zeta_{80}^{29}$ | $-\sqrt{3-\sqrt{5}}$ |
| $[\text{FR}_4^{6,1,0}]_{4,3,2}$ | | $\phi^2$ | $\zeta_{80}^{29}$ | $\sqrt{3-\sqrt{5}}$ |
| $[\text{FR}_4^{6,1,0}]_{4,4,1}$ | | $\phi^2$ | $\zeta_{80}^{61}$ | $-\sqrt{3-\sqrt{5}}$ |
| $[\text{FR}_4^{6,1,0}]_{4,4,2}$ | | $\phi^2$ | $\zeta_{80}^{61}$ | $\sqrt{3-\sqrt{5}}$ |
| $[\text{FR}_4^{6,1,0}]_{4,5,1}$ | | $\phi^2$ | $\zeta_{80}^{69}$ | $-\sqrt{3-\sqrt{5}}$ |
| $[\text{FR}_4^{6,1,0}]_{4,5,2}$ | | $\phi^2$ | $\zeta_{80}^{69}$ | $\sqrt{3-\sqrt{5}}$ |
| $[\text{FR}_4^{6,1,0}]_{4,6,1}$ | | $\phi^2$ | $\zeta_{80}^{21}$ | $-\sqrt{3-\sqrt{5}}$ |
| $[\text{FR}_4^{6,1,0}]_{4,6,2}$ | | $\phi^2$ | $\zeta_{80}^{21}$ | $\sqrt{3-\sqrt{5}}$ |
| $[\text{FR}_4^{6,1,0}]_{4,7,1}$ | | $\phi^2$ | $\zeta_{80}^{19}$ | $-\sqrt{3-\sqrt{5}}$ |
| $[\text{FR}_4^{6,1,0}]_{4,7,2}$ | | $\phi^2$ | $\zeta_{80}^{19}$ | $\sqrt{3-\sqrt{5}}$ |
| $[\text{FR}_4^{6,1,0}]_{4,8,1}$ | | $\phi^2$ | $\zeta_{80}^{51}$ | $-\sqrt{3-\sqrt{5}}$ |
| $[\text{FR}_4^{6,1,0}]_{4,8,2}$ | | $\phi^2$ | $\zeta_{80}^{51}$ | $\sqrt{3-\sqrt{5}}$ |

Table 56: Numeric invariants

| Name | Properties | $[F_1^{656}]_6^6[F_6^{666}]_5^5$ | $R_1^{66}$ | $d_6^L$ |
|---|---|---|---|---|
| $[\text{FR}_4^{6,1,0}]_{1,1,1}$ | | $-0.437$ | $0.972 + 0.233i$ | $2.288$ |
| $[\text{FR}_4^{6,1,0}]_{1,1,2}$ | | $-0.437$ | $0.972 + 0.233i$ | $-2.288$ |
| $[\text{FR}_4^{6,1,0}]_{1,2,1}$ | | $-0.437$ | $0.522 - 0.853i$ | $2.288$ |
| $[\text{FR}_4^{6,1,0}]_{1,2,2}$ | | $-0.437$ | $0.522 - 0.853i$ | $-2.288$ |
| $[\text{FR}_4^{6,1,0}]_{1,3,1}$ | | $-0.437$ | $-0.522 - 0.853i$ | $2.288$ |
| $[\text{FR}_4^{6,1,0}]_{1,3,2}$ | | $-0.437$ | $-0.522 - 0.853i$ | $-2.288$ |
| $[\text{FR}_4^{6,1,0}]_{1,4,1}$ | | $-0.437$ | $-0.972 + 0.233i$ | $2.288$ |
| $[\text{FR}_4^{6,1,0}]_{1,4,2}$ | | $-0.437$ | $-0.972 + 0.233i$ | $-2.288$ |
| $[\text{FR}_4^{6,1,0}]_{1,5,1}$ | | $-0.437$ | $0.522 + 0.853i$ | $2.288$ |
| $[\text{FR}_4^{6,1,0}]_{1,5,2}$ | | $-0.437$ | $0.522 + 0.853i$ | $-2.288$ |
| $[\text{FR}_4^{6,1,0}]_{1,6,1}$ | | $-0.437$ | $0.972 - 0.233i$ | $2.288$ |
| $[\text{FR}_4^{6,1,0}]_{1,6,2}$ | | $-0.437$ | $0.972 - 0.233i$ | $-2.288$ |
| $[\text{FR}_4^{6,1,0}]_{1,7,1}$ | | $-0.437$ | $-0.972 - 0.233i$ | $2.288$ |
| $[\text{FR}_4^{6,1,0}]_{1,7,2}$ | | $-0.437$ | $-0.972 - 0.233i$ | $-2.288$ |
| $[\text{FR}_4^{6,1,0}]_{1,8,1}$ | | $-0.437$ | $-0.522 + 0.853i$ | $2.288$ |
| $[\text{FR}_4^{6,1,0}]_{1,8,2}$ | | $-0.437$ | $-0.522 + 0.853i$ | $-2.288$ |

Table 56: Numeric invariants (Continued)

| Name | Properties | $[F_1^{656}]_6^6[F_6^{666}]_5^5$ | $R_1^{66}$ | $d_6^L$ |
|---|---|---|---|---|
| $[FR_4^{6,1,0}]_{2,1,1}$ | | 0.437 | $0.853 - 0.522i$ | 2.288 |
| $[FR_4^{6,1,0}]_{2,1,2}$ | | 0.437 | $0.853 - 0.522i$ | $-2.288$ |
| $[FR_4^{6,1,0}]_{2,2,1}$ | | 0.437 | $-0.233 - 0.972i$ | 2.288 |
| $[FR_4^{6,1,0}]_{2,2,2}$ | | 0.437 | $-0.233 - 0.972i$ | $-2.288$ |
| $[FR_4^{6,1,0}]_{2,3,1}$ | | 0.437 | $0.233 - 0.972i$ | 2.288 |
| $[FR_4^{6,1,0}]_{2,3,2}$ | | 0.437 | $0.233 - 0.972i$ | $-2.288$ |
| $[FR_4^{6,1,0}]_{2,4,1}$ | | 0.437 | $-0.853 - 0.522i$ | 2.288 |
| $[FR_4^{6,1,0}]_{2,4,2}$ | | 0.437 | $-0.853 - 0.522i$ | $-2.288$ |
| $[FR_4^{6,1,0}]_{2,5,1}$ | | 0.437 | $-0.233 + 0.972i$ | 2.288 |
| $[FR_4^{6,1,0}]_{2,5,2}$ | | 0.437 | $-0.233 + 0.972i$ | $-2.288$ |
| $[FR_4^{6,1,0}]_{2,6,1}$ | | 0.437 | $0.853 + 0.522i$ | 2.288 |
| $[FR_4^{6,1,0}]_{2,6,2}$ | | 0.437 | $0.853 + 0.522i$ | $-2.288$ |
| $[FR_4^{6,1,0}]_{2,7,1}$ | | 0.437 | $-0.853 + 0.522i$ | 2.288 |
| $[FR_4^{6,1,0}]_{2,7,2}$ | | 0.437 | $-0.853 + 0.522i$ | $-2.288$ |
| $[FR_4^{6,1,0}]_{2,8,1}$ | | 0.437 | $0.233 + 0.972i$ | 2.288 |
| $[FR_4^{6,1,0}]_{2,8,2}$ | | 0.437 | $0.233 + 0.972i$ | $-2.288$ |
| $[FR_4^{6,1,0}]_{3,1,1}$ | | $-1.144$ | $-0.760 - 0.649i$ | $-0.874$ |
| $[FR_4^{6,1,0}]_{3,1,2}$ | | $-1.144$ | $-0.760 - 0.649i$ | 0.874 |
| $[FR_4^{6,1,0}]_{3,2,1}$ | | $-1.144$ | $0.997 + 0.078i$ | $-0.874$ |
| $[FR_4^{6,1,0}]_{3,2,2}$ | | $-1.144$ | $0.997 + 0.078i$ | 0.874 |
| $[FR_4^{6,1,0}]_{3,3,1}$ | | $-1.144$ | $-0.997 + 0.078i$ | $-0.874$ |
| $[FR_4^{6,1,0}]_{3,3,2}$ | | $-1.144$ | $-0.997 + 0.078i$ | 0.874 |
| $[FR_4^{6,1,0}]_{3,4,1}$ | | $-1.144$ | $0.760 - 0.649i$ | $-0.874$ |
| $[FR_4^{6,1,0}]_{3,4,2}$ | | $-1.144$ | $0.760 - 0.649i$ | 0.874 |
| $[FR_4^{6,1,0}]_{3,5,1}$ | | $-1.144$ | $0.997 - 0.078i$ | $-0.874$ |
| $[FR_4^{6,1,0}]_{3,5,2}$ | | $-1.144$ | $0.997 - 0.078i$ | 0.874 |
| $[FR_4^{6,1,0}]_{3,6,1}$ | | $-1.144$ | $-0.760 + 0.649i$ | $-0.874$ |
| $[FR_4^{6,1,0}]_{3,6,2}$ | | $-1.144$ | $-0.760 + 0.649i$ | 0.874 |
| $[FR_4^{6,1,0}]_{3,7,1}$ | | $-1.144$ | $0.760 + 0.649i$ | $-0.874$ |
| $[FR_4^{6,1,0}]_{3,7,2}$ | | $-1.144$ | $0.760 + 0.649i$ | 0.874 |
| $[FR_4^{6,1,0}]_{3,8,1}$ | | $-1.144$ | $-0.997 - 0.078i$ | $-0.874$ |
| $[FR_4^{6,1,0}]_{3,8,2}$ | | $-1.144$ | $-0.997 - 0.078i$ | 0.874 |
| $[FR_4^{6,1,0}]_{4,1,1}$ | | 1.144 | $-0.078 - 0.997i$ | $-0.874$ |

Table 56: Numeric invariants (Continued)

| Name | Properties | $[F_1^{656}]_6^6[F_6^{666}]_5^5$ | $R_1^{66}$ | $d_6^L$ |
|---|---|---|---|---|
| $[FR_4^{6,1,0}]_{4,1,2}$ | 🧬 🌐 🎗 ⚭ | 1.144 | $-0.078 - 0.997i$ | 0.874 |
| $[FR_4^{6,1,0}]_{4,2,1}$ | 🧬 🌐 🎗 ⚭ | 1.144 | $0.649 + 0.760i$ | $-0.874$ |
| $[FR_4^{6,1,0}]_{4,2,2}$ | 🧬 🌐 🎗 ⚭ | 1.144 | $0.649 + 0.760i$ | 0.874 |
| $[FR_4^{6,1,0}]_{4,3,1}$ | 🧬 🌐 🎗 ⚭ | 1.144 | $-0.649 + 0.760i$ | $-0.874$ |
| $[FR_4^{6,1,0}]_{4,3,2}$ | 🧬 🌐 🎗 ⚭ | 1.144 | $-0.649 + 0.760i$ | 0.874 |
| $[FR_4^{6,1,0}]_{4,4,1}$ | 🧬 🌐 🎗 ⚭ | 1.144 | $0.078 - 0.997i$ | $-0.874$ |
| $[FR_4^{6,1,0}]_{4,4,2}$ | 🧬 🌐 🎗 ⚭ | 1.144 | $0.078 - 0.997i$ | 0.874 |
| $[FR_4^{6,1,0}]_{4,5,1}$ | 🧬 🌐 🎗 ⚭ | 1.144 | $0.649 - 0.760i$ | $-0.874$ |
| $[FR_4^{6,1,0}]_{4,5,2}$ | 🧬 🌐 🎗 ⚭ | 1.144 | $0.649 - 0.760i$ | 0.874 |
| $[FR_4^{6,1,0}]_{4,6,1}$ | 🧬 🌐 🎗 ⚭ | 1.144 | $-0.078 + 0.997i$ | $-0.874$ |
| $[FR_4^{6,1,0}]_{4,6,2}$ | 🧬 🌐 🎗 ⚭ | 1.144 | $-0.078 + 0.997i$ | 0.874 |
| $[FR_4^{6,1,0}]_{4,7,1}$ | 🧬 🌐 🎗 ⚭ | 1.144 | $0.078 + 0.997i$ | $-0.874$ |
| $[FR_4^{6,1,0}]_{4,7,2}$ | 🧬 🌐 🎗 ⚭ | 1.144 | $0.078 + 0.997i$ | 0.874 |
| $[FR_4^{6,1,0}]_{4,8,1}$ | 🧬 🌐 🎗 ⚭ | 1.144 | $-0.649 - 0.760i$ | $-0.874$ |
| $[FR_4^{6,1,0}]_{4,8,2}$ | 🧬 🌐 🎗 ⚭ | 1.144 | $-0.649 - 0.760i$ | 0.874 |

### 4.5.30   $FR_5^{6,1,0}$: **Fib** ⊗ **Rep**($D_3$)

For the fusion ring, the following multiplication table is used.

| 1 | 2 | 3 | 4 | 5 | 6 |
|---|---|---|---|---|---|
| 2 | 1 | 4 | 3 | 5 | 6 |
| 3 | 4 | $1+4$ | $2+3$ | 6 | $5+6$ |
| 4 | 3 | $2+3$ | $1+4$ | 6 | $5+6$ |
| 5 | 5 | 6 | 6 | $1+2+5$ | $3+4+6$ |
| 6 | 6 | $5+6$ | $5+6$ | $3+4+6$ | $1+2+3+4+5+6$ |

Only the trivial permutation leaves the fusion rules invariant.

The following table lists a small set of invariants whose values completely distinguish between all MFPBFCs and MFPNBFCs with the given fusion rules.

Table 57: Symbolic invariants

| Name | Properties | $[F_6^{363}]_6^6$ | $R_1^{66}$ |
|---|---|---|---|
| $[FR_5^{6,1,0}]_{1,1,1}$ | 🧬 ✝ 🌐 🎗 | $\overline{\phi}$ | $\zeta_5^3$ |
| $[FR_5^{6,1,0}]_{1,2,1}$ | 🧬 ✝ 🌐 🎗 | $\overline{\phi}$ | $\zeta_{15}^{14}$ |
| $[FR_5^{6,1,0}]_{1,3,1}$ | 🧬 ✝ 🌐 🎗 | $\overline{\phi}$ | $\zeta_{15}^{4}$ |
| $[FR_5^{6,1,0}]_{1,4,1}$ | 🧬 ✝ 🌐 🎗 | $\overline{\phi}$ | $\zeta_5^2$ |

Table 57: Symbolic invariants (Continued)

| Name | Properties | $[F_6^{363}]_6^6$ | $R_1^{66}$ |
|---|---|---|---|
| $[FR_5^{6,1,0}]_{1,5,1}$ | ✕ ✝ 🌐 🎗 | $\overline{\phi}$ | $\zeta_{15}^{11}$ |
| $[FR_5^{6,1,0}]_{1,6,1}$ | ✕ ✝ 🌐 🎗 | $\overline{\phi}$ | $\zeta_{15}$ |
| $[FR_5^{6,1,0}]_{2,1,1}$ | ✕ 🌐 🎗 | $\phi$ | $\zeta_5$ |
| $[FR_5^{6,1,0}]_{2,2,1}$ | ✕ 🌐 🎗 | $\phi$ | $\zeta_{15}^{13}$ |
| $[FR_5^{6,1,0}]_{2,3,1}$ | ✕ 🌐 🎗 | $\phi$ | $\zeta_{15}^{8}$ |
| $[FR_5^{6,1,0}]_{2,4,1}$ | ✕ 🌐 🎗 | $\phi$ | $\zeta_5^4$ |
| $[FR_5^{6,1,0}]_{2,5,1}$ | ✕ 🌐 🎗 | $\phi$ | $\zeta_{15}^{7}$ |
| $[FR_5^{6,1,0}]_{2,6,1}$ | ✕ 🌐 🎗 | $\phi$ | $\zeta_{15}^{2}$ |

Table 58: Numeric invariants

| Name | Properties | $[F_6^{363}]_6^6$ | $R_1^{66}$ |
|---|---|---|---|
| $[FR_5^{6,1,0}]_{1,1,1}$ | ✕ ✝ 🌐 🎗 | $-0.618$ | $-0.809 - 0.588i$ |
| $[FR_5^{6,1,0}]_{1,2,1}$ | ✕ ✝ 🌐 🎗 | $-0.618$ | $0.914 - 0.407i$ |
| $[FR_5^{6,1,0}]_{1,3,1}$ | ✕ ✝ 🌐 🎗 | $-0.618$ | $-0.105 + 0.995i$ |
| $[FR_5^{6,1,0}]_{1,4,1}$ | ✕ ✝ 🌐 🎗 | $-0.618$ | $-0.809 + 0.588i$ |
| $[FR_5^{6,1,0}]_{1,5,1}$ | ✕ ✝ 🌐 🎗 | $-0.618$ | $-0.105 - 0.995i$ |
| $[FR_5^{6,1,0}]_{1,6,1}$ | ✕ ✝ 🌐 🎗 | $-0.618$ | $0.914 + 0.407i$ |
| $[FR_5^{6,1,0}]_{2,1,1}$ | ✕ 🌐 🎗 | $1.618$ | $0.309 + 0.951i$ |
| $[FR_5^{6,1,0}]_{2,2,1}$ | ✕ 🌐 🎗 | $1.618$ | $0.669 - 0.743i$ |
| $[FR_5^{6,1,0}]_{2,3,1}$ | ✕ 🌐 🎗 | $1.618$ | $-0.978 - 0.208i$ |
| $[FR_5^{6,1,0}]_{2,4,1}$ | ✕ 🌐 🎗 | $1.618$ | $0.309 - 0.951i$ |
| $[FR_5^{6,1,0}]_{2,5,1}$ | ✕ 🌐 🎗 | $1.618$ | $-0.978 + 0.208i$ |
| $[FR_5^{6,1,0}]_{2,6,1}$ | ✕ 🌐 🎗 | $1.618$ | $0.669 + 0.743i$ |

### 4.5.31 $FR_6^{6,1,0}$: $SU(2)_5$

For the fusion ring, the following multiplication table is used.

| 1 | 2 | 3 | 4 | 5 | 6 |
|---|---|---|---|---|---|
| 2 | 1 | 4 | 3 | 6 | 5 |
| 3 | 4 | $1+5$ | $2+6$ | $3+5$ | $4+6$ |
| 4 | 3 | $2+6$ | $1+5$ | $4+6$ | $3+5$ |
| 5 | 6 | $3+5$ | $4+6$ | $1+3+5$ | $2+4+6$ |
| 6 | 5 | $4+6$ | $3+5$ | $2+4+6$ | $1+3+5$ |

Only the trivial permutation leaves the fusion rules invariant.

The following table lists a small set of invariants whose values completely distinguish between all MFPBFCs and MFPNBFCs with the given fusion rules.

Table 59: Symbolic invariants

| Name | Properties | $[F_4^{444}]_1^1$ | $R_1^{44}$ | $d_6^L$ |
|---|---|---|---|---|
| $[FR_6^{6,1,0}]_{1,1,1}$ | | $-\zeta_7^5 - \zeta_7^2 - 1$ | $\zeta_{28}^{25}$ | $+\zeta_7^6 + \zeta + 1$ |
| $[FR_6^{6,1,0}]_{1,1,2}$ | | $-\zeta_7^5 - \zeta_7^2 - 1$ | $\zeta_{28}^{25}$ | $-\zeta_7^6 - \zeta - 1$ |
| $[FR_6^{6,1,0}]_{1,2,1}$ | | $-\zeta_7^5 - \zeta_7^2 - 1$ | $\zeta_{28}^{3}$ | $+\zeta_7^6 + \zeta + 1$ |
| $[FR_6^{6,1,0}]_{1,2,2}$ | | $-\zeta_7^5 - \zeta_7^2 - 1$ | $\zeta_{28}^{3}$ | $-\zeta_7^6 - \zeta - 1$ |
| $[FR_6^{6,1,0}]_{1,3,1}$ | | $-\zeta_7^5 - \zeta_7^2 - 1$ | $\zeta_{28}^{17}$ | $+\zeta_7^6 + \zeta + 1$ |
| $[FR_6^{6,1,0}]_{1,3,2}$ | | $-\zeta_7^5 - \zeta_7^2 - 1$ | $\zeta_{28}^{17}$ | $-\zeta_7^6 - \zeta - 1$ |
| $[FR_6^{6,1,0}]_{1,4,1}$ | | $-\zeta_7^5 - \zeta_7^2 - 1$ | $\zeta_{28}^{11}$ | $+\zeta_7^6 + \zeta + 1$ |
| $[FR_6^{6,1,0}]_{1,4,2}$ | | $-\zeta_7^5 - \zeta_7^2 - 1$ | $\zeta_{28}^{11}$ | $-\zeta_7^6 - \zeta - 1$ |
| $[FR_6^{6,1,0}]_{2,1,1}$ | | $\zeta_7^5 + \zeta_7^2 + 1$ | $\zeta_7^6$ | $+\zeta_7^6 + \zeta + 1$ |
| $[FR_6^{6,1,0}]_{2,1,2}$ | | $\zeta_7^5 + \zeta_7^2 + 1$ | $\zeta_7^6$ | $-\zeta_7^6 - \zeta - 1$ |
| $[FR_6^{6,1,0}]_{2,2,1}$ | | $\zeta_7^5 + \zeta_7^2 + 1$ | $\zeta_7$ | $+\zeta_7^6 + \zeta + 1$ |
| $[FR_6^{6,1,0}]_{2,2,2}$ | | $\zeta_7^5 + \zeta_7^2 + 1$ | $\zeta_7$ | $-\zeta_7^6 - \zeta - 1$ |
| $[FR_6^{6,1,0}]_{2,3,1}$ | | $\zeta_7^5 + \zeta_7^2 + 1$ | $\zeta_{14}^{9}$ | $+\zeta_7^6 + \zeta + 1$ |
| $[FR_6^{6,1,0}]_{2,3,2}$ | | $\zeta_7^5 + \zeta_7^2 + 1$ | $\zeta_{14}^{9}$ | $-\zeta_7^6 - \zeta - 1$ |
| $[FR_6^{6,1,0}]_{2,4,1}$ | | $\zeta_7^5 + \zeta_7^2 + 1$ | $\zeta_{14}^{5}$ | $+\zeta_7^6 + \zeta + 1$ |
| $[FR_6^{6,1,0}]_{2,4,2}$ | | $\zeta_7^5 + \zeta_7^2 + 1$ | $\zeta_{14}^{5}$ | $-\zeta_7^6 - \zeta - 1$ |
| $[FR_6^{6,1,0}]_{3,1,1}$ | | $-\zeta_7^4 - \zeta_7^3 - 1$ | $\zeta_{28}^{13}$ | $\zeta_7^5 + \zeta_7^2 + 1$ |
| $[FR_6^{6,1,0}]_{3,1,2}$ | | $-\zeta_7^4 - \zeta_7^3 - 1$ | $\zeta_{28}^{13}$ | $-\zeta_7^5 - \zeta_7^2 - 1$ |
| $[FR_6^{6,1,0}]_{3,2,1}$ | | $-\zeta_7^4 - \zeta_7^3 - 1$ | $\zeta_{28}^{15}$ | $\zeta_7^5 + \zeta_7^2 + 1$ |
| $[FR_6^{6,1,0}]_{3,2,2}$ | | $-\zeta_7^4 - \zeta_7^3 - 1$ | $\zeta_{28}^{15}$ | $-\zeta_7^5 - \zeta_7^2 - 1$ |
| $[FR_6^{6,1,0}]_{3,3,1}$ | | $-\zeta_7^4 - \zeta_7^3 - 1$ | $\zeta_{28}$ | $\zeta_7^5 + \zeta_7^2 + 1$ |
| $[FR_6^{6,1,0}]_{3,3,2}$ | | $-\zeta_7^4 - \zeta_7^3 - 1$ | $\zeta_{28}$ | $-\zeta_7^5 - \zeta_7^2 - 1$ |
| $[FR_6^{6,1,0}]_{3,4,1}$ | | $-\zeta_7^4 - \zeta_7^3 - 1$ | $\zeta_{28}^{27}$ | $\zeta_7^5 + \zeta_7^2 + 1$ |
| $[FR_6^{6,1,0}]_{3,4,2}$ | | $-\zeta_7^4 - \zeta_7^3 - 1$ | $\zeta_{28}^{27}$ | $-\zeta_7^5 - \zeta_7^2 - 1$ |
| $[FR_6^{6,1,0}]_{4,1,1}$ | | $-\zeta_7^6 - \zeta - 1$ | $\zeta_{28}^{9}$ | $\zeta_7^4 + \zeta_7^3 + 1$ |
| $[FR_6^{6,1,0}]_{4,1,2}$ | | $-\zeta_7^6 - \zeta - 1$ | $\zeta_{28}^{9}$ | $-\zeta_7^4 - \zeta_7^3 - 1$ |
| $[FR_6^{6,1,0}]_{4,2,1}$ | | $-\zeta_7^6 - \zeta - 1$ | $\zeta_{28}^{19}$ | $\zeta_7^4 + \zeta_7^3 + 1$ |
| $[FR_6^{6,1,0}]_{4,2,2}$ | | $-\zeta_7^6 - \zeta - 1$ | $\zeta_{28}^{19}$ | $-\zeta_7^4 - \zeta_7^3 - 1$ |
| $[FR_6^{6,1,0}]_{4,3,1}$ | | $-\zeta_7^6 - \zeta - 1$ | $\zeta_{28}^{5}$ | $\zeta_7^4 + \zeta_7^3 + 1$ |
| $[FR_6^{6,1,0}]_{4,3,2}$ | | $-\zeta_7^6 - \zeta - 1$ | $\zeta_{28}^{5}$ | $-\zeta_7^4 - \zeta_7^3 - 1$ |
| $[FR_6^{6,1,0}]_{4,4,1}$ | | $-\zeta_7^6 - \zeta - 1$ | $\zeta_{28}^{23}$ | $\zeta_7^4 + \zeta_7^3 + 1$ |

Table 59: Symbolic invariants (Continued)

| Name | Properties | $[F_4^{444}]_1^1$ | $R_1^{44}$ | $d_6^L$ |
|---|---|---|---|---|
| $[FR_6^{6,1,0}]_{4,4,2}$ | ✗ 🌐🎗️⚭ | $-\zeta_7^6 - \zeta_7 - 1$ | $\zeta_{28}^{23}$ | $-\zeta_7^4 - \zeta_7^3 - 1$ |
| $[FR_6^{6,1,0}]_{5,1,1}$ | ✗ 🌐🎗️ | $\zeta_7^4 + \zeta_7^3 + 1$ | $\zeta_7^2$ | $\zeta_7^5 + \zeta_7^2 + 1$ |
| $[FR_6^{6,1,0}]_{5,1,2}$ | ✗ 🌐🎗️ | $\zeta_7^4 + \zeta_7^3 + 1$ | $\zeta_7^2$ | $-\zeta_7^5 - \zeta_7^2 - 1$ |
| $[FR_6^{6,1,0}]_{5,2,1}$ | ✗ 🌐🎗️ | $\zeta_7^4 + \zeta_7^3 + 1$ | $\zeta_7^5$ | $\zeta_7^5 + \zeta_7^2 + 1$ |
| $[FR_6^{6,1,0}]_{5,2,2}$ | ✗ 🌐🎗️ | $\zeta_7^4 + \zeta_7^3 + 1$ | $\zeta_7^5$ | $-\zeta_7^5 - \zeta_7^2 - 1$ |
| $[FR_6^{6,1,0}]_{5,3,1}$ | ✗ 🌐🎗️ | $\zeta_7^4 + \zeta_7^3 + 1$ | $\zeta_{14}^3$ | $\zeta_7^5 + \zeta_7^2 + 1$ |
| $[FR_6^{6,1,0}]_{5,3,2}$ | ✗ 🌐🎗️ | $\zeta_7^4 + \zeta_7^3 + 1$ | $\zeta_{14}^3$ | $-\zeta_7^5 - \zeta_7^2 - 1$ |
| $[FR_6^{6,1,0}]_{5,4,1}$ | ✗ 🌐🎗️ | $\zeta_7^4 + \zeta_7^3 + 1$ | $\zeta_{14}^{11}$ | $\zeta_7^5 + \zeta_7^2 + 1$ |
| $[FR_6^{6,1,0}]_{5,4,2}$ | ✗ 🌐🎗️ | $\zeta_7^4 + \zeta_7^3 + 1$ | $\zeta_{14}^{11}$ | $-\zeta_7^5 - \zeta_7^2 - 1$ |
| $[FR_6^{6,1,0}]_{6,1,1}$ | ✗ 🌐🎗️ | $\zeta_7^6 + \zeta_7 + 1$ | $\zeta_7^3$ | $\zeta_7^4 + \zeta_7^3 + 1$ |
| $[FR_6^{6,1,0}]_{6,1,2}$ | ✗ 🌐🎗️ | $\zeta_7^6 + \zeta_7 + 1$ | $\zeta_7^3$ | $-\zeta_7^4 - \zeta_7^3 - 1$ |
| $[FR_6^{6,1,0}]_{6,2,1}$ | ✗ 🌐🎗️ | $\zeta_7^6 + \zeta_7 + 1$ | $\zeta_7^4$ | $\zeta_7^4 + \zeta_7^3 + 1$ |
| $[FR_6^{6,1,0}]_{6,2,2}$ | ✗ 🌐🎗️ | $\zeta_7^6 + \zeta_7 + 1$ | $\zeta_7^4$ | $-\zeta_7^4 - \zeta_7^3 - 1$ |
| $[FR_6^{6,1,0}]_{6,3,1}$ | ✗ 🌐🎗️ | $\zeta_7^6 + \zeta_7 + 1$ | $\zeta_{14}$ | $\zeta_7^4 + \zeta_7^3 + 1$ |
| $[FR_6^{6,1,0}]_{6,3,2}$ | ✗ 🌐🎗️ | $\zeta_7^6 + \zeta_7 + 1$ | $\zeta_{14}$ | $-\zeta_7^4 - \zeta_7^3 - 1$ |
| $[FR_6^{6,1,0}]_{6,4,1}$ | ✗ 🌐🎗️ | $\zeta_7^6 + \zeta_7 + 1$ | $\zeta_{14}^{13}$ | $\zeta_7^4 + \zeta_7^3 + 1$ |
| $[FR_6^{6,1,0}]_{6,4,2}$ | ✗ 🌐🎗️ | $\zeta_7^6 + \zeta_7 + 1$ | $\zeta_{14}^{13}$ | $-\zeta_7^4 - \zeta_7^3 - 1$ |

Table 60: Numeric invariants

| Name | Properties | $[F_4^{444}]_1^1$ | $R_1^{44}$ | $d_4^L$ |
|---|---|---|---|---|
| $[FR_6^{6,1,0}]_{1,1,1}$ | ✗ † 🌐🎗️⚭ | $-0.555$ | $0.782 - 0.623i$ | $1.802$ |
| $[FR_6^{6,1,0}]_{1,1,2}$ | ✗ 🌐🎗️⚭ | $-0.555$ | $0.782 - 0.623i$ | $-1.802$ |
| $[FR_6^{6,1,0}]_{1,2,1}$ | ✗ † 🌐🎗️⚭ | $-0.555$ | $0.782 + 0.623i$ | $1.802$ |
| $[FR_6^{6,1,0}]_{1,2,2}$ | ✗ 🌐🎗️⚭ | $-0.555$ | $0.782 + 0.623i$ | $-1.802$ |
| $[FR_6^{6,1,0}]_{1,3,1}$ | ✗ † 🌐🎗️⚭ | $-0.555$ | $-0.782 - 0.623i$ | $1.802$ |
| $[FR_6^{6,1,0}]_{1,3,2}$ | ✗ 🌐🎗️⚭ | $-0.555$ | $-0.782 - 0.623i$ | $-1.802$ |
| $[FR_6^{6,1,0}]_{1,4,1}$ | ✗ † 🌐🎗️⚭ | $-0.555$ | $-0.782 + 0.623i$ | $1.802$ |
| $[FR_6^{6,1,0}]_{1,4,2}$ | ✗ 🌐🎗️⚭ | $-0.555$ | $-0.782 + 0.623i$ | $-1.802$ |
| $[FR_6^{6,1,0}]_{2,1,1}$ | ✗ † 🌐🎗️ | $0.555$ | $0.623 - 0.782i$ | $1.802$ |
| $[FR_6^{6,1,0}]_{2,1,2}$ | ✗ 🌐🎗️ | $0.555$ | $0.623 - 0.782i$ | $-1.802$ |
| $[FR_6^{6,1,0}]_{2,2,1}$ | ✗ † 🌐🎗️ | $0.555$ | $0.623 + 0.782i$ | $1.802$ |
| $[FR_6^{6,1,0}]_{2,2,2}$ | ✗ 🌐🎗️ | $0.555$ | $0.623 + 0.782i$ | $-1.802$ |

Table 60: Numeric invariants (Continued)

| Name | Properties | $[F_4^{444}]_1^1$ | $R_1^{44}$ | $d_4^L$ |
|---|---|---|---|---|
| $[FR_6^{6,1,0}]_{2,3,1}$ | ✕ † 🌐 🎗 | 0.555 | $-0.623 - 0.782i$ | 1.802 |
| $[FR_6^{6,1,0}]_{2,3,2}$ | ✕ 🌐 🎗 | 0.555 | $-0.623 - 0.782i$ | $-1.802$ |
| $[FR_6^{6,1,0}]_{2,4,1}$ | ✕ † 🌐 🎗 | 0.555 | $-0.623 + 0.782i$ | 1.802 |
| $[FR_6^{6,1,0}]_{2,4,2}$ | ✕ 🌐 🎗 | 0.555 | $-0.623 + 0.782i$ | $-1.802$ |
| $[FR_6^{6,1,0}]_{3,1,1}$ | ✕ 🌐 🎗 ⭕ | 0.802 | $-0.975 + 0.223i$ | $-1.247$ |
| $[FR_6^{6,1,0}]_{3,1,2}$ | ✕ 🌐 🎗 ⭕ | 0.802 | $-0.975 + 0.223i$ | 1.247 |
| $[FR_6^{6,1,0}]_{3,2,1}$ | ✕ 🌐 🎗 ⭕ | 0.802 | $-0.975 - 0.223i$ | $-1.247$ |
| $[FR_6^{6,1,0}]_{3,2,2}$ | ✕ 🌐 🎗 ⭕ | 0.802 | $-0.975 - 0.223i$ | 1.247 |
| $[FR_6^{6,1,0}]_{3,3,1}$ | ✕ 🌐 🎗 ⭕ | 0.802 | $0.975 + 0.223i$ | $-1.247$ |
| $[FR_6^{6,1,0}]_{3,3,2}$ | ✕ 🌐 🎗 ⭕ | 0.802 | $0.975 + 0.223i$ | 1.247 |
| $[FR_6^{6,1,0}]_{3,4,1}$ | ✕ 🌐 🎗 ⭕ | 0.802 | $0.975 - 0.223i$ | $-1.247$ |
| $[FR_6^{6,1,0}]_{3,4,2}$ | ✕ 🌐 🎗 ⭕ | 0.802 | $0.975 - 0.223i$ | 1.247 |
| $[FR_6^{6,1,0}]_{4,1,1}$ | ✕ 🌐 🎗 ⭕ | $-2.247$ | $-0.434 + 0.901i$ | 0.445 |
| $[FR_6^{6,1,0}]_{4,1,2}$ | ✕ 🌐 🎗 ⭕ | $-2.247$ | $-0.434 + 0.901i$ | $-0.445$ |
| $[FR_6^{6,1,0}]_{4,2,1}$ | ✕ 🌐 🎗 ⭕ | $-2.247$ | $-0.434 - 0.901i$ | 0.445 |
| $[FR_6^{6,1,0}]_{4,2,2}$ | ✕ 🌐 🎗 ⭕ | $-2.247$ | $-0.434 - 0.901i$ | $-0.445$ |
| $[FR_6^{6,1,0}]_{4,3,1}$ | ✕ 🌐 🎗 ⭕ | $-2.247$ | $0.434 + 0.901i$ | 0.445 |
| $[FR_6^{6,1,0}]_{4,3,2}$ | ✕ 🌐 🎗 ⭕ | $-2.247$ | $0.434 + 0.901i$ | $-0.445$ |
| $[FR_6^{6,1,0}]_{4,4,1}$ | ✕ 🌐 🎗 ⭕ | $-2.247$ | $0.434 - 0.901i$ | 0.445 |
| $[FR_6^{6,1,0}]_{4,4,2}$ | ✕ 🌐 🎗 ⭕ | $-2.247$ | $0.434 - 0.901i$ | $-0.445$ |
| $[FR_6^{6,1,0}]_{5,1,1}$ | ✕ 🌐 🎗 | $-0.802$ | $-0.223 + 0.975i$ | $-1.247$ |
| $[FR_6^{6,1,0}]_{5,1,2}$ | ✕ 🌐 🎗 | $-0.802$ | $-0.223 + 0.975i$ | 1.247 |
| $[FR_6^{6,1,0}]_{5,2,1}$ | ✕ 🌐 🎗 | $-0.802$ | $-0.223 - 0.975i$ | $-1.247$ |
| $[FR_6^{6,1,0}]_{5,2,2}$ | ✕ 🌐 🎗 | $-0.802$ | $-0.223 - 0.975i$ | 1.247 |
| $[FR_6^{6,1,0}]_{5,3,1}$ | ✕ 🌐 🎗 | $-0.802$ | $0.223 + 0.975i$ | $-1.247$ |
| $[FR_6^{6,1,0}]_{5,3,2}$ | ✕ 🌐 🎗 | $-0.802$ | $0.223 + 0.975i$ | 1.247 |
| $[FR_6^{6,1,0}]_{5,4,1}$ | ✕ 🌐 🎗 | $-0.802$ | $0.223 - 0.975i$ | $-1.247$ |
| $[FR_6^{6,1,0}]_{5,4,2}$ | ✕ 🌐 🎗 | $-0.802$ | $0.223 - 0.975i$ | 1.247 |
| $[FR_6^{6,1,0}]_{6,1,1}$ | ✕ 🌐 🎗 | 2.247 | $-0.901 + 0.434i$ | 0.445 |
| $[FR_6^{6,1,0}]_{6,1,2}$ | ✕ 🌐 🎗 | 2.247 | $-0.901 + 0.434i$ | $-0.445$ |
| $[FR_6^{6,1,0}]_{6,2,1}$ | ✕ 🌐 🎗 | 2.247 | $-0.901 - 0.434i$ | 0.445 |
| $[FR_6^{6,1,0}]_{6,2,2}$ | ✕ 🌐 🎗 | 2.247 | $-0.901 - 0.434i$ | $-0.445$ |
| $[FR_6^{6,1,0}]_{6,3,1}$ | ✕ 🌐 🎗 | 2.247 | $0.901 + 0.434i$ | 0.445 |

Table 60: Numeric invariants (Continued)

| Name | Properties | $[F_4^{444}]_1^1$ | $R_1^{44}$ | $d_4^L$ |
|---|---|---|---|---|
| $[FR_6^{6,1,0}]_{6,3,2}$ | ✗ 🌐 🎀 | 2.247 | $0.901 + 0.434i$ | $-0.445$ |
| $[FR_6^{6,1,0}]_{6,4,1}$ | ✗ 🌐 🎀 | 2.247 | $0.901 - 0.434i$ | $0.445$ |
| $[FR_6^{6,1,0}]_{6,4,2}$ | ✗ 🌐 🎀 | 2.247 | $0.901 - 0.434i$ | $-0.445$ |

**4.5.32** $FR_7^{6,1,0}$: $\mathbf{Rep}(\mathbb{Z}_3 \rtimes D_3)$

For the fusion ring, the following multiplication table is used.

| 1 | 2 | 3 | 4 | 5 | 6 |
|---|---|---|---|---|---|
| 2 | 1 | 3 | 4 | 5 | 6 |
| 3 | 3 | $1+2+6$ | $5+6$ | $4+5$ | $3+4$ |
| 4 | 4 | $5+6$ | $1+2+4$ | $3+6$ | $3+5$ |
| 5 | 5 | $4+5$ | $3+6$ | $1+2+3$ | $4+6$ |
| 6 | 6 | $3+4$ | $3+5$ | $4+6$ | $1+2+5$ |

The following is the group of all non-trivial permutations that leave the fusion rules invariant:

$$S = \{(), (3\ 5\ 6), (3\ 6\ 5)\}.$$

$$X_1 = S\left(\frac{[F_4^{334}]_6^1 [F_4^{343}]_5^6 [F_1^{535}]_5^5}{[F_4^{333}]_6^6 [F_4^{334}]_5^1 [F_5^{343}]_5^5}\right), \tag{61}$$

$$X_2 = S\left(R_1^{33}\right). \tag{62}$$

The following table lists a small set of invariants whose values completely distinguish between all MFPBFCs and MFPNBFCs with the given fusion rules.

Table 61: Symbolic invariants

| Name | Properties | $X_1$ | $X_2$ |
|---|---|---|---|
| $[FR_7^{6,1,0}]_{1,1,1}$ | ✗ † 🌐 🎀 | $\{1\}$ | $\{1\}$ |
| $[FR_7^{6,1,0}]_{1,2,1}$ | ✗ † 🌐 🎀 | $\{1\}$ | $\{\zeta_3\}$ |
| $[FR_7^{6,1,0}]_{1,3,1}$ | ✗ † 🌐 🎀 | $\{1\}$ | $\{\zeta_3^2\}$ |
| $[FR_7^{6,1,0}]_{1,4,1}$ | ✗ † 🌐 🎀 | $\{1\}$ | $\{\zeta_9^5, \zeta_9^2, \zeta_9^8\}$ |
| $[FR_7^{6,1,0}]_{1,5,1}$ | ✗ † 🌐 🎀 | $\{1\}$ | $\{\zeta_9^4, \zeta_9^7, \zeta_9\}$ |
| $[FR_7^{6,1,0}]_{2,0,1}$ | † 🌐 | $\{\zeta_3^2\}$ | - |
| $[FR_7^{6,1,0}]_{3,0,1}$ | † 🌐 | $\{\zeta_3\}$ | - |
| $[FR_7^{6,1,0}]_{4,0,1}$ | † 🌐 | $\{\zeta_9^5, \zeta_9^2, \zeta_9^8\}$ | - |
| $[FR_7^{6,1,0}]_{5,0,1}$ | † 🌐 | $\{\zeta_9^4, \zeta_9^7, \zeta_9\}$ | - |

Table 62: Numeric invariants

| Name | Properties | $X_1$ | $X_2$ |
|---|---|---|---|
| $[FR_7^{6,1,0}]_{1,1,1}$ | 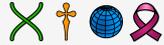 | $\{1\}$ | $\{1\}$ |
| $[FR_7^{6,1,0}]_{1,2,1}$ | 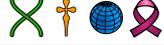 | $\{1\}$ | $\{-0.5 + 0.866i\}$ |
| $[FR_7^{6,1,0}]_{1,3,1}$ | 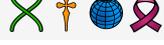 | $\{1\}$ | $\{-0.5 - 0.866i\}$ |
| $[FR_7^{6,1,0}]_{1,4,1}$ | 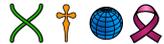 | $\{1\}$ | $\{-0.940 - 0.342i,\ 0.174 + 0.985i,\ 0.766 - 0.643i\}$ |
| $[FR_7^{6,1,0}]_{1,5,1}$ | 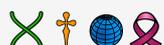 | $\{1\}$ | $\{-0.940 + 0.342i,\ 0.174 - 0.985i,\ 0.766 + 0.643i\}$ |
| $[FR_7^{6,1,0}]_{2,0,1}$ | 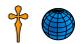 | $\{-0.5 - 0.866i\}$ | - |
| $[FR_7^{6,1,0}]_{3,0,1}$ | 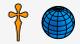 | $\{-0.5 + 0.866i\}$ | - |
| $[FR_7^{6,1,0}]_{4,0,1}$ | 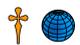 | $\{-0.940 - 0.342i,\ 0.174 + 0.985i,\ 0.766 - 0.643i\}$ | - |
| $[FR_7^{6,1,0}]_{5,0,1}$ | 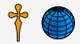 | $\{-0.940 + 0.342i,\ 0.174 - 0.985i,\ 0.766 + 0.643i\}$ | - |

### 4.5.33 $FR_8^{6,1,0}$: $\mathbf{Rep}(D_9)$

For the fusion ring, the following multiplication table is used.

| | 1 | 2 | 3 | 4 | 5 | 6 |
|---|---|---|---|---|---|---|
| | 2 | 1 | 3 | 4 | 5 | 6 |
| | 3 | 3 | $1+2+3$ | $5+6$ | $4+6$ | $4+5$ |
| | 4 | 4 | $5+6$ | $1+2+4$ | $3+6$ | $3+5$ |
| | 5 | 5 | $4+6$ | $3+6$ | $1+2+5$ | $3+4$ |
| | 6 | 6 | $4+5$ | $3+5$ | $3+4$ | $1+2+6$ |

The following is the group of all non-trivial permutations that leave the fusion rules invariant:

$$S = \{(), (3\ 4), (3\ 5), (3\ 6), (4\ 5), (4\ 6), (5\ 6),$$
$$(3\ 4\ 5), (3\ 4\ 6), (3\ 5\ 4), (3\ 5\ 6), (3\ 6\ 4), (3\ 6\ 5), (4\ 5\ 6),\ (4\ 6\ 5),$$
$$(3\ 4\ 5\ 6), (3\ 4\ 6\ 5), (3\ 5\ 4\ 6), (3\ 5\ 6\ 4), (3\ 6\ 4\ 5), (3\ 6\ 5\ 4),$$
$$(3\ 4)(5\ 6), (3\ 5)(4\ 6), (3\ 6)(4\ 5)\}.$$

Let

$$X_1 = S\left([F_1^{333}]_3^3\right), X_2 = S\left(R_1^{33}\right).$$

The following table lists a small set of invariants whose values completely distinguish between all MFPBFCs and MFPNBFCs with the given fusion rules.

Table 63: Symbolic invariants

| Name | Properties | $X_1$ | $X_2$ |
|---|---|---|---|
| $[FR_8^{6,1,0}]_{1,1,1}$ | 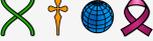 | $\{1\}$ | $\{\zeta_3^2, \zeta_3\}$ |
| $[FR_8^{6,1,0}]_{1,2,1}$ | 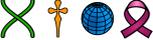 | $\{1\}$ | $\{1, \zeta_3^2, \zeta_3\}$ |
| $[FR_8^{6,1,0}]_{1,3,1}$ | 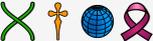 | $\{1\}$ | $\{1, \zeta_3^2\}$ |
| $[FR_8^{6,1,0}]_{1,4,1}$ | 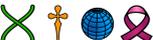 | $\{1\}$ | $\{1\}$ |
| $[FR_8^{6,1,0}]_{1,5,1}$ | 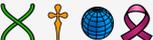 | $\{1\}$ | $\{1, \zeta_3\}$ |
| $[FR_8^{6,1,0}]_{2,0,1}$ | 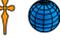 | $\{1, \zeta_3\}$ | - |
| $[FR_8^{6,1,0}]_{3,0,1}$ | 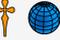 | $\{1, \zeta_3^2, \zeta_3\}$ | - |
| $[FR_8^{6,1,0}]_{4,0,1}$ | 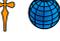 | $\{1, \zeta_3^2\}$ | - |
| $[FR_8^{6,1,0}]_{5,0,1}$ | 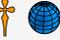 | $\{\zeta_3^2, \zeta_3\}$ | - |

Table 64: Numeric invariants

| Name | Properties | $X_1$ | $X_2$ |
|---|---|---|---|
| $[FR_8^{6,1,0}]_{1,1,1}$ | 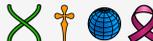 | $\{1\}$ | $\{-0.5 - 0.866i, -0.5 + 0.866i\}$ |
| $[FR_8^{6,1,0}]_{1,2,1}$ | 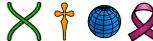 | $\{1\}$ | $\{-0.5 - 0.866i, -0.5 + 0.866i, 1\}$ |
| $[FR_8^{6,1,0}]_{1,3,1}$ | 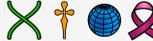 | $\{1\}$ | $\{-0.5 - 0.866i, 1\}$ |
| $[FR_8^{6,1,0}]_{1,4,1}$ | 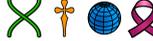 | $\{1\}$ | $\{1\}$ |
| $[FR_8^{6,1,0}]_{1,5,1}$ | 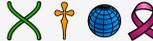 | $\{1\}$ | $\{-0.5 + 0.866i, 1\}$ |
| $[FR_8^{6,1,0}]_{2,0,1}$ | 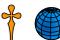 | $\{-0.5 + 0.866i, 1\}$ | - |
| $[FR_8^{6,1,0}]_{3,0,1}$ | 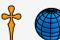 | $\{-0.5 - 0.866i, -0.5 + 0.866i, 1\}$ | - |
| $[FR_8^{6,1,0}]_{4,0,1}$ | 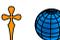 | $\{-0.5 - 0.866i, 1\}$ | - |
| $[FR_8^{6,1,0}]_{5,0,1}$ | 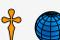 | $\{-0.5 - 0.866i, -0.5 + 0.866i\}$ | - |

### 4.5.34 $FR_9^{6,1,0}$: $SO(5)_2$

For the fusion ring, the following multiplication table is used.

|   |   |   |   |   |   |
|---|---|---|---|---|---|
| 1 | 2 | 3 | 4 | 5 | 6 |
| 2 | 1 | 3 | 4 | 6 | 5 |
| 3 | 3 | 1+2+4 | 3+4 | 5+6 | 5+6 |
| 4 | 4 | 3+4 | 1+2+3 | 5+6 | 5+6 |
| 5 | 6 | 5+6 | 5+6 | 1+3+4 | 2+3+4 |
| 6 | 5 | 5+6 | 5+6 | 2+3+4 | 1+3+4 |

The following is the group of all non-trivial permutations that leave the fusion rules invariant:

$$S = \{(), (3\ 4), (5\ 6), (3\ 4)(5\ 6)\}.$$

Let

$$X_1 = S\left([F_2^{525}]_6^6 [F_6^{565}]_2^2\right),$$
$$X_2 = S\left([F_5^{353}]_5^5\right),$$
$$X_3 = S\left(d_5^L\right).$$

The following table lists a small set of invariants whose values completely distinguish between all MFPBFCs and MFPNBFCs with the given fusion rules.

Table 65: Symbolic invariants

| Name | Properties | $X_1$ | $X_2$ | $X_3$ |
|---|---|---|---|---|
| $[FR_9^{6,1,0}]_{1,1,1}$ | ✂ † 🌐 🎗 ⦾ | $\left\{\frac{1}{\sqrt{5}}\right\}$ | $\left\{\frac{-\phi}{2}\right\}$ | $\left\{\sqrt{5}\right\}$ |
| $[FR_9^{6,1,0}]_{1,1,2}$ | ✂ 🌐 🎗 ⦾ | $\left\{\frac{1}{\sqrt{5}}\right\}$ | $\left\{\frac{-\phi}{2}\right\}$ | $\left\{-\sqrt{5}\right\}$ |
| $[FR_9^{6,1,0}]_{2,1,1}$ | ✂ † 🌐 🎗 ⦾ | $\left\{\frac{1}{\sqrt{5}}\right\}$ | $\left\{-\frac{\overline{\phi}}{2}\right\}$ | $\left\{\sqrt{5}\right\}$ |
| $[FR_9^{6,1,0}]_{2,1,2}$ | ✂ 🌐 🎗 ⦾ | $\left\{\frac{1}{\sqrt{5}}\right\}$ | $\left\{-\frac{\overline{\phi}}{2}\right\}$ | $\left\{-\sqrt{5}\right\}$ |
| $[FR_9^{6,1,0}]_{3,1,1}$ | ✂ † 🌐 🎗 ⦾ | $\left\{-\frac{1}{\sqrt{5}}\right\}$ | $\left\{\frac{-\phi}{2}\right\}$ | $\left\{\sqrt{5}\right\}$ |
| $[FR_9^{6,1,0}]_{3,1,2}$ | ✂ 🌐 🎗 ⦾ | $\left\{-\frac{1}{\sqrt{5}}\right\}$ | $\left\{\frac{-\phi}{2}\right\}$ | $\left\{-\sqrt{5}\right\}$ |
| $[FR_9^{6,1,0}]_{4,1,1}$ | ✂ † 🌐 🎗 ⦾ | $\left\{-\frac{1}{\sqrt{5}}\right\}$ | $\left\{-\frac{\overline{\phi}}{2}\right\}$ | $\left\{\sqrt{5}\right\}$ |
| $[FR_9^{6,1,0}]_{4,1,2}$ | ✂ 🌐 🎗 ⦾ | $\left\{-\frac{1}{\sqrt{5}}\right\}$ | $\left\{-\frac{\overline{\phi}}{2}\right\}$ | $\left\{-\sqrt{5}\right\}$ |

Table 66: Numeric invariants

| Name | Properties | $X_1$ | $X_2$ | $X_3$ |
|---|---|---|---|---|
| $[FR_9^{6,1,0}]_{1,1,1}$ | ✂ † 🌐 🎗 ⦾ | $\{0.447\}$ | $\{-0.809\}$ | $\{2.236\}$ |
| $[FR_9^{6,1,0}]_{1,1,2}$ | ✂ 🌐 🎗 ⦾ | $\{0.447\}$ | $\{-0.809\}$ | $\{-2.236\}$ |
| $[FR_9^{6,1,0}]_{2,1,1}$ | ✂ † 🌐 🎗 ⦾ | $\{0.447\}$ | $\{0.309\}$ | $\{2.236\}$ |
| $[FR_9^{6,1,0}]_{2,1,2}$ | ✂ 🌐 🎗 ⦾ | $\{0.447\}$ | $\{0.309\}$ | $\{-2.236\}$ |

Table 66: Numeric invariants (Continued)

| Name | Properties | $X_1$ | $X_2$ | $X_3$ |
|---|---|---|---|---|
| $[\mathrm{FR}_9^{6,1,0}]_{3,1,1}$ | ✂ ✝ 🌐 🎗 ⚭ | $\{-0.447\}$ | $\{-0.809\}$ | $\{2.236\}$ |
| $[\mathrm{FR}_9^{6,1,0}]_{3,1,2}$ | ✂ 🌐 🎗 ⚭ | $\{-0.447\}$ | $\{-0.809\}$ | $\{-2.236\}$ |
| $[\mathrm{FR}_9^{6,1,0}]_{4,1,1}$ | ✂ ✝ 🌐 🎗 ⚭ | $\{-0.447\}$ | $\{0.309\}$ | $\{2.236\}$ |
| $[\mathrm{FR}_9^{6,1,0}]_{4,1,2}$ | ✂ 🌐 🎗 ⚭ | $\{-0.447\}$ | $\{0.309\}$ | $\{-2.236\}$ |

## 4.5.35 $\mathrm{FR}_{14}^{6,1,0}$: Fib ⊗ PSU(2)$_5$

For the fusion ring, the following multiplication table is used.

| 1 | 2 | 3 | 4 | 5 | 6 |
|---|---|---|---|---|---|
| 2 | 1 + 2 | 5 | 6 | 3 + 5 | 4 + 6 |
| 3 | 5 | 1 + 4 | 3 + 4 | 2 + 6 | 5 + 6 |
| 4 | 6 | 3 + 4 | 1 + 3 + 4 | 5 + 6 | 2 + 5 + 6 |
| 5 | 3 + 5 | 2 + 6 | 5 + 6 | 1 + 2 + 4 + 6 | 3 + 4 + 5 + 6 |
| 6 | 4 + 6 | 5 + 6 | 2 + 5 + 6 | 3 + 4 + 5 + 6 | 1 + 2 + 3 + 4 + 5 + 6 |

Only the trivial permutation leaves the fusion rules invariant.

The following table lists a small set of invariants whose values completely distinguish between all MFPBFCs and MFPNBFCs with the given fusion rules.

Table 67: Distinguishing invariants for categories with $\mathrm{FR}_{14}^{6,1,0}$ fusion rules.

| Name | Properties | $[F_6^{666}]_6^6$ | $R_1^{55}$ |
|---|---|---|---|
| $[\mathrm{FR}_{14}^{6,1,0}]_{1,1,1}$ | ✂ ✝ 🌐 🎗 ⚭ | $\overline{\phi}\left(\zeta_7^4 + \zeta_7^3 + 2\right)$ | $\zeta_{35}^{16}$ |
| $[\mathrm{FR}_{14}^{6,1,0}]_{1,2,1}$ | ✂ ✝ 🌐 🎗 ⚭ | $\overline{\phi}\left(\zeta_7^4 + \zeta_7^3 + 2\right)$ | $\zeta_{35}^{26}$ |
| $[\mathrm{FR}_{14}^{6,1,0}]_{1,3,1}$ | ✂ ✝ 🌐 🎗 ⚭ | $\overline{\phi}\left(\zeta_7^4 + \zeta_7^3 + 2\right)$ | $\zeta_{35}^{9}$ |
| $[\mathrm{FR}_{14}^{6,1,0}]_{1,4,1}$ | ✂ ✝ 🌐 🎗 ⚭ | $\overline{\phi}\left(\zeta_7^4 + \zeta_7^3 + 2\right)$ | $\zeta_{35}^{19}$ |
| $[\mathrm{FR}_{14}^{6,1,0}]_{2,1,1}$ | ✂ 🌐 🎗 ⚭ | $\phi\left(\zeta_7^5 + \zeta_7^2 + 2\right)$ | $\zeta_{35}^{13}$ |
| $[\mathrm{FR}_{14}^{6,1,0}]_{2,2,1}$ | ✂ 🌐 🎗 ⚭ | $\phi\left(\zeta_7^5 + \zeta_7^2 + 2\right)$ | $\zeta_{35}^{8}$ |
| $[\mathrm{FR}_{14}^{6,1,0}]_{2,3,1}$ | ✂ 🌐 🎗 ⚭ | $\phi\left(\zeta_7^5 + \zeta_7^2 + 2\right)$ | $\zeta_{35}^{27}$ |
| $[\mathrm{FR}_{14}^{6,1,0}]_{2,4,1}$ | ✂ 🌐 🎗 ⚭ | $\phi\left(\zeta_7^5 + \zeta_7^2 + 2\right)$ | $\zeta_{35}^{22}$ |
| $[\mathrm{FR}_{14}^{6,1,0}]_{3,1,1}$ | ✂ 🌐 🎗 ⚭ | $\overline{\phi}\left(\zeta_7^5 + \zeta_7^2 + 2\right)$ | $\zeta_{35}^{6}$ |
| $[\mathrm{FR}_{14}^{6,1,0}]_{3,2,1}$ | ✂ 🌐 🎗 ⚭ | $\overline{\phi}\left(\zeta_7^5 + \zeta_7^2 + 2\right)$ | $\zeta_{35}$ |
| $[\mathrm{FR}_{14}^{6,1,0}]_{3,3,1}$ | ✂ 🌐 🎗 ⚭ | $\overline{\phi}\left(\zeta_7^5 + \zeta_7^2 + 2\right)$ | $\zeta_{35}^{34}$ |
| $[\mathrm{FR}_{14}^{6,1,0}]_{3,4,1}$ | ✂ 🌐 🎗 ⚭ | $\overline{\phi}\left(\zeta_7^5 + \zeta_7^2 + 2\right)$ | $\zeta_{35}^{29}$ |
| $[\mathrm{FR}_{14}^{6,1,0}]_{4,1,1}$ | ✂ 🌐 🎗 ⚭ | $\phi\left(\zeta_7^4 + \zeta_7^3 + 2\right)$ | $\zeta_{35}^{23}$ |

Table 67: Distinguishing invariants for categories with $FR_{14}^{6,1,0}$ fusion rules. (Continued)

| Name | Properties | $[F_6^{666}]_6^6$ | $R_1^{55}$ |
|---|---|---|---|
| $[FR_{14}^{6,1,0}]_{4,2,1}$ | ✗ 🌐🎗️◎ | $\phi\left(\zeta_7^4 + \zeta_7^3 + 2\right)$ | $\zeta_{35}^{33}$ |
| $[FR_{14}^{6,1,0}]_{4,3,1}$ | ✗ 🌐🎗️◎ | $\phi\left(\zeta_7^4 + \zeta_7^3 + 2\right)$ | $\zeta_{35}^{2}$ |
| $[FR_{14}^{6,1,0}]_{4,4,1}$ | ✗ 🌐🎗️◎ | $\phi\left(\zeta_7^4 + \zeta_7^3 + 2\right)$ | $\zeta_{35}^{12}$ |
| $[FR_{14}^{6,1,0}]_{5,1,1}$ | ✗ 🌐🎗️◎ | $\phi\left(\zeta_7^6 + \zeta_7 + 2\right)$ | $\zeta_{35}^{3}$ |
| $[FR_{14}^{6,1,0}]_{5,2,1}$ | ✗ 🌐🎗️◎ | $\phi\left(\zeta_7^6 + \zeta_7 + 2\right)$ | $\zeta_{35}^{18}$ |
| $[FR_{14}^{6,1,0}]_{5,3,1}$ | ✗ 🌐🎗️◎ | $\phi\left(\zeta_7^6 + \zeta_7 + 2\right)$ | $\zeta_{35}^{17}$ |
| $[FR_{14}^{6,1,0}]_{5,4,1}$ | ✗ 🌐🎗️◎ | $\phi\left(\zeta_7^6 + \zeta_7 + 2\right)$ | $\zeta_{35}^{32}$ |
| $[FR_{14}^{6,1,0}]_{6,1,1}$ | ✗ 🌐🎗️◎ | $\overline{\phi}\left(\zeta_7^6 + \zeta_7 + 2\right)$ | $\zeta_{35}^{31}$ |
| $[FR_{14}^{6,1,0}]_{6,2,1}$ | ✗ 🌐🎗️◎ | $\overline{\phi}\left(\zeta_7^6 + \zeta_7 + 2\right)$ | $\zeta_{35}^{11}$ |
| $[FR_{14}^{6,1,0}]_{6,3,1}$ | ✗ 🌐🎗️◎ | $\overline{\phi}\left(\zeta_7^6 + \zeta_7 + 2\right)$ | $\zeta_{35}^{24}$ |
| $[FR_{14}^{6,1,0}]_{6,4,1}$ | ✗ 🌐🎗️◎ | $\overline{\phi}\left(\zeta_7^6 + \zeta_7 + 2\right)$ | $\zeta_{35}^{4}$ |

Table 68: Numeric invariants

| Name | Properties | $[F_6^{666}]_6^6$ | $R_1^{55}$ |
|---|---|---|---|
| $[FR_{14}^{6,1,0}]_{1,1,1}$ | ✗ †🌐🎗️◎ | $-0.122$ | $-0.964 + 0.266i$ |
| $[FR_{14}^{6,1,0}]_{1,2,1}$ | ✗ †🌐🎗️◎ | $-0.122$ | $-0.045 - 0.999i$ |
| $[FR_{14}^{6,1,0}]_{1,3,1}$ | ✗ †🌐🎗️◎ | $-0.122$ | $-0.045 + 0.999i$ |
| $[FR_{14}^{6,1,0}]_{1,4,1}$ | ✗ †🌐🎗️◎ | $-0.122$ | $-0.964 - 0.266i$ |
| $[FR_{14}^{6,1,0}]_{2,1,1}$ | ✗ 🌐🎗️◎ | $2.516$ | $-0.691 + 0.723i$ |
| $[FR_{14}^{6,1,0}]_{2,2,1}$ | ✗ 🌐🎗️◎ | $2.516$ | $0.134 + 0.991i$ |
| $[FR_{14}^{6,1,0}]_{2,3,1}$ | ✗ 🌐🎗️◎ | $2.516$ | $0.134 - 0.991i$ |
| $[FR_{14}^{6,1,0}]_{2,4,1}$ | ✗ 🌐🎗️◎ | $2.516$ | $-0.691 - 0.723i$ |
| $[FR_{14}^{6,1,0}]_{3,1,1}$ | ✗ 🌐🎗️◎ | $-0.961$ | $0.474 + 0.881i$ |
| $[FR_{14}^{6,1,0}]_{3,2,1}$ | ✗ 🌐🎗️◎ | $-0.961$ | $0.984 + 0.179i$ |
| $[FR_{14}^{6,1,0}]_{3,3,1}$ | ✗ 🌐🎗️◎ | $-0.961$ | $0.984 - 0.179i$ |
| $[FR_{14}^{6,1,0}]_{3,4,1}$ | ✗ 🌐🎗️◎ | $-0.961$ | $0.474 - 0.881i$ |
| $[FR_{14}^{6,1,0}]_{4,1,1}$ | ✗ 🌐🎗️◎ | $0.320$ | $-0.551 - 0.835i$ |
| $[FR_{14}^{6,1,0}]_{4,2,1}$ | ✗ 🌐🎗️◎ | $0.320$ | $0.936 - 0.351i$ |
| $[FR_{14}^{6,1,0}]_{4,3,1}$ | ✗ 🌐🎗️◎ | $0.320$ | $0.936 + 0.351i$ |
| $[FR_{14}^{6,1,0}]_{4,4,1}$ | ✗ 🌐🎗️◎ | $0.320$ | $-0.551 + 0.835i$ |

Table 68: Numeric invariants (Continued)

| Name | Properties | $[F_6^{666}]_6^6$ | $R_1^{55}$ |
|---|---|---|---|
| $[FR_{14}^{6,1,0}]_{5,1,1}$ | ✕ 🌐🎗️⭕ | 5.254 | $0.858 + 0.513i$ |
| $[FR_{14}^{6,1,0}]_{5,2,1}$ | ✕ 🌐🎗️⭕ | 5.254 | $-0.996 - 0.090i$ |
| $[FR_{14}^{6,1,0}]_{5,3,1}$ | ✕ 🌐🎗️⭕ | 5.254 | $-0.996 + 0.090i$ |
| $[FR_{14}^{6,1,0}]_{5,4,1}$ | ✕ 🌐🎗️⭕ | 5.254 | $0.858 - 0.513i$ |
| $[FR_{14}^{6,1,0}]_{6,1,1}$ | ✕ 🌐🎗️⭕ | $-2.007$ | $0.753 - 0.658i$ |
| $[FR_{14}^{6,1,0}]_{6,2,1}$ | ✕ 🌐🎗️⭕ | $-2.007$ | $-0.393 + 0.920i$ |
| $[FR_{14}^{6,1,0}]_{6,3,1}$ | ✕ 🌐🎗️⭕ | $-2.007$ | $-0.393 - 0.920i$ |
| $[FR_{14}^{6,1,0}]_{6,4,1}$ | ✕ 🌐🎗️⭕ | $-2.007$ | $0.753 + 0.658i$ |

### 4.5.36 $FR_{16}^{6,1,0}$: $PSU(2)_{10}$

For the fusion ring, the following multiplication table is used.

| 1 | 2 | 3 | 4 | 5 | 6 |
|---|---|---|---|---|---|
| 2 | 1 | 4 | 3 | 6 | 5 |
| 3 | 4 | $1+4+6$ | $2+3+5$ | $4+5+6$ | $3+5+6$ |
| 4 | 3 | $2+3+5$ | $1+4+6$ | $3+5+6$ | $4+5+6$ |
| 5 | 6 | $4+5+6$ | $3+5+6$ | $1+3+4+5+6$ | $2+3+4+5+6$ |
| 6 | 5 | $3+5+6$ | $4+5+6$ | $2+3+4+5+6$ | $1+3+4+5+6$ |

Only the trivial permutation leaves the fusion rules invariant.

The following table lists a small set of invariants whose values completely distinguish between all MFPBFCs and MFPNBFCs with the given fusion rules.

Table 69: Symbolic invariants

| Name | Properties | $[F_5^{353}]_5^5$ | $R_4^{33}$ |
|---|---|---|---|
| $[FR_{16}^{6,1,0}]_{1,1,1}$ | ✕ † 🌐🎗️ | $\frac{1}{2}(2-\sqrt{3})$ | $\zeta_{12}$ |
| $[FR_{16}^{6,1,0}]_{1,2,1}$ | ✕ † 🌐🎗️ | $\frac{1}{2}(2-\sqrt{3})$ | $\zeta_{12}^{11}$ |
| $[FR_{16}^{6,1,0}]_{2,1,1}$ | ✕ 🌐🎗️ | $\frac{1}{2}(2+\sqrt{3})$ | $\zeta_{12}^{7}$ |
| $[FR_{16}^{6,1,0}]_{2,2,1}$ | ✕ 🌐🎗️ | $\frac{1}{2}(2+\sqrt{3})$ | $\zeta_{12}^{5}$ |

Table 70: Numeric invariants

| Name | Properties | $[F_5^{353}]_5^5$ | $R_4^{33}$ |
|---|---|---|---|
| $[FR_{16}^{6,1,0}]_{1,1,1}$ | ✕ † 🌐🎗️ | 0.134 | $0.866 + 0.5i$ |
| $[FR_{16}^{6,1,0}]_{1,2,1}$ | ✕ † 🌐🎗️ | 0.134 | $0.866 - 0.5i$ |
| $[FR_{16}^{6,1,0}]_{2,1,1}$ | ✕ 🌐🎗️ | 1.866 | $-0.866 - 0.5i$ |

Table 70: Numeric invariants (Continued)

| Name | Properties | $[F_5^{353}]_5^5$ | $R_4^{33}$ |
|---|---|---|---|
| $[\mathrm{FR}_{16}^{6,1,0}]_{2,2,1}$ | | 1.866 | $-0.866 + 0.5i$ |

### 4.5.37 $\mathrm{FR}_{18}^{6,1,0}$: $\mathrm{PSU}(2)_{11}$

For the fusion ring, the following multiplication table is used.

| 1 | 2 | 3 | 4 | 5 | 6 |
|---|---|---|---|---|---|
| 2 | 1+3 | 2+4 | 3+5 | 4+6 | 5+6 |
| 3 | 2+4 | 1+3+5 | 2+4+6 | 3+5+6 | 4+5+6 |
| 4 | 3+5 | 2+4+6 | 1+3+5+6 | 2+4+5+6 | 3+4+5+6 |
| 5 | 4+6 | 3+5+6 | 2+4+5+6 | 1+3+4+5+6 | 2+3+4+5+6 |
| 6 | 5+6 | 4+5+6 | 3+4+5+6 | 2+3+4+5+6 | 1+2+3+4+5+6 |

Only the trivial permutation leaves the fusion rules invariant.

The following table lists a small set of invariants whose values completely distinguish between all MFPBFCs and MFPNBFCs with the given fusion rules.

Table 71: Symbolic invariants

| Name | Properties | $[F_3^{333}]_3^3$ | $R_1^{44}$ |
|---|---|---|---|
| $[\mathrm{FR}_{18}^{6,1,0}]_{1,1,1}$ | | $\zeta_{13}^{11} + \zeta_{13}^{10} + \zeta_{13}^{7} + \zeta_{13}^{6} + \zeta_{13}^{3} + \zeta_{13}^{2} + 1$ | $\zeta_{13}^{7}$ |
| $[\mathrm{FR}_{18}^{6,1,0}]_{1,2,1}$ | | $\zeta_{13}^{11} + \zeta_{13}^{10} + \zeta_{13}^{7} + \zeta_{13}^{6} + \zeta_{13}^{3} + \zeta_{13}^{2} + 1$ | $\zeta_{13}^{6}$ |
| $[\mathrm{FR}_{18}^{6,1,0}]_{2,1,1}$ | | $-\zeta_{13}^{11} - \zeta_{13}^{10} - \zeta_{13}^{8} - \zeta_{13}^{5} - \zeta_{13}^{3} - \zeta_{13}^{2}$ | $\zeta_{13}$ |
| $[\mathrm{FR}_{18}^{6,1,0}]_{2,2,1}$ | | $-\zeta_{13}^{11} - \zeta_{13}^{10} - \zeta_{13}^{8} - \zeta_{13}^{5} - \zeta_{13}^{3} - \zeta_{13}^{2}$ | $\zeta_{13}^{12}$ |
| $[\mathrm{FR}_{18}^{6,1,0}]_{3,1,1}$ | | $-\zeta_{13}^{10} - \zeta_{13}^{9} - \zeta_{13}^{7} - \zeta_{13}^{6} - \zeta_{13}^{4} - \zeta_{13}^{3}$ | $\zeta_{13}^{11}$ |
| $[\mathrm{FR}_{18}^{6,1,0}]_{3,2,1}$ | | $-\zeta_{13}^{10} - \zeta_{13}^{9} - \zeta_{13}^{7} - \zeta_{13}^{6} - \zeta_{13}^{4} - \zeta_{13}^{3}$ | $\zeta_{13}^{2}$ |
| $[\mathrm{FR}_{18}^{6,1,0}]_{4,1,1}$ | | $-\zeta_{13}^{11} - \zeta_{13}^{9} - \zeta_{13}^{7} - \zeta_{13}^{6} - \zeta_{13}^{4} - \zeta_{13}^{2}$ | $\zeta_{13}^{3}$ |
| $[\mathrm{FR}_{18}^{6,1,0}]_{4,2,1}$ | | $-\zeta_{13}^{11} - \zeta_{13}^{9} - \zeta_{13}^{7} - \zeta_{13}^{6} - \zeta_{13}^{4} - \zeta_{13}^{2}$ | $\zeta_{13}^{10}$ |
| $[\mathrm{FR}_{18}^{6,1,0}]_{5,1,1}$ | | $\zeta_{13}^{11} + \zeta_{13}^{10} + \zeta_{13}^{9} + \zeta_{13}^{4} + \zeta_{13}^{3} + \zeta_{13}^{2} + 1$ | $\zeta_{13}^{9}$ |
| $[\mathrm{FR}_{18}^{6,1,0}]_{5,2,1}$ | | $\zeta_{13}^{11} + \zeta_{13}^{10} + \zeta_{13}^{9} + \zeta_{13}^{4} + \zeta_{13}^{3} + \zeta_{13}^{2} + 1$ | $\zeta_{13}^{4}$ |
| $[\mathrm{FR}_{18}^{6,1,0}]_{6,1,1}$ | | $\zeta_{13}^{9} + \zeta_{13}^{8} + \zeta_{13}^{7} + \zeta_{13}^{6} + \zeta_{13}^{5} + \zeta_{13}^{4} + 1$ | $\zeta_{13}^{8}$ |
| $[\mathrm{FR}_{18}^{6,1,0}]_{6,2,1}$ | | $\zeta_{13}^{9} + \zeta_{13}^{8} + \zeta_{13}^{7} + \zeta_{13}^{6} + \zeta_{13}^{5} + \zeta_{13}^{4} + 1$ | $\zeta_{13}^{5}$ |

Table 72: Numeric invariants

| Name | Properties | $[F_3^{333}]_3^3$ | $R_1^{44}$ |
|---|---|---|---|
| $[\mathrm{FR}_{18}^{6,1,0}]_{1,1,1}$ | | 0.435 | $-0.971 - 0.239i$ |
| $[\mathrm{FR}_{18}^{6,1,0}]_{1,2,1}$ | | 0.435 | $-0.971 + 0.239i$ |

Table 72: Numeric invariants (Continued)

| Name | Properties | $[F_3^{333}]_3^3$ | $R_1^{44}$ |
|---|---|---|---|
| $[FR_{18}^{6,1,0}]_{2,1,1}$ | ✕ 🌐🎗⭕ | 0.120 | $0.885 + 0.465i$ |
| $[FR_{18}^{6,1,0}]_{2,2,1}$ | ✕ 🌐🎗⭕ | 0.120 | $0.885 - 0.465i$ |
| $[FR_{18}^{6,1,0}]_{3,1,1}$ | ✕ 🌐🎗⭕ | 2.410 | $0.568 - 0.823i$ |
| $[FR_{18}^{6,1,0}]_{3,2,1}$ | ✕ 🌐🎗⭕ | 2.410 | $0.568 + 0.823i$ |
| $[FR_{18}^{6,1,0}]_{4,1,1}$ | ✕ 🌐🎗⭕ | 1.515 | $0.121 + 0.993i$ |
| $[FR_{18}^{6,1,0}]_{4,2,1}$ | ✕ 🌐🎗⭕ | 1.515 | $0.121 - 0.993i$ |
| $[FR_{18}^{6,1,0}]_{5,1,1}$ | ✕ 🌐🎗⭕ | 1.668 | $-0.355 - 0.935i$ |
| $[FR_{18}^{6,1,0}]_{5,2,1}$ | ✕ 🌐🎗⭕ | 1.668 | $-0.355 + 0.935i$ |
| $[FR_{18}^{6,1,0}]_{6,1,1}$ | ✕ 🌐🎗⭕ | $-3.148$ | $-0.749 - 0.663i$ |
| $[FR_{18}^{6,1,0}]_{6,2,1}$ | ✕ 🌐🎗⭕ | $-3.148$ | $-0.749 + 0.663i$ |

### 4.5.38  $FR_1^{6,1,2}$: $D_3$

For the fusion ring, the following multiplication table is used.

| 1 | 2 | 3 | 4 | 5 | 6 |
|---|---|---|---|---|---|
| 2 | 1 | 6 | 5 | 4 | 3 |
| 3 | 5 | 1 | 6 | 2 | 4 |
| 4 | 6 | 5 | 1 | 3 | 2 |
| 5 | 3 | 4 | 2 | 6 | 1 |
| 6 | 4 | 2 | 3 | 1 | 5 |

The following is the group of all non-trivial permutations that leave the fusion rules invariant:

$$S = \{(), (2\ 3\ 4), (2\ 4\ 3), (2\ 3)(5\ 6), (2\ 4)(5\ 6), (3\ 4)(5\ 6)\}.$$

Let

$$X_1 = S\left(\frac{[F_5^{262}]_4^3 [F_2^{332}]_5^1 [F_4^{342}]_6^6 [F_5^{426}]_3^6 [F_2^{434}]_6^5}{[F_2^{324}]_5^5 [F_4^{334}]_6^1 [F_5^{346}]_2^6}\right), \tag{63}$$

$$X_2 = S\left(d_2^L\right). \tag{64}$$

The following table lists a small set of invariants whose values completely distinguish between all MFPBFCs and MFPNBFCs with the given fusion rules.

Table 73: Symbolic invariants

| Name | Properties | $X_1$ | $X_2$ |
|---|---|---|---|
| $[FR_1^{6,1,2}]_{1,0,1}$ | † 🌐 | $\{1\}$ | $\{1\}$ |
| $[FR_1^{6,1,2}]_{1,0,2}$ | 🌐 | $\{1\}$ | $\{-1\}$ |
| $[FR_1^{6,1,2}]_{2,0,1}$ | † 🌐 | $\{\zeta_3\}$ | $\{1\}$ |

Table 73: Symbolic invariants (Continued)

| Name | Properties | $X_1$ | $X_2$ |
|---|---|---|---|
| $[FR_1^{6,1,2}]_{2,0,2}$ | 🌐 | $\{\zeta_3\}$ | $\{-1\}$ |
| $[FR_1^{6,1,2}]_{3,0,1}$ | †🌐 | $\{\zeta_3^2\}$ | $\{1\}$ |
| $[FR_1^{6,1,2}]_{3,0,2}$ | 🌐 | $\{\zeta_3^2\}$ | $\{-1\}$ |
| $[FR_1^{6,1,2}]_{4,0,1}$ | †🌐 | $\{\zeta_6\}$ | $\{1\}$ |
| $[FR_1^{6,1,2}]_{4,0,2}$ | 🌐 | $\{\zeta_6\}$ | $\{-1\}$ |
| $[FR_1^{6,1,2}]_{5,0,1}$ | †🌐 | $\{-1\}$ | $\{1\}$ |
| $[FR_1^{6,1,2}]_{5,0,2}$ | 🌐 | $\{-1\}$ | $\{-1\}$ |
| $[FR_1^{6,1,2}]_{6,0,1}$ | †🌐 | $\{\zeta_6^5\}$ | $\{1\}$ |
| $[FR_1^{6,1,2}]_{6,0,2}$ | 🌐 | $\{\zeta_6^5\}$ | $\{-1\}$ |

Table 74: Numeric invariants

| Name | Properties | $X_1$ | $X_2$ |
|---|---|---|---|
| $[FR_1^{6,1,2}]_{1,0,1}$ | †🌐 | $\{1\}$ | $\{1\}$ |
| $[FR_1^{6,1,2}]_{1,0,2}$ | 🌐 | $\{1\}$ | $\{-1\}$ |
| $[FR_1^{6,1,2}]_{2,0,1}$ | †🌐 | $\{-0.5 + 0.866i\}$ | $\{1\}$ |
| $[FR_1^{6,1,2}]_{2,0,2}$ | 🌐 | $\{-0.5 + 0.866i\}$ | $\{-1\}$ |
| $[FR_1^{6,1,2}]_{3,0,1}$ | †🌐 | $\{-0.5 - 0.866i\}$ | $\{1\}$ |
| $[FR_1^{6,1,2}]_{3,0,2}$ | 🌐 | $\{-0.5 - 0.866i\}$ | $\{-1\}$ |
| $[FR_1^{6,1,2}]_{4,0,1}$ | †🌐 | $\{0.5 + 0.866i, 0.5 + 0.866i\}$ | $\{1\}$ |
| $[FR_1^{6,1,2}]_{4,0,2}$ | 🌐 | $\{0.5 + 0.866i, 0.5 + 0.866i\}$ | $\{-1\}$ |
| $[FR_1^{6,1,2}]_{5,0,1}$ | †🌐 | $\{-1\}$ | $\{1\}$ |
| $[FR_1^{6,1,2}]_{5,0,2}$ | 🌐 | $\{-1\}$ | $\{-1\}$ |
| $[FR_1^{6,1,2}]_{6,0,1}$ | †🌐 | $\{0.5 - 0.866i, 0.5 - 0.866i\}$ | $\{1\}$ |
| $[FR_1^{6,1,2}]_{6,0,2}$ | 🌐 | $\{0.5 - 0.866i, 0.5 - 0.866i\}$ | $\{-1\}$ |

## 4.5.39 $FR_2^{6,1,2}$: $[\mathbb{Z}_2 \triangleleft \mathbb{Z}_4]_{1|0}^{\text{Id}}$

For the fusion ring, the following multiplication table is used.

| 1 | 2 | 3 | 4 | 5 | 6 |
|---|---|---|---|---|---|
| 2 | 1 | 4 | 3 | 5 | 6 |
| 3 | 4 | 2 | 1 | 6 | 5 |
| 4 | 3 | 1 | 2 | 6 | 5 |
| 5 | 5 | 6 | 6 | 1+2 | 3+4 |
| 6 | 6 | 5 | 5 | 3+4 | 1+2 |

The following is the group of all non-trivial permutations that leave the fusion rules invariant:

$$S = \{(), (3\ 4), (5\ 6), (3\ 4)(5\ 6)\}.$$

Let

$$X_1 = S\left(\left([F_5^{353}]_6^6[F_6^{363}]_5^5, [F_3^{535}]_6^6[F_6^{565}]_3^3, d_3^L, d_5^L\right)\right). \tag{65}$$

The following table lists a small set of invariants whose values completely distinguish between all MFPBFCs and MFPNBFCs with the given fusion rules.

Table 75: Symbolic invariants

| Name | Properties | $X_1$ |
|---|---|---|
| $[FR_2^{6,1,2}]_{1,0,1}$ | †  | $\left\{\left(-i, \frac{1}{\sqrt{2}}, 1, \sqrt{2}\right)\right\}$ |
| $[FR_2^{6,1,2}]_{1,0,2}$ |  | $\left\{\left(-i, \frac{1}{\sqrt{2}}, -1, -\sqrt{2}\right), \left(-i, \frac{1}{\sqrt{2}}, -1, \sqrt{2}\right)\right\}$ |
| $[FR_2^{6,1,2}]_{1,0,3}$ |  | $\left\{\left(-i, \frac{1}{\sqrt{2}}, 1, -\sqrt{2}\right)\right\}$ |
| $[FR_2^{6,1,2}]_{2,0,1}$ | †  | $\left\{\left(i, \frac{1}{\sqrt{2}}, 1, \sqrt{2}\right)\right\}$ |
| $[FR_2^{6,1,2}]_{2,0,2}$ |  | $\left\{\left(i, \frac{1}{\sqrt{2}}, -1, -\sqrt{2}\right), \left(i, \frac{1}{\sqrt{2}}, -1, \sqrt{2}\right)\right\}$ |
| $[FR_2^{6,1,2}]_{2,0,3}$ |  | $\left\{\left(i, \frac{1}{\sqrt{2}}, 1, -\sqrt{2}\right)\right\}$ |
| $[FR_2^{6,1,2}]_{3,0,1}$ | †  | $\left\{\left(-i, -\frac{1}{\sqrt{2}}, 1, \sqrt{2}\right), \left(-i, \frac{1}{\sqrt{2}}, 1, \sqrt{2}\right)\right\}$ |
| $[FR_2^{6,1,2}]_{3,0,2}$ |  | $\left\{\left(-i, -\frac{1}{\sqrt{2}}, -1, \sqrt{2}\right), \left(-i, \frac{1}{\sqrt{2}}, -1, -\sqrt{2}\right)\right\}$ |
| $[FR_2^{6,1,2}]_{3,0,3}$ |  | $\left\{\left(-i, -\frac{1}{\sqrt{2}}, -1, -\sqrt{2}\right), \left(-i, \frac{1}{\sqrt{2}}, -1, \sqrt{2}\right)\right\}$ |
| $[FR_2^{6,1,2}]_{3,0,4}$ |  | $\left\{\left(-i, -\frac{1}{\sqrt{2}}, 1, -\sqrt{2}\right), \left(-i, \frac{1}{\sqrt{2}}, 1, -\sqrt{2}\right)\right\}$ |
| $[FR_2^{6,1,2}]_{4,0,1}$ | †  | $\left\{\left(i, -\frac{1}{\sqrt{2}}, 1, \sqrt{2}\right), \left(i, \frac{1}{\sqrt{2}}, 1, \sqrt{2}\right)\right\}$ |
| $[FR_2^{6,1,2}]_{4,0,2}$ |  | $\left\{\left(i, -\frac{1}{\sqrt{2}}, -1, \sqrt{2}\right), \left(i, \frac{1}{\sqrt{2}}, -1, \sqrt{2}\right)\right\}$ |
| $[FR_2^{6,1,2}]_{4,0,3}$ |  | $\left\{\left(i, -\frac{1}{\sqrt{2}}, -1, -\sqrt{2}\right), \left(i, \frac{1}{\sqrt{2}}, -1, \sqrt{2}\right)\right\}$ |
| $[FR_2^{6,1,2}]_{5,0,1}$ |  | $\left\{\left(i, -\frac{1}{\sqrt{2}}, 1, -\sqrt{2}\right), \left(i, \frac{1}{\sqrt{2}}, 1, -\sqrt{2}\right)\right\}$ |
| $[FR_2^{6,1,2}]_{5,0,2}$ | †  | $\left\{\left(-i, -\frac{1}{\sqrt{2}}, 1, \sqrt{2}\right)\right\}$ |
| $[FR_2^{6,1,2}]_{5,0,3}$ |  | $\left\{\left(-i, -\frac{1}{\sqrt{2}}, -1, -\sqrt{2}\right), \left(-i, -\frac{1}{\sqrt{2}}, -1, \sqrt{2}\right)\right\}$ |
| $[FR_2^{6,1,2}]_{6,0,1}$ |  | $\left\{\left(-i, -\frac{1}{\sqrt{2}}, 1, -\sqrt{2}\right)\right\}$ |
| $[FR_2^{6,1,2}]_{6,0,2}$ | †  | $\left\{\left(i, -\frac{1}{\sqrt{2}}, 1, \sqrt{2}\right)\right\}$ |
| $[FR_2^{6,1,2}]_{6,0,3}$ |  | $\left\{\left(i, -\frac{1}{\sqrt{2}}, -1, -\sqrt{2}\right), \left(i, -\frac{1}{\sqrt{2}}, -1, \sqrt{2}\right)\right\}$ |
| $[FR_2^{6,1,2}]_{6,0,4}$ |  | $\left\{\left(i, -\frac{1}{\sqrt{2}}, 1, -\sqrt{2}\right)\right\}$ |

Table 76: Numeric invariants

| Name | Properties | $X_1$ |
|---|---|---|
| $[FR_2^{6,1,2}]_{1,0,1}$ |  | $\{(-i, 0.707, 1, 1.414)\}$ |
| $[FR_2^{6,1,2}]_{1,0,2}$ |  | $\{(-i, 0.707, -1, -1.414), (-i, 0.707, -1, 1.414)\}$ |
| $[FR_2^{6,1,2}]_{1,0,3}$ |  | $\{(-i, 0.707, 1, -1.414)\}$ |
| $[FR_2^{6,1,2}]_{2,0,1}$ |  | $\{(i, 0.707, 1, 1.414)\}$ |
| $[FR_2^{6,1,2}]_{2,0,2}$ |  | $\{(i, 0.707, -1, -1.414), (i, 0.707, -1, 1.414)\}$ |
| $[FR_2^{6,1,2}]_{2,0,3}$ |  | $\{(i, 0.707, 1, -1.414)\}$ |
| $[FR_2^{6,1,2}]_{3,0,1}$ |  | $\{(-i, -0.707, 1, 1.414), (-i, 0.707, 1, 1.414)\}$ |
| $[FR_2^{6,1,2}]_{3,0,2}$ |  | $\{(-i, -0.707, -1, 1.414), (-i, 0.707, -1, -1.414)\}$ |
| $[FR_2^{6,1,2}]_{3,0,3}$ |  | $\{(-i, -0.707, -1, -1.414), (-i, 0.707, -1, 1.414)\}$ |
| $[FR_2^{6,1,2}]_{3,0,4}$ |  | $\{(-i, -0.707, 1, -1.414), (-i, 0.707, 1, -1.414)\}$ |
| $[FR_2^{6,1,2}]_{4,0,1}$ |  | $\{(i, -0.707, 1, 1.414), (i, 0.707, 1, 1.414)\}$ |
| $[FR_2^{6,1,2}]_{4,0,2}$ |  | $\{(i, -0.707, -1, 1.414), (i, 0.707, -1, -1.414)\}$ |
| $[FR_2^{6,1,2}]_{4,0,3}$ |  | $\{(i, -0.707, -1, -1.414), (i, 0.707, -1, 1.414)\}$ |
| $[FR_2^{6,1,2}]_{5,0,1}$ |  | $\{(i, -0.707, 1, -1.414), (i, 0.707, 1, -1.414)\}$ |
| $[FR_2^{6,1,2}]_{5,0,2}$ |  | $\{(-i, -0.707, 1, 1.414)\}$ |
| $[FR_2^{6,1,2}]_{5,0,3}$ |  | $\{(-i, -0.707, -1, -1.414), (-i, -0.707, -1, 1.414)\}$ |
| $[FR_2^{6,1,2}]_{6,0,1}$ |  | $\{(-i, -0.707, 1, -1.414)\}$ |
| $[FR_2^{6,1,2}]_{6,0,2}$ |  | $\{(i, -0.707, 1, 1.414)\}$ |
| $[FR_2^{6,1,2}]_{6,0,3}$ |  | $\{(i, -0.707, -1, -1.414), (i, -0.707, -1, 1.414)\}$ |
| $[FR_2^{6,1,2}]_{6,0,4}$ |  | $\{(i, -0.707, 1, -1.414)\}$ |

## 4.5.40 $FR_3^{6,1,2}$: $[\mathbb{Z}_2 \triangleleft \mathbb{Z}_2 \otimes \mathbb{Z}_2]_{3|0}^{\text{Id}}$

For the fusion ring, the following multiplication table is used.

$$\begin{vmatrix} 1 & 2 & 3 & 4 & 5 & 6 \\ 2 & 1 & 4 & 3 & 6 & 5 \\ 3 & 4 & 1 & 2 & 6 & 5 \\ 4 & 3 & 2 & 1 & 5 & 6 \\ 5 & 6 & 6 & 5 & 2+3 & 1+4 \\ 6 & 5 & 5 & 6 & 1+4 & 2+3 \end{vmatrix}$$

The following is the group of all non-trivial permutations that leave the fusion rules invariant:

$$S = \{(), (2\ 3), (5\ 6), (2\ 3)(5\ 6)\}.$$

Let

$$X_1 = S\left([F_3^{232}]_4^4[F_4^{242}]_3^3\right), \tag{66}$$
$$X_2 = S\left(R_2^{55}\right), \tag{67}$$
$$X_3 = S\left(d_5^L\right). \tag{68}$$

The following table lists a small set of invariants whose values completely distinguish between all MFPBFCs and MFPNBFCs with the given fusion rules.

Table 77: Symbolic invariants

| Name | Properties | $X_1$ | $X_2$ | $X_3$ |
|---|---|---|---|---|
| $[FR_3^{6,1,2}]_{1,1,1}$ | ✂ † 🌐 🎗 | $\{1\}$ | $\{\zeta_{16}^7, \zeta_{16}^{11}\}$ | $\{\sqrt{2}\}$ |
| $[FR_3^{6,1,2}]_{1,1,2}$ | ✂ 🌐 🎗 | $\{1\}$ | $\{\zeta_{16}^7, \zeta_{16}^{11}\}$ | $\{-\sqrt{2}\}$ |
| $[FR_3^{6,1,2}]_{1,1,3}$ | ✂ | $\{1\}$ | $\{\zeta_{16}^7, \zeta_{16}^{11}\}$ | $\{-i\sqrt{2}, i\sqrt{2}\}$ |
| $[FR_3^{6,1,2}]_{1,2,1}$ | ✂ † 🌐 🎗 | $\{1\}$ | $\{\zeta_{16}^7, \zeta_{16}^3\}$ | $\{\sqrt{2}\}$ |
| $[FR_3^{6,1,2}]_{1,2,2}$ | ✂ 🌐 🎗 | $\{1\}$ | $\{\zeta_{16}^7, \zeta_{16}^3\}$ | $\{-\sqrt{2}\}$ |
| $[FR_3^{6,1,2}]_{1,2,3}$ | ✂ | $\{1\}$ | $\{\zeta_{16}^7, \zeta_{16}^3\}$ | $\{-i\sqrt{2}, i\sqrt{2}\}$ |
| $[FR_3^{6,1,2}]_{1,3,1}$ | ✂ † 🌐 🎗 | $\{1\}$ | $\{\zeta_{16}^{11}, \zeta_{16}^{15}\}$ | $\{\sqrt{2}\}$ |
| $[FR_3^{6,1,2}]_{1,3,2}$ | ✂ 🌐 🎗 | $\{1\}$ | $\{\zeta_{16}^{11}, \zeta_{16}^{15}\}$ | $\{-\sqrt{2}\}$ |
| $[FR_3^{6,1,2}]_{1,3,3}$ | ✂ | $\{1\}$ | $\{\zeta_{16}^{11}, \zeta_{16}^{15}\}$ | $\{-i\sqrt{2}, i\sqrt{2}\}$ |
| $[FR_3^{6,1,2}]_{1,4,1}$ | ✂ † 🌐 🎗 | $\{1\}$ | $\{\zeta_{16}^3, \zeta_{16}^{15}\}$ | $\{\sqrt{2}\}$ |
| $[FR_3^{6,1,2}]_{1,4,2}$ | ✂ 🌐 🎗 | $\{1\}$ | $\{\zeta_{16}^3, \zeta_{16}^{15}\}$ | $\{-\sqrt{2}\}$ |
| $[FR_3^{6,1,2}]_{1,4,3}$ | ✂ | $\{1\}$ | $\{\zeta_{16}^3, \zeta_{16}^{15}\}$ | $\{-i\sqrt{2}, i\sqrt{2}\}$ |
| $[FR_3^{6,1,2}]_{1,5,1}$ | ✂ † 🌐 🎗 | $\{1\}$ | $\{\zeta_{16}^9, \zeta_{16}^5\}$ | $\{\sqrt{2}\}$ |
| $[FR_3^{6,1,2}]_{1,5,2}$ | ✂ 🌐 🎗 | $\{1\}$ | $\{\zeta_{16}^9, \zeta_{16}^5\}$ | $\{-\sqrt{2}\}$ |
| $[FR_3^{6,1,2}]_{1,5,3}$ | ✂ | $\{1\}$ | $\{\zeta_{16}^9, \zeta_{16}^5\}$ | $\{-i\sqrt{2}, i\sqrt{2}\}$ |
| $[FR_3^{6,1,2}]_{1,6,1}$ | ✂ † 🌐 🎗 | $\{1\}$ | $\{\zeta_{16}^9, \zeta_{16}^{13}\}$ | $\{\sqrt{2}\}$ |

Table 77: Symbolic invariants (Continued)

| Name | Properties | $X_1$ | $X_2$ | $X_3$ |
|---|---|---|---|---|
| $[FR_3^{6,1,2}]_{1,6,2}$ | ✗ 🌐 🎗 | $\{1\}$ | $\{\zeta_{16}^9, \zeta_{16}^{13}\}$ | $\{-\sqrt{2}\}$ |
| $[FR_3^{6,1,2}]_{1,6,3}$ | ✗ | $\{1\}$ | $\{\zeta_{16}^9, \zeta_{16}^{13}\}$ | $\{-i\sqrt{2}, i\sqrt{2}\}$ |
| $[FR_3^{6,1,2}]_{1,7,1}$ | ✗ † 🌐 🎗 | $\{1\}$ | $\{\zeta_{16}^5, \zeta_{16}\}$ | $\{\sqrt{2}\}$ |
| $[FR_3^{6,1,2}]_{1,7,2}$ | ✗ 🌐 🎗 | $\{1\}$ | $\{\zeta_{16}^5, \zeta_{16}\}$ | $\{-\sqrt{2}\}$ |
| $[FR_3^{6,1,2}]_{1,7,3}$ | ✗ | $\{1\}$ | $\{\zeta_{16}^5, \zeta_{16}\}$ | $\{-i\sqrt{2}, i\sqrt{2}\}$ |
| $[FR_3^{6,1,2}]_{1,8,1}$ | ✗ † 🌐 🎗 | $\{1\}$ | $\{\zeta_{16}^{13}, \zeta_{16}\}$ | $\{\sqrt{2}\}$ |
| $[FR_3^{6,1,2}]_{1,8,2}$ | ✗ 🌐 🎗 | $\{1\}$ | $\{\zeta_{16}^{13}, \zeta_{16}\}$ | $\{-\sqrt{2}\}$ |
| $[FR_3^{6,1,2}]_{1,8,3}$ | ✗ | $\{1\}$ | $\{\zeta_{16}^{13}, \zeta_{16}\}$ | $\{-i\sqrt{2}, i\sqrt{2}\}$ |
| $[FR_3^{6,1,2}]_{2,0,1}$ | † 🌐 | $\{-1\}$ | - | $\{\sqrt{2}\}$ |
| $[FR_3^{6,1,2}]_{2,0,2}$ | 🌐 | $\{-1\}$ | - | $\{-\sqrt{2}\}$ |
| $[FR_3^{6,1,2}]_{2,0,3}$ |  | $\{-1\}$ | - | $\{-i\sqrt{2}, i\sqrt{2}\}$ |

Table 78: Numeric invariants

| Name | Properties | $X_1$ | $X_2$ | $X_3$ |
|---|---|---|---|---|
| $[FR_3^{6,1,2}]_{1,1,1}$ | ✗ † 🌐 🎗 | $\{1\}$ | $\{-0.924 + 0.383i,\ -0.383 - 0.924i\}$ | $\{1.414\}$ |
| $[FR_3^{6,1,2}]_{1,1,2}$ | ✗ 🌐 🎗 | $\{1\}$ | $\{-0.924 + 0.383i,\ -0.383 - 0.924i\}$ | $\{-1.414\}$ |
| $[FR_3^{6,1,2}]_{1,1,3}$ | ✗ | $\{1\}$ | $\{-0.924 + 0.383i,\ -0.383 - 0.924i\}$ | $\{-1.414i,\ 1.414i\}$ |
| $[FR_3^{6,1,2}]_{1,2,1}$ | ✗ † 🌐 🎗 | $\{1\}$ | $\{-0.924 + 0.383i,\ 0.383 + 0.924i\}$ | $\{1.414\}$ |
| $[FR_3^{6,1,2}]_{1,2,2}$ | ✗ 🌐 🎗 | $\{1\}$ | $\{-0.924 + 0.383i,\ 0.383 + 0.924i\}$ | $\{-1.414\}$ |
| $[FR_3^{6,1,2}]_{1,2,3}$ | ✗ | $\{1\}$ | $\{-0.924 + 0.383i,\ 0.383 + 0.924i\}$ | $\{-1.414i,\ 1.414i\}$ |
| $[FR_3^{6,1,2}]_{1,3,1}$ | ✗ † 🌐 🎗 | $\{1\}$ | $\{-0.383 - 0.924i,\ 0.924 - 0.383i\}$ | $\{1.414\}$ |
| $[FR_3^{6,1,2}]_{1,3,2}$ | ✗ 🌐 🎗 | $\{1\}$ | $\{-0.383 - 0.924i,\ 0.924 - 0.383i\}$ | $\{-1.414\}$ |
| $[FR_3^{6,1,2}]_{1,3,3}$ | ✗ | $\{1\}$ | $\{-0.383 - 0.924i,\ 0.924 - 0.383i\}$ | $\{-1.414i,\ 1.414i\}$ |
| $[FR_3^{6,1,2}]_{1,4,1}$ | ✗ † 🌐 🎗 | $\{1\}$ | $\{0.383 + 0.924i,\ 0.924 - 0.383i\}$ | $\{1.414\}$ |

Table 78: Numeric invariants (Continued)

| Name | Properties | $X_1$ | $X_2$ | $X_3$ |
|---|---|---|---|---|
| $[FR_3^{6,1,2}]_{1,4,2}$ | ✗ 🌐 🎗 | $\{1\}$ | $\left\{\begin{array}{l}0.383 + 0.924i, \\ 0.924 - 0.383i\end{array}\right\}$ | $\{-1.414\}$ |
| $[FR_3^{6,1,2}]_{1,4,3}$ | ✗ | $\{1\}$ | $\left\{\begin{array}{l}0.383 + 0.924i, \\ 0.924 - 0.383i\end{array}\right\}$ | $\left\{\begin{array}{l}-1.414i, \\ 1.414i\end{array}\right\}$ |
| $[FR_3^{6,1,2}]_{1,5,1}$ | ✗ † 🌐 🎗 | $\{1\}$ | $\left\{\begin{array}{l}-0.924 - 0.383i, \\ -0.383 + 0.924i\end{array}\right\}$ | $\{1.414\}$ |
| $[FR_3^{6,1,2}]_{1,5,2}$ | ✗ 🌐 🎗 | $\{1\}$ | $\left\{\begin{array}{l}-0.924 - 0.383i, \\ -0.383 + 0.924i\end{array}\right\}$ | $\{-1.414\}$ |
| $[FR_3^{6,1,2}]_{1,5,3}$ | ✗ | $\{1\}$ | $\left\{\begin{array}{l}-0.924 - 0.383i, \\ -0.383 + 0.924i\end{array}\right\}$ | $\left\{\begin{array}{l}-1.414i, \\ 1.414i\end{array}\right\}$ |
| $[FR_3^{6,1,2}]_{1,6,1}$ | ✗ † 🌐 🎗 | $\{1\}$ | $\left\{\begin{array}{l}-0.924 - 0.383i, \\ 0.383 - 0.924i\end{array}\right\}$ | $\{1.414\}$ |
| $[FR_3^{6,1,2}]_{1,6,2}$ | ✗ 🌐 🎗 | $\{1\}$ | $\left\{\begin{array}{l}-0.924 - 0.383i, \\ 0.383 - 0.924i\end{array}\right\}$ | $\{-1.414\}$ |
| $[FR_3^{6,1,2}]_{1,6,3}$ | ✗ | $\{1\}$ | $\left\{\begin{array}{l}-0.924 - 0.383i, \\ 0.383 - 0.924i\end{array}\right\}$ | $\left\{\begin{array}{l}-1.414i, \\ 1.414i\end{array}\right\}$ |
| $[FR_3^{6,1,2}]_{1,7,1}$ | ✗ † 🌐 🎗 | $\{1\}$ | $\left\{\begin{array}{l}-0.383 + 0.924i, \\ 0.924 + 0.383i\end{array}\right\}$ | $\{1.414\}$ |
| $[FR_3^{6,1,2}]_{1,7,2}$ | ✗ 🌐 🎗 | $\{1\}$ | $\left\{\begin{array}{l}-0.383 + 0.924i, \\ 0.924 + 0.383i\end{array}\right\}$ | $\{-1.414\}$ |
| $[FR_3^{6,1,2}]_{1,7,3}$ | ✗ | $\{1\}$ | $\left\{\begin{array}{l}-0.383 + 0.924i, \\ 0.924 + 0.383i\end{array}\right\}$ | $\left\{\begin{array}{l}-1.414i, \\ 1.414i\end{array}\right\}$ |
| $[FR_3^{6,1,2}]_{1,8,1}$ | ✗ † 🌐 🎗 | $\{1\}$ | $\left\{\begin{array}{l}0.383 - 0.924i, \\ 0.924 + 0.383i\end{array}\right\}$ | $\{1.414\}$ |
| $[FR_3^{6,1,2}]_{1,8,2}$ | ✗ 🌐 🎗 | $\{1\}$ | $\left\{\begin{array}{l}0.383 - 0.924i, \\ 0.924 + 0.383i\end{array}\right\}$ | $\{-1.414\}$ |
| $[FR_3^{6,1,2}]_{1,8,3}$ | ✗ | $\{1\}$ | $\left\{\begin{array}{l}0.383 - 0.924i, \\ 0.924 + 0.383i\end{array}\right\}$ | $\left\{\begin{array}{l}-1.414i, \\ 1.414i\end{array}\right\}$ |
| $[FR_3^{6,1,2}]_{2,0,1}$ | † 🌐 | $\{-1\}$ | - | $\{1.414i\}$ |
| $[FR_3^{6,1,2}]_{2,0,2}$ | 🌐 | $\{-1\}$ | - | $\{-1.414\}$ |
| $[FR_3^{6,1,2}]_{2,0,3}$ |  | $\{-1\}$ | - | $\left\{\begin{array}{l}-1.414i, \\ 1.414i\end{array}\right\}$ |

### 4.5.41 $FR_4^{6,1,2}$: $\mathbf{Rep}(\mathbf{Dic}_{12})$

For the fusion ring, the following multiplication table is used.

$$\begin{array}{|cccccc|}
\hline
1 & 2 & 3 & 4 & 5 & 6 \\
2 & 1 & 4 & 3 & 5 & 6 \\
3 & 4 & 2 & 1 & 6 & 5 \\
4 & 3 & 1 & 2 & 6 & 5 \\
5 & 5 & 6 & 6 & 1+2+6 & 3+4+5 \\
6 & 6 & 5 & 5 & 3+4+5 & 1+2+6 \\
\hline
\end{array}$$

The following is the group of all non-trivial permutations that leave the fusion rules invariant:

$$S = \{(), (3\ 4)\}.$$

Let

$$X_1 = S\left([F_5^{353}]_6^6 [F_6^{363}]_5^5\right), \tag{69}$$
$$X_2 = S\left(d_3^L\right). \tag{70}$$

The following table lists a small set of invariants whose values completely distinguish between all MFPBFCs and MFPNBFCs with the given fusion rules.

Table 79: Symbolic invariants

| Name | Properties | $[F_1^{666}]_6^6$ | $X_1$ | $R_1^{55}$ | $X_2$ |
|---|---|---|---|---|---|
| $[FR_4^{6,1,2}]_{1,1,1}$ | ✗ ✝ 🌐 🎗 | 1 | $\{-1\}$ | $-i$ | $\{1\}$ |
| $[FR_4^{6,1,2}]_{1,1,2}$ | ✗ 🌐 🎗 | 1 | $\{-1\}$ | $-i$ | $\{-1\}$ |
| $[FR_4^{6,1,2}]_{1,2,1}$ | ✗ ✝ 🌐 🎗 | 1 | $\{-1\}$ | $\zeta_{12}^5$ | $\{1\}$ |
| $[FR_4^{6,1,2}]_{1,2,2}$ | ✗ 🌐 🎗 | 1 | $\{-1\}$ | $\zeta_{12}^5$ | $\{-1\}$ |
| $[FR_4^{6,1,2}]_{1,3,1}$ | ✗ ✝ 🌐 🎗 | 1 | $\{-1\}$ | $\zeta_{12}$ | $\{1\}$ |
| $[FR_4^{6,1,2}]_{1,3,2}$ | ✗ 🌐 🎗 | 1 | $\{-1\}$ | $\zeta_{12}$ | $\{-1\}$ |
| $[FR_4^{6,1,2}]_{1,4,1}$ | ✗ ✝ 🌐 🎗 | 1 | $\{-1\}$ | $i$ | $\{1\}$ |
| $[FR_4^{6,1,2}]_{1,4,2}$ | ✗ 🌐 🎗 | 1 | $\{-1\}$ | $i$ | $\{-1\}$ |
| $[FR_4^{6,1,2}]_{1,5,1}$ | ✗ ✝ 🌐 🎗 | 1 | $\{-1\}$ | $\zeta_{12}^{11}$ | $\{1\}$ |
| $[FR_4^{6,1,2}]_{1,5,2}$ | ✗ 🌐 🎗 | 1 | $\{-1\}$ | $\zeta_{12}^{11}$ | $\{-1\}$ |
| $[FR_4^{6,1,2}]_{1,6,1}$ | ✗ ✝ 🌐 🎗 | 1 | $\{-1\}$ | $\zeta_{12}^7$ | $\{1\}$ |
| $[FR_4^{6,1,2}]_{1,6,2}$ | ✗ 🌐 🎗 | 1 | $\{-1\}$ | $\zeta_{12}^7$ | $\{-1\}$ |
| $[FR_4^{6,1,2}]_{2,1,1}$ | ✗ ✝ 🌐 🎗 | 1 | $\{1\}$ | $-1$ | $\{1\}$ |
| $[FR_4^{6,1,2}]_{2,1,2}$ | ✗ 🌐 🎗 | 1 | $\{1\}$ | $-1$ | $\{-1\}$ |
| $[FR_4^{6,1,2}]_{2,2,1}$ | ✗ ✝ 🌐 🎗 | 1 | $\{1\}$ | $\zeta_6$ | $\{1\}$ |
| $[FR_4^{6,1,2}]_{2,2,2}$ | ✗ 🌐 🎗 | 1 | $\{1\}$ | $\zeta_6$ | $\{-1\}$ |
| $[FR_4^{6,1,2}]_{2,3,1}$ | ✗ ✝ 🌐 🎗 | 1 | $\{1\}$ | $\zeta_6^5$ | $\{1\}$ |
| $[FR_4^{6,1,2}]_{2,3,2}$ | ✗ 🌐 🎗 | 1 | $\{1\}$ | $\zeta_6^5$ | $\{-1\}$ |

Table 79: Symbolic invariants (Continued)

| Name | Properties | $[F_1^{666}]_6^6$ | $X_1$ | $R_1^{55}$ | $X_2$ |
|---|---|---|---|---|---|
| $[FR_4^{6,1,2}]_{2,4,1}$ | | 1 | {1} | 1 | {1} |
| $[FR_4^{6,1,2}]_{2,4,2}$ | | 1 | {1} | 1 | {−1} |
| $[FR_4^{6,1,2}]_{2,5,1}$ | | 1 | {1} | $\zeta_3^2$ | {1} |
| $[FR_4^{6,1,2}]_{2,5,2}$ | | 1 | {1} | $\zeta_3^2$ | {−1} |
| $[FR_4^{6,1,2}]_{2,6,1}$ | | 1 | {1} | $\zeta_3$ | {1} |
| $[FR_4^{6,1,2}]_{2,6,2}$ | | 1 | {1} | $\zeta_3$ | {−1} |
| $[FR_4^{6,1,2}]_{3,0,1}$ | | $\zeta_3^2$ | {−1} | − | {1} |
| $[FR_4^{6,1,2}]_{3,0,2}$ | | $\zeta_3^2$ | {−1} | − | {−1} |
| $[FR_4^{6,1,2}]_{4,0,1}$ | | $\zeta_3^2$ | {1} | − | {1} |
| $[FR_4^{6,1,2}]_{4,0,2}$ | | $\zeta_3^2$ | {1} | − | {−1} |
| $[FR_4^{6,1,2}]_{5,0,1}$ | | $\zeta_3$ | {−1} | − | {1} |
| $[FR_4^{6,1,2}]_{5,0,2}$ | | $\zeta_3$ | {−1} | − | {−1} |
| $[FR_4^{6,1,2}]_{6,0,1}$ | | $\zeta_3$ | {1} | − | {1} |
| $[FR_4^{6,1,2}]_{6,0,2}$ | | $\zeta_3$ | {1} | − | {−1} |

Table 80: Numeric invariants

| Name | Properties | $[F_1^{666}]_6^6$ | $X_1$ | $R_1^{55}$ | $X_2$ |
|---|---|---|---|---|---|
| $[FR_4^{6,1,2}]_{1,1,1}$ | | 1 | {−1} | $-i$ | {1} |
| $[FR_4^{6,1,2}]_{1,1,2}$ | | 1 | {−1} | $-i$ | {−1} |
| $[FR_4^{6,1,2}]_{1,2,1}$ | | 1 | {−1} | $-0.866 + 0.5i$ | {1} |
| $[FR_4^{6,1,2}]_{1,2,2}$ | | 1 | {−1} | $-0.866 + 0.5i$ | {−1} |
| $[FR_4^{6,1,2}]_{1,3,1}$ | | 1 | {−1} | $0.866 + 0.5i$ | {1} |
| $[FR_4^{6,1,2}]_{1,3,2}$ | | 1 | {−1} | $0.866 + 0.5i$ | {−1} |
| $[FR_4^{6,1,2}]_{1,4,1}$ | | 1 | {−1} | $i$ | {1} |
| $[FR_4^{6,1,2}]_{1,4,2}$ | | 1 | {−1} | $i$ | {−1} |
| $[FR_4^{6,1,2}]_{1,5,1}$ | | 1 | {−1} | $0.866 - 0.5i$ | {1} |
| $[FR_4^{6,1,2}]_{1,5,2}$ | | 1 | {−1} | $0.866 - 0.5i$ | {−1} |
| $[FR_4^{6,1,2}]_{1,6,1}$ | | 1 | {−1} | $-0.866 - 0.5i$ | {1} |
| $[FR_4^{6,1,2}]_{1,6,2}$ | | 1 | {−1} | $-0.866 - 0.5i$ | {−1} |
| $[FR_4^{6,1,2}]_{2,1,1}$ | | 1 | {1} | $-1$ | {1} |
| $[FR_4^{6,1,2}]_{2,1,2}$ | | 1 | {1} | $-1$ | {−1} |
| $[FR_4^{6,1,2}]_{2,2,1}$ | | 1 | {1} | $0.5 + 0.866i$ | {1} |

Table 80: Numeric invariants (Continued)

| Name | Properties | $[F_1^{666}]_6^6$ | $X_1$ | $R_1^{55}$ | $X_2$ |
|---|---|---|---|---|---|
| $[FR_4^{6,1,2}]_{2,2,2}$ | ✗ 🌐 🎗 | 1 | {1} | $0.5 + 0.866i$ | {−1} |
| $[FR_4^{6,1,2}]_{2,3,1}$ | ✗ † 🌐 🎗 | 1 | {1} | $0.5 − 0.866i$ | {1} |
| $[FR_4^{6,1,2}]_{2,3,2}$ | ✗ 🌐 🎗 | 1 | {1} | $0.5 − 0.866i$ | {−1} |
| $[FR_4^{6,1,2}]_{2,4,1}$ | ✗ † 🌐 🎗 | 1 | {1} | 1 | {1} |
| $[FR_4^{6,1,2}]_{2,4,2}$ | ✗ 🌐 🎗 | 1 | {1} | 1 | {−1} |
| $[FR_4^{6,1,2}]_{2,5,1}$ | ✗ † 🌐 🎗 | 1 | {1} | $−0.5 − 0.866i$ | {1} |
| $[FR_4^{6,1,2}]_{2,5,2}$ | ✗ 🌐 🎗 | 1 | {1} | $−0.5 − 0.866i$ | {−1} |
| $[FR_4^{6,1,2}]_{2,6,1}$ | ✗ † 🌐 🎗 | 1 | {1} | $−0.5 + 0.866i$ | {1} |
| $[FR_4^{6,1,2}]_{2,6,2}$ | ✗ 🌐 🎗 | 1 | {1} | $−0.5 + 0.866i$ | {−1} |
| $[FR_4^{6,1,2}]_{3,0,1}$ | † 🌐 | $−0.5 − 0.866i$ | {−1} | - | {1} |
| $[FR_4^{6,1,2}]_{3,0,2}$ | 🌐 | $−0.5 − 0.866i$ | {−1} | - | {−1} |
| $[FR_4^{6,1,2}]_{4,0,1}$ | † 🌐 | $−0.5 − 0.866i$ | {1} | - | {1} |
| $[FR_4^{6,1,2}]_{4,0,2}$ | 🌐 | $−0.5 − 0.866i$ | {1} | - | {−1} |
| $[FR_4^{6,1,2}]_{5,0,1}$ | † 🌐 | $−0.5 + 0.866i$ | {−1} | - | {1} |
| $[FR_4^{6,1,2}]_{5,0,2}$ | 🌐 | $−0.5 + 0.866i$ | {−1} | - | {−1} |
| $[FR_4^{6,1,2}]_{6,0,1}$ | † 🌐 | $−0.5 + 0.866i$ | {1} | - | {1} |
| $[FR_4^{6,1,2}]_{6,0,2}$ | 🌐 | $−0.5 + 0.866i$ | {1} | - | {−1} |

### 4.5.42 $FR_7^{6,1,2}$: Pseudo $SO(5)_2$

For the fusion ring, the following multiplication table is used.

| 1 | 2 | 3 | 4 | 5 | 6 |
|---|---|---|---|---|---|
| 2 | 1 | 3 | 4 | 6 | 5 |
| 3 | 3 | $1+2+4$ | $3+4$ | $5+6$ | $5+6$ |
| 4 | 4 | $3+4$ | $1+2+3$ | $5+6$ | $5+6$ |
| 5 | 6 | $5+6$ | $5+6$ | $2+3+4$ | $1+3+4$ |
| 6 | 5 | $5+6$ | $5+6$ | $1+3+4$ | $2+3+4$ |

The following is the group of all non-trivial permutations that leave the fusion rules invariant:

$$S = \{(), (3\ 4), (5\ 6), (3\ 4)(5\ 6)\}.$$

Let

$$X_1 = S\left([F_5^{353}]_5^5\right), \tag{71}$$

$$X_2 = S\left(d_5^L\right). \tag{72}$$

The following table lists a small set of invariants whose values completely distinguish between all MFPBFCs and MFPNBFCs with the given fusion rules.

Table 81: Symbolic invariants

| Name | Properties | $X_1$ | $X_2$ |
|---|---|---|---|
| $[FR_7^{6,1,2}]_{1,0,1}$ | 🗡️🌐 | $\left\{\frac{-\phi}{2}\right\}$ | $\{\sqrt{5}\}$ |
| $[FR_7^{6,1,2}]_{1,0,2}$ | 🌐 | $\left\{\frac{-\phi}{2}\right\}$ | $\{-\sqrt{5}\}$ |
| $[FR_7^{6,1,2}]_{2,0,1}$ | 🗡️🌐 | $\left\{-\frac{\bar\phi}{2}\right\}$ | $\{\sqrt{5}\}$ |
| $[FR_7^{6,1,2}]_{2,0,2}$ | 🌐 | $\left\{-\frac{\bar\phi}{2}\right\}$ | $\{-\sqrt{5}\}$ |

Table 82: Numeric invariants

| Name | Properties | $X_1$ | $X_2$ |
|---|---|---|---|
| $[FR_7^{6,1,2}]_{1,0,1}$ | 🗡️🌐 | $\{-0.809\}$ | $\{2.236\}$ |
| $[FR_7^{6,1,2}]_{1,0,2}$ | 🌐 | $\{-0.809\}$ | $\{-2.236\}$ |
| $[FR_7^{6,1,2}]_{2,0,1}$ | 🗡️🌐 | $\{0.309\}$ | $\{2.236\}$ |
| $[FR_7^{6,1,2}]_{2,0,2}$ | 🌐 | $\{0.309\}$ | $\{-2.236\}$ |

### 4.5.43 $FR_8^{6,1,2}$: $HI(\mathbb{Z}_3)$

For the fusion ring, the following multiplication table is used.

```
1  2  3  4            5            6
2  3  1  5            6            4
3  1  2  6            4            5
4  6  5  1+4+5+6      3+4+5+6      2+4+5+6
5  4  6  2+4+5+6      1+4+5+6      3+4+5+6
6  5  4  3+4+5+6      2+4+5+6      1+4+5+6
```

The following is the group of all non-trivial permutations that leave the fusion rules invariant:

$$S = \{(), (4\ 5\ 6), (4\ 6\ 5), (2\ 3)(4\ 5), (2\ 3)(4\ 6), (2\ 3)(5\ 6)\}.$$

Let

$$X_1 = S\left([F_4^{444}]_4^4\right). \tag{73}$$

The following table lists a small set of invariants whose values completely distinguish between all MFPBFCs and MFPNBFCs with the given fusion rules.

Table 83: Symbolic invariants

| Name | Properties | $X_1$ |
|---|---|---|
| $[FR_8^{6,1,2}]_{1,0,1}$ | 🗡️🌐 | $\left\{\frac{1}{3}\left(2-\sqrt{13}\right)\right\}$ |

Table 83: Symbolic invariants (Continued)

| Name | Properties | $X_1$ |
|---|---|---|
| $[FR_8^{6,1,2}]_{2,0,1}$ | † 🌐 | $\{\frac{1}{6}(7-\sqrt{13})\}$ |
| $[FR_8^{6,1,2}]_{3,0,1}$ | 🌐 | $\{\frac{1}{6}(7+\sqrt{13})\}$ |
| $[FR_8^{6,1,2}]_{4,0,1}$ | 🌐 | $\{\frac{1}{3}(2+\sqrt{13})\}$ |

Table 84: Numeric invariants

| Name | Properties | $X_1$ |
|---|---|---|
| $[FR_8^{6,1,2}]_{1,0,1}$ | † 🌐 | $\{-0.535\}$ |
| $[FR_8^{6,1,2}]_{2,0,1}$ | † 🌐 | $\{0.566\}$ |
| $[FR_8^{6,1,2}]_{3,0,1}$ | 🌐 | $\{1.768\}$ |
| $[FR_8^{6,1,2}]_{4,0,1}$ | 🌐 | $\{1.869\}$ |

### 4.5.44 $FR_1^{6,1,4}$: $\mathbb{Z}_6$

For the fusion ring, the following multiplication table is used.

| 1 | 2 | 3 | 4 | 5 | 6 |
|---|---|---|---|---|---|
| 2 | 1 | 6 | 5 | 4 | 3 |
| 3 | 6 | 5 | 1 | 2 | 4 |
| 4 | 5 | 1 | 6 | 3 | 2 |
| 5 | 4 | 2 | 3 | 6 | 1 |
| 6 | 3 | 4 | 2 | 1 | 5 |

The following is the group of all non-trivial permutations that leave the fusion rules invariant:

$$S = \{(), (3\ 4)(5\ 6)\}.$$

Let

$$X_1 = S\left(\frac{[F_3^{223}]_6^1 [F_4^{233}]_5^6 [F_3^{322}]_1^6 [F_6^{335}]_2^5 [F_4^{355}]_6^2}{[F_3^{232}]_6^6 [F_4^{323}]_6^6}\right), \tag{74}$$

$$X_2 = S\left(R_5^{33}\right), \tag{75}$$

$$X_3 = S\left(d_3^L\right). \tag{76}$$

The following table lists a small set of invariants whose values completely distinguish between all MFPBFCs and MFPNBFCs with the given fusion rules.

Table 85: Symbolic invariants

| Name | Properties | $X_1$ | $X_2$ | $X_3$ |
|---|---|---|---|---|
| $[FR_1^{6,1,4}]_{1,1,1}$ | ✗ † 🌐 🎀 ⊚ | $\{-1\}$ | $\{\zeta_{12}^5\}$ | $\{1\}$ |

Table 85: Symbolic invariants (Continued)

| Name | Properties | $X_1$ | $X_2$ | $X_3$ |
|---|---|---|---|---|
| $[\text{FR}_1^{6,1,4}]_{1,1,2}$ | ✻ 🌐 🎗 ⚭ | $\{-1\}$ | $\{\zeta_{12}^5\}$ | $\{-1\}$ |
| $[\text{FR}_1^{6,1,4}]_{1,1,3}$ | ✻ | $\{-1\}$ | $\{\zeta_{12}^5\}$ | $\{\zeta_3^2, \zeta_3\}$ |
| $[\text{FR}_1^{6,1,4}]_{1,1,4}$ | ✻ | $\{-1\}$ | $\{\zeta_{12}^5\}$ | $\{\zeta_6^5, \zeta_6\}$ |
| $[\text{FR}_1^{6,1,4}]_{1,2,1}$ | ✻ † 🌐 🎗 ⚭ | $\{-1\}$ | $\{\zeta_{12}^{11}\}$ | $\{1\}$ |
| $[\text{FR}_1^{6,1,4}]_{1,2,2}$ | ✻ 🌐 🎗 ⚭ | $\{-1\}$ | $\{\zeta_{12}^{11}\}$ | $\{-1\}$ |
| $[\text{FR}_1^{6,1,4}]_{1,2,3}$ | ✻ | $\{-1\}$ | $\{\zeta_{12}^{11}\}$ | $\{\zeta_3^2, \zeta_3\}$ |
| $[\text{FR}_1^{6,1,4}]_{1,2,4}$ | ✻ | $\{-1\}$ | $\{\zeta_{12}^{11}\}$ | $\{\zeta_6^5, \zeta_6\}$ |
| $[\text{FR}_1^{6,1,4}]_{1,3,1}$ | ✻ † 🌐 🎗 ⚭ | $\{-1\}$ | $\{\zeta_{12}\}$ | $\{1\}$ |
| $[\text{FR}_1^{6,1,4}]_{1,3,2}$ | ✻ 🌐 🎗 ⚭ | $\{-1\}$ | $\{\zeta_{12}\}$ | $\{-1\}$ |
| $[\text{FR}_1^{6,1,4}]_{1,3,3}$ | ✻ | $\{-1\}$ | $\{\zeta_{12}\}$ | $\{\zeta_3^2, \zeta_3\}$ |
| $[\text{FR}_1^{6,1,4}]_{1,3,4}$ | ✻ | $\{-1\}$ | $\{\zeta_{12}\}$ | $\{\zeta_6^5, \zeta_6\}$ |
| $[\text{FR}_1^{6,1,4}]_{1,4,1}$ | ✻ † 🌐 🎗 ⚭ | $\{-1\}$ | $\{\zeta_{12}^7\}$ | $\{1\}$ |
| $[\text{FR}_1^{6,1,4}]_{1,4,2}$ | ✻ 🌐 🎗 ⚭ | $\{-1\}$ | $\{\zeta_{12}^7\}$ | $\{-1\}$ |
| $[\text{FR}_1^{6,1,4}]_{1,4,3}$ | ✻ | $\{-1\}$ | $\{\zeta_{12}^7\}$ | $\{\zeta_3^2, \zeta_3\}$ |
| $[\text{FR}_1^{6,1,4}]_{1,4,4}$ | ✻ | $\{-1\}$ | $\{\zeta_{12}^7\}$ | $\{\zeta_6^5, \zeta_6\}$ |
| $[\text{FR}_1^{6,1,4}]_{1,5,1}$ | ✻ † 🌐 🎗 | $\{-1\}$ | $\{-i\}$ | $\{1\}$ |
| $[\text{FR}_1^{6,1,4}]_{1,5,2}$ | ✻ 🌐 🎗 | $\{-1\}$ | $\{-i\}$ | $\{-1\}$ |
| $[\text{FR}_1^{6,1,4}]_{1,5,3}$ | ✻ | $\{-1\}$ | $\{-i\}$ | $\{\zeta_3^2, \zeta_3\}$ |
| $[\text{FR}_1^{6,1,4}]_{1,5,4}$ | ✻ | $\{-1\}$ | $\{-i\}$ | $\{\zeta_6^5, \zeta_6\}$ |
| $[\text{FR}_1^{6,1,4}]_{1,6,1}$ | ✻ † 🌐 🎗 | $\{-1\}$ | $\{i\}$ | $\{1\}$ |
| $[\text{FR}_1^{6,1,4}]_{1,6,2}$ | ✻ 🌐 🎗 | $\{-1\}$ | $\{i\}$ | $\{-1\}$ |
| $[\text{FR}_1^{6,1,4}]_{1,6,3}$ | ✻ | $\{-1\}$ | $\{i\}$ | $\{\zeta_3^2, \zeta_3\}$ |
| $[\text{FR}_1^{6,1,4}]_{1,6,4}$ | ✻ | $\{-1\}$ | $\{i\}$ | $\{\zeta_6^5, \zeta_6\}$ |
| $[\text{FR}_1^{6,1,4}]_{2,1,1}$ | ✻ † 🌐 🎗 | $\{1\}$ | $\{1\}$ | $\{1\}$ |
| $[\text{FR}_1^{6,1,4}]_{2,1,2}$ | ✻ 🌐 🎗 | $\{1\}$ | $\{1\}$ | $\{-1\}$ |
| $[\text{FR}_1^{6,1,4}]_{2,1,3}$ | ✻ | $\{1\}$ | $\{1\}$ | $\{\zeta_3^2, \zeta_3\}$ |
| $[\text{FR}_1^{6,1,4}]_{2,1,4}$ | ✻ | $\{1\}$ | $\{1\}$ | $\{\zeta_6^5, \zeta_6\}$ |
| $[\text{FR}_1^{6,1,4}]_{2,2,1}$ | ✻ † 🌐 🎗 | $\{1\}$ | $\{-1\}$ | $\{1\}$ |
| $[\text{FR}_1^{6,1,4}]_{2,2,2}$ | ✻ 🌐 🎗 | $\{1\}$ | $\{-1\}$ | $\{-1\}$ |
| $[\text{FR}_1^{6,1,4}]_{2,2,3}$ | ✻ | $\{1\}$ | $\{-1\}$ | $\{\zeta_3^2, \zeta_3\}$ |
| $[\text{FR}_1^{6,1,4}]_{2,2,4}$ | ✻ | $\{1\}$ | $\{-1\}$ | $\{\zeta_6^5, \zeta_6\}$ |
| $[\text{FR}_1^{6,1,4}]_{2,3,1}$ | ✻ † 🌐 🎗 | $\{1\}$ | $\{\zeta_3^2\}$ | $\{1\}$ |

Table 85: Symbolic invariants (Continued)

| Name | Properties | $X_1$ | $X_2$ | $X_3$ |
|---|---|---|---|---|
| $[FR_1^{6,1,4}]_{2,3,2}$ | ✖ 🌐 🎗 | $\{1\}$ | $\{\zeta_3^2\}$ | $\{-1\}$ |
| $[FR_1^{6,1,4}]_{2,3,3}$ | ✖ | $\{1\}$ | $\{\zeta_3^2\}$ | $\{\zeta_3^2, \zeta_3\}$ |
| $[FR_1^{6,1,4}]_{2,3,4}$ | ✖ | $\{1\}$ | $\{\zeta_3^2\}$ | $\{\zeta_6^5, \zeta_6\}$ |
| $[FR_1^{6,1,4}]_{2,4,1}$ | ✖ † 🌐 🎗 | $\{1\}$ | $\{\zeta_6\}$ | $\{1\}$ |
| $[FR_1^{6,1,4}]_{2,4,2}$ | ✖ 🌐 🎗 | $\{1\}$ | $\{\zeta_6\}$ | $\{-1\}$ |
| $[FR_1^{6,1,4}]_{2,4,3}$ | ✖ | $\{1\}$ | $\{\zeta_6\}$ | $\{\zeta_3^2, \zeta_3\}$ |
| $[FR_1^{6,1,4}]_{2,4,4}$ | ✖ | $\{1\}$ | $\{\zeta_6\}$ | $\{\zeta_6^5, \zeta_6\}$ |
| $[FR_1^{6,1,4}]_{2,5,1}$ | ✖ † 🌐 🎗 | $\{1\}$ | $\{\zeta_3\}$ | $\{1\}$ |
| $[FR_1^{6,1,4}]_{2,5,2}$ | ✖ 🌐 🎗 | $\{1\}$ | $\{\zeta_3\}$ | $\{-1\}$ |
| $[FR_1^{6,1,4}]_{2,5,3}$ | ✖ | $\{1\}$ | $\{\zeta_3\}$ | $\{\zeta_3^2, \zeta_3\}$ |
| $[FR_1^{6,1,4}]_{2,5,4}$ | ✖ | $\{1\}$ | $\{\zeta_3\}$ | $\{\zeta_6^5, \zeta_6\}$ |
| $[FR_1^{6,1,4}]_{2,6,1}$ | ✖ † 🌐 🎗 | $\{1\}$ | $\{\zeta_6^5\}$ | $\{1\}$ |
| $[FR_1^{6,1,4}]_{2,6,2}$ | ✖ 🌐 🎗 | $\{1\}$ | $\{\zeta_6^5\}$ | $\{-1\}$ |
| $[FR_1^{6,1,4}]_{2,6,3}$ | ✖ | $\{1\}$ | $\{\zeta_6^5\}$ | $\{\zeta_3^2, \zeta_3\}$ |
| $[FR_1^{6,1,4}]_{2,6,4}$ | ✖ | $\{1\}$ | $\{\zeta_6^5\}$ | $\{\zeta_6^5, \zeta_6\}$ |
| $[FR_1^{6,1,4}]_{3,0,1}$ | † 🌐 | $\{\zeta_3\}$ | – | $\{1\}$ |
| $[FR_1^{6,1,4}]_{3,0,2}$ | 🌐 | $\{\zeta_3\}$ | – | $\{-1\}$ |
| $[FR_1^{6,1,4}]_{3,0,3}$ | | $\{\zeta_3\}$ | – | $\{\zeta_3^2, \zeta_3\}$ |
| $[FR_1^{6,1,4}]_{3,0,4}$ | | $\{\zeta_3\}$ | – | $\{\zeta_6^5, \zeta_6\}$ |
| $[FR_1^{6,1,4}]_{4,0,1}$ | † 🌐 | $\{\zeta_6^5\}$ | – | $\{1\}$ |
| $[FR_1^{6,1,4}]_{4,0,2}$ | 🌐 | $\{\zeta_6^5\}$ | – | $\{-1\}$ |
| $[FR_1^{6,1,4}]_{4,0,3}$ | | $\{\zeta_6^5\}$ | – | $\{\zeta_3^2, \zeta_3\}$ |
| $[FR_1^{6,1,4}]_{4,0,4}$ | | $\{\zeta_6^5\}$ | – | $\{\zeta_6^5, \zeta_6\}$ |
| $[FR_1^{6,1,4}]_{5,0,1}$ | † 🌐 | $\{\zeta_6\}$ | – | $\{1\}$ |
| $[FR_1^{6,1,4}]_{5,0,2}$ | 🌐 | $\{\zeta_6\}$ | – | $\{-1\}$ |
| $[FR_1^{6,1,4}]_{5,0,3}$ | | $\{\zeta_6\}$ | – | $\{\zeta_3^2, \zeta_3\}$ |
| $[FR_1^{6,1,4}]_{5,0,4}$ | | $\{\zeta_6\}$ | – | $\{\zeta_6^5, \zeta_6\}$ |
| $[FR_1^{6,1,4}]_{6,0,1}$ | † 🌐 | $\{\zeta_3^2\}$ | – | $\{1\}$ |
| $[FR_1^{6,1,4}]_{6,0,2}$ | 🌐 | $\{\zeta_3^2\}$ | – | $\{-1\}$ |
| $[FR_1^{6,1,4}]_{6,0,3}$ | | $\{\zeta_3^2\}$ | – | $\{\zeta_3^2, \zeta_3\}$ |
| $[FR_1^{6,1,4}]_{6,0,4}$ | | $\{\zeta_3^2\}$ | – | $\{\zeta_6^5, \zeta_6\}$ |

Table 86: Numeric invariants

| Name | Properties | $X_1$ | $X_2$ | $X_3$ |
|---|---|---|---|---|
| $[FR_1^{6,1,4}]_{1,1,1}$ | ✕ † 🌐 🎗 ⊚ | $\{-1\}$ | $\{-0.866 + 0.5i\}$ | $\{1\}$ |
| $[FR_1^{6,1,4}]_{1,1,2}$ | ✕ 🌐 🎗 ⊚ | $\{-1\}$ | $\{-0.866 + 0.5i\}$ | $\{-1\}$ |
| $[FR_1^{6,1,4}]_{1,1,3}$ | ✕ | $\{-1\}$ | $\{-0.866 + 0.5i\}$ | $\left\{\begin{array}{l}-0.5 - 0.866i,\\ -0.5 + 0.866i\end{array}\right\}$ |
| $[FR_1^{6,1,4}]_{1,1,4}$ | ✕ | $\{-1\}$ | $\{-0.866 + 0.5i\}$ | $\left\{\begin{array}{l}0.5 - 0.866i,\\ 0.5 + 0.866i\end{array}\right\}$ |
| $[FR_1^{6,1,4}]_{1,2,1}$ | ✕ † 🌐 🎗 ⊚ | $\{-1\}$ | $\{0.866 - 0.5i\}$ | $\{1\}$ |
| $[FR_1^{6,1,4}]_{1,2,2}$ | ✕ 🌐 🎗 ⊚ | $\{-1\}$ | $\{0.866 - 0.5i\}$ | $\{-1\}$ |
| $[FR_1^{6,1,4}]_{1,2,3}$ | ✕ | $\{-1\}$ | $\{0.866 - 0.5i\}$ | $\left\{\begin{array}{l}-0.5 - 0.866i,\\ -0.5 + 0.866i\end{array}\right\}$ |
| $[FR_1^{6,1,4}]_{1,2,4}$ | ✕ | $\{-1\}$ | $\{0.866 - 0.5i\}$ | $\left\{\begin{array}{l}0.5 - 0.866i,\\ 0.5 + 0.866i\end{array}\right\}$ |
| $[FR_1^{6,1,4}]_{1,3,1}$ | ✕ † 🌐 🎗 ⊚ | $\{-1\}$ | $\{0.866 + 0.5i\}$ | $\{1\}$ |
| $[FR_1^{6,1,4}]_{1,3,2}$ | ✕ 🌐 🎗 ⊚ | $\{-1\}$ | $\{0.866 + 0.5i\}$ | $\{-1\}$ |
| $[FR_1^{6,1,4}]_{1,3,3}$ | ✕ | $\{-1\}$ | $\{0.866 + 0.5i\}$ | $\left\{\begin{array}{l}-0.5 - 0.866i,\\ -0.5 + 0.866i\end{array}\right\}$ |
| $[FR_1^{6,1,4}]_{1,3,4}$ | ✕ | $\{-1\}$ | $\{0.866 + 0.5i\}$ | $\left\{\begin{array}{l}0.5 - 0.866i,\\ 0.5 + 0.866i\end{array}\right\}$ |
| $[FR_1^{6,1,4}]_{1,4,1}$ | ✕ † 🌐 🎗 ⊚ | $\{-1\}$ | $\{-0.866 - 0.5i\}$ | $\{1\}$ |
| $[FR_1^{6,1,4}]_{1,4,2}$ | ✕ 🌐 🎗 ⊚ | $\{-1\}$ | $\{-0.866 - 0.5i\}$ | $\{-1\}$ |
| $[FR_1^{6,1,4}]_{1,4,3}$ | ✕ | $\{-1\}$ | $\{-0.866 - 0.5i\}$ | $\left\{\begin{array}{l}-0.5 - 0.866i,\\ -0.5 + 0.866i\end{array}\right\}$ |
| $[FR_1^{6,1,4}]_{1,4,4}$ | ✕ | $\{-1\}$ | $\{-0.866 - 0.5i\}$ | $\left\{\begin{array}{l}0.5 - 0.866i,\\ 0.5 + 0.866i\end{array}\right\}$ |
| $[FR_1^{6,1,4}]_{1,5,1}$ | ✕ † 🌐 🎗 | $\{-1\}$ | $\{-i\}$ | $\{1\}$ |
| $[FR_1^{6,1,4}]_{1,5,2}$ | ✕ 🌐 🎗 | $\{-1\}$ | $\{-i\}$ | $\{-1\}$ |
| $[FR_1^{6,1,4}]_{1,5,3}$ | ✕ | $\{-1\}$ | $\{-i\}$ | $\left\{\begin{array}{l}-0.5 - 0.866i,\\ -0.5 + 0.866i\end{array}\right\}$ |
| $[FR_1^{6,1,4}]_{1,5,4}$ | ✕ | $\{-1\}$ | $\{-i\}$ | $\left\{\begin{array}{l}0.5 - 0.866i,\\ 0.5 + 0.866i\end{array}\right\}$ |
| $[FR_1^{6,1,4}]_{1,6,1}$ | ✕ † 🌐 🎗 | $\{-1\}$ | $\{i\}$ | $\{1\}$ |
| $[FR_1^{6,1,4}]_{1,6,2}$ | ✕ 🌐 🎗 | $\{-1\}$ | $\{i\}$ | $\{-1\}$ |
| $[FR_1^{6,1,4}]_{1,6,3}$ | ✕ | $\{-1\}$ | $\{i\}$ | $\left\{\begin{array}{l}-0.5 - 0.866i,\\ -0.5 + 0.866i\end{array}\right\}$ |
| $[FR_1^{6,1,4}]_{1,6,4}$ | ✕ | $\{-1\}$ | $\{i\}$ | $\left\{\begin{array}{l}0.5 - 0.866i,\\ 0.5 + 0.866i\end{array}\right\}$ |

Table 86: Numeric invariants (Continued)

| Name | Properties | $X_1$ | $X_2$ | $X_3$ |
|---|---|---|---|---|
| $[FR_1^{6,1,4}]_{2,1,1}$ | ✗ † 🌐 🎗 | $\{1\}$ | $\{1\}$ | $\{1\}$ |
| $[FR_1^{6,1,4}]_{2,1,2}$ | ✗ 🌐 🎗 | $\{1\}$ | $\{1\}$ | $\{-1\}$ |
| $[FR_1^{6,1,4}]_{2,1,3}$ | ✗ | $\{1\}$ | $\{1\}$ | $\{-0.5 - 0.866i, -0.5 + 0.866i\}$ |
| $[FR_1^{6,1,4}]_{2,1,4}$ | ✗ | $\{1\}$ | $\{1\}$ | $\{0.5 - 0.866i, 0.5 + 0.866i\}$ |
| $[FR_1^{6,1,4}]_{2,2,1}$ | ✗ † 🌐 🎗 | $\{1\}$ | $\{-1\}$ | $\{1\}$ |
| $[FR_1^{6,1,4}]_{2,2,2}$ | ✗ 🌐 🎗 | $\{1\}$ | $\{-1\}$ | $\{-1\}$ |
| $[FR_1^{6,1,4}]_{2,2,3}$ | ✗ | $\{1\}$ | $\{-1\}$ | $\{-0.5 - 0.866i, -0.5 + 0.866i\}$ |
| $[FR_1^{6,1,4}]_{2,2,4}$ | ✗ | $\{1\}$ | $\{-1\}$ | $\{0.5 - 0.866i, 0.5 + 0.866i\}$ |
| $[FR_1^{6,1,4}]_{2,3,1}$ | ✗ † 🌐 🎗 | $\{1\}$ | $\{-0.5 - 0.866i\}$ | $\{1\}$ |
| $[FR_1^{6,1,4}]_{2,3,2}$ | ✗ 🌐 🎗 | $\{1\}$ | $\{-0.5 - 0.866i\}$ | $\{-1\}$ |
| $[FR_1^{6,1,4}]_{2,3,3}$ | ✗ | $\{1\}$ | $\{-0.5 - 0.866i\}$ | $\{-0.5 - 0.866i, -0.5 + 0.866i\}$ |
| $[FR_1^{6,1,4}]_{2,3,4}$ | ✗ | $\{1\}$ | $\{-0.5 - 0.866i\}$ | $\{0.5 - 0.866i, 0.5 + 0.866i\}$ |
| $[FR_1^{6,1,4}]_{2,4,1}$ | ✗ † 🌐 🎗 | $\{1\}$ | $\{0.5 + 0.866i\}$ | $\{1\}$ |
| $[FR_1^{6,1,4}]_{2,4,2}$ | ✗ 🌐 🎗 | $\{1\}$ | $\{0.5 + 0.866i\}$ | $\{-1\}$ |
| $[FR_1^{6,1,4}]_{2,4,3}$ | ✗ | $\{1\}$ | $\{0.5 + 0.866i\}$ | $\{-0.5 - 0.866i, -0.5 + 0.866i\}$ |
| $[FR_1^{6,1,4}]_{2,4,4}$ | ✗ | $\{1\}$ | $\{0.5 + 0.866i\}$ | $\{0.5 - 0.866i, 0.5 + 0.866i\}$ |
| $[FR_1^{6,1,4}]_{2,5,1}$ | ✗ † 🌐 🎗 | $\{1\}$ | $\{-0.5 + 0.866i\}$ | $\{1\}$ |
| $[FR_1^{6,1,4}]_{2,5,2}$ | ✗ 🌐 🎗 | $\{1\}$ | $\{-0.5 + 0.866i\}$ | $\{-1\}$ |
| $[FR_1^{6,1,4}]_{2,5,3}$ | ✗ | $\{1\}$ | $\{-0.5 + 0.866i\}$ | $\{-0.5 - 0.866i, -0.5 + 0.866i\}$ |
| $[FR_1^{6,1,4}]_{2,5,4}$ | ✗ | $\{1\}$ | $\{-0.5 + 0.866i\}$ | $\{0.5 - 0.866i, 0.5 + 0.866i\}$ |
| $[FR_1^{6,1,4}]_{2,6,1}$ | ✗ † 🌐 🎗 | $\{1\}$ | $\{0.5 - 0.866i\}$ | $\{1\}$ |
| $[FR_1^{6,1,4}]_{2,6,2}$ | ✗ 🌐 🎗 | $\{1\}$ | $\{0.5 - 0.866i\}$ | $\{-1\}$ |
| $[FR_1^{6,1,4}]_{2,6,3}$ | ✗ | $\{1\}$ | $\{0.5 - 0.866i\}$ | $\{-0.5 - 0.866i, -0.5 + 0.866i\}$ |
| $[FR_1^{6,1,4}]_{2,6,4}$ | ✗ | $\{1\}$ | $\{0.5 - 0.866i\}$ | $\{0.5 - 0.866i, 0.5 + 0.866i\}$ |

Table 86: Numeric invariants (Continued)

| Name | Properties | $X_1$ | $X_2$ | $X_3$ |
|---|---|---|---|---|
| $[FR_1^{6,1,4}]_{3,0,1}$ |  | $\{-0.5 + 0.866i\}$ | - | $\{1, 1i\}$ |
| $[FR_1^{6,1,4}]_{3,0,2}$ |  | $\{-0.5 + 0.866i\}$ | - | $\{-1, -1\}$ |
| $[FR_1^{6,1,4}]_{3,0,3}$ | | $\{-0.5 + 0.866i\}$ | - | $\left\{\begin{array}{l}-0.5 - 0.866i,\\ -0.5 + 0.866i\end{array}\right\}$ |
| $[FR_1^{6,1,4}]_{3,0,4}$ | | $\{-0.5 + 0.866i\}$ | - | $\left\{\begin{array}{l}0.5 - 0.866i,\\ 0.5 + 0.866i\end{array}\right\}$ |
| $[FR_1^{6,1,4}]_{4,0,1}$ |  | $\{0.5 - 0.866i\}$ | - | $\{1\}$ |
| $[FR_1^{6,1,4}]_{4,0,2}$ |  | $\{0.5 - 0.866i\}$ | - | $\{-1\}$ |
| $[FR_1^{6,1,4}]_{4,0,3}$ | | $\{0.5 - 0.866i\}$ | - | $\left\{\begin{array}{l}-0.5 - 0.866i,\\ -0.5 + 0.866i\end{array}\right\}$ |
| $[FR_1^{6,1,4}]_{4,0,4}$ | | $\{0.5 - 0.866i\}$ | - | $\left\{\begin{array}{l}0.5 - 0.866i,\\ 0.5 + 0.866i\end{array}\right\}$ |
| $[FR_1^{6,1,4}]_{5,0,1}$ |  | $\{0.5 + 0.866i\}$ | - | $\{1\}$ |
| $[FR_1^{6,1,4}]_{5,0,2}$ |  | $\{0.5 + 0.866i\}$ | - | $\{-1\}$ |
| $[FR_1^{6,1,4}]_{5,0,3}$ | | $\{0.5 + 0.866i\}$ | - | $\left\{\begin{array}{l}-0.5 - 0.866i,\\ -0.5 + 0.866i\end{array}\right\}$ |
| $[FR_1^{6,1,4}]_{5,0,4}$ | | $\{0.5 + 0.866i\}$ | - | $\left\{\begin{array}{l}0.5 - 0.866i,\\ 0.5 + 0.866i\end{array}\right\}$ |
| $[FR_1^{6,1,4}]_{6,0,1}$ |  | $\{-0.5 - 0.866i\}$ | - | $\{1, 1i\}$ |
| $[FR_1^{6,1,4}]_{6,0,2}$ |  | $\{-0.5 - 0.866i\}$ | - | $\{-1, -1\}$ |
| $[FR_1^{6,1,4}]_{6,0,3}$ | | $\{-0.5 - 0.866i\}$ | - | $\left\{\begin{array}{l}-0.5 - 0.866i,\\ -0.5 + 0.866i\end{array}\right\}$ |
| $[FR_1^{6,1,4}]_{6,0,4}$ | | $\{-0.5 - 0.866i\}$ | - | $\left\{\begin{array}{l}0.5 - 0.866i,\\ 0.5 + 0.866i\end{array}\right\}$ |

### 4.5.45 $FR_2^{6,1,4}$: $\mathbf{MR}_6$

For the fusion ring, the following multiplication table is used.

| 1 | 2 | 3 | 4 | 5 | 6 |
|---|---|---|---|---|---|
| 2 | 1 | 4 | 3 | 5 | 6 |
| 3 | 4 | 2 | 1 | 6 | 5 |
| 4 | 3 | 1 | 2 | 6 | 5 |
| 5 | 5 | 6 | 6 | $3+4$ | $1+2$ |
| 6 | 6 | 5 | 5 | $1+2$ | $3+4$ |

The following is the group of all non-trivial permutations that leave the fusion rules invariant:

$$S = \{(), (3\ 4), (5\ 6), (3\ 4)(5\ 6)\}.$$

Let
$$X_1 = S\left(\frac{[F_5^{246}]_5^3[F_1^{466}]_3^5[F_3^{536}]_5^6[F_2^{553}]_6^3}{[F_5^{225}]_5^1[F_2^{243}]_1^3[F_1^{256}]_2^5}\right), \tag{77}$$
$$X_2 = S\left(d_5^L\right). \tag{78}$$

The following table lists a small set of invariants whose values completely distinguish between all MFPBFCs and MFPNBFCs with the given fusion rules.

Table 87: Symbolic invariants

| Name | Properties | $X_1$ | $X_2$ |
|---|---|---|---|
| $[FR_2^{6,1,4}]_{1,0,1}$ |  | $\{\zeta_8^3\}$ | $\{\sqrt{2}\}$ |
| $[FR_2^{6,1,4}]_{1,0,2}$ |  | $\{\zeta_8^3\}$ | $\{-\sqrt{2}\}$ |
| $[FR_2^{6,1,4}]_{1,0,3}$ | | $\{\zeta_8^3\}$ | $\{-i\sqrt{2}, i\sqrt{2}\}$ |
| $[FR_2^{6,1,4}]_{2,0,1}$ |  | $\{\zeta_8^7\}$ | $\{\sqrt{2}\}$ |
| $[FR_2^{6,1,4}]_{2,0,2}$ |  | $\{\zeta_8^7\}$ | $\{-\sqrt{2}\}$ |
| $[FR_2^{6,1,4}]_{2,0,3}$ | | $\{\zeta_8^7\}$ | $\{-i\sqrt{2}, i\sqrt{2}\}$ |
| $[FR_2^{6,1,4}]_{3,0,1}$ |  | $\{\zeta_8\}$ | $\{\sqrt{2}\}$ |
| $[FR_2^{6,1,4}]_{3,0,2}$ |  | $\{\zeta_8\}$ | $\{-\sqrt{2}\}$ |
| $[FR_2^{6,1,4}]_{3,0,3}$ | | $\{\zeta_8\}$ | $\{-i\sqrt{2}, i\sqrt{2}\}$ |
| $[FR_2^{6,1,4}]_{4,0,1}$ |  | $\{\zeta_8^5\}$ | $\{\sqrt{2}\}$ |
| $[FR_2^{6,1,4}]_{4,0,2}$ |  | $\{\zeta_8^5\}$ | $\{-\sqrt{2}\}$ |
| $[FR_2^{6,1,4}]_{4,0,3}$ | | $\{\zeta_8^5\}$ | $\{-i\sqrt{2}, i\sqrt{2}\}$ |

Table 88: Numeric invariants

| Name | Properties | $X_1$ | $X_2$ |
|---|---|---|---|
| $[FR_2^{6,1,4}]_{1,0,1}$ |  | $\{-0.707 + 0.707i\}$ | $\{1.414\}$ |
| $[FR_2^{6,1,4}]_{1,0,2}$ |  | $\{-0.707 + 0.707i\}$ | $\{-1.414\}$ |
| $[FR_2^{6,1,4}]_{1,0,3}$ | | $\{-0.707 + 0.707i\}$ | $\{-1.414i, 1.414i\}$ |
| $[FR_2^{6,1,4}]_{2,0,1}$ |  | $\{0.707 - 0.707i\}$ | $\{1.414\}$ |
| $[FR_2^{6,1,4}]_{2,0,2}$ |  | $\{0.707 - 0.707i\}$ | $\{-1.414\}$ |
| $[FR_2^{6,1,4}]_{2,0,3}$ | | $\{0.707 - 0.707i\}$ | $\{-1.414i, 1.414i\}$ |
| $[FR_2^{6,1,4}]_{3,0,1}$ |  | $\{0.707 + 0.707i\}$ | $\{1.414\}$ |
| $[FR_2^{6,1,4}]_{3,0,2}$ |  | $\{0.707 + 0.707i\}$ | $\{-1.414\}$ |
| $[FR_2^{6,1,4}]_{3,0,3}$ | | $\{0.707 + 0.707i\}$ | $\{-1.414i, 1.414i\}$ |

Table 88: Numeric invariants (Continued)

| Name | Properties | $X_1$ | $X_2$ |
|---|---|---|---|
| $[FR_2^{6,1,4}]_{4,0,1}$ |  | $\{-0.707 - 0.707i\}$ | $\{1.414\}$ |
| $[FR_2^{6,1,4}]_{4,0,2}$ |  | $\{-0.707 - 0.707i\}$ | $\{-1.414\}$ |
| $[FR_2^{6,1,4}]_{4,0,3}$ | | $\{-0.707 - 0.707i\}$ | $\{-1.414i, 1.414i\}$ |

### 4.5.46 $FR_3^{6,1,4}$: $TY(\mathbb{Z}_5)$

For the fusion ring, the following multiplication table is used.

| 1 | 2 | 3 | 4 | 5 | 6 |
|---|---|---|---|---|---|
| 2 | 5 | 1 | 3 | 4 | 6 |
| 3 | 1 | 4 | 5 | 2 | 6 |
| 4 | 3 | 5 | 2 | 1 | 6 |
| 5 | 4 | 2 | 1 | 3 | 6 |
| 6 | 6 | 6 | 6 | 6 | $1+2+3+4+5$ |

The following is the group of all non-trivial permutations that leave the fusion rules invariant:

$$S = \{(), (2\ 4\ 3\ 5), (2\ 5\ 3\ 4), (2\ 3)(4\ 5)\}.$$

$$X_1 = S\left([F_1^{626}]_6^6 [F_6^{666}]_2^2\right) \tag{79}$$

$$\tag{80}$$

The following table lists a small set of invariants whose values completely distinguish between all MFPBFCs and MFPNBFCs with the given fusion rules.

Table 89: Symbolic invariants

| Name | Properties | $X_1$ | $d_6^L$ |
|---|---|---|---|
| $[FR_3^{6,1,4}]_{1,0,1}$ |  | $\left\{\frac{\zeta_5^4}{\sqrt{5}}, \frac{\zeta_5}{\sqrt{5}}\right\}$ | $\sqrt{5}$ |
| $[FR_3^{6,1,4}]_{1,0,2}$ |  | $\left\{\frac{\zeta_5^4}{\sqrt{5}}, \frac{\zeta_5}{\sqrt{5}}\right\}$ | $-\sqrt{5}$ |
| $[FR_3^{6,1,4}]_{2,0,1}$ |  | $\left\{\frac{\zeta_{10}^7}{\sqrt{5}}, \frac{\zeta_{10}^3}{\sqrt{5}}\right\}$ | $\sqrt{5}$ |
| $[FR_3^{6,1,4}]_{2,0,2}$ |  | $\left\{\frac{\zeta_{10}^7}{\sqrt{5}}, \frac{\zeta_{10}^3}{\sqrt{5}}\right\}$ | $-\sqrt{5}$ |
| $[FR_3^{6,1,4}]_{3,0,1}$ |  | $\left\{\frac{\zeta_5^3}{\sqrt{5}}, \frac{\zeta_5^2}{\sqrt{5}}\right\}$ | $\sqrt{5}$ |
| $[FR_3^{6,1,4}]_{3,0,2}$ |  | $\left\{\frac{\zeta_5^3}{\sqrt{5}}, \frac{\zeta_5^2}{\sqrt{5}}\right\}$ | $-\sqrt{5}$ |
| $[FR_3^{6,1,4}]_{4,0,1}$ |  | $\left\{\frac{\zeta_{10}^9}{\sqrt{5}}, \frac{\zeta_{10}}{\sqrt{5}}\right\}$ | $\sqrt{5}$ |

Table 89: Symbolic invariants (Continued)

| Name | Properties | $X_1$ | $d_6^L$ |
|---|---|---|---|
| $[FR_3^{6,1,4}]_{4,0,2}$ |  | $\left\{\frac{\zeta_{10}^9}{\sqrt{5}}, \frac{\zeta_{10}}{\sqrt{5}}\right\}$ | $-\sqrt{5}$ |

Table 90: Numeric invariants

| Name | Properties | $X_1$ | $d_6^L$ |
|---|---|---|---|
| $[FR_3^{6,1,4}]_{1,0,1}$ |  | $\{0.138 - 0.425i, 0.138 + 0.425i\}$ | $2.236$ |
| $[FR_3^{6,1,4}]_{1,0,2}$ |  | $\{0.138 - 0.425i, 0.138 + 0.425i\}$ | $-2.236$ |
| $[FR_3^{6,1,4}]_{2,0,1}$ |  | $\{-0.138 - 0.425i, -0.138 + 0.425i\}$ | $2.236$ |
| $[FR_3^{6,1,4}]_{2,0,2}$ |  | $\{-0.138 - 0.425i, -0.138 + 0.425i\}$ | $-2.236$ |
| $[FR_3^{6,1,4}]_{3,0,1}$ |  | $\{-0.362 - 0.263i, -0.362 + 0.263i\}$ | $2.236$ |
| $[FR_3^{6,1,4}]_{3,0,2}$ |  | $\{-0.362 - 0.263i, -0.362 + 0.263i\}$ | $-2.236$ |
| $[FR_3^{6,1,4}]_{4,0,1}$ |  | $\{0.362 - 0.263i, 0.362 + 0.263i\}$ | $2.236$ |
| $[FR_3^{6,1,4}]_{4,0,2}$ |  | $\{0.362 - 0.263i, 0.362 + 0.263i\}$ | $-2.236$ |

### 4.5.47 $FR_5^{6,1,4}$: **Fib** ⊗ $\mathbb{Z}_3$

For the fusion ring, the following multiplication table is used.

| 1 | 2 | 3 | 4 | 5 | 6 |
|---|---|---|---|---|---|
| 2 | 3 | 1 | 6 | 4 | 5 |
| 3 | 1 | 2 | 5 | 6 | 4 |
| 4 | 6 | 5 | 1+4 | 3+5 | 2+6 |
| 5 | 4 | 6 | 3+5 | 2+6 | 1+4 |
| 6 | 5 | 4 | 2+6 | 1+4 | 3+5 |

The following is the group of all non-trivial permutations that leave the fusion rules invariant:

$$S = \{(), (2\ 3)(5\ 6)\}.$$

Let

$$X_1 = S\left(R_2^{55}\right), \tag{81}$$
$$X_2 = S\left(d_5^L\right). \tag{82}$$

The following table lists a small set of invariants whose values completely distinguish between all MFPBFCs and MFPNBFCs with the given fusion rules.

Table 91: Symbolic invariants

| Name | Properties | $[F_4^{555}]_6^6[F_4^{666}]_5^5$ | $X_1$ | $X_2$ |
|---|---|---|---|---|
| $[FR_5^{6,1,4}]_{1,1,1}$ | ✂ † 🌐 🎗 ⚭ | $\overline{\phi}^2$ | $\{\zeta_{15}^4\}$ | $\{\phi\}$ |
| $[FR_5^{6,1,4}]_{1,1,2}$ | ✂ | $\overline{\phi}^2$ | $\{\zeta_{15}^4\}$ | $\{\phi\zeta_3^2, \phi\zeta_3\}$ |
| $[FR_5^{6,1,4}]_{1,2,1}$ | ✂ † 🌐 🎗 ⚭ | $\overline{\phi}^2$ | $\{\zeta_{15}^{14}\}$ | $\{\phi\}$ |
| $[FR_5^{6,1,4}]_{1,2,2}$ | ✂ | $\overline{\phi}^2$ | $\{\zeta_{15}^{14}\}$ | $\{\phi\zeta_3^2, \phi\zeta_3\}$ |
| $[FR_5^{6,1,4}]_{1,3,1}$ | ✂ † 🌐 🎗 ⚭ | $\overline{\phi}^2$ | $\{\zeta_{15}\}$ | $\{\phi\}$ |
| $[FR_5^{6,1,4}]_{1,3,2}$ | ✂ | $\overline{\phi}^2$ | $\{\zeta_{15}\}$ | $\{\phi\zeta_3^2, \phi\zeta_3\}$ |
| $[FR_5^{6,1,4}]_{1,4,1}$ | ✂ † 🌐 🎗 ⚭ | $\overline{\phi}^2$ | $\{\zeta_{15}^{11}\}$ | $\{\phi\}$ |
| $[FR_5^{6,1,4}]_{1,4,2}$ | ✂ | $\overline{\phi}^2$ | $\{\zeta_{15}^{11}\}$ | $\{\phi\zeta_3^2, \phi\zeta_3\}$ |
| $[FR_5^{6,1,4}]_{1,5,1}$ | ✂ † 🌐 🎗 | $\overline{\phi}^2$ | $\{\zeta_5^3\}$ | $\{\phi\}$ |
| $[FR_5^{6,1,4}]_{1,5,2}$ | ✂ | $\overline{\phi}^2$ | $\{\zeta_5^3\}$ | $\{\phi\zeta_3^2, \phi\zeta_3\}$ |
| $[FR_5^{6,1,4}]_{1,6,1}$ | ✂ † 🌐 🎗 | $\overline{\phi}^2$ | $\{\zeta_5^2\}$ | $\{\phi\}$ |
| $[FR_5^{6,1,4}]_{1,6,2}$ | ✂ | $\overline{\phi}^2$ | $\{\zeta_5^2\}$ | $\{\phi\zeta_3^2, \phi\zeta_3\}$ |
| $[FR_5^{6,1,4}]_{2,1,1}$ | ✂ 🌐 🎗 ⚭ | $\phi^2$ | $\{\zeta_{15}^7\}$ | $\{\overline{\phi}\}$ |
| $[FR_5^{6,1,4}]_{2,1,2}$ | ✂ | $\phi^2$ | $\{\zeta_{15}^7\}$ | $\{-\overline{\phi}\zeta_6^5, -\overline{\phi}\zeta_6\}$ |
| $[FR_5^{6,1,4}]_{2,2,1}$ | ✂ 🌐 🎗 ⚭ | $\phi^2$ | $\{\zeta_{15}^2\}$ | $\{\overline{\phi}\}$ |
| $[FR_5^{6,1,4}]_{2,2,2}$ | ✂ | $\phi^2$ | $\{\zeta_{15}^2\}$ | $\{-\overline{\phi}\zeta_6^5, -\overline{\phi}\zeta_6\}$ |
| $[FR_5^{6,1,4}]_{2,3,1}$ | ✂ 🌐 🎗 ⚭ | $\phi^2$ | $\{\zeta_{15}^{13}\}$ | $\{\overline{\phi}\}$ |
| $[FR_5^{6,1,4}]_{2,3,2}$ | ✂ | $\phi^2$ | $\{\zeta_{15}^{13}\}$ | $\{-\overline{\phi}\zeta_6^5, -\overline{\phi}\zeta_6\}$ |
| $[FR_5^{6,1,4}]_{2,4,1}$ | ✂ 🌐 🎗 ⚭ | $\phi^2$ | $\{\zeta_{15}^8\}$ | $\{\overline{\phi}\}$ |
| $[FR_5^{6,1,4}]_{2,4,2}$ | ✂ | $\phi^2$ | $\{\zeta_{15}^8\}$ | $\{-\overline{\phi}\zeta_6^5, -\overline{\phi}\zeta_6\}$ |
| $[FR_5^{6,1,4}]_{2,5,1}$ | ✂ 🌐 🎗 | $\phi^2$ | $\{\zeta_5^4\}$ | $\{\overline{\phi}\}$ |
| $[FR_5^{6,1,4}]_{2,5,2}$ | ✂ | $\phi^2$ | $\{\zeta_5^4\}$ | $\{-\overline{\phi}\zeta_6^5, -\overline{\phi}\zeta_6\}$ |
| $[FR_5^{6,1,4}]_{2,6,1}$ | ✂ 🌐 🎗 | $\phi^2$ | $\{\zeta_5\}$ | $\{\overline{\phi}\}$ |
| $[FR_5^{6,1,4}]_{2,6,2}$ | ✂ | $\phi^2$ | $\{\zeta_5\}$ | $\{-\overline{\phi}\zeta_6^5, -\overline{\phi}\zeta_6\}$ |
| $[FR_5^{6,1,4}]_{3,0,1}$ | † 🌐 | $\overline{\phi}^2\zeta_3^2$ | - | $\{\phi\}$ |
| $[FR_5^{6,1,4}]_{3,0,2}$ | | $\overline{\phi}^2\zeta_3^2$ | - | $\{\phi\zeta_3^2, \phi\zeta_3\}$ |
| $[FR_5^{6,1,4}]_{4,0,1}$ | † 🌐 | $\overline{\phi}^2\zeta_3$ | - | $\{\phi\}$ |
| $[FR_5^{6,1,4}]_{4,0,2}$ | | $\overline{\phi}^2\zeta_3$ | - | $\{\phi\zeta_3^2, \phi\zeta_3\}$ |
| $[FR_5^{6,1,4}]_{5,0,1}$ | 🌐 | $\phi^2\zeta_3^2$ | - | $\{\overline{\phi}\}$ |
| $[FR_5^{6,1,4}]_{5,0,2}$ | | $\phi^2\zeta_3^2$ | - | $\{-\overline{\phi}\zeta_6^5, -\overline{\phi}\zeta_6\}$ |
| $[FR_5^{6,1,4}]_{6,0,1}$ | 🌐 | $\phi^2\zeta_3$ | - | $\{\overline{\phi}\}$ |

Table 91: Symbolic invariants (Continued)

| Name | Properties | $[F_4^{555}]_6^6[F_4^{666}]_5^5$ | $X_1$ | $X_2$ |
|---|---|---|---|---|
| $[\mathrm{FR}_5^{6,1,4}]_{6,0,2}$ | | $\phi^2\zeta_3$ | - | $\{-\overline{\phi}\zeta_6^5, -\overline{\phi}\zeta_6\}$ |

Table 92: Numeric invariants

| Name | Properties | $[F_4^{555}]_6^6[F_4^{666}]_5^5$ | $X_1$ | $X_2$ |
|---|---|---|---|---|
| $[\mathrm{FR}_5^{6,1,4}]_{1,1,1}$ | ✗ † 🌐 🎗 ⚭ | 0.382 | $\{-0.105 + 0.995i\}$ | $\{1.618\}$ |
| $[\mathrm{FR}_5^{6,1,4}]_{1,1,2}$ | ✗ | 0.382 | $\{-0.105 + 0.995i\}$ | $\{-0.809 - 1.401i, -0.809 + 1.401i\}$ |
| $[\mathrm{FR}_5^{6,1,4}]_{1,2,1}$ | ✗ † 🌐 🎗 ⚭ | 0.382 | $\{0.914 - 0.407i\}$ | $\{1.618\}$ |
| $[\mathrm{FR}_5^{6,1,4}]_{1,2,2}$ | ✗ | 0.382 | $\{0.914 - 0.407i\}$ | $\{-0.809 - 1.401i, -0.809 + 1.401i\}$ |
| $[\mathrm{FR}_5^{6,1,4}]_{1,3,1}$ | ✗ † 🌐 🎗 ⚭ | 0.382 | $\{0.914 + 0.407i\}$ | $\{1.618\}$ |
| $[\mathrm{FR}_5^{6,1,4}]_{1,3,2}$ | ✗ | 0.382 | $\{0.914 + 0.407i\}$ | $\{-0.809 - 1.401i, -0.809 + 1.401i\}$ |
| $[\mathrm{FR}_5^{6,1,4}]_{1,4,1}$ | ✗ † 🌐 🎗 ⚭ | 0.382 | $\{-0.105 - 0.995i\}$ | $\{1.618\}$ |
| $[\mathrm{FR}_5^{6,1,4}]_{1,4,2}$ | ✗ | 0.382 | $\{-0.105 - 0.995i\}$ | $\{-0.809 - 1.401i, -0.809 + 1.401i\}$ |
| $[\mathrm{FR}_5^{6,1,4}]_{1,5,1}$ | ✗ † 🌐 🎗 | 0.382 | $\{-0.809 - 0.588i\}$ | $\{1.618\}$ |
| $[\mathrm{FR}_5^{6,1,4}]_{1,5,2}$ | ✗ | 0.382 | $\{-0.809 - 0.588i\}$ | $\{-0.809 - 1.401i, -0.809 + 1.401i\}$ |
| $[\mathrm{FR}_5^{6,1,4}]_{1,6,1}$ | ✗ † 🌐 🎗 | 0.382 | $\{-0.809 + 0.588i\}$ | $\{1.618\}$ |
| $[\mathrm{FR}_5^{6,1,4}]_{1,6,2}$ | ✗ | 0.382 | $\{-0.809 + 0.588i\}$ | $\{-0.809 - 1.401i, -0.809 + 1.401i\}$ |
| $[\mathrm{FR}_5^{6,1,4}]_{2,1,1}$ | ✗ 🌐 🎗 ⚭ | 2.618 | $\{-0.978 + 0.208i\}$ | $\{-0.618\}$ |
| $[\mathrm{FR}_5^{6,1,4}]_{2,1,2}$ | ✗ | 2.618 | $\{-0.978 + 0.208i\}$ | $\{0.309 - 0.535i, 0.309 + 0.535i\}$ |
| $[\mathrm{FR}_5^{6,1,4}]_{2,2,1}$ | ✗ 🌐 🎗 ⚭ | 2.618 | $\{0.669 + 0.743i\}$ | $\{-0.618\}$ |
| $[\mathrm{FR}_5^{6,1,4}]_{2,2,2}$ | ✗ | 2.618 | $\{0.669 + 0.743i\}$ | $\{0.309 - 0.535i, 0.309 + 0.535i\}$ |
| $[\mathrm{FR}_5^{6,1,4}]_{2,3,1}$ | ✗ 🌐 🎗 ⚭ | 2.618 | $\{0.669 - 0.743i\}$ | $\{-0.618\}$ |
| $[\mathrm{FR}_5^{6,1,4}]_{2,3,2}$ | ✗ | 2.618 | $\{0.669 - 0.743i\}$ | $\{0.309 - 0.535i, 0.309 + 0.535i\}$ |
| $[\mathrm{FR}_5^{6,1,4}]_{2,4,1}$ | ✗ 🌐 🎗 ⚭ | 2.618 | $\{-0.978 - 0.208i\}$ | $\{-0.618\}$ |
| $[\mathrm{FR}_5^{6,1,4}]_{2,4,2}$ | ✗ | 2.618 | $\{-0.978 - 0.208i\}$ | $\{0.309 - 0.535i, 0.309 + 0.535i\}$ |
| $[\mathrm{FR}_5^{6,1,4}]_{2,5,1}$ | ✗ 🌐 🎗 | 2.618 | $\{0.309 - 0.951i\}$ | $\{-0.618\}$ |

Table 92: Numeric invariants (Continued)

| Name | Properties | $[F_4^{555}]_6^6[F_4^{666}]_5^5$ | $X_1$ | $X_2$ |
|---|---|---|---|---|
| $[FR_5^{6,1,4}]_{2,5,2}$ | ✗ | 2.618 | $\{0.309 - 0.951i\}$ | $\{0.309 - 0.535i,\ 0.309 + 0.535i\}$ |
| $[FR_5^{6,1,4}]_{2,6,1}$ | ✗ 🌐 🎗 | 2.618 | $\{0.309 + 0.951i\}$ | $\{-0.618\}$ |
| $[FR_5^{6,1,4}]_{2,6,2}$ | ✗ | 2.618 | $\{0.309 + 0.951i\}$ | $\{0.309 - 0.535i,\ 0.309 + 0.535i\}$ |
| $[FR_5^{6,1,4}]_{3,0,1}$ | † 🌐 | $-0.191 - 0.331i$ | - | $\{1.618\}$ |
| $[FR_5^{6,1,4}]_{3,0,2}$ | | $-0.191 - 0.331i$ | - | $\{-0.809 - 1.401i,\ -0.809 + 1.401i\}$ |
| $[FR_5^{6,1,4}]_{4,0,1}$ | † 🌐 | $-0.191 + 0.331i$ | - | $\{1.618\}$ |
| $[FR_5^{6,1,4}]_{4,0,2}$ | | $-0.191 + 0.331i$ | - | $\{-0.809 - 1.401i,\ -0.809 + 1.401i\}$ |
| $[FR_5^{6,1,4}]_{5,0,1}$ | 🌐 | $-1.309 - 2.267i$ | - | $\{-0.618\}$ |
| $[FR_5^{6,1,4}]_{5,0,2}$ | | $-1.309 - 2.267i$ | - | $\{0.309 - 0.535i,\ 0.309 + 0.535i\}$ |
| $[FR_5^{6,1,4}]_{6,0,1}$ | 🌐 | $-1.309 + 2.267i$ | - | $\{-0.618\}$ |
| $[FR_5^{6,1,4}]_{6,0,2}$ | | $-1.309 + 2.267i$ | - | $\{.309 - 0.535i,\ 0.309 + 0.535i\}$ |

### 4.5.48  $FR_1^{7,1,0}$: **Adj(SO(16)$_2$)**

For the fusion ring, the following multiplication table is used.

| 1 | 2 | 3 | 4 | 5 | 6 | 7 |
|---|---|---|---|---|---|---|
| 2 | 1 | 4 | 3 | 7 | 6 | 5 |
| 3 | 4 | 1 | 2 | 7 | 6 | 5 |
| 4 | 3 | 2 | 1 | 5 | 6 | 7 |
| 5 | 7 | 7 | 5 | $1+4+6$ | $5+7$ | $2+3+6$ |
| 6 | 6 | 6 | 6 | $5+7$ | $1+2+3+4$ | $5+7$ |
| 7 | 5 | 5 | 7 | $2+3+6$ | $5+7$ | $1+4+6$ |

The following is the group of all non-trivial permutations that leave the fusion rules invariant:

$$S = \{(), (2\ 3), (5\ 7), (2\ 3)(5\ 7)\}.$$

Let

$$X_1 = S\left([F_5^{557}]_6^6[F_6^{567}]_5^5\right), \tag{83}$$

$$X_2 = S\left(\frac{[F_5^{522}]_1^7[F_5^{555}]_6^1[F_1^{556}]_5^6[F_5^{766}]_2^7}{[F_7^{256}]_5^7[F_1^{266}]_2^6[F_5^{526}]_6^7}\right), \tag{84}$$

$$X_3 = S\left(R_6^{55}\right), \tag{85}$$

$$X_4 = S\left(d_5^L\right). \tag{86}$$

The following table lists a small set of invariants whose values completely distinguish between all MFPBFCs and MFPNBFCs with the given fusion rules.

Table 93: Symbolic invariants

| Name | Properties | $X_1$ | $X_2$ | $X_3$ | $X_4$ |
|---|---|---|---|---|---|
| $[FR_1^{7,1,0}]_{1,1,1}$ | ✂ † 🌐 🎗 | $\{1\}$ | $\{\frac{1}{2}\}$ | $\{1\}$ | $\{2\}$ |
| $[FR_1^{7,1,0}]_{1,1,2}$ | ✂ 🌐 🎗 | $\{1\}$ | $\{\frac{1}{2}\}$ | $\{1\}$ | $\{-2\}$ |
| $[FR_1^{7,1,0}]_{1,2,1}$ | ✂ † 🌐 🎗 | $\{1\}$ | $\{\frac{1}{2}\}$ | $\{-1\}$ | $\{2\}$ |
| $[FR_1^{7,1,0}]_{1,2,2}$ | ✂ 🌐 🎗 | $\{1\}$ | $\{\frac{1}{2}\}$ | $\{-1\}$ | $\{-2\}$ |
| $[FR_1^{7,1,0}]_{1,3,1}$ | ✂ † 🌐 🎗 | $\{1\}$ | $\{\frac{1}{2}\}$ | $\{-i\}$ | $\{2\}$ |
| $[FR_1^{7,1,0}]_{1,3,2}$ | ✂ 🌐 🎗 | $\{1\}$ | $\{\frac{1}{2}\}$ | $\{-i\}$ | $\{-2\}$ |
| $[FR_1^{7,1,0}]_{1,4,1}$ | ✂ † 🌐 🎗 | $\{1\}$ | $\{\frac{1}{2}\}$ | $\{i\}$ | $\{2\}$ |
| $[FR_1^{7,1,0}]_{1,4,2}$ | ✂ 🌐 🎗 | $\{1\}$ | $\{\frac{1}{2}\}$ | $\{i\}$ | $\{-2\}$ |
| $[FR_1^{7,1,0}]_{1,5,1}$ | ✂ † 🌐 🎗 | $\{1\}$ | $\{\frac{1}{2}\}$ | $\{\zeta_8^7\}$ | $\{2\}$ |
| $[FR_1^{7,1,0}]_{1,5,2}$ | ✂ 🌐 🎗 | $\{1\}$ | $\{\frac{1}{2}\}$ | $\{\zeta_8^7\}$ | $\{-2\}$ |
| $[FR_1^{7,1,0}]_{1,6,1}$ | ✂ † 🌐 🎗 | $\{1\}$ | $\{\frac{1}{2}\}$ | $\{\zeta_8^3\}$ | $\{2\}$ |
| $[FR_1^{7,1,0}]_{1,6,2}$ | ✂ 🌐 🎗 | $\{1\}$ | $\{\frac{1}{2}\}$ | $\{\zeta_8^3\}$ | $\{-2\}$ |
| $[FR_1^{7,1,0}]_{1,7,1}$ | ✂ † 🌐 🎗 | $\{1\}$ | $\{\frac{1}{2}\}$ | $\{\zeta_8\}$ | $\{2\}$ |
| $[FR_1^{7,1,0}]_{1,7,2}$ | ✂ 🌐 🎗 | $\{1\}$ | $\{\frac{1}{2}\}$ | $\{\zeta_8\}$ | $\{-2\}$ |
| $[FR_1^{7,1,0}]_{1,8,1}$ | ✂ † 🌐 🎗 | $\{1\}$ | $\{\frac{1}{2}\}$ | $\{\zeta_8^5\}$ | $\{2\}$ |
| $[FR_1^{7,1,0}]_{1,8,2}$ | ✂ 🌐 🎗 | $\{1\}$ | $\{\frac{1}{2}\}$ | $\{\zeta_8^5\}$ | $\{-2\}$ |
| $[FR_1^{7,1,0}]_{2,1,1}$ | ✂ † 🌐 🎗 | $\{1\}$ | $\{-\frac{1}{2}\}$ | $\{\zeta_{16}^5, \zeta_{16}^{13}\}$ | $\{2\}$ |
| $[FR_1^{7,1,0}]_{2,1,2}$ | ✂ 🌐 🎗 | $\{1\}$ | $\{-\frac{1}{2}\}$ | $\{\zeta_{16}^5, \zeta_{16}^{13}\}$ | $\{-2\}$ |
| $[FR_1^{7,1,0}]_{2,2,1}$ | ✂ † 🌐 🎗 | $\{1\}$ | $\{-\frac{1}{2}\}$ | $\{\zeta_{16}^{11}, \zeta_{16}^3\}$ | $\{2\}$ |
| $[FR_1^{7,1,0}]_{2,2,2}$ | ✂ 🌐 🎗 | $\{1\}$ | $\{-\frac{1}{2}\}$ | $\{\zeta_{16}^{11}, \zeta_{16}^3\}$ | $\{-2\}$ |
| $[FR_1^{7,1,0}]_{2,3,1}$ | ✂ † 🌐 🎗 | $\{1\}$ | $\{-\frac{1}{2}\}$ | $\{\zeta_{16}^7, \zeta_{16}^{15}\}$ | $\{2\}$ |
| $[FR_1^{7,1,0}]_{2,3,2}$ | ✂ 🌐 🎗 | $\{1\}$ | $\{-\frac{1}{2}\}$ | $\{\zeta_{16}^7, \zeta_{16}^{15}\}$ | $\{-2\}$ |
| $[FR_1^{7,1,0}]_{2,4,1}$ | ✂ † 🌐 🎗 | $\{1\}$ | $\{-\frac{1}{2}\}$ | $\{\zeta_{16}^9, \zeta_{16}\}$ | $\{2\}$ |
| $[FR_1^{7,1,0}]_{2,4,2}$ | ✂ 🌐 🎗 | $\{1\}$ | $\{-\frac{1}{2}\}$ | $\{\zeta_{16}^9, \zeta_{16}\}$ | $\{-2\}$ |
| $[FR_1^{7,1,0}]_{2,5,1}$ | ✂ † 🌐 🎗 | $\{-1\}$ | $\{-\frac{1}{2}\}$ | $\{i\}$ | $\{2\}$ |
| $[FR_1^{7,1,0}]_{2,5,2}$ | ✂ 🌐 🎗 | $\{-1\}$ | $\{-\frac{1}{2}\}$ | $\{i\}$ | $\{-2\}$ |
| $[FR_1^{7,1,0}]_{2,6,1}$ | ✂ † 🌐 🎗 | $\{-1\}$ | $\{-\frac{1}{2}\}$ | $\{-i\}$ | $\{2\}$ |
| $[FR_1^{7,1,0}]_{2,6,2}$ | ✂ 🌐 🎗 | $\{-1\}$ | $\{-\frac{1}{2}\}$ | $\{-i\}$ | $\{-2\}$ |
| $[FR_1^{7,1,0}]_{2,7,1}$ | ✂ † 🌐 🎗 | $\{-1\}$ | $\{-\frac{1}{2}\}$ | $\{-1\}$ | $\{2\}$ |

Table 93: Symbolic invariants (Continued)

| Name | Properties | $X_1$ | $X_2$ | $X_3$ | $X_4$ |
|---|---|---|---|---|---|
| $[FR_1^{7,1,0}]_{2,7,2}$ | ✗ 🌐 🎗 | $\{-1\}$ | $\{-\frac{1}{2}\}$ | $\{-1\}$ | $\{-2\}$ |
| $[FR_1^{7,1,0}]_{2,8,1}$ | ✗ † 🌐 🎗 | $\{-1\}$ | $\{-\frac{1}{2}\}$ | $\{1\}$ | $\{2\}$ |
| $[FR_1^{7,1,0}]_{2,8,2}$ | ✗ 🌐 🎗 | $\{-1\}$ | $\{-\frac{1}{2}\}$ | $\{1\}$ | $\{-2\}$ |
| $[FR_1^{7,1,0}]_{3,1,1}$ | ✗ † 🌐 🎗 | $\{-1\}$ | $\{-\frac{1}{2}\}$ | $\{\zeta_8^5\}$ | $\{2\}$ |
| $[FR_1^{7,1,0}]_{3,1,2}$ | ✗ 🌐 🎗 | $\{-1\}$ | $\{-\frac{1}{2}\}$ | $\{\zeta_8^5\}$ | $\{-2\}$ |
| $[FR_1^{7,1,0}]_{3,2,1}$ | ✗ † 🌐 🎗 | $\{-1\}$ | $\{-\frac{1}{2}\}$ | $\{\zeta_8\}$ | $\{2\}$ |
| $[FR_1^{7,1,0}]_{3,2,2}$ | ✗ 🌐 🎗 | $\{-1\}$ | $\{-\frac{1}{2}\}$ | $\{\zeta_8\}$ | $\{-2\}$ |
| $[FR_1^{7,1,0}]_{3,3,1}$ | ✗ † 🌐 🎗 | $\{-1\}$ | $\{-\frac{1}{2}\}$ | $\{\zeta_8^3\}$ | $\{2\}$ |
| $[FR_1^{7,1,0}]_{3,3,2}$ | ✗ 🌐 🎗 | $\{-1\}$ | $\{-\frac{1}{2}\}$ | $\{\zeta_8^3\}$ | $\{-2\}$ |
| $[FR_1^{7,1,0}]_{3,4,1}$ | ✗ † 🌐 🎗 | $\{-1\}$ | $\{-\frac{1}{2}\}$ | $\{\zeta_8^7\}$ | $\{2\}$ |
| $[FR_1^{7,1,0}]_{3,4,2}$ | ✗ 🌐 🎗 | $\{-1\}$ | $\{-\frac{1}{2}\}$ | $\{\zeta_8^7\}$ | $\{-2\}$ |
| $[FR_1^{7,1,0}]_{4,1,1}$ | ✗ † 🌐 🎗 | $\{-1\}$ | $\{\frac{1}{2}\}$ | $\{\zeta_{16}^9, \zeta_{16}\}$ | $\{2\}$ |
| $[FR_1^{7,1,0}]_{4,1,2}$ | ✗ 🌐 🎗 | $\{-1\}$ | $\{\frac{1}{2}\}$ | $\{\zeta_{16}^9, \zeta_{16}\}$ | $\{-2\}$ |
| $[FR_1^{7,1,0}]_{4,2,1}$ | ✗ † 🌐 🎗 | $\{-1\}$ | $\{\frac{1}{2}\}$ | $\{\zeta_{16}^7, \zeta_{16}^{15}\}$ | $\{2\}$ |
| $[FR_1^{7,1,0}]_{4,2,2}$ | ✗ 🌐 🎗 | $\{-1\}$ | $\{\frac{1}{2}\}$ | $\{\zeta_{16}^7, \zeta_{16}^{15}\}$ | $\{-2\}$ |
| $[FR_1^{7,1,0}]_{4,3,1}$ | ✗ † 🌐 🎗 | $\{-1\}$ | $\{\frac{1}{2}\}$ | $\{\zeta_{16}^{11}, \zeta_{16}^3\}$ | $\{2\}$ |
| $[FR_1^{7,1,0}]_{4,3,2}$ | ✗ 🌐 🎗 | $\{-1\}$ | $\{\frac{1}{2}\}$ | $\{\zeta_{16}^{11}, \zeta_{16}^3\}$ | $\{-2\}$ |
| $[FR_1^{7,1,0}]_{4,4,1}$ | ✗ † 🌐 🎗 | $\{-1\}$ | $\{\frac{1}{2}\}$ | $\{\zeta_{16}^5, \zeta_{16}^{13}\}$ | $\{2\}$ |
| $[FR_1^{7,1,0}]_{4,4,2}$ | ✗ 🌐 🎗 | $\{-1\}$ | $\{\frac{1}{2}\}$ | $\{\zeta_{16}^5, \zeta_{16}^{13}\}$ | $\{-2\}$ |
| $[FR_1^{7,1,0}]_{5,0,1}$ | † 🌐 | $\{1\}$ | $\{-\frac{i}{2}\}$ | - | $\{2\}$ |
| $[FR_1^{7,1,0}]_{5,0,2}$ | 🌐 | $\{1\}$ | $\{-\frac{i}{2}\}$ | - | $\{-2\}$ |
| $[FR_1^{7,1,0}]_{6,0,1}$ | † 🌐 | $\{1\}$ | $\{\frac{i}{2}\}$ | - | $\{2\}$ |
| $[FR_1^{7,1,0}]_{6,0,2}$ | 🌐 | $\{1\}$ | $\{\frac{i}{2}\}$ | - | $\{-2\}$ |
| $[FR_1^{7,1,0}]_{7,0,1}$ | † 🌐 | $\{-1\}$ | $\{\frac{i}{2}\}$ | - | $\{2\}$ |
| $[FR_1^{7,1,0}]_{7,0,2}$ | 🌐 | $\{-1\}$ | $\{\frac{i}{2}\}$ | - | $\{-2\}$ |
| $[FR_1^{7,1,0}]_{8,0,1}$ | † 🌐 | $\{-1\}$ | $\{-\frac{i}{2}\}$ | - | $\{2\}$ |
| $[FR_1^{7,1,0}]_{8,0,2}$ | 🌐 | $\{-1\}$ | $\{-\frac{i}{2}\}$ | - | $\{-2\}$ |

Table 94: Numeric invariants

| Name | Properties | $X_1$ | $X_2$ | $X_3$ | $X_4$ |
|---|---|---|---|---|---|
| $[FR_1^{7,1,0}]_{1,1,1}$ | ✗ † 🌐 🎗 | $\{1\}$ | $\{0.5\}$ | $\{1\}$ | $\{2\}$ |

Table 94: Numeric invariants (Continued)

| Name | Properties | $X_1$ | $X_2$ | $X_3$ | $X_4$ |
|---|---|---|---|---|---|
| $[\mathrm{FR}_1^{7,1,0}]_{1,1,2}$ | | $\{1\}$ | $\{0.5\}$ | $\{1\}$ | $\{-2\}$ |
| $[\mathrm{FR}_1^{7,1,0}]_{1,2,1}$ | | $\{1\}$ | $\{0.5\}$ | $\{-1\}$ | $\{2\}$ |
| $[\mathrm{FR}_1^{7,1,0}]_{1,2,2}$ | | $\{1\}$ | $\{0.5\}$ | $\{-1\}$ | $\{-2\}$ |
| $[\mathrm{FR}_1^{7,1,0}]_{1,3,1}$ | | $\{1\}$ | $\{0.5\}$ | $\{-i\}$ | $\{2\}$ |
| $[\mathrm{FR}_1^{7,1,0}]_{1,3,2}$ | | $\{1\}$ | $\{0.5\}$ | $\{-i\}$ | $\{-2\}$ |
| $[\mathrm{FR}_1^{7,1,0}]_{1,4,1}$ | | $\{1\}$ | $\{0.5\}$ | $\{i\}$ | $\{2\}$ |
| $[\mathrm{FR}_1^{7,1,0}]_{1,4,2}$ | | $\{1\}$ | $\{0.5\}$ | $\{i\}$ | $\{-2\}$ |
| $[\mathrm{FR}_1^{7,1,0}]_{1,5,1}$ | | $\{1\}$ | $\{0.5\}$ | $\{0.707 - 0.707i\}$ | $\{2\}$ |
| $[\mathrm{FR}_1^{7,1,0}]_{1,5,2}$ | | $\{1\}$ | $\{0.5\}$ | $\{0.707 - 0.707i\}$ | $\{-2\}$ |
| $[\mathrm{FR}_1^{7,1,0}]_{1,6,1}$ | | $\{1\}$ | $\{0.5\}$ | $\{-0.707 + 0.707i\}$ | $\{2\}$ |
| $[\mathrm{FR}_1^{7,1,0}]_{1,6,2}$ | | $\{1\}$ | $\{0.5\}$ | $\{-0.707 + 0.707i\}$ | $\{-2\}$ |
| $[\mathrm{FR}_1^{7,1,0}]_{1,7,1}$ | | $\{1\}$ | $\{0.5\}$ | $\{0.707 + 0.707i\}$ | $\{2\}$ |
| $[\mathrm{FR}_1^{7,1,0}]_{1,7,2}$ | | $\{1\}$ | $\{0.5\}$ | $\{0.707 + 0.707i\}$ | $\{-2\}$ |
| $[\mathrm{FR}_1^{7,1,0}]_{1,8,1}$ | | $\{1\}$ | $\{0.5\}$ | $\{-0.707 - 0.707i\}$ | $\{2\}$ |
| $[\mathrm{FR}_1^{7,1,0}]_{1,8,2}$ | | $\{1\}$ | $\{0.5\}$ | $\{-0.707 - 0.707i\}$ | $\{-2\}$ |
| $[\mathrm{FR}_1^{7,1,0}]_{2,1,1}$ | | $\{1\}$ | $\{-0.5\}$ | $\{-0.383 + 0.924i, 0.383 - 0.924i\}$ | $\{2\}$ |
| $[\mathrm{FR}_1^{7,1,0}]_{2,1,2}$ | | $\{1\}$ | $\{-0.5\}$ | $\{-0.383 + 0.924i, 0.383 - 0.924i\}$ | $\{-2\}$ |
| $[\mathrm{FR}_1^{7,1,0}]_{2,2,1}$ | | $\{1\}$ | $\{-0.5\}$ | $\{-0.383 - 0.924i, 0.383 + 0.924i\}$ | $\{2\}$ |
| $[\mathrm{FR}_1^{7,1,0}]_{2,2,2}$ | | $\{1\}$ | $\{-0.5\}$ | $\{-0.383 - 0.924i, 0.383 + 0.924i\}$ | $\{-2\}$ |
| $[\mathrm{FR}_1^{7,1,0}]_{2,3,1}$ | | $\{1\}$ | $\{-0.5\}$ | $\{-0.924 + 0.383i, 0.924 - 0.383i\}$ | $\{2\}$ |
| $[\mathrm{FR}_1^{7,1,0}]_{2,3,2}$ | | $\{1\}$ | $\{-0.5\}$ | $\{-0.924 + 0.383i, 0.924 - 0.383i\}$ | $\{-2\}$ |
| $[\mathrm{FR}_1^{7,1,0}]_{2,4,1}$ | | $\{1\}$ | $\{-0.5\}$ | $\{-0.924 - 0.383i, 0.924 + 0.383i\}$ | $\{2\}$ |
| $[\mathrm{FR}_1^{7,1,0}]_{2,4,2}$ | | $\{1\}$ | $\{-0.5\}$ | $\{-0.924 - 0.383i, 0.924 + 0.383i\}$ | $\{-2\}$ |
| $[\mathrm{FR}_1^{7,1,0}]_{2,5,1}$ | | $\{-1\}$ | $\{-0.5\}$ | $\{i\}$ | $\{2\}$ |
| $[\mathrm{FR}_1^{7,1,0}]_{2,5,2}$ | | $\{-1\}$ | $\{-0.5\}$ | $\{i\}$ | $\{-2\}$ |
| $[\mathrm{FR}_1^{7,1,0}]_{2,6,1}$ | | $\{-1\}$ | $\{-0.5\}$ | $\{-i\}$ | $\{2\}$ |
| $[\mathrm{FR}_1^{7,1,0}]_{2,6,2}$ | | $\{-1\}$ | $\{-0.5\}$ | $\{-i\}$ | $\{-2\}$ |
| $[\mathrm{FR}_1^{7,1,0}]_{2,7,1}$ | | $\{-1\}$ | $\{-0.5\}$ | $\{-1\}$ | $\{2\}$ |
| $[\mathrm{FR}_1^{7,1,0}]_{2,7,2}$ | | $\{-1\}$ | $\{-0.5\}$ | $\{-1\}$ | $\{-2\}$ |
| $[\mathrm{FR}_1^{7,1,0}]_{2,8,1}$ | | $\{-1\}$ | $\{-0.5\}$ | $\{1\}$ | $\{2\}$ |
| $[\mathrm{FR}_1^{7,1,0}]_{2,8,2}$ | | $\{-1\}$ | $\{-0.5\}$ | $\{1\}$ | $\{-2\}$ |
| $[\mathrm{FR}_1^{7,1,0}]_{3,1,1}$ | | $\{-1\}$ | $\{-0.5\}$ | $\{-0.707 - 0.707i\}$ | $\{2\}$ |
| $[\mathrm{FR}_1^{7,1,0}]_{3,1,2}$ | | $\{-1\}$ | $\{-0.5\}$ | $\{-0.707 - 0.707i\}$ | $\{-2\}$ |

Table 94: Numeric invariants (Continued)

| Name | Properties | $X_1$ | $X_2$ | $X_3$ | $X_4$ |
|------|------------|-------|-------|-------|-------|
| $[FR_1^{7,1,0}]_{3,2,1}$ | | $\{-1\}$ | $\{-0.5\}$ | $\{0.707 + 0.707i\}$ | $\{2\}$ |
| $[FR_1^{7,1,0}]_{3,2,2}$ | | $\{-1\}$ | $\{-0.5\}$ | $\{0.707 + 0.707i\}$ | $\{-2\}$ |
| $[FR_1^{7,1,0}]_{3,3,1}$ | | $\{-1\}$ | $\{-0.5\}$ | $\{-0.707 + 0.707i\}$ | $\{2\}$ |
| $[FR_1^{7,1,0}]_{3,3,2}$ | | $\{-1\}$ | $\{-0.5\}$ | $\{-0.707 + 0.707i\}$ | $\{-2\}$ |
| $[FR_1^{7,1,0}]_{3,4,1}$ | | $\{-1\}$ | $\{-0.5\}$ | $\{0.707 - 0.707i\}$ | $\{2\}$ |
| $[FR_1^{7,1,0}]_{3,4,2}$ | | $\{-1\}$ | $\{-0.5\}$ | $\{0.707 - 0.707i\}$ | $\{-2\}$ |
| $[FR_1^{7,1,0}]_{4,1,1}$ | | $\{-1\}$ | $\{0.5\}$ | $\{-0.924 - 0.383i, 0.924 + 0.383i\}$ | $\{2\}$ |
| $[FR_1^{7,1,0}]_{4,1,2}$ | | $\{-1\}$ | $\{0.5\}$ | $\{-0.924 - 0.383i, 0.924 + 0.383i\}$ | $\{-2\}$ |
| $[FR_1^{7,1,0}]_{4,2,1}$ | | $\{-1\}$ | $\{0.5\}$ | $\{-0.924 + 0.383i, 0.924 - 0.383i\}$ | $\{2\}$ |
| $[FR_1^{7,1,0}]_{4,2,2}$ | | $\{-1\}$ | $\{0.5\}$ | $\{-0.924 + 0.383i, 0.924 - 0.383i\}$ | $\{-2\}$ |
| $[FR_1^{7,1,0}]_{4,3,1}$ | | $\{-1\}$ | $\{0.5\}$ | $\{-0.383 - 0.924i, 0.383 + 0.924i\}$ | $\{2\}$ |
| $[FR_1^{7,1,0}]_{4,3,2}$ | | $\{-1\}$ | $\{0.5\}$ | $\{-0.383 - 0.924i, 0.383 + 0.924i\}$ | $\{-2\}$ |
| $[FR_1^{7,1,0}]_{4,4,1}$ | | $\{-1\}$ | $\{0.5\}$ | $\{-0.383 + 0.924i, 0.383 - 0.924i\}$ | $\{2\}$ |
| $[FR_1^{7,1,0}]_{4,4,2}$ | | $\{-1\}$ | $\{0.5\}$ | $\{-0.383 + 0.924i, 0.383 - 0.924i\}$ | $\{-2\}$ |
| $[FR_1^{7,1,0}]_{5,0,1}$ | | $\{1\}$ | $\{-0.5i\}$ | - | $\{2\}$ |
| $[FR_1^{7,1,0}]_{5,0,2}$ | | $\{1\}$ | $\{-0.5i\}$ | - | $\{-2\}$ |
| $[FR_1^{7,1,0}]_{6,0,1}$ | | $\{1\}$ | $\{0.5i\}$ | - | $\{2\}$ |
| $[FR_1^{7,1,0}]_{6,0,2}$ | | $\{1\}$ | $\{0.5i\}$ | - | $\{-2\}$ |
| $[FR_1^{7,1,0}]_{7,0,1}$ | | $\{-1\}$ | $\{0.5i\}$ | - | $\{2\}$ |
| $[FR_1^{7,1,0}]_{7,0,2}$ | | $\{-1\}$ | $\{0.5i\}$ | - | $\{-2\}$ |
| $[FR_1^{7,1,0}]_{8,0,1}$ | | $\{-1\}$ | $\{-0.5i\}$ | - | $\{2\}$ |
| $[FR_1^{7,1,0}]_{8,0,2}$ | | $\{-1\}$ | $\{-0.5i\}$ | - | $\{-2\}$ |

### 4.5.49 $FR_6^{7,1,0}$: Adj(SO(11)$_2$)

For the fusion ring, the following multiplication table is used.

| | 1 | 2 | 3 | 4 | 5 | 6 | 7 |
|---|---|---|---|---|---|---|---|
| 1 | | 2 | 3 | 4 | 5 | 6 | 7 |
| 2 | 1 | | 3 | 4 | 5 | 6 | 7 |
| 3 | 3 | 3 | 1+2+7 | 6+7 | 5+6 | 4+5 | 3+4 |
| 4 | 4 | 4 | 6+7 | 1+2+5 | 4+7 | 3+6 | 3+5 |
| 5 | 5 | 5 | 5+6 | 4+7 | 1+2+3 | 3+7 | 4+6 |
| 6 | 6 | 6 | 4+5 | 3+6 | 3+7 | 1+2+4 | 5+7 |
| 7 | 7 | 7 | 3+4 | 3+5 | 4+6 | 5+7 | 1+2+6 |

The following is the group of all non-trivial permutations that leave the fusion rules invariant:

$$S = \{(), (3\ 4\ 7\ 5\ 6), (3\ 5\ 4\ 6\ 7), (3\ 6\ 5\ 7\ 4), (3\ 7\ 6\ 4\ 5)\}.$$

Let

$$X_1 = S\left(\frac{[F_7^{437}]_3^6[F_3^{557}]_6^3[F_4^{567}]_7^3}{[F_6^{543}]_7^4[F_4^{557}]_4^3}\right), \quad (87)$$

$$X_2 = S\left(R_1^{33}\right). \quad (88)$$

The following table lists a small set of invariants whose values completely distinguish between all MFPBFCs and MFPNBFCs with the given fusion rules.

Table 95: Symbolic invariants

| Name | Properties | $X_1$ | $X_2$ |
|---|---|---|---|
| $[FR_6^{7,1,0}]_{1,1,1}$ | 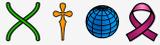 | $\{1\}$ | $\{1\}$ |
| $[FR_6^{7,1,0}]_{1,2,1}$ | 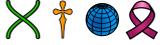 | $\{1\}$ | $\{\zeta_{11}^6, \zeta_{11}^7, \zeta_{11}^8, \zeta_{11}^2, \zeta_{11}^{10}\}$ |
| $[FR_6^{7,1,0}]_{1,3,1}$ | 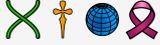 | $\{1\}$ | $\{\zeta_{11}^5, \zeta_{11}^4, \zeta_{11}^3, \zeta_{11}^9, \zeta_{11}\}$ |
| $[FR_6^{7,1,0}]_{2,0,1}$ | 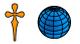 | $\{\zeta_{11}^5, \zeta_{11}^4, \zeta_{11}^3, \zeta_{11}^9, \zeta_{11}\}$ | - |
| $[FR_6^{7,1,0}]_{3,0,1}$ | 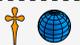 | $\{\zeta_{11}^6, \zeta_{11}^7, \zeta_{11}^8, \zeta_{11}^2, \zeta_{11}^{10}\}$ | - |

Table 96: Numeric invariants

| Name | Properties | $X_1$ | $X_2$ |
|---|---|---|---|
| $[FR_6^{7,1,0}]_{1,1,1}$ | 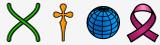 | $\{1\}$ | $\{1\}$ |
| $[FR_6^{7,1,0}]_{1,2,1}$ | 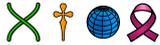 | $\{1\}$ | $\begin{Bmatrix} -0.959 - 0.282i, \\ -0.655 - 0.756i, \\ -0.142 - 0.990i, \\ 0.415 + 0.910i, \\ 0.841 - 0.541i \end{Bmatrix}$ |
| $[FR_6^{7,1,0}]_{1,3,1}$ | 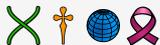 | $\{1\}$ | $\begin{Bmatrix} -0.959 + 0.282i, \\ -0.655 + 0.756i, \\ -0.142 + 0.990i, \\ 0.415 - 0.910i, \\ 0.841 + 0.541i \end{Bmatrix}$ |
| $[FR_6^{7,1,0}]_{2,0,1}$ | 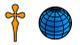 | $\begin{Bmatrix} -0.959 + 0.282i, \\ -0.655 + 0.756i, \\ -0.142 + 0.990i, \\ 0.415 - 0.910i, \\ 0.841 + 0.541i \end{Bmatrix}$ | - |

Table 96: Numeric invariants (Continued)

| Name | Properties | $X_1$ | $X_2$ |
|---|---|---|---|
| $[FR_6^{7,1,0}]_{3,0,1}$ |  | $\left\{\begin{array}{l} -0.959 - 0.282i, \\ -0.655 - 0.756i, \\ -0.142 - 0.990i, \\ 0.415 + 0.910i, \\ 0.841 - 0.541i \end{array}\right\}$ | - |

### 4.5.50 $FR_7^{7,1,0}$: $SU(2)_6$

For the fusion ring, the following multiplication table is used.

| 1 | 2 | 3 | 4 | 5 | 6 | 7 |
|---|---|---|---|---|---|---|
| 2 | 1 | 4 | 3 | 6 | 5 | 7 |
| 3 | 4 | 1+6 | 2+5 | 4+7 | 3+7 | 5+6 |
| 4 | 3 | 2+5 | 1+6 | 3+7 | 4+7 | 5+6 |
| 5 | 6 | 4+7 | 3+7 | 1+5+6 | 2+5+6 | 3+4+7 |
| 6 | 5 | 3+7 | 4+7 | 2+5+6 | 1+5+6 | 3+4+7 |
| 7 | 7 | 5+6 | 5+6 | 3+4+7 | 3+4+7 | 1+2+5+6 |

The following is the group of all non-trivial permutations that leave the fusion rules invariant:

$$S = \{(), (3\ 4)\}.$$

Let

$$X_1 = S\left(R_1^{33}\right), \tag{89}$$
$$X_2 = S\left(d_3^L\right). \tag{90}$$

The following table lists a small set of invariants whose values completely distinguish between all MFPBFCs and MFPNBFCs with the given fusion rules.

Table 97: Symbolic invariants

| Name | Properties | $[F_5^{555}]_5^5$ | $X_1$ | $X_2$ |
|---|---|---|---|---|
| $[FR_7^{7,1,0}]_{1,1,1}$ |  | $\frac{1}{2}(2-\sqrt{2})$ | $\{\zeta_{32}^{13}\}$ | $\{\sqrt{2+\sqrt{2}}\}$ |
| $[FR_7^{7,1,0}]_{1,1,2}$ |  | $\frac{1}{2}(2-\sqrt{2})$ | $\{\zeta_{32}^{13}\}$ | $\{-\sqrt{2+\sqrt{2}}\}$ |
| $[FR_7^{7,1,0}]_{1,2,1}$ |  | $\frac{1}{2}(2-\sqrt{2})$ | $\{\zeta_{32}^{19}\}$ | $\{\sqrt{2+\sqrt{2}}\}$ |
| $[FR_7^{7,1,0}]_{1,2,2}$ |  | $\frac{1}{2}(2-\sqrt{2})$ | $\{\zeta_{32}^{19}\}$ | $\{-\sqrt{2+\sqrt{2}}\}$ |
| $[FR_7^{7,1,0}]_{1,3,1}$ |  | $\frac{1}{2}(2-\sqrt{2})$ | $\{\zeta_{32}^{3}\}$ | $\{\sqrt{2+\sqrt{2}}\}$ |
| $[FR_7^{7,1,0}]_{1,3,2}$ |  | $\frac{1}{2}(2-\sqrt{2})$ | $\{\zeta_{32}^{3}\}$ | $\{-\sqrt{2+\sqrt{2}}\}$ |

Table 97: Symbolic invariants (Continued)

| Name | Properties | $[F_5^{555}]_5^5$ | $X_1$ | $X_2$ |
|---|---|---|---|---|
| $[FR_7^{7,1,0}]_{1,4,1}$ | ✗ ✝ 🌐 🎗 ⭕ | $\frac{1}{2}(2-\sqrt{2})$ | $\{\zeta_{32}^{29}\}$ | $\{\sqrt{2+\sqrt{2}}\}$ |
| $[FR_7^{7,1,0}]_{1,4,2}$ | ✗ 🌐 🎗 ⭕ | $\frac{1}{2}(2-\sqrt{2})$ | $\{\zeta_{32}^{29}\}$ | $\{-\sqrt{2+\sqrt{2}}\}$ |
| $[FR_7^{7,1,0}]_{2,1,1}$ | ✗ 🌐 🎗 ⭕ | $\frac{1}{2}(2+\sqrt{2})$ | $\{\zeta_{32}^{9}\}$ | $\{\sqrt{2-\sqrt{2}}\}$ |
| $[FR_7^{7,1,0}]_{2,1,2}$ | ✗ 🌐 🎗 ⭕ | $\frac{1}{2}(2+\sqrt{2})$ | $\{\zeta_{32}^{9}\}$ | $\{-\sqrt{2-\sqrt{2}}\}$ |
| $[FR_7^{7,1,0}]_{2,2,1}$ | ✗ 🌐 🎗 ⭕ | $\frac{1}{2}(2+\sqrt{2})$ | $\{\zeta_{32}^{23}\}$ | $\{\sqrt{2-\sqrt{2}}\}$ |
| $[FR_7^{7,1,0}]_{2,2,2}$ | ✗ 🌐 🎗 ⭕ | $\frac{1}{2}(2+\sqrt{2})$ | $\{\zeta_{32}^{23}\}$ | $\{-\sqrt{2-\sqrt{2}}\}$ |
| $[FR_7^{7,1,0}]_{2,3,1}$ | ✗ 🌐 🎗 ⭕ | $\frac{1}{2}(2+\sqrt{2})$ | $\{\zeta_{32}^{7}\}$ | $\{\sqrt{2-\sqrt{2}}\}$ |
| $[FR_7^{7,1,0}]_{2,3,2}$ | ✗ 🌐 🎗 ⭕ | $\frac{1}{2}(2+\sqrt{2})$ | $\{\zeta_{32}^{7}\}$ | $\{-\sqrt{2-\sqrt{2}}\}$ |
| $[FR_7^{7,1,0}]_{2,4,1}$ | ✗ 🌐 🎗 ⭕ | $\frac{1}{2}(2+\sqrt{2})$ | $\{\zeta_{32}^{25}\}$ | $\{\sqrt{2-\sqrt{2}}\}$ |
| $[FR_7^{7,1,0}]_{2,4,2}$ | ✗ 🌐 🎗 ⭕ | $\frac{1}{2}(2+\sqrt{2})$ | $\{\zeta_{32}^{25}\}$ | $\{-\sqrt{2-\sqrt{2}}\}$ |

Table 98: Numeric invariants

| Name | Properties | $[F_5^{555}]_5^5$ | $X_1$ | $X_2$ |
|---|---|---|---|---|
| $[FR_7^{7,1,0}]_{1,1,1}$ | ✗ ✝ 🌐 🎗 ⭕ | 0.293 | $\{-0.831 + 0.556i\}$ | $\{1.848\}$ |
| $[FR_7^{7,1,0}]_{1,1,2}$ | ✗ 🌐 🎗 ⭕ | 0.293 | $\{-0.831 + 0.556i\}$ | $\{-1.848\}$ |
| $[FR_7^{7,1,0}]_{1,2,1}$ | ✗ ✝ 🌐 🎗 ⭕ | 0.293 | $\{-0.831 - 0.556i\}$ | $\{1.848\}$ |
| $[FR_7^{7,1,0}]_{1,2,2}$ | ✗ 🌐 🎗 ⭕ | 0.293 | $\{-0.831 - 0.556i\}$ | $\{-1.848\}$ |
| $[FR_7^{7,1,0}]_{1,3,1}$ | ✗ ✝ 🌐 🎗 ⭕ | 0.293 | $\{0.831 + 0.556i\}$ | $\{1.848\}$ |
| $[FR_7^{7,1,0}]_{1,3,2}$ | ✗ 🌐 🎗 ⭕ | 0.293 | $\{0.831 + 0.556i\}$ | $\{-1.848\}$ |
| $[FR_7^{7,1,0}]_{1,4,1}$ | ✗ ✝ 🌐 🎗 ⭕ | 0.293 | $\{0.831 - 0.556i\}$ | $\{1.848\}$ |
| $[FR_7^{7,1,0}]_{1,4,2}$ | ✗ 🌐 🎗 ⭕ | 0.293 | $\{0.831 - 0.556i\}$ | $\{-1.848\}$ |
| $[FR_7^{7,1,0}]_{2,1,1}$ | ✗ 🌐 🎗 ⭕ | 1.707 | $\{-0.195 + 0.981i\}$ | $\{0.765\}$ |
| $[FR_7^{7,1,0}]_{2,1,2}$ | ✗ 🌐 🎗 ⭕ | 1.707 | $\{-0.195 + 0.981i\}$ | $\{-0.765\}$ |
| $[FR_7^{7,1,0}]_{2,2,1}$ | ✗ 🌐 🎗 ⭕ | 1.707 | $\{-0.195 - 0.981i\}$ | $\{0.765\}$ |
| $[FR_7^{7,1,0}]_{2,2,2}$ | ✗ 🌐 🎗 ⭕ | 1.707 | $\{-0.195 - 0.981i\}$ | $\{-0.765\}$ |
| $[FR_7^{7,1,0}]_{2,3,1}$ | ✗ 🌐 🎗 ⭕ | 1.707 | $\{0.195 + 0.981i\}$ | $\{0.765\}$ |
| $[FR_7^{7,1,0}]_{2,3,2}$ | ✗ 🌐 🎗 ⭕ | 1.707 | $\{0.195 + 0.981i\}$ | $\{-0.765\}$ |
| $[FR_7^{7,1,0}]_{2,4,1}$ | ✗ 🌐 🎗 ⭕ | 1.707 | $\{0.195 - 0.981i\}$ | $\{0.765\}$ |
| $[FR_7^{7,1,0}]_{2,4,2}$ | ✗ 🌐 🎗 ⭕ | 1.707 | $\{0.195 - 0.981i\}$ | $\{-0.765\}$ |

## 4.5.51 $FR_8^{7,1,0} : SO_7(2)$

For the fusion ring, the following multiplication table is used.

| 1 | 2 | 3 | 4 | 5 | 6 | 7 |
|---|---|---|---|---|---|---|
| 2 | 1 | 3 | 4 | 5 | 7 | 6 |
| 3 | 3 | 1+2+5 | 4+5 | 3+4 | 6+7 | 6+7 |
| 4 | 4 | 4+5 | 1+2+3 | 3+5 | 6+7 | 6+7 |
| 5 | 5 | 3+4 | 3+5 | 1+2+4 | 6+7 | 6+7 |
| 6 | 7 | 6+7 | 6+7 | 6+7 | 1+3+4+5 | 2+3+4+5 |
| 7 | 6 | 6+7 | 6+7 | 6+7 | 2+3+4+5 | 1+3+4+5 |

The following is the group of all non-trivial permutations that leave the fusion rules invariant:

$$S = \{(), (6\ 7), (3\ 4\ 5), (3\ 5\ 4), (3\ 4\ 5)(6\ 7), (3\ 5\ 4)(6\ 7)\}.$$

Let

$$X_1 = S\left([F_2^{626}]_7^7 [F_7^{676}]_2^2\right), \tag{91}$$

$$X_2 = S\left(R_1^{66}\right), \tag{92}$$

$$X_3 = S\left(d_6^L\right). \tag{93}$$

The following table lists a small set of invariants whose values completely distinguish between all MFPBFCs and MFPNBFCs with the given fusion rules.

Table 99: Symbolic invariants

| Name | Properties | $X_1$ | $X_2$ | $X_3$ |
|---|---|---|---|---|
| $[FR_8^{7,1,0}]_{1,1,1}$ | ✂ † 🌐 🎗 ⊚ | $\left\{-\frac{1}{\sqrt{7}}\right\}$ | $\{\zeta_8^3, \zeta_8^7\}$ | $\{\sqrt{7}\}$ |
| $[FR_8^{7,1,0}]_{1,1,2}$ | ✂ 🌐 🎗 ⊚ | $\left\{-\frac{1}{\sqrt{7}}\right\}$ | $\{\zeta_8^3, \zeta_8^7\}$ | $\{-\sqrt{7}\}$ |
| $[FR_8^{7,1,0}]_{1,2,1}$ | ✂ † 🌐 🎗 ⊚ | $\left\{-\frac{1}{\sqrt{7}}\right\}$ | $\{\zeta_8^5, \zeta_8\}$ | $\{\sqrt{7}\}$ |
| $[FR_8^{7,1,0}]_{1,2,2}$ | ✂ 🌐 🎗 ⊚ | $\left\{-\frac{1}{\sqrt{7}}\right\}$ | $\{\zeta_8^5, \zeta_8\}$ | $\{-\sqrt{7}\}$ |
| $[FR_8^{7,1,0}]_{2,1,1}$ | ✂ † 🌐 🎗 ⊚ | $\left\{\frac{1}{\sqrt{7}}\right\}$ | $\{\zeta_8^3, \zeta_8^7\}$ | $\{\sqrt{7}\}$ |
| $[FR_8^{7,1,0}]_{2,1,2}$ | ✂ 🌐 🎗 ⊚ | $\left\{\frac{1}{\sqrt{7}}\right\}$ | $\{\zeta_8^3, \zeta_8^7\}$ | $\{-\sqrt{7}\}$ |
| $[FR_8^{7,1,0}]_{2,2,1}$ | ✂ † 🌐 🎗 ⊚ | $\left\{\frac{1}{\sqrt{7}}\right\}$ | $\{\zeta_8^5, \zeta_8\}$ | $\{\sqrt{7}\}$ |
| $[FR_8^{7,1,0}]_{2,2,2}$ | ✂ 🌐 🎗 ⊚ | $\left\{\frac{1}{\sqrt{7}}\right\}$ | $\{\zeta_8^5, \zeta_8\}$ | $\{-\sqrt{7}\}$ |

Table 100: Numeric invariants

| Name | Properties | $X_1$ | $X_2$ | $X_3$ |
|---|---|---|---|---|
| $[FR_8^{7,1,0}]_{1,1,1}$ | ✂ † 🌐 🎗 ⊚ | $\{-0.378\}$ | $\{-0.707 + 0.707i, 0.707 - 0.707i\}$ | $\{2.646\}$ |
| $[FR_8^{7,1,0}]_{1,1,2}$ | ✂ 🌐 🎗 ⊚ | $\{-0.378\}$ | $\{-0.707 + 0.707i, 0.707 - 0.707i\}$ | $\{-2.646\}$ |

Table 100: Numeric invariants (Continued)

| Name | Properties | $X_1$ | $X_2$ | $X_3$ |
|---|---|---|---|---|
| $[FR_8^{7,1,0}]_{1,2,1}$ | ✗ † 🌐 🎗 ⚭ | $\{-0.378\}$ | $\{-0.707 - 0.707i, 0.707 + 0.707i\}$ | $\{2.646\}$ |
| $[FR_8^{7,1,0}]_{1,2,2}$ | ✗ 🌐 🎗 ⚭ | $\{-0.378\}$ | $\{-0.707 - 0.707i, 0.707 + 0.707i\}$ | $\{-2.646\}$ |
| $[FR_8^{7,1,0}]_{2,1,1}$ | ✗ † 🌐 🎗 ⚭ | $\{0.378\}$ | $\{-0.707 + 0.707i, 0.707 - 0.707i\}$ | $\{2.646\}$ |
| $[FR_8^{7,1,0}]_{2,1,2}$ | ✗ 🌐 🎗 ⚭ | $\{0.378\}$ | $\{-0.707 + 0.707i, 0.707 - 0.707i\}$ | $\{-2.646\}$ |
| $[FR_8^{7,1,0}]_{2,2,1}$ | ✗ † 🌐 🎗 ⚭ | $\{0.378\}$ | $\{-0.707 - 0.707i, 0.707 + 0.707i\}$ | $\{2.646\}$ |
| $[FR_8^{7,1,0}]_{2,2,2}$ | ✗ 🌐 🎗 ⚭ | $\{0.378\}$ | $\{-0.707 - 0.707i, 0.707 + 0.707i\}$ | $\{-2.646\}$ |

### 4.5.52 $FR_{14}^{7,1,0}$: $PSU(2)_{12}$

For the fusion ring, the following multiplication table is used.

| | | | | | | |
|---|---|---|---|---|---|---|
| 1 | 2 | 3 | 4 | 5 | 6 | 7 |
| 2 | 1 | 4 | 3 | 6 | 5 | 7 |
| 3 | 4 | $1+4+6$ | $2+3+5$ | $4+6+7$ | $3+5+7$ | $5+6+7$ |
| 4 | 3 | $2+3+5$ | $1+4+6$ | $3+5+7$ | $4+6+7$ | $5+6+7$ |
| 5 | 6 | $4+6+7$ | $3+5+7$ | $1+4+5+6+7$ | $2+3+5+6+7$ | $3+4+5+6+7$ |
| 6 | 5 | $3+5+7$ | $4+6+7$ | $2+3+5+6+7$ | $1+4+5+6+7$ | $3+4+5+6+7$ |
| 7 | 7 | $5+6+7$ | $5+6+7$ | $3+4+5+6+7$ | $3+4+5+6+7$ | $1+2+3+4+5+6+7$ |

Only the trivial permutation leaves the fusion rules invariant.

The following table lists a small set of invariants whose values completely distinguish between all MFPBFCs and MFPNBFCs with the given fusion rules.

Table 101: Symbolic invariants

| Name | Properties | $[F_4^{444}]_4^4$ | $R_1^{33}$ |
|---|---|---|---|
| $[FR_{14}^{7,1,0}]_{1,1,1}$ | ✗ † 🌐 🎗 | $\zeta_{14}^3 - \zeta_{14}^4$ | $\zeta_7$ |
| $[FR_{14}^{7,1,0}]_{1,2,1}$ | ✗ † 🌐 🎗 | $\zeta_{14}^3 - \zeta_{14}^4$ | $\zeta_7^6$ |
| $[FR_{14}^{7,1,0}]_{2,1,1}$ | ✗ 🌐 🎗 | $\zeta_{14} - \zeta_{14}^6$ | $\zeta_7^5$ |
| $[FR_{14}^{7,1,0}]_{2,2,1}$ | ✗ 🌐 🎗 | $\zeta_{14} - \zeta_{14}^6$ | $\zeta_7^2$ |
| $[FR_{14}^{7,1,0}]_{3,1,1}$ | ✗ 🌐 🎗 | $\zeta_{14}^5 - \zeta_{14}^2$ | $\zeta_7^3$ |
| $[FR_{14}^{7,1,0}]_{3,2,1}$ | ✗ 🌐 🎗 | $\zeta_{14}^5 - \zeta_{14}^2$ | $\zeta_7^4$ |

Table 102: Numeric invariants

| Name | Properties | $[F_4^{444}]_4^4$ | $R_1^{33}$ |
|---|---|---|---|
| $[FR_{14}^{7,1,0}]_{1,1,1}$ | ✗ † 🌐 🎗 | 0.445 | $0.623 + 0.782i$ |
| $[FR_{14}^{7,1,0}]_{1,2,1}$ | ✗ † 🌐 🎗 | 0.445 | $0.623 - 0.782i$ |

Table 102: Numeric invariants (Continued)

| Name | Properties | $[F_4^{444}]_4^4$ | $R_1^{33}$ |
|---|---|---|---|
| $[FR_{14}^{7,1,0}]_{2,1,1}$ | ✗ 🌐 🎗 | 1.802 | $-0.223 - 0.975i$ |
| $[FR_{14}^{7,1,0}]_{2,2,1}$ | ✗ 🌐 🎗 | 1.802 | $-0.223 + 0.975i$ |
| $[FR_{14}^{7,1,0}]_{3,1,1}$ | ✗ 🌐 🎗 | $-1.247$ | $-0.901 + 0.434i$ |
| $[FR_{14}^{7,1,0}]_{3,2,1}$ | ✗ 🌐 🎗 | $-1.247$ | $-0.901 - 0.434i$ |

### 4.5.53 $FR_{17}^{7,1,0}$: $PSU(2)_{13}$

For the fusion ring, the following multiplication table is used.

| 1 | 2 | 3 | 4 | 5 | 6 | 7 |
|---|---|---|---|---|---|---|
| 2 | 1+3 | 2+4 | 3+5 | 4+6 | 5+7 | 6+7 |
| 3 | 2+4 | 1+3+5 | 2+4+6 | 3+5+7 | 4+6+7 | 5+6+7 |
| 4 | 3+5 | 2+4+6 | 1+3+5+7 | 2+4+6+7 | 3+5+6+7 | 4+5+6+7 |
| 5 | 4+6 | 3+5+7 | 2+4+6+7 | 1+3+5+6+7 | 2+4+5+6+7 | 3+4+5+6+7 |
| 6 | 5+7 | 4+6+7 | 3+5+6+7 | 2+4+5+6+7 | 1+3+4+5+6+7 | 2+3+4+5+6+7 |
| 7 | 6+7 | 5+6+7 | 4+5+6+7 | 3+4+5+6+7 | 2+3+4+5+6+7 | 1+2+3+4+5+6+7 |

Only the trivial permutation leaves the fusion rules invariant.

The following table lists a small set of invariants whose values completely distinguish between all MFPBFCs and MFPNBFCs with the given fusion rules.

Table 103: Symbolic invariants

| Name | Properties | $[F_3^{333}]_3^3$ | $R_1^{33}$ |
|---|---|---|---|
| $[FR_{17}^{7,1,0}]_{1,1,1}$ | ✗ † 🌐 🎗 ⭕ | $\zeta_{15}^7 - \zeta_{15}^6 - \zeta_{15}^5 + 2\zeta_{15}^4 - \zeta_{15}^2 + 1$ | $\zeta_{15}^2$ |
| $[FR_{17}^{7,1,0}]_{1,2,1}$ | ✗ † 🌐 🎗 ⭕ | $\zeta_{15}^7 - \zeta_{15}^6 - \zeta_{15}^5 + 2\zeta_{15}^4 - \zeta_{15}^2 + 1$ | $\zeta_{15}^{13}$ |
| $[FR_{17}^{7,1,0}]_{2,1,1}$ | ✗ 🌐 🎗 ⭕ | $-\zeta_{15}^6 + \zeta_{15}^5 - \zeta_{15}^3 + 2\zeta_{15}^2 - 2\zeta_{15} + 2$ | $\zeta_{15}^{14}$ |
| $[FR_{17}^{7,1,0}]_{2,2,1}$ | ✗ 🌐 🎗 ⭕ | $-\zeta_{15}^6 + \zeta_{15}^5 - \zeta_{15}^3 + 2\zeta_{15}^2 - 2\zeta_{15} + 2$ | $\zeta_{15}$ |
| $[FR_{17}^{7,1,0}]_{3,1,1}$ | ✗ 🌐 🎗 ⭕ | $2\zeta_{15}^7 + \zeta_{15}^6 - \zeta_{15}^5 - \zeta_{15}^3 + 2\zeta_{15} + 1$ | $\zeta_{15}^4$ |
| $[FR_{17}^{7,1,0}]_{3,2,1}$ | ✗ 🌐 🎗 ⭕ | $2\zeta_{15}^7 + \zeta_{15}^6 - \zeta_{15}^5 - \zeta_{15}^3 + 2\zeta_{15} + 1$ | $\zeta_{15}^{11}$ |
| $[FR_{17}^{7,1,0}]_{4,1,1}$ | ✗ 🌐 🎗 ⭕ | $-3\zeta_{15}^7 + \zeta_{15}^6 + \zeta_{15}^5 - 2\zeta_{15}^4 + 2\zeta_{15}^3 - \zeta_{15}^2 + 4$ | $\zeta_{15}^7$ |
| $[FR_{17}^{7,1,0}]_{4,2,1}$ | ✗ 🌐 🎗 ⭕ | $-3\zeta_{15}^7 + \zeta_{15}^6 + \zeta_{15}^5 - 2\zeta_{15}^4 + 2\zeta_{15}^3 - \zeta_{15}^2 + 4$ | $\zeta_{15}^8$ |

Table 104: Numeric invariants

| Name | Properties | $[F_3^{333}]_3^3$ | $R_1^{33}$ |
|---|---|---|---|
| $[FR_{17}^{7,1,0}]_{1,1,1}$ | ✗ † 🌐 🎗 ⭕ | 0.453 | $0.669 + 0.743i$ |
| $[FR_{17}^{7,1,0}]_{1,2,1}$ | ✗ † 🌐 🎗 ⭕ | 0.453 | $0.669 - 0.743i$ |

Table 104: Numeric invariants (Continued)

| Name | Properties | $[F_3^{333}]_3^3$ | $R_1^{33}$ |
|---|---|---|---|
| $[FR_{17}^{7,1,0}]_{2,1,1}$ | ✕ 🌐🎗⚭ | 1.511 | $0.914 - 0.407i$ |
| $[FR_{17}^{7,1,0}]_{2,2,1}$ | ✕ 🌐🎗⚭ | 1.511 | $0.914 + 0.407i$ |
| $[FR_{17}^{7,1,0}]_{3,1,1}$ | ✕ 🌐🎗⚭ | 0.253 | $-0.105 + 0.995i$ |
| $[FR_{17}^{7,1,0}]_{3,2,1}$ | ✕ 🌐🎗⚭ | 0.253 | $-0.105 - 0.995i$ |
| $[FR_{17}^{7,1,0}]_{4,1,1}$ | ✕ 🌐🎗⚭ | 5.783 | $-0.978 + 0.208i$ |
| $[FR_{17}^{7,1,0}]_{4,2,1}$ | ✕ 🌐🎗⚭ | 5.783 | $-0.978 - 0.208i$ |

### 4.5.54 $FR_3^{7,1,2}$

For the fusion ring, the following multiplication table is used.

| 1 | 2 | 3 | 4 | 5 | 6 | 7 |
|---|---|---|---|---|---|---|
| 2 | 1 | 4 | 3 | 5 | 6 | 7 |
| 3 | 4 | 2 | 1 | 7 | 6 | 5 |
| 4 | 3 | 1 | 2 | 7 | 6 | 5 |
| 5 | 5 | 7 | 7 | $1+2+6$ | $5+7$ | $3+4+6$ |
| 6 | 6 | 6 | 6 | $5+7$ | $1+2+3+4$ | $5+7$ |
| 7 | 7 | 5 | 5 | $3+4+6$ | $5+7$ | $1+2+6$ |

The following is the group of all non-trivial permutations that leave the fusion rules invariant:

$$S = \{(), (3\ 4), (5\ 7), (3\ 4)(5\ 7)\}.$$

Let

$$X_1 = S\left([F_5^{557}]_6^6 [F_6^{567}]_5^5\right), \tag{94}$$

$$X_2 = S\left(\frac{[F_6^{336}]_6^2 [F_5^{356}]_7^7 [F_5^{522}]_1^5 [F_5^{533}]_2^7 [F_5^{555}]_6^1 [F_1^{556}]_5^6 [F_5^{566}]_2^7}{[F_1^{266}]_2^6 [F_5^{353}]_7^7 \left([F_5^{536}]_6^7\right)^2}\right), \tag{95}$$

$$X_3 = S\left(d_5^L\right). \tag{96}$$

The following table lists a small set of invariants whose values completely distinguish between all MFPBFCs and MFPNBFCs with the given fusion rules.

Table 105: Symbolic invariants

| Name | Properties | $X_1$ | $X_2$ | $X_3$ |
|---|---|---|---|---|
| $[FR_3^{7,1,2}]_{1,0,1}$ | 🗡🌐 | $\{1\}$ | $\left\{\frac{\zeta_8^5}{2}\right\}$ | $\{2\}$ |
| $[FR_3^{7,1,2}]_{1,0,2}$ | 🌐 | $\{1\}$ | $\left\{\frac{\zeta_8^5}{2}\right\}$ | $\{-2\}$ |
| $[FR_3^{7,1,2}]_{2,0,1}$ | 🗡🌐 | $\{1\}$ | $\left\{\frac{\zeta_8^7}{2}\right\}$ | $\{2\}$ |

Table 105: Symbolic invariants (Continued)

| Name | Properties | $X_1$ | $X_2$ | $X_3$ |
|---|---|---|---|---|
| $[FR_3^{7,1,2}]_{2,0,2}$ |  | $\{1\}$ | $\left\{\frac{\zeta_8^7}{2}\right\}$ | $\{-2\}$ |
| $[FR_3^{7,1,2}]_{3,0,1}$ |  | $\{1\}$ | $\left\{\frac{\zeta_8}{2}\right\}$ | $\{2\}$ |
| $[FR_3^{7,1,2}]_{3,0,2}$ |  | $\{1\}$ | $\left\{\frac{\zeta_8}{2}\right\}$ | $\{-2\}$ |
| $[FR_3^{7,1,2}]_{4,0,1}$ |  | $\{1\}$ | $\left\{\frac{\zeta_8^3}{2}\right\}$ | $\{2\}$ |
| $[FR_3^{7,1,2}]_{4,0,2}$ |  | $\{1\}$ | $\left\{\frac{\zeta_8^3}{2}\right\}$ | $\{-2\}$ |
| $[FR_3^{7,1,2}]_{5,0,1}$ |  | $\{-1\}$ | $\left\{\frac{\zeta_8}{2}\right\}$ | $\{2\}$ |
| $[FR_3^{7,1,2}]_{5,0,2}$ |  | $\{-1\}$ | $\left\{\frac{\zeta_8}{2}\right\}$ | $\{-2\}$ |
| $[FR_3^{7,1,2}]_{6,0,1}$ |  | $\{-1\}$ | $\left\{\frac{\zeta_8^3}{2}\right\}$ | $\{2\}$ |
| $[FR_3^{7,1,2}]_{6,0,2}$ |  | $\{-1\}$ | $\left\{\frac{\zeta_8^3}{2}\right\}$ | $\{-2\}$ |
| $[FR_3^{7,1,2}]_{7,0,1}$ |  | $\{-1\}$ | $\left\{\frac{\zeta_8^5}{2}\right\}$ | $\{2\}$ |
| $[FR_3^{7,1,2}]_{7,0,2}$ |  | $\{-1\}$ | $\left\{\frac{\zeta_8^5}{2}\right\}$ | $\{-2\}$ |
| $[FR_3^{7,1,2}]_{8,0,1}$ |  | $\{-1\}$ | $\left\{\frac{\zeta_8^7}{2}\right\}$ | $\{2\}$ |
| $[FR_3^{7,1,2}]_{8,0,2}$ |  | $\{-1\}$ | $\left\{\frac{\zeta_8^7}{2}\right\}$ | $\{-2\}$ |

Table 106: Numeric invariants

| Name | Properties | $X_1$ | $X_2$ | $X_3$ |
|---|---|---|---|---|
| $[FR_3^{7,1,2}]_{1,0,1}$ |  | $\{1\}$ | $\{-0.354 - 0.354i\}$ | $\{2\}$ |
| $[FR_3^{7,1,2}]_{1,0,2}$ |  | $\{1\}$ | $\{-0.354 - 0.354i\}$ | $\{-2\}$ |
| $[FR_3^{7,1,2}]_{2,0,1}$ |  | $\{1\}$ | $\{0.354 - 0.354i\}$ | $\{2\}$ |
| $[FR_3^{7,1,2}]_{2,0,2}$ |  | $\{1\}$ | $\{0.354 - 0.354i\}$ | $\{-2\}$ |
| $[FR_3^{7,1,2}]_{3,0,1}$ |  | $\{1\}$ | $\{0.354 + 0.354i\}$ | $\{2\}$ |
| $[FR_3^{7,1,2}]_{3,0,2}$ |  | $\{1\}$ | $\{0.354 + 0.354i\}$ | $\{-2\}$ |
| $[FR_3^{7,1,2}]_{4,0,1}$ |  | $\{1\}$ | $\{-0.354 + 0.354i\}$ | $\{2\}$ |
| $[FR_3^{7,1,2}]_{4,0,2}$ |  | $\{1\}$ | $\{-0.354 + 0.354i\}$ | $\{-2\}$ |
| $[FR_3^{7,1,2}]_{5,0,1}$ |  | $\{-1\}$ | $\{0.354 + 0.354i\}$ | $\{2\}$ |
| $[FR_3^{7,1,2}]_{5,0,2}$ |  | $\{-1\}$ | $\{0.354 + 0.354i\}$ | $\{-2\}$ |
| $[FR_3^{7,1,2}]_{6,0,1}$ |  | $\{-1\}$ | $\{-0.354 + 0.354i\}$ | $\{2\}$ |
| $[FR_3^{7,1,2}]_{6,0,2}$ |  | $\{-1\}$ | $\{-0.354 + 0.354i\}$ | $\{-2\}$ |

Table 106: Numeric invariants (Continued)

| Name | Properties | $X_1$ | $X_2$ | $X_3$ |
|---|---|---|---|---|
| $[FR_3^{7,1,2}]_{7,0,1}$ |   | $\{-1\}$ | $\{-0.354 - 0.354i\}$ | $\{2\}$ |
| $[FR_3^{7,1,2}]_{7,0,2}$ |  | $\{-1\}$ | $\{-0.354 - 0.354i\}$ | $\{-2\}$ |
| $[FR_3^{7,1,2}]_{8,0,1}$ |   | $\{-1\}$ | $\{0.354 - 0.354i\}$ | $\{2\}$ |
| $[FR_3^{7,1,2}]_{8,0,2}$ |  | $\{-1\}$ | $\{0.354 - 0.354i\}$ | $\{-2\}$ |

## 4.5.55 $FR_4^{7,1,2}$ : $\mathbf{Rep}(SD_{16})$

For the fusion ring, the following multiplication table is used.

| 1 | 2 | 3 | 4 | 5 | 6 | 7 |
|---|---|---|---|---|---|---|
| 2 | 1 | 4 | 3 | 5 | 7 | 6 |
| 3 | 4 | 1 | 2 | 5 | 7 | 6 |
| 4 | 3 | 2 | 1 | 5 | 6 | 7 |
| 5 | 5 | 5 | 5 | $1+2+3+4$ | $6+7$ | $6+7$ |
| 6 | 7 | 7 | 6 | $6+7$ | $2+3+5$ | $1+4+5$ |
| 7 | 6 | 6 | 7 | $6+7$ | $1+4+5$ | $2+3+5$ |

The following is the group of all non-trivial permutations that leave the fusion rules invariant:

$$S = \{(), (2\ 3), (6\ 7), (2\ 3)(6\ 7)\}.$$

Let

$$X_1 = S\left([F_2^{656}]_6^6 [F_6^{666}]_5^5\right), \tag{97}$$
$$X_2 = S\left(R_5^{66}\right), \tag{98}$$
$$X_3 = S\left(d_6^L\right). \tag{99}$$

The following table lists a small set of invariants whose values completely distinguish between all MFPBFCs and MFPNBFCs with the given fusion rules.

Table 107: Symbolic invariants

| Name | Properties | $X_1$ | $X_2$ | $X_3$ |
|---|---|---|---|---|
| $[FR_4^{7,1,2}]_{1,1,1}$ |     | $\{-1, 1\}$ | $\{i\}$ | $\{2\}$ |
| $[FR_4^{7,1,2}]_{1,1,2}$ |    | $\{-1, 1\}$ | $\{i\}$ | $\{-2\}$ |
| $[FR_4^{7,1,2}]_{1,2,1}$ |     | $\{-1, 1\}$ | $\{-i\}$ | $\{2\}$ |
| $[FR_4^{7,1,2}]_{1,2,2}$ |    | $\{-1, 1\}$ | $\{-i\}$ | $\{-2\}$ |
| $[FR_4^{7,1,2}]_{1,3,1}$ |     | $\{-1, 1\}$ | $\{1\}$ | $\{2\}$ |
| $[FR_4^{7,1,2}]_{1,3,2}$ |    | $\{-1, 1\}$ | $\{1\}$ | $\{-2\}$ |
| $[FR_4^{7,1,2}]_{1,4,1}$ |     | $\{-1, 1\}$ | $\{-1\}$ | $\{2\}$ |

Table 107: Symbolic invariants (Continued)

| Name | Properties | $X_1$ | $X_2$ | $X_3$ |
|---|---|---|---|---|
| $[\mathrm{FR}_4^{7,1,2}]_{1,4,2}$ | ✕ 🌐 🎗 | $\{-1,1\}$ | $\{-1\}$ | $\{-2\}$ |
| $[\mathrm{FR}_4^{7,1,2}]_{1,5,1}$ | ✕ † 🌐 🎗 | $\{-1,1\}$ | $\{\zeta_8^3\}$ | $\{2\}$ |
| $[\mathrm{FR}_4^{7,1,2}]_{1,5,2}$ | ✕ 🌐 🎗 | $\{-1,1\}$ | $\{\zeta_8^3\}$ | $\{-2\}$ |
| $[\mathrm{FR}_4^{7,1,2}]_{1,6,1}$ | ✕ † 🌐 🎗 | $\{-1,1\}$ | $\{\zeta_8\}$ | $\{2\}$ |
| $[\mathrm{FR}_4^{7,1,2}]_{1,6,2}$ | ✕ 🌐 🎗 | $\{-1,1\}$ | $\{\zeta_8\}$ | $\{-2\}$ |
| $[\mathrm{FR}_4^{7,1,2}]_{1,7,1}$ | ✕ † 🌐 🎗 | $\{-1,1\}$ | $\{\zeta_8^5\}$ | $\{2\}$ |
| $[\mathrm{FR}_4^{7,1,2}]_{1,7,2}$ | ✕ 🌐 🎗 | $\{-1,1\}$ | $\{\zeta_8^5\}$ | $\{-2\}$ |
| $[\mathrm{FR}_4^{7,1,2}]_{1,8,1}$ | ✕ † 🌐 🎗 | $\{-1,1\}$ | $\{\zeta_8^7\}$ | $\{2\}$ |
| $[\mathrm{FR}_4^{7,1,2}]_{1,8,2}$ | ✕ 🌐 🎗 | $\{-1,1\}$ | $\{\zeta_8^7\}$ | $\{-2\}$ |
| $[\mathrm{FR}_4^{7,1,2}]_{2,0,1}$ | † 🌐 | $\{\zeta_8^3, \zeta_8^7\}$ | - | $\{2\}$ |
| $[\mathrm{FR}_4^{7,1,2}]_{2,0,2}$ | 🌐 | $\{\zeta_8^3, \zeta_8^7\}$ | - | $\{-2\}$ |
| $[\mathrm{FR}_4^{7,1,2}]_{3,0,1}$ | † 🌐 | $\{-i, i\}$ | - | $\{2\}$ |
| $[\mathrm{FR}_4^{7,1,2}]_{3,0,2}$ | 🌐 | $\{-i, i\}$ | - | $\{-2\}$ |
| $[\mathrm{FR}_4^{7,1,2}]_{4,0,1}$ | † 🌐 | $\{\zeta_8^5, \zeta_8\}$ | - | $\{2\}$ |
| $[\mathrm{FR}_4^{7,1,2}]_{4,0,2}$ | 🌐 | $\{\zeta_8^5, \zeta_8\}$ | - | $\{-2\}$ |

Table 108: Numeric invariants

| Name | Properties | $X_1$ | $X_2$ | $X_3$ |
|---|---|---|---|---|
| $[\mathrm{FR}_4^{7,1,2}]_{1,1,1}$ | ✕ † 🌐 🎗 | $\{-1,1\}$ | $\{i\}$ | $\{2\}$ |
| $[\mathrm{FR}_4^{7,1,2}]_{1,1,2}$ | ✕ 🌐 🎗 | $\{-1,1\}$ | $\{i\}$ | $\{-2\}$ |
| $[\mathrm{FR}_4^{7,1,2}]_{1,2,1}$ | ✕ † 🌐 🎗 | $\{-1,1\}$ | $\{-i\}$ | $\{2\}$ |
| $[\mathrm{FR}_4^{7,1,2}]_{1,2,2}$ | ✕ 🌐 🎗 | $\{-1,1\}$ | $\{-i\}$ | $\{-2\}$ |
| $[\mathrm{FR}_4^{7,1,2}]_{1,3,1}$ | ✕ † 🌐 🎗 | $\{-1,1\}$ | $\{1\}$ | $\{2\}$ |
| $[\mathrm{FR}_4^{7,1,2}]_{1,3,2}$ | ✕ 🌐 🎗 | $\{-1,1\}$ | $\{1\}$ | $\{-2\}$ |
| $[\mathrm{FR}_4^{7,1,2}]_{1,4,1}$ | ✕ † 🌐 🎗 | $\{-1,1\}$ | $\{-1\}$ | $\{2\}$ |
| $[\mathrm{FR}_4^{7,1,2}]_{1,4,2}$ | ✕ 🌐 🎗 | $\{-1,1\}$ | $\{-1\}$ | $\{-2\}$ |
| $[\mathrm{FR}_4^{7,1,2}]_{1,5,1}$ | ✕ † 🌐 🎗 | $\{-1,1\}$ | $\{-0.707 + 0.707i\}$ | $\{2\}$ |
| $[\mathrm{FR}_4^{7,1,2}]_{1,5,2}$ | ✕ 🌐 🎗 | $\{-1,1\}$ | $\{-0.707 + 0.707i\}$ | $\{-2\}$ |
| $[\mathrm{FR}_4^{7,1,2}]_{1,6,1}$ | ✕ † 🌐 🎗 | $\{-1,1\}$ | $\{0.707 + 0.707i\}$ | $\{2\}$ |
| $[\mathrm{FR}_4^{7,1,2}]_{1,6,2}$ | ✕ 🌐 🎗 | $\{-1,1\}$ | $\{0.707 + 0.707i\}$ | $\{-2\}$ |
| $[\mathrm{FR}_4^{7,1,2}]_{1,7,1}$ | ✕ † 🌐 🎗 | $\{-1,1\}$ | $\{-0.707 - 0.707i\}$ | $\{2\}$ |
| $[\mathrm{FR}_4^{7,1,2}]_{1,7,2}$ | ✕ 🌐 🎗 | $\{-1,1\}$ | $\{-0.707 - 0.707i\}$ | $\{-2\}$ |

Table 108: Numeric invariants (Continued)

| Name | Properties | $X_1$ | $X_2$ | $X_3$ |
|---|---|---|---|---|
| $[FR_4^{7,1,2}]_{1,8,1}$ | ✗ † 🌐 🎗 | $\{-1, 1\}$ | $\{0.707 - 0.707i\}$ | $\{2\}$ |
| $[FR_4^{7,1,2}]_{1,8,2}$ | ✗ 🌐 🎗 | $\{-1, 1\}$ | $\{0.707 - 0.707i\}$ | $\{-2\}$ |
| $[FR_4^{7,1,2}]_{2,0,1}$ | † 🌐 | $\{-0.707 + 0.707i, 0.707 - 0.707i\}$ | - | $\{2\}$ |
| $[FR_4^{7,1,2}]_{2,0,2}$ | 🌐 | $\{-0.707 + 0.707i, 0.707 - 0.707i\}$ | - | $\{-2\}$ |
| $[FR_4^{7,1,2}]_{3,0,1}$ | † 🌐 | $\{-i, i\}$ | - | $\{2\}$ |
| $[FR_4^{7,1,2}]_{3,0,2}$ | 🌐 | $\{-i, i\}$ | - | $\{-2\}$ |
| $[FR_4^{7,1,2}]_{4,0,1}$ | † 🌐 | $\{-0.707 - 0.707i, 0.707 + 0.707i\}$ | - | $\{2\}$ |
| $[FR_4^{7,1,2}]_{4,0,2}$ | 🌐 | $\{-0.707 - 0.707i, 0.707 + 0.707i\}$ | - | $\{-2\}$ |

### 4.5.56 $FR_{12}^{7,1,2}$

For the fusion ring, the following multiplication table is used.

| 1 | 2 | 3 | 4 | 5 | 6 | 7 |
|---|---|---|---|---|---|---|
| 2 | 1 | 3 | 4 | 5 | 7 | 6 |
| 3 | 3 | $1+2+5$ | $4+5$ | $3+4$ | $6+7$ | $6+7$ |
| 4 | 4 | $4+5$ | $1+2+3$ | $3+5$ | $6+7$ | $6+7$ |
| 5 | 5 | $3+4$ | $3+5$ | $1+2+4$ | $6+7$ | $6+7$ |
| 6 | 7 | $6+7$ | $6+7$ | $6+7$ | $2+3+4+5$ | $1+3+4+5$ |
| 7 | 6 | $6+7$ | $6+7$ | $6+7$ | $1+3+4+5$ | $2+3+4+5$ |

The following is the group of all non-trivial permutations that leave the fusion rules invariant:

$$S = \{(), (6\ 7), (3\ 4\ 5), (3\ 5\ 4), (3\ 4\ 5)(6\ 7), (3\ 5\ 4)(6\ 7)\}.$$

Let

$$X_1 = S\left([F_2^{636}]_6^6 [F_6^{666}]_3^3\right), \tag{100}$$

$$X_2 = S\left(d_6^L\right), \tag{101}$$

$$r_1 = \frac{1}{7}\left(3\zeta_{28}^{11} - 2\zeta_{28}^9 + 4\zeta_{28}^7 - 2\zeta_{28}^5 + 3\zeta_{28}^3\right), \tag{102}$$

$$r_2 = \frac{1}{7}\left(-2\zeta_{28}^{11} - \zeta_{28}^9 + 2\zeta_{28}^7 - \zeta_{28}^5 - 2\zeta_{28}^3\right), \tag{103}$$

$$r_3 = \frac{1}{7}\left(-\zeta_{28}^{11} + 3\zeta_{28}^9 + \zeta_{28}^7 + 3\zeta_{28}^5 - \zeta_{28}^3\right). \tag{104}$$

The following table lists a small set of invariants whose values completely distinguish between all MFPBFCs and MFPNBFCs with the given fusion rules.

Table 109: Symbolic invariants

| Name | Properties | $X_1$ | $X_2$ |
|---|---|---|---|
| $[FR_{12}^{7,1,2}]_{1,0,1}$ | † 🌐 | $\{r_1, r_2, r_3\}$ | $\{\sqrt{7}\}$ |

Table 109: Symbolic invariants (Continued)

| Name | Properties | $X_1$ | $X_2$ |
|---|---|---|---|
| $[\text{FR}_{12}^{7,1,2}]_{1,0,2}$ | 🌐 | $\{r_1, r_2, r_3\}$ | $\{-\sqrt{7}\}$ |
| $[\text{FR}_{12}^{7,1,2}]_{2,0,1}$ | 🗡🌐 | $\{-r_1, -r_2, -r_3\}$ | $\{\sqrt{7}\}$ |
| $[\text{FR}_{12}^{7,1,2}]_{2,0,2}$ | 🌐 | $\{-r_1, -r_2, -r_3\}$ | $\{-\sqrt{7}\}$ |

Table 110: Numeric invariants

| Name | Properties | $X_1$ | $X_2$ |
|---|---|---|---|
| $[\text{FR}_{12}^{7,1,2}]_{1,0,1}$ | 🗡🌐 | $\{-0.328i, 0.591i, 0.737i\}$ | $\{2.646\}$ |
| $[\text{FR}_{12}^{7,1,2}]_{1,0,2}$ | 🌐 | $\{-0.328i, 0.591i, 0.737i\}$ | $\{-2.646\}$ |
| $[\text{FR}_{12}^{7,1,2}]_{2,0,1}$ | 🗡🌐 | $\{0.328i, -0.591i, -0.737i\}$ | $\{2.646\}$ |
| $[\text{FR}_{12}^{7,1,2}]_{2,0,2}$ | 🌐 | $\{0.328i, -0.591i, -0.737i\}$ | $\{-2.646\}$ |

### 4.5.57 $\text{FR}_1^{7,1,4}$: $\text{TY}(\mathbb{Z}_2 \otimes \mathbb{Z}_3)$

For the fusion ring, the following multiplication table is used.

```
1 2 3 4 5 6 7
2 1 6 5 4 3 7
3 6 4 1 2 5 7
4 5 1 3 6 2 7
5 4 2 6 3 1 7
6 3 5 2 1 4 7
7 7 7 7 7 7 1+2+3+4+5+6
```

The following is the group of all non-trivial permutations that leave the fusion rules invariant:

$$S = \{(), (3\ 4)(5\ 6)\}.$$

Let

$$X_1 = S\left([F_1^{757}]_7^7 [F_7^{777}]_5^5\right). \tag{105}$$

The following table lists a small set of invariants whose values completely distinguish between all MFPBFCs and MFPNBFCs with the given fusion rules.

Table 111: Symbolic invariants

| Name | Properties | $X_1$ | $d_7^L$ |
|---|---|---|---|
| $[\text{FR}_1^{7,1,4}]_{1,0,1}$ | 🗡🌐 | $\left\{\frac{\zeta_6^5}{\sqrt{6}}\right\}$ | $\sqrt{6}$ |
| $[\text{FR}_1^{7,1,4}]_{1,0,2}$ | 🌐 | $\left\{\frac{\zeta_6^5}{\sqrt{6}}\right\}$ | $-\sqrt{6}$ |

Table 111: Symbolic invariants (Continued)

| Name | Properties | $X_1$ | $d_7^L$ |
|---|---|---|---|
| $[FR_1^{7,1,4}]_{2,0,1}$ | † 🌐 | $\left\{\frac{\zeta_3}{\sqrt{6}}\right\}$ | $\sqrt{6}$ |
| $[FR_1^{7,1,4}]_{2,0,2}$ | 🌐 | $\left\{\frac{\zeta_3}{\sqrt{6}}\right\}$ | $-\sqrt{6}$ |
| $[FR_1^{7,1,4}]_{3,0,1}$ | † 🌐 | $\left\{\frac{\zeta_6}{\sqrt{6}}\right\}$ | $\sqrt{6}$ |
| $[FR_1^{7,1,4}]_{3,0,2}$ | 🌐 | $\left\{\frac{\zeta_6}{\sqrt{6}}\right\}$ | $-\sqrt{6}$ |
| $[FR_1^{7,1,4}]_{4,0,1}$ | † 🌐 | $\left\{\frac{\zeta_3^2}{\sqrt{6}}\right\}$ | $\sqrt{6}$ |
| $[FR_1^{7,1,4}]_{4,0,2}$ | 🌐 | $\left\{\frac{\zeta_3^2}{\sqrt{6}}\right\}$ | $-\sqrt{6}$ |

Table 112: Numeric invariants

| Name | Properties | $X_1$ | $d_7^L$ |
|---|---|---|---|
| $[FR_1^{7,1,4}]_{1,0,1}$ | † 🌐 | $\{0.204 - 0.354i\}$ | $2.449$ |
| $[FR_1^{7,1,4}]_{1,0,2}$ | 🌐 | $\{0.204 - 0.354i\}$ | $-2.449$ |
| $[FR_1^{7,1,4}]_{2,0,1}$ | † 🌐 | $\{-0.204 + 0.354i\}$ | $2.449$ |
| $[FR_1^{7,1,4}]_{2,0,2}$ | 🌐 | $\{-0.204 + 0.354i\}$ | $-2.449$ |
| $[FR_1^{7,1,4}]_{3,0,1}$ | † 🌐 | $\{0.204 + 0.354i\}$ | $2.449$ |
| $[FR_1^{7,1,4}]_{3,0,2}$ | 🌐 | $\{0.204 + 0.354i\}$ | $-2.449$ |
| $[FR_1^{7,1,4}]_{4,0,1}$ | † 🌐 | $\{-0.204 - 0.354i\}$ | $2.449$ |
| $[FR_1^{7,1,4}]_{4,0,2}$ | 🌐 | $\{-0.204 - 0.354i\}$ | $-2.449$ |

### 4.5.58 $FR_3^{7,1,4}$

For the fusion ring, the following multiplication table is used.

| 1 | 2 | 3 | 4 | 5 | 6 | 7 |
|---|---|---|---|---|---|---|
| 2 | 1 | 4 | 3 | 5 | 6 | 7 |
| 3 | 4 | 2 | 1 | 5 | 7 | 6 |
| 4 | 3 | 1 | 2 | 5 | 7 | 6 |
| 5 | 5 | 5 | 5 | $1+2+3+4$ | $6+7$ | $6+7$ |
| 6 | 6 | 7 | 7 | $6+7$ | $3+4+5$ | $1+2+5$ |
| 7 | 7 | 6 | 6 | $6+7$ | $1+2+5$ | $3+4+5$ |

The following is the group of all non-trivial permutations that leave the fusion rules invariant:

$$S = \{(), (3\ 4), (6\ 7), (3\ 4)(6\ 7)\}.$$

Let

$$X_1 = S\left([F_3^{656}]_6^6 [F_6^{666}]_5^5\right). \tag{106}$$
$$X_2 = S\left(d_6^L\right) \tag{107}$$

The following table lists a small set of invariants whose values completely distinguish between all MFPBFCs and MFPNBFCs with the given fusion rules.

Table 113: Symbolic invariants

| Name | Properties | $X_1$ | $X_2$ |
|---|---|---|---|
| $[FR_3^{7,1,4}]_{1,0,1}$ | † 🌐 | $\{\zeta_{16}^5, \zeta_{16}^{13}\}$ | $\{2\}$ |
| $[FR_3^{7,1,4}]_{1,0,2}$ | 🌐 | $\{\zeta_{16}^5, \zeta_{16}^{13}\}$ | $\{-2\}$ |
| $[FR_3^{7,1,4}]_{2,0,1}$ | † 🌐 | $\{\zeta_{16}^9, \zeta_{16}\}$ | $\{2\}$ |
| $[FR_3^{7,1,4}]_{2,0,2}$ | 🌐 | $\{\zeta_{16}^9, \zeta_{16}\}$ | $\{-2\}$ |
| $[FR_3^{7,1,4}]_{3,0,1}$ | † 🌐 | $\{\zeta_{16}^7, \zeta_{16}^{15}\}$ | $\{2\}$ |
| $[FR_3^{7,1,4}]_{3,0,2}$ | 🌐 | $\{\zeta_{16}^7, \zeta_{16}^{15}\}$ | $\{-2\}$ |
| $[FR_3^{7,1,4}]_{4,0,1}$ | † 🌐 | $\{\zeta_{16}^{11}, \zeta_{16}^3\}$ | $\{2\}$ |
| $[FR_3^{7,1,4}]_{4,0,2}$ | 🌐 | $\{\zeta_{16}^{11}, \zeta_{16}^3\}$ | $\{-2\}$ |

Table 114: Numeric invariants

| Name | Properties | $X_1$ | $X_2$ |
|---|---|---|---|
| $[FR_3^{7,1,4}]_{1,0,1}$ | † 🌐 | $\{-0.383 + 0.924i, 0.383 - 0.924i\}$ | $\{2\}$ |
| $[FR_3^{7,1,4}]_{1,0,2}$ | 🌐 | $\{-0.383 + 0.924i, 0.383 - 0.924i\}$ | $\{-2\}$ |
| $[FR_3^{7,1,4}]_{2,0,1}$ | † 🌐 | $\{-0.924 - 0.383i, 0.924 + 0.383i\}$ | $\{2\}$ |
| $[FR_3^{7,1,4}]_{2,0,2}$ | 🌐 | $\{-0.924 - 0.383i, 0.924 + 0.383i\}$ | $\{-2\}$ |
| $[FR_3^{7,1,4}]_{3,0,1}$ | † 🌐 | $\{-0.924 + 0.383i, 0.924 - 0.383i\}$ | $\{2\}$ |
| $[FR_3^{7,1,4}]_{3,0,2}$ | 🌐 | $\{-0.924 + 0.383i, 0.924 - 0.383i\}$ | $\{-2\}$ |
| $[FR_3^{7,1,4}]_{4,0,1}$ | † 🌐 | $\{-0.383 - 0.924i, 0.383 + 0.924i\}$ | $\{2\}$ |
| $[FR_3^{7,1,4}]_{4,0,2}$ | 🌐 | $\{-0.383 - 0.924i, 0.383 + 0.924i\}$ | $\{-2\}$ |

### 4.5.59 $FR_1^{7,1,6}$: $\mathbb{Z}_7$

For the fusion ring, the following multiplication table is used.

| 1 | 2 | 3 | 4 | 5 | 6 | 7 |
|---|---|---|---|---|---|---|
| 2 | 7 | 1 | 6 | 4 | 3 | 5 |
| 3 | 1 | 6 | 5 | 7 | 4 | 2 |
| 4 | 6 | 5 | 2 | 1 | 7 | 3 |
| 5 | 4 | 7 | 1 | 3 | 2 | 6 |
| 6 | 3 | 4 | 7 | 2 | 5 | 1 |
| 7 | 5 | 2 | 3 | 6 | 1 | 4 |

The following is the group of all non-trivial permutations that leave the fusion rules invariant:

$$S = \{(), (2\ 5\ 7\ 3\ 4\ 6), (2\ 6\ 4\ 3\ 7\ 5), (2\ 4\ 7)(3\ 5\ 6), (2\ 7\ 4)(3\ 6\ 5), (2\ 3)(4\ 5)(6\ 7)\}$$

Let

$$X_1 = S\left(\left([F_5^{222}]_7^7[F_4^{432}]_1^5[F_5^{447}]_3^2[F_2^{452}]_4^1[F_1^{472}]_5^3, R_2^{44}, d_4^L\right)\right) \tag{108}$$

The following table lists a small set of invariants whose values completely distinguish between all MFPBFCs and MFPNBFCs with the given fusion rules.

Table 115: Symbolic invariants

| Name | Properties | $X_1$ |
|---|---|---|
| $[\mathrm{FR}_1^{7,1,6}]_{1,1,1}$ | ✗ † 🌐 🎗 ⦿ | $\{(1, \zeta_7^3, 1), (1, \zeta_7^5, 1), (1, \zeta_7^6, 1)\}$ |
| $[\mathrm{FR}_1^{7,1,6}]_{1,1,2}$ | ✗ | $\{(1, \zeta_7^3, \zeta_7^4), (1, \zeta_7^3, \zeta_7^3), (1, \zeta_7^5, \zeta_7^6), (1, \zeta_7^5, \zeta_7), (1, \zeta_7^6, \zeta_7^5), (1, \zeta_7^6, \zeta_7^2)\}$ |
| $[\mathrm{FR}_1^{7,1,6}]_{1,1,3}$ | ✗ | $\{(1, \zeta_7^3, \zeta_7^6), (1, \zeta_7^3, \zeta_7), (1, \zeta_7^5, \zeta_7^5), (1, \zeta_7^5, \zeta_7^2), (1, \zeta_7^6, \zeta_7^4), (1, \zeta_7^6, \zeta_7^3)\}$ |
| $[\mathrm{FR}_1^{7,1,6}]_{1,1,4}$ | ✗ | $\{(1, \zeta_7^3, \zeta_7^5), (1, \zeta_7^3, \zeta_7^2), (1, \zeta_7^5, \zeta_7^4), (1, \zeta_7^5, \zeta_7^3), (1, \zeta_7^6, \zeta_7^6), (1, \zeta_7^6, \zeta_7)\}$ |
| $[\mathrm{FR}_1^{7,1,6}]_{1,2,1}$ | ✗ † 🌐 🎗 ⦿ | $\{(1, \zeta_7^4, 1), (1, \zeta_7^2, 1), (1, \zeta_7, 1)\}$ |
| $[\mathrm{FR}_1^{7,1,6}]_{1,2,2}$ | ✗ | $\{(1, \zeta_7^4, \zeta_7^6), (1, \zeta_7^4, \zeta_7), (1, \zeta_7^2, \zeta_7^5), (1, \zeta_7^2, \zeta_7^2), (1, \zeta_7, \zeta_7^4), (1, \zeta_7, \zeta_7^3)\}$ |
| $[\mathrm{FR}_1^{7,1,6}]_{1,2,3}$ | ✗ | $\{(1, \zeta_7^4, \zeta_7^5), (1, \zeta_7^4, \zeta_7^2), (1, \zeta_7^2, \zeta_7^4), (1, \zeta_7^2, \zeta_7^3), (1, \zeta_7, \zeta_7^6), (1, \zeta_7, \zeta_7)\}$ |
| $[\mathrm{FR}_1^{7,1,6}]_{1,2,4}$ | ✗ | $\{(1, \zeta_7^4, \zeta_7^4), (1, \zeta_7^4, \zeta_7^3), (1, \zeta_7^2, \zeta_7^6), (1, \zeta_7^2, \zeta_7), (1, \zeta_7, \zeta_7^5), (1, \zeta_7, \zeta_7^2)\}$ |
| $[\mathrm{FR}_1^{7,1,6}]_{1,3,1}$ | ✗ † 🌐 🎗 | $\{(1, 1, 1)\}$ |
| $[\mathrm{FR}_1^{7,1,6}]_{1,3,2}$ | ✗ | $\{(1, 1, \zeta_7^4), (1, 1, \zeta_7^3), (1, 1, \zeta_7^5), (1, 1, \zeta_7^2), (1, 1, \zeta_7^6), (1, 1, \zeta_7)\}$ |
| $[\mathrm{FR}_1^{7,1,6}]_{2,0,1}$ | † 🌐 | $\{(\zeta_7^4, -, 1), (\zeta_7^2, -, 1), (\zeta_7, -, 1)\}$ |
| $[\mathrm{FR}_1^{7,1,6}]_{2,0,2}$ | | $\{(\zeta_7^4, -, \zeta_7^3), (\zeta_7^4, -, \zeta_7^4), (\zeta_7^2, -, \zeta_7), (\zeta_7^2, -, \zeta_7^6), (\zeta_7, -, \zeta_7^2), (\zeta_7, -, \zeta_7^5)\}$ |
| $[\mathrm{FR}_1^{7,1,6}]_{2,0,3}$ | | $\{(\zeta_7^4, -, \zeta_7), (\zeta_7^4, -, \zeta_7^6), (\zeta_7^2, -, \zeta_7^5), (\zeta_7^2, -, \zeta_7^2), (\zeta_7, -, \zeta_7^3), (\zeta_7, -, \zeta_7^4)\}$ |
| $[\mathrm{FR}_1^{7,1,6}]_{2,0,4}$ | | $\{(\zeta_7^4, -, \zeta_7^2), (\zeta_7^4, -, \zeta_7^5), (\zeta_7^2, -, \zeta_7^3), (\zeta_7^2, -, \zeta_7^4), (\zeta_7, -, \zeta_7^6), (\zeta_7, -, \zeta_7)\}$ |
| $[\mathrm{FR}_1^{7,1,6}]_{3,0,1}$ | † 🌐 | $\{(\zeta_7^3, -, 1), (\zeta_7^5, -, 1), (\zeta_7^6, -, 1)\}$ |
| $[\mathrm{FR}_1^{7,1,6}]_{3,0,2}$ | | $\{(\zeta_7^3, -, \zeta_7^2), (\zeta_7^3, -, \zeta_7^5), (\zeta_7^5, -, \zeta_7^3), (\zeta_7^5, -, \zeta_7^4), (\zeta_7^6, -, \zeta_7), (\zeta_7^6, -, \zeta_7^6)\}$ |
| $[\mathrm{FR}_1^{7,1,6}]_{3,0,3}$ | | $\{(\zeta_7^3, -, \zeta_7^3), (\zeta_7^3, -, \zeta_7^4), (\zeta_7^5, -, \zeta_7), (\zeta_7^5, -, \zeta_7^6), (\zeta_7^6, -, \zeta_7^5), (\zeta_7^6, -, \zeta_7^2)\}$ |
| $[\mathrm{FR}_1^{7,1,6}]_{3,0,4}$ | | $\{(\zeta_7^3, -, \zeta_7^6), (\zeta_7^3, -, \zeta_7), (\zeta_7^5, -, \zeta_7^2), (\zeta_7^5, -, \zeta_7^5), (\zeta_7^6, -, \zeta_7^3), (\zeta_7^6, -, \zeta_7^4)\}$ |

Table 116: Numeric invariants

| Name | Properties | $X_1$ |
|---|---|---|
| $[\mathrm{FR}_1^{7,1,6}]_{1,1,1}$ | ✗ † 🌐 🎗 ⦿ | $\left\{\begin{array}{rrr}(1, & -0.901 + 0.434i, & 1), \\ (1, & -0.223 - 0.975i, & 1), \\ (1, & 0.623 - 0.782i, & 1)\end{array}\right\}$ |

| Name | Properties | $X_1$ |
|---|---|---|
| $[FR_1^{7,1,6}]_{1,1,2}$ | ✗ | $\left\{\begin{array}{l}(1,\ -0.901+0.434i,\ -0.901-0.434i),\\(1,\ -0.901+0.434i,\ -0.901+0.434i),\\(1,\ -0.223-0.975i,\ 0.623-0.782i),\\(1,\ -0.223-0.975i,\ 0.623+0.782i),\\(1,\ 0.623-0.782i,\ -0.223-0.975i),\\(1,\ 0.623-0.782i,\ -0.223+0.975i)\end{array}\right\}$ |
| $[FR_1^{7,1,6}]_{1,1,3}$ | ✗ | $\left\{\begin{array}{l}(1,\ -0.901+0.434i,\ 0.623-0.782i),\\(1,\ -0.901+0.434i,\ 0.623+0.782i),\\(1,\ -0.223-0.975i,\ -0.223-0.975i),\\(1,\ -0.223-0.975i,\ -0.223+0.975i),\\(1,\ 0.623-0.782i,\ -0.901-0.434i),\\(1,\ 0.623-0.782i,\ -0.901+0.434i)\end{array}\right\}$ |
| $[FR_1^{7,1,6}]_{1,1,4}$ | ✗ | $\left\{\begin{array}{l}(1,\ -0.901+0.434i,\ -0.223-0.975i),\\(1,\ -0.901+0.434i,\ -0.223+0.975i),\\(1,\ -0.223-0.975i,\ -0.901-0.434i),\\(1,\ -0.223-0.975i,\ -0.901+0.434i),\\(1,\ 0.623-0.782i,\ 0.623-0.782i),\\(1,\ 0.623-0.782i,\ 0.623+0.782i)\end{array}\right\}$ |
| $[FR_1^{7,1,6}]_{1,2,1}$ | ✗ ✝ 🌐 🎗 ⭕ | $\left\{\begin{array}{l}(1,\ -0.901-0.434i,\ 1),\\(1,\ -0.223+0.975i,\ 1),\\(1,\ 0.623+0.782i,\ 1)\end{array}\right\}$ |
| $[FR_1^{7,1,6}]_{1,2,2}$ | ✗ | $\left\{\begin{array}{l}(1,\ -0.901-0.434i,\ 0.623-0.782i),\\(1,\ -0.901-0.434i,\ 0.623+0.782i),\\(1,\ -0.223+0.975i,\ -0.223-0.975i),\\(1,\ -0.223+0.975i,\ -0.223+0.975i),\\(1,\ 0.623+0.782i,\ -0.901-0.434i),\\(1,\ 0.623+0.782i,\ -0.901+0.434i)\end{array}\right\}$ |
| $[FR_1^{7,1,6}]_{1,2,3}$ | ✗ | $\left\{\begin{array}{l}(1,\ -0.901-0.434i,\ -0.223-0.975i),\\(1,\ -0.901-0.434i,\ -0.223+0.975i),\\(1,\ -0.223+0.975i,\ -0.901-0.434i),\\(1,\ -0.223+0.975i,\ -0.901+0.434i),\\(1,\ 0.623+0.782i,\ 0.623-0.782i),\\(1,\ 0.623+0.782i,\ 0.623+0.782i)\end{array}\right\}$ |
| $[FR_1^{7,1,6}]_{1,2,4}$ | ✗ | $\left\{\begin{array}{l}(1,\ -0.901-0.434i,\ -0.901-0.434i),\\(1,\ -0.901-0.434i,\ -0.901+0.434i),\\(1,\ -0.223+0.975i,\ 0.623-0.782i),\\(1,\ -0.223+0.975i,\ 0.623+0.782i),\\(1,\ 0.623+0.782i,\ -0.223-0.975i),\\(1,\ 0.623+0.782i,\ -0.223+0.975i)\end{array}\right\}$ |
| $[FR_1^{7,1,6}]_{1,3,1}$ | ✗ ✝ 🌐 🎗 | $\{(1,\ 1,\ 1)\}$ |

Table 116: Numeric invariants (Continued)

| Name | Properties | $X_1$ |
|---|---|---|
| $[FR_1^{7,1,6}]_{1,3,2}$ | 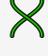 | $\left\{\begin{array}{rrr}(1, & 1, & -0.901 - 0.434i), \\ (1, & 1, & -0.901 + 0.434i), \\ (1, & 1, & -0.223 - 0.975i), \\ (1, & 1, & -0.223 + 0.975i), \\ (1, & 1, & 0.623 - 0.782i), \\ (1, & 1, & 0.623 + 0.782i)\end{array}\right\}$ |
| $[FR_1^{7,1,6}]_{2,0,1}$ | 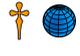 | $\left\{\begin{array}{rrr}(-0.901 - 0.434i, & -, & 1), \\ (-0.223 + 0.975i, & -, & 1), \\ (0.623 + 0.782i, & -, & 1),\end{array}\right\}$ |
| $[FR_1^{7,1,6}]_{2,0,2}$ | | $\left\{\begin{array}{rrr}(-0.901 - 0.434i, & -, & -0.901 + 0.434i), \\ (-0.901 - 0.434i, & -, & -0.901 - 0.434i), \\ (-0.223 + 0.975i, & -, & 0.623 + 0.782i), \\ (-0.223 + 0.975i, & -, & 0.623 - 0.782i), \\ (0.623 + 0.782i, & -, & -0.223 + 0.975i), \\ (0.623 + 0.782i, & -, & -0.223 - 0.975i)\end{array}\right\}$ |
| $[FR_1^{7,1,6}]_{2,0,3}$ | | $\left\{\begin{array}{rrr}(-0.901 - 0.434i, & -, & 0.623 + 0.782i), \\ (-0.901 - 0.434i, & -, & 0.623 - 0.782i), \\ (-0.223 + 0.975i, & -, & -0.223 - 0.975i), \\ (-0.223 + 0.975i, & -, & -0.223 + 0.975i), \\ (0.623 + 0.782i, & -, & -0.901 + 0.434i), \\ (0.623 + 0.782i, & -, & -0.901 - 0.434i)\end{array}\right\}$ |
| $[FR_1^{7,1,6}]_{2,0,4}$ | | $\left\{\begin{array}{rrr}(-0.901 - 0.434i, & -, & -0.223 + 0.975i), \\ (-0.901 - 0.434i, & -, & -0.223 - 0.975i), \\ (-0.223 + 0.975i, & -, & -0.901 + 0.434i), \\ (-0.223 + 0.975i, & -, & -0.901 - 0.434i), \\ (0.623 + 0.782i, & -, & 0.623 - 0.782i), \\ (0.623 + 0.782i, & -, & 0.623 + 0.782i)\end{array}\right\}$ |
| $[FR_1^{7,1,6}]_{3,0,1}$ | 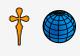 | $\left\{\begin{array}{rrr}(-0.901 + 0.434i, & -, & 1), \\ (-0.223 - 0.975i, & -, & 1), \\ (0.623 - 0.782i, & -, & 1)\end{array}\right\}$ |
| $[FR_1^{7,1,6}]_{3,0,2}$ | | $\left\{\begin{array}{rrr}(-0.901 + 0.434i, & -, & -0.223 + 0.975i), \\ (-0.901 + 0.434i, & -, & -0.223 - 0.975i), \\ (-0.223 - 0.975i, & -, & -0.901 + 0.434i), \\ (-0.223 - 0.975i, & -, & -0.901 - 0.434i), \\ (0.623 - 0.782i, & -, & 0.623 + 0.782i), \\ (0.623 - 0.782i, & -, & 0.623 - 0.782i)\end{array}\right\}$ |

Table 116: Numeric invariants (Continued)

| Name | Properties | $X_1$ |
|---|---|---|
| $[FR_1^{7,1,6}]_{3,0,3}$ | | $\left\{\begin{array}{rrr}(-0.901 + 0.434i, & -, & -0.901 + 0.434i), \\ (-0.901 + 0.434i, & -, & -0.901 - 0.434i), \\ (-0.223 - 0.975i, & -, & 0.623 + 0.782i), \\ (-0.223 - 0.975i, & -, & 0.623 - 0.782i), \\ (0.623 - 0.782i, & -, & -0.223 - 0.975i), \\ (0.623 - 0.782i, & -, & -0.223 + 0.975i)\end{array}\right\}$ |
| $[FR_1^{7,1,6}]_{3,0,4}$ | | $\left\{\begin{array}{rrr}(-0.901 + 0.434i, & -, & 0.623 - 0.782i), \\ (-0.901 + 0.434i, & -, & 0.623 + 0.782i), \\ (-0.223 - 0.975i, & -, & -0.223 + 0.975i), \\ (-0.223 - 0.975i, & -, & -0.223 - 0.975i), \\ (0.623 - 0.782i, & -, & -0.901 + 0.434i), \\ (0.623 - 0.782i, & -, & -0.901 - 0.434i)\end{array}\right\}$ |